\documentclass[hyper,12pt,letterpaper]{JHEP3}

\usepackage{graphicx}
\usepackage{latexsym,amsmath,amsfonts,amssymb}
\usepackage{mathrsfs}
\usepackage[makeroom]{cancel}
\usepackage{bbm}
\usepackage{bm}
\usepackage{subfigure}
\usepackage{paralist}
\usepackage{url}

\newcommand{\vol}{\mathrm{vol}}

\newcommand{\slashed}{{\bf\not}}

\newcommand{\ie}{\textit{i.e.}}

\newcommand{\vev}[1]{\langle #1 \rangle}

\numberwithin{equation}{section}

\newcommand{\nn}{\nonumber}
\newcommand{\mat}[1]{\begin{pmatrix} #1 \end{pmatrix}}

\newcommand{\be}{\begin{equation}} 
\newcommand{\ee}{\end{equation}}
\newcommand{\bea}{\begin{equation} \begin{aligned}} \newcommand{\eea}{\end{aligned} \end{equation}}

\newcommand{\bit}{\begin{itemize}} 
\newcommand{\eit}{\end{itemize}}

\newcommand{\cA}{\mathcal{A}}

\newcommand{\cC}{\mathcal{C}}
\newcommand{\cD}{\mathcal{D}}

\newcommand{\cH}{\mathcal{H}}
\newcommand{\cI}{\mathcal{I}}
\newcommand{\cJ}{\mathcal{J}}
\newcommand{\cK}{\mathcal{K}}

\newcommand{\cN}{\mathcal{N}}
\newcommand{\cO}{\mathcal{O}}

\newcommand{\cV}{\mathcal{V}}

\newcommand{\cZ}{\mathcal{Z}}

\newcommand{\bC}{\mathbb{C}}

\newcommand{\bR}{\mathbb{R}}

\newcommand{\bZ}{\mathbb{Z}}

\newcommand{\Z}{\mathbb{Z}}
\newcommand{\C}{\mathbb{C}}
\newcommand{\R}{\mathbb{R}}

\renewcommand{\t}{\widetilde }
\renewcommand{\d}{\partial }
\renewcommand{\b}{\bar }

\newcommand{\half}{{1\over 2}}

\newcommand{\bz}{{\b z}}

\newcommand{\CA}{\mathcal{A}}
\newcommand{\CB}{\mathcal{B}}
\newcommand{\CC}{\mathcal{C}}
\newcommand{\CD}{\mathcal{D}}

\newcommand{\CF}{\mathcal{F}}

\newcommand{\CH}{\mathcal{H}}

\newcommand{\CK}{\mathcal{K}}
\newcommand{\CL}{\mathcal{L}}
\newcommand{\CM}{\mathcal{M}}
\newcommand{\CN}{\mathcal{N}}
\newcommand{\CO}{\mathcal{O}}

\newcommand{\CQ}{\mathcal{Q}}
\newcommand{\CR}{\mathcal{R}}
\newcommand{\CS}{\mathcal{S}}

\newcommand{\CV}{\mathcal{V}}

\newcommand{\CZ}{\mathcal{Z}}

\newcommand{\FR}{\mathfrak{R}}
\newcommand{\Fg}{\mathfrak{g}}
\newcommand{\Fh}{\mathfrak{h}}

\newcommand{\GG}{\mathbf{G}}
\newcommand{\GF}{\mathbf{F}}
\newcommand{\GH}{\mathbf{H}}

\newcommand{\rk}{{{\rm rk}(\GG)}}

\newcommand{\epsdef}{{ \boldsymbol\epsilon_\Omega}}
\newcommand{\h}{\hat}

\newcommand{\s}{\sigma}
\newcommand{\hs}{\hat{\sigma}}
\newcommand{\hS}{\hat{\Sigma}}
\newcommand{\oneloop}{\text{1-loop}}

\DeclareMathOperator{\Tr}{Tr}
\DeclareMathOperator{\tr}{tr}

\newcommand{\eps}{\epsilon}

\def\res{\mathop{\mathrm{Res}}}

\newcommand{\rs}{{\bf r}}

\newcommand{\SL}{{\mathscr L}}

\newcommand{\p}{\partial}
\newcommand{\eq}[1]{(\ref{#1})}
\newcommand{\ov}{\over}
\newcommand{\ZS}{{\bf Z}_{S^2_\Omega}}
\newcommand{\ra}{\rightarrow}

\newcommand{\e}{{\rm e}}
\newcommand{\g}{{\rm g}}
\newcommand{\hD}{{\hat D}}
\newcommand{\eff}{\text{eff}}
\newcommand{\pos}{\text{pos}}
\newcommand{\fM}{\mathfrak{M}}
\newcommand{\tfM}{{\widetilde{\mathfrak{M}}}}

\newcommand{\idom}{{\widetilde{\mathfrak{M}} \setminus \Delta_\epsilon}}
\newcommand{\De}{{\Delta_\epsilon}}
\newcommand{\Dek}{{\Delta_{\epsilon,k}}}
\newcommand{\imag}{{\rm Im\,}}
\newcommand{\setcond}[2]{\{\,#1\,:\,#2\,\}}

\title{
The equivariant A-twist and gauged linear sigma models on the two-sphere
}

\author{Cyril~Closset,$^{\flat}$  Stefano~Cremonesi,$^\sharp$   Daniel~S.~Park$^{\flat}$\\

{}$^{\flat}$ Simons Center for Geometry and Physics\\ 
State University of New York, Stony Brook, NY 11794, USA\\

{}$^{\sharp}$ Department of Mathematics, King's College London \\
The Strand, London WC2R 2LS, United Kingdom

}

\preprint{}

\abstract{We study two-dimensional $\CN=(2,2)$ supersymmetric gauged linear sigma models (GLSM) on the $\Omega$-deformed sphere, $S^2_\Omega$, which is a one-parameter deformation of the $A$-twisted sphere. We provide an exact formula for the $S^2_\Omega$ supersymmetric correlation functions using supersymmetric localization. The contribution of each instanton sector is given in terms of a Jeffrey-Kirwan residue on the Coulomb branch. In the limit of vanishing $\Omega$-deformation, the  localization formula greatly simplifies the computation of $A$-twisted correlation functions, and leads to new results for non-abelian theories.  We discuss a number of examples and comment on  the $\epsdef$-deformation of the quantum cohomology relations. Finally, we present a complementary Higgs branch localization scheme in the special case of abelian gauge groups.
}

\begin{document}

\tableofcontents

\section{Introduction}
The study of supersymmetric quantum field theories on curved manifolds often leads to exact non-perturbative results,  by effectively isolating interesting supersymmetric sub-sectors. One of the simplest examples of this approach is the ($A$- or $B$-type) topological twist in two dimensions \cite{Witten:1988xj}, defined by ``twisting'' the spin by the (vector-like or axial-like) $R$-charge. It preserves two scalar supercharges  $\CQ, \t\CQ$ such that $\CQ^2=\t\CQ^2=0$, on any orientable Riemann surface $\Sigma$. The supersymmetric sector it isolates corresponds to the twisted chiral  operators%
~\footnote{Note that there are two distinct uses of the term ``twisted'' here. The first one refers to the $A$-twist and the corresponding ``twisting'' of the spin  by the $R$-charge, while the other refers to the ``twisted multiplets'', which are representations of the $\CN=(2,2)$ supersymmetry algebra. We shall distinguish between the two acceptations by writing ``$A$-twisted'' and ``twisted'', respectively.}
  in the case of the $A$-twist (or to the chiral operators in the case of the $B$-twist).

Let us consider two-dimensional $\CN=(2,2)$ theories with a vector-like $R$-symmetry, $U(1)_R$. Any supersymmetric background on a closed orientable Riemann surface $\Sigma$ can be understood as an {\it off-shell supergravity} background  \cite{Festuccia:2011ws,Closset:2014pda}
\be\label{sugra mult}
\left(\Sigma~; \, g_{\mu\nu}~,\,  A_\mu^{(R)}~,\, C_\mu~,\, \t C_\mu\right)~,
\ee
where, in addition to a metric $g_{\mu\nu}$ on $\Sigma$, we have a background $R$-symmetry gauge field $A_\mu^{(R)}$ and a complex graviphoton $C_\mu, \t C_\mu$ coupling to the conserved current for the central charge $Z, \t Z$. Supersymmetry imposes particular relations between the fields in \eqref{sugra mult}  \cite{Closset:2014pda}. In this work, we study the $\Omega$-deformed sphere, which we denote $S^2_\Omega$. It corresponds to $\Sigma= \C \mathbb{P}^1$ with a $U(1)$ isometry generated by  the Killing vector 
\be\label{def KV V}
V= i z\d_z - i \bz \d_\bz~,
\ee
using the usual complex coordinate on the sphere. Note that $V$ has fixed points at the north and south poles, $z=0$ and $z=\infty$, respectively. The supergravity background is further characterized by one unit of flux for the $U(1)_R$ gauge field,
\be
{1\over 2\pi} \int_{S^2} dA^{(R)}= -1~,
\ee 
and by the following background for the graviphoton:
\be\label{vev C}
C_\mu = i{  \epsdef \over 2} V_\mu~, \qquad \t C_\mu =0~,
\ee
with $\epsdef \in\C$ a constant of mass dimension $1$. The supersymmetry algebra on $S^2_\Omega$ is
\be\label{susy algebra A twist}
\CQ^2=0~, \qquad \t\CQ^2=0~, \qquad \{\CQ, \t\CQ\} = - 2i\left(Z  - \epsdef  J_V \right)~,
\ee
where $Z$ is the holomorphic central charge of the flat-space $\CN=(2,2)$ algebra that commutes with $U(1)_R$, and $J_V$ is the generator of rotations along \eqref{def KV V}. The $S^2_\Omega$ background is  a $J_V$-equivariant deformation of the $A$-twist on the sphere---the $A$-twist itself  corresponds to $\epsdef=0$.

In this paper, we consider general gauged linear sigma models (GLSM) \cite{Witten:1993yc} on the $\Omega$-deformed sphere. 
A GLSM  is a two-dimensional gauge theory consisting of a vector multiplet for some gauge group $\GG$ and of some matter fields in chiral multiplets charged under $\GG$.  The chiral multiplets can interact through a $U(1)_R$-preserving superpotential. 
If $\GG$ includes some $U(1)$ factors, 
\be
\GG \supset \prod_{I=1}^n U(1)_I~, \qquad  1\leq n \leq {\rm rank}(\GG)~, 
\ee
we can turn on Fayet-Iliopoulos (FI) couplings and $\theta$-angles:~\footnote{In the notation of this paper, the actual FI Lagrangian is given in \eqref{FI Lag expanded}, which includes an idiosyncratic redefinition of the $D$-term.} 

\be\label{FI Lagrangian}
\SL_{\rm FI} =- \sum_I \xi^I \tr_I(D) + {i\over 2\pi} \sum_I \theta^I  \tr_I\left(2 i f_{1\b 1}\right)~.
\ee
Here  $\tr_I$ denotes the generator of $U(1)_I$ inside  $\GG$.
The couplings $\xi^I, \theta^I$ are paired by supersymmetry into the complex combination
\be\label{def tauI}
\tau^I = {\theta^I\over 2\pi} + i \xi^I~.
\ee
The Lagrangian \eqref{FI Lagrangian} descends from a  twisted superpotential
 $\h W(\sigma)$ linear in $\sigma$, with $\sigma$ the complex scalar in the vector multiplet:
\be\label{Wh linear}
\h W(\sigma) = \sum_{I=1}^n \tau^I \tr_I\left(\sigma\right)~.
\ee
Much of the interest in this class of theories is that it  provides renormalizable ultraviolet (UV) completions of interesting strongly-interacting field theories in the infrared (IR). In particular, GLSMs can UV-complete non-linear sigma models (NLSM) on  K\"ahler---and in particular Calabi-Yau (CY)---manifolds, as well as superconformal field theories (SCFT) with no geometric description.

The computable quantities of interest on $S^2_\Omega$ are the correlation functions of gauge-invariant polynomials in $\sigma$ inserted at the fixed points of the isometry \eqref{def KV V}. Consider the two operators $\CO^{(N)}(\sigma)$ and $\CO^{(S)}(\sigma)$ inserted at the north and south poles,  respectively. The main result of this paper is an exact formula for their correlation function, of the schematic form
\be\label{formula intro 1}
\left\langle  \CO^{(N)}(\sigma_N) \CO^{(S)}(\sigma_S)\right\rangle ={1\over |W|}\sum_k q^k\,  \widetilde{\bf Z}_k(\CO)~,
\ee
where $\sigma=\sigma_N$, $\sigma=\sigma_S$ stands for the north and south pole insertions,  and
\be\label{formula intro 2}
\widetilde{\bf Z}_k(\CO)=   \oint_{{\rm JK}(\xi_{\rm eff}^{\rm UV})}  \Big(\prod_{a=1}^\rk {d \h\sigma_a\over 2\pi i}\Big) \; \cZ^\oneloop_k(\h\sigma; \epsdef)   \,\CO^{(N)}\left(\h\sigma-\epsdef {k\over 2}\right)  \CO^{(S)}\left(\h\sigma+\epsdef {k\over 2}\right)~.
\ee
The sum over fluxes $k$ in \eqref{formula intro 1} runs over the weight lattice $\Gamma_{\GG^\vee}$ of the GNO (or Langlands) dual group $\GG^\vee$ of $\GG$,~
\footnote{$\GG^\vee$ is the group whose weights $k$ satisfy the Dirac quantization condition $e^{2\pi i k} = 1_{\GG}$ \cite{Goddard:1976qe,Englert:1976ng}.} 
weighted by an instanton factor of the form $q= e^{2\pi i \tau}$ for the couplings \eqref{def tauI}.  
$|W|$ is the order of the Weyl group of $\GG$. 
The $k$-instanton factor  \eqref{formula intro 2} is a multi-dimensional contour integral on the  ``Coulomb branch'' spanned by the constant vacuum expectation values (VEV)  $\sigma= {\rm diag}(\h\sigma_a)$. More precisely,  the contour integral  in  \eqref{formula intro 2} is a  particular residue operation known as Jeffrey-Kirwan (JK) residue \cite{JK1995, 1999math......3178B,2004InMat.158..453S}, the definition of which depends on the effective FI parameter in the UV, $\xi_{\rm eff}^{\rm UV}$.
Finally, the integrand of \eqref{formula intro 2} consists of a one-loop determinant $Z^\oneloop_k(\h\sigma; \epsdef)$ from massive fields on the Coulomb branch, and of the operator insertions themselves (including an important $\epsdef$-dependent shift of $\h\sigma_a$).~\footnote{We determine the integrand up to an overall sign ambiguity. We shall give an {\it ad hoc} prescription to fix this sign, consistent with all the examples.} An executive summary of this formula is provided in section \ref{subsec: summary}.

While we shall spend much time deriving and explaining  \eqref{formula intro 1}-\eqref{formula intro 2} in the following, a few important remarks should be made from the outset:

 \begin{itemize}

\item The result  \eqref{formula intro 1} is holomorphic in the parameters $q$ and $\epsdef$. In the presence of a flavor symmetry group, it is also holomorphic in any twisted mass $m^F$ that can be turned on. The $q$ parameters have an interpretation as coordinates on the K\"ahler moduli space of the geometry the GLSM engineers. They are referred to as ``algebraic coordinates" in \cite{Morrison:1994fr}, and are identified with complex coefficients of superpotential terms in the mirror theory \cite{Candelas:1990rm,Aspinwall:1993rj,Hori:2000kt}.

\item This ``Coulomb branch'' formula is obtained by supersymmetric localization in the UV. Thus the term ``Coulomb branch'' should be taken with a grain of salt, as we do not sum over flat-space infrared vacua. Rather, we use the $S^2_\Omega$ supersymmetry to force the path integral into saddles that mimic a Coulomb branch. As we will show in section \ref{section: Higgs branch loc} (in the abelian case), a different localization computation can lead to a complementary understanding of \eqref{formula intro 1} as a sum over ``Higgs branch'' configurations,  which are supersymmetric vortices corresponding to the residues picked by \eqref{formula intro 2}. This point of view is closer in spirit  to the seminal work of Morrison and Plesser \cite{Morrison:1994fr}. 

\item The formula is valid in any of the chambers in FI-parameter space---the famous GLSM {\it phases} \cite{Witten:1993yc}---except on the chamber walls. The JK residue  prescription only depends on a choice of chamber.  In any given chamber, only some particular set of fluxes $k$ contributes to the JK residue, and the sum \eqref{formula intro 1} is convergent.

\item One  should be  careful about the meaning of \eqref{formula intro 1}-\eqref{formula intro 2}  when the FI-parameters run under RG flow. In that case, one can write  the formula in a RG-invariant way in terms of dynamical scales.  Moreover, many of the classical GLSM chambers are lifted at one-loop. This is reflected in the JK residue prescription in \eqref{formula intro 2}, which depends on the one-loop UV effective FI parameter $\xi_{\rm eff}^{\rm UV}$.

\item The formula \eqref{formula intro 1}-\eqref{formula intro 2}  provides a direct way of computing various correlators in the $\epsdef$-deformed theory. The $A$-model correlators can be recovered from these correlators by sending $\epsdef$ to zero. For example, we obtain
\be
\begin{split}
\langle  \sigma^n_N \rangle &= 0  \qquad\qquad (n=0,1,2)\\
\langle  \sigma^3_N \rangle &= \frac{5}{1+5^5 q}\\
\langle  \sigma^4_N \rangle &= \epsdef \frac{2\cdot 5^6 q}{(1+5^5 q)^2}\\
\langle  \sigma^5_N \rangle &= \epsdef^2 \frac{5^5 q(-17 + 13\cdot 5^5 q)}{(1+5^5 q)^3}\\
& \vdots
\end{split}
\ee
for the quintic threefold, where $\sigma$ is the lowest component of the unique twisted chiral field in the theory. Notice that the celebrated triple-intersection formula \cite{Candelas:1990rm} is reproduced.

\item The formula \eqref{formula intro 1}-\eqref{formula intro 2} applies straightforwardly to non-abelian gauge theories. In particular, it is applicable to the computation of correlators constructed out of higher Casimir operators of twisted chiral fields. We are thus able to compute some correlators in (submanifolds of) non-abelian K\"ahler quotient manifolds that have not been computed before, to our knowledge (see section \ref{sec:_Amodel_examples}).

\item  Setting $\epsdef=0$, we obtain a simple formula for (genus zero) correlations functions $\langle \CO(\sigma)\rangle_0$ in the $A$-twisted GLSM:
\be\label{formula Amodel intro}
\langle \CO(\sigma)\rangle_0 = {1\over |W|}\sum_k q^k\,  \oint_{{\rm JK}(\xi_{\rm eff}^{\rm UV})}  \Big(\prod_{a=1}^\rk {d \h\sigma_a\over 2\pi i}\Big) \; \cZ^\oneloop_k(\h\sigma)  \, \CO(\h \sigma)~.
\ee
 This includes in particular the holomorphic Yukawa couplings of CY string phenomenology, and some non-abelian generalizations thereof. The results we obtain from \eqref{formula Amodel intro} can be compared to various results in the literature, whether obtained from mirror symmetry or from direct GLSM  computations  \cite{Morrison:1994fr, Losev:1999nt, Melnikov:2005tk, Guffin:2008kt, Nekrasov:2014xaa}.
When $\GG$ is abelian, the ``Coulomb branch'' formula significantly simplifies the toric geometry computations of  \cite{Morrison:1994fr}.  In fact,  this particular use of the JK residue was first introduced in \cite{2004InMat.158..453S} from a mathematical point of view.  In a different approach, a related Coulomb branch formula has also been introduced in   \cite{Melnikov:2005tk, Nekrasov:2014xaa}, which can be recovered from \eqref{formula Amodel intro}  in the appropriate regime of validity.

\item The formula \eqref{formula Amodel intro} can be viewed as a generalization of Vafa's formula for $A$-twisted Landau-Ginzburg theories of twisted chiral multiplets \cite{Vafa:1990mu} to the case of gauge fields.

\item In the $\epsdef=0$ case, it is relatively easy to show from  \eqref{formula Amodel intro} that the quantum chiral ring relations (also known as quantum cohomology relations) are realized by the correlation functions, given a technical assumption about the integrand. This assumption  corresponds physically to the absence of certain dangerous gauge invariant operators which could take any VEV. (In geometric models, such a situation occurs on non-compact geometries with all mass terms set to zero. This clarifies some observations made in \cite{Melnikov:2005tk}.)

\item For $\epsdef\neq 0$, one can derive recursion relations for the $\epsdef$-dependence of correlations functions, generalizing the quantum cohomology relations of the $A$-twisted theory. These recursion relations simplify many explicit computations, and have deep relationships to enumerative geometry.%
~\footnote{The recursion relations have been derived for the computation of Gromov-Witten invariants of complete intersections inside toric manifolds in the mathematical literature, for example, in the works of Givental \cite{GiventalEquivariant,GiventalToric,GiventalQuintic}. These recursion relations and their relation to Picard-Fuchs equations have been noticed \cite{Honda:2013uca,recursion} in the context of correlators on the supersymmetric hemisphere \cite{Honda:2013uca,Hori:2013ika,Sugishita:2013jca} and two-sphere \cite{Benini:2012ui,Doroud:2012xw}. Some related relations, that translate into difference equations of ``holomorphic blocks" \cite{Pasquetti:2011fj,Beem:2012mb} have been studied in \cite{Bullimore:2014awa}.}
It is interesting to observe that the one-loop determinant $\cZ^\oneloop$ we have computed can be identified with the densities computed in \cite{GiventalEquivariant, GiventalToric, GiventalQuintic, 2012arXiv1201.6350C} that are integrated over moduli spaces of curves to obtain certain geometric invariants, once $\epsdef$ is identified with the ``equivariant parameter" $\hbar$ of these works.

\end{itemize}

To conclude this introduction,  let us briefly compare our setup to similar localization results for $\CN=(2,2)$ theories obtained in recent  years. The authors of \cite{Benini:2012ui, Doroud:2012xw} localized $\CN=(2,2)$ gauge theories on a different $S^2$ background with {\it vanishing $U(1)_R$ flux} and some unit flux for the graviphotons $C_\mu, \t C_\mu$. Thus, in that background the $R$-charges can be arbitrary while the central charge $Z$ is quantized.~\footnote{More precisely, the imaginary part (or real part, depending on conventions) of $Z$ is quantized. This leads to a  Coulomb branch integral over the real part of $\sigma$, as opposed to our integral formula which is holomorphic in $\sigma$.} That $S^2$ background corresponds to a supersymmetric fusion of the $A$- and $\b A$-twists  on two hemispheres \cite{Cecotti:1991me, Jockers:2012dk, Gomis:2012wy}, while this work considers a deformation of the $A$-twist on $S^2$. (See \cite{Sugishita:2013jca,Honda:2013uca,Hori:2013ika} for the localization of 2d $\CN=(2,2)$ theories on hemispheres.) 

Another closely related localization result is the computation of the  $\CN=(2,2)$ elliptic genus \cite{Benini:2013nda,  Gadde:2013dda, Benini:2013xpa}---the $T^2$ partition function---which was found to be given in terms of a JK residue on the space of flat connections \cite{Benini:2013xpa}. Finally, the recent computation  of the 1d supersymmetric index in \cite{Hori:2014tda} was very influential to our derivation of the Coulomb branch formula \eqref{formula intro 1}. Some other partially related recent works, in the context of topologically twisted 4d $\cN=2$ theories, are \cite{Bawane:2014uka, Sinamuli:2014lma,Rodriguez-Gomez:2014eza}. 

This paper is organized as follows.  In section \ref{sec: susy}, we expound on supersymmetry on $S^2_\Omega$. In section \ref{sec: GLSM and observ}, we give some relevant background material on GLSM and we discuss the supersymmetric observables we are set on computing. We present the  derivation of the Coulomb branch formula in sections \ref{section: der1} and \ref{section: derivation}. Section \ref{section: derivation} is more technical, and might be skipped on first reading. In section \ref{section: Q Cohomology}, we discuss the quantum cohomology relations and their $\epsdef$-deformations. In sections \ref{sec: Examples 1} and \ref{sec:_Amodel_examples}, we present several instructive examples. In section \ref{section: Higgs branch loc}, we discuss the Higgs branch localization. Several appendices summarize our conventions and provide some useful technical results.
 
\vskip0.2cm
\noindent
{\it Note added:}  As this paper was being completed, a related work \cite{Benini:2015noa} appeared on the arXiv which contains some overlapping material.



\section{Supersymmetry on the $\Omega$-deformed sphere}\label{sec: susy}

In this section we study off-shell supersymmetry on the $S^2_\Omega$  background,  we discuss various supersymmetric Lagrangians which will be used in the following, and we present the equations satisfied by any supersymmetric configuration of vector and chiral multiplets.

\subsection{Supersymmetric background on $S^2_\Omega$}
In the ``curved-space supersymmetry'' formalism we are using, a general supersymmetric background is an off-shell supergravity background \eqref{sugra mult} which preserves  some generalized Killing spinors $\zeta_\pm$ or $\t\zeta_\pm$ \cite{Closset:2014pda}.  
We have the following generalized Killing spinor equations for $\zeta_\pm$,
\bea\label{KSE i}
& (\nabla_z -i A_z) \zeta_- =0~, \qquad \qquad\;\; \qquad\qquad  (\nabla_\bz -i A_\bz) \zeta_- = \half\CH\, e^{\b 1}_\bz\, \zeta_+~,\cr
& (\nabla_z -i A_z)  \zeta_+ = \half\t\CH\, e^{ 1}_z \,\zeta_-~,  \qquad\quad\qquad\;   (\nabla_\bz -i A_\bz) \zeta_+ = 0~,
\eea
and  for $\t\zeta_\pm$,
\bea\label{KSE ii}
& (\nabla_z +i A_z) \t\zeta_- =0~, \qquad \qquad\;\; \qquad\qquad  (\nabla_\bz +i A_\bz) \t\zeta_- = \half\t\CH\, e^{\b 1}_\bz\,\t\zeta_+~,\cr
& (\nabla_z +i A_z) \t \zeta_+ = \half\CH\, e^{ 1}_z \,\t\zeta_-~,  \qquad\quad\qquad\;   (\nabla_\bz +i A_\bz) \t\zeta_+ = 0~.
\eea
 Note that $\zeta_\pm$ and $\t\zeta_\pm$ have $U(1)_R$ charge $\pm 1$, respectively. Here we introduced the graviphoton dual field strengths
\be\label{def CH}
\CH = -i \epsilon^{\mu\nu}\d_\mu C_\nu~, \qquad  \quad
\t\CH = -i \epsilon^{\mu\nu}\d_\mu \t C_\nu~
\ee
and the canonical complex frame $\{e^1=e^1_z dz = g^{1\over 4} dz, e^{\b 1}=e^{\b 1}_\bz d\bz=g^{1\over 4}d\bz\}$. 

Let us consider a sphere with metric
\be\label{S2 metric}
ds^2 = 2 g_{z\b z}(|z|^2)dz d\bz = \sqrt{g}d z d \bz = e^1e^{\b 1}~,
\ee
with a $U(1)$ isometry generated by the real Killing vector $V$ in \eqref{def KV V}. 
The supersymmetric background $S^2_\Omega$ is given by \eqref{S2 metric} together with
\be\label{sugra background}
A_\mu = \half \omega_\mu~, \qquad  \ \CH =  {\epsdef\over 2}\, \epsilon^{\mu\nu}\d_\mu V_\nu~, \qquad\quad \t\CH=0~,
\ee
where  $\omega_\mu$ is the spin connection. The graviphotons are given by \eqref{vev C}. The supergravity  fluxes are: 
\be
{1\over 2\pi} \int_{S^2} dA^{(R)}= -1~,\qquad  {1\over 2\pi} \int_{S^2} d C= {1\over 2\pi} \int_{S^2} d \t C= 0~.
\ee 
In consequence, the $R$-charges of all the fields  must be integer by Dirac quantization, while the value of the central charge $Z, \t Z$ is unconstrained.  One can check that the background \eqref{S2 metric}-\eqref{sugra background} gives a solution of \eqref{KSE i}-\eqref{KSE ii} with the  Killing spinors
\be\label{KSEquivAtwist}
\zeta = \mat{\zeta_- \cr \zeta_+}=  \mat{i \epsdef \, V_{1}\cr 1 }~,     \qquad\qquad  
\t\zeta= \mat{\t\zeta_- \cr \t\zeta_+}= \mat{1 \cr - i \epsdef \, V_{\b 1} }~.
\ee
In the language of the $A$-twist, and in keeping with the formalism of \cite{Closset:2014pda}, the components $\zeta_+$ and $\t\zeta_-$ transform as scalars while $\zeta_-$ and $\t\zeta_+$ are naturally sections of $\overline{\CO(2)}$ and $\CO(2)$, respectively.

The Killing spinors \eqref{KSEquivAtwist} can be used to provide an explicit map between the usual flat-space variables and the more convenient $A$-twisted variables---see appendix \ref{app: conventions}. By construction, $A$-twisted variables are fields of vanishing $R$-charge and spin $s= s_0 +{r\over 2}$, where $s_0$ and $r$ are the original flat-space spin and $R$-charge. By abuse of terminology, we always refer to $r$ as the $R$-charge even when the field is technically $U(1)_R$-neutral after the $A$-twist.

Denoting by $\delta$ and $\t\delta$ the supersymmetry variations along $\zeta$ and $\t\zeta$, respectively, the $S^2_\Omega$ supersymmetry algebra \eqref{susy algebra A twist} is realized on $A$-twisted fields as \cite{Closset:2014pda}:
\be\label{susy algebra on fields}
\delta^2=0~, \qquad \t\delta^2=0~, \qquad 
\{\delta,\t\delta\}=- 2i  \left(Z+ i\epsdef \CL_V \right)~,
\ee
where $\CL_V$ denotes the Lie derivative along $V$. This gives
a $U(1)$-equivariant deformation of the topological $A$-twist algebra with equivariant parameter $\epsdef$, also known as  $\Omega$-deformation \cite{Nekrasov:2002qd, Nekrasov:2003rj, Shadchin:2006yz}. 

A more familiar description of the $\Omega$-background is in terms of a fibration of space-time over a torus \cite{Nekrasov:2003rj}. An $\CN=(2,2)$ supersymmetric theory on $S^2_\Omega$ can be naturally uplifted to a 4d $\CN=1$ theory on an $S^2 \times T^2$ supersymmetric background.~\footnote{At least classically;  4d gauge anomalies forbid many matter contents that are allowed in 2d.} As explained in \cite{Dumitrescu:2012ha, Closset:2013vra, Closset:2013sxa}, one can consider a two-parameter family of complex structures on $S^2 \times T^2$ while preserving two supercharges of opposite chiralities. The complex structure moduli are denoted by $\tau^{S^2\times T^2}$ and $\sigma^{S^2\times T^2}$ in \cite{Closset:2013sxa}, where $\tau^{S^2\times T^2}$ is the complex structure modulus of the $T^2$ factor, while  $\sigma^{S^2\times T^2}$ governs a (topologically but not holomorphically trivial) fibration of $S^2$ over $T^2$. One can show that the supersymmetric uplift of $S^2_\Omega$ to $S^2\times T^2$ is precisely the  background of \cite{Closset:2013sxa} with the identification
\be\label{eps eq sigmaS2T2}
  \epsdef=\sigma^{S^2\times T^2}~.
\ee
This relationship to $4d$ $\CN=1$ is another motivation to study $S^2_\Omega$ in detail. The $T^2\times S^2$ partition function should be given by an elliptic uplift of our $S^2_\Omega$ results (see \cite{Benini:2015noa,Honda:2015yha} for some recent progress  in that direction).

\subsection{Supersymmetric multiplets}
Let us consider the vector and chiral multiplets, which are the building blocks of the GLSM. We shall also discuss the twisted chiral multiplet, which is important to understand the vector multiplet itself. 
 We discuss all multiplets in $A$-twisted notation, as summarized in appendix \ref{app: conventions}.

\subsubsection{Vector multiplet $\CV$} 
Consider a vector multiplet $\CV$ with gauge group $\GG$, and denote ${\frak g}={\rm Lie}(\GG)$. In  Wess-Zumino (WZ) gauge, $\CV$ has components:
\be\label{V components}
\CV= \left(a_\mu~,\, \sigma~, \,\t\sigma~, \,\Lambda_1~, \,\lambda~, \, \t\Lambda_{\b 1}~,\, \t\lambda~,\, D\right)~.
\ee
All the fields are valued in the adjoint representation of $\Fg$. Let us define the field strength
\be
f_{1\b 1} = \d_1 a_{\b 1} - \d_{\b1}a_1 -i [a_1, a_{\b1}]~.
\ee
The covariant derivative $D_\mu$ is taken to be gauge-covariant, and we denote by $\CL_V^{(a)}$ the gauge-covariant version of the Lie derivative along $V$. 
The supersymmetry transformations of \eqref{V components} are
\bea\label{susyVector twisted i}
&\delta a_1 = 0~,\cr
&\delta a_{\b 1} = i \t\Lambda_{\b 1}~,\cr
&\delta\sigma = 2 i \epsdef V_1 \t\Lambda_{\b 1}~,\cr
&\delta\t\sigma = -2\t \lambda~, \cr
& \delta \Lambda_1 = -\epsdef V_1 (4 i f_{1\b 1}) + 2 i D_1 \sigma~,\cr
&\delta \t\Lambda_{\b1} =0~,\cr
&\delta \lambda = i \left(D - 2 i f_{1\b 1}-\half [\sigma, \t\sigma] \right) - 2  \epsdef V_1 D_{\b 1} \t\sigma \cr
& \delta\t\lambda =0~,\cr
&\delta D= - 2 D_1 \t\Lambda_{\b 1} + 4 i \epsdef V_1 D_{\b 1} \t\lambda  -[\sigma, \t\lambda] + i \epsdef V_1[\t\sigma, \t\Lambda_{\b1}]~,
\eea
for the supersymmetry $\CQ$, and
\bea\label{susyVector twisted ii}
& \t\delta a_1 = - i \Lambda_1~,\cr
& \t\delta a_{\b 1} =0~,\cr
&  \t\delta \sigma = -2 i \epsdef V_{\b 1} \Lambda_1~,\cr
& \t\delta \t\sigma =-2 \lambda~,\cr
& \t\delta \Lambda_1=0~,\cr
&\t\delta \t\Lambda_{\b1}= -\epsdef V_{\b1} (4 i f_{1\b 1}) - 2 i D_{\b1}\sigma~,\cr
& \t\delta \lambda=0~,\cr
&\t\delta\t\lambda = -i \left(D- 2 i f_{1\b 1}+\half [\sigma, \t\sigma]\right)- 2 \epsdef V_{\b1}D_1 \t\sigma~,\cr
&\t\delta D= -2 D_{\b 1} \Lambda_1 - 4 i \epsdef V_{\b 1}D_1 \lambda +[\sigma, \lambda]+ i\epsdef V_{\b1}[\t\sigma, \Lambda_1]~,
\eea
for the supersymmetry $\t \CQ$.
 These transformations realize a gauge-covariant version of the supersymmetry algebra  \eqref{susy algebra on fields}. One has
\be
\delta^2\varphi =0~, \qquad \t\delta^2 \varphi=0~, \qquad \{\delta, \t\delta\}\varphi= - 2 i \left(-  [\sigma,\varphi]  + i \epsdef \CL_V^{(a)}\varphi \right)
\ee
on every ${\frak g}$-covariant field  $\varphi$ in $\CV$, while for the gauge field $a_\mu$ one has
\be
 \{\delta, \t\delta\}a_\mu=  2\epsdef  \CL_V a_\mu- 2 \epsdef V^\nu\left(\d_\mu a_\nu - i [a_\mu, a_\nu] \right)+  2 D_\mu\sigma~.
\ee
Note that $\sigma$ enters the supersymmetry algebra similarly to a central charge $Z= -\sigma$, as a result of the WZ gauge fixing.  This is expected from the dimensional reduction of 4d $\cN=1$ to 2d $\cN=(2,2)$ supersymmetry, where $Z\propto P_3+i P_4$ and $\sigma\propto a_3+i a_4$.

\subsubsection{Charged chiral multiplet $\Phi$}
Consider a chiral multiplet $\Phi$ of $R$-charge $r$, transforming in a representation $\FR$ of $\Fg$. In $A$-twisted notation (see appendix \ref{app: conventions}), we denote the components of $\Phi$ by
\be\label{compo Phi}
\Phi= \left(\CA~,\, \CB~,\, \CC~,\, \CF\right)~.
\ee
They are sections of appropriate powers of the canonical line bundle:
\be
\CA~,\,  \CB \,\in \Gamma(\CK^{r\over 2})~, \qquad \CC~,\, \CF \in \,\Gamma(\CK^{r\over 2}\otimes\b \CK)~.
\ee
The supersymmetry transformations are given by
\bea\label{susytranfoPhitwistBis}
& \delta \CA = \CB~, \qquad\qquad  && \t\delta \CA=0~,\cr
& \delta \CB=0~, \qquad \qquad && \t\delta \CB= -2i\big(- \sigma +i \epsdef \CL_V^{(a)}\big)\CA~,\cr
& \delta \CC=\CF~, \qquad \qquad && \t\delta \CC= 2i D_\bz\CA~,\cr
& \delta \CF=0~, \qquad \qquad && \t\delta \CF=- 2i \big(-\sigma +i \epsdef \CL_V^{(a)}\big)\CC -2i D_\bz\CB -2i \t\Lambda_\bz\CA~,
\eea
where $D_\mu$ is appropriately gauge-covariant and $\sigma$ and $\t\Lambda_{\bz}$ act in the representation $\FR$.
Similarly, the charge-conjugate antichiral multiplet $\t\Phi$ of $R$-charge $-r$ in the representation  $\b\FR$ has components
\be\label{compo tPhi}
\t\Phi= \left(\t\CA~,\, \t\CB~,\, \t\CC~,\, \t\CF\right)~,\qquad \;
 \t\CA~,\, \t\CB \,\in \Gamma(\b\CK^{r\over 2})~,\qquad  \t\CC~,\,\t \CF\, \in \Gamma(\b\CK^{r\over 2}\otimes \CK)~.
\ee
Its supersymmetry transformations are
\bea\label{susytranfotPhitwistBis}
& \delta \t\CA = 0~, \qquad\qquad\qquad\qquad\qquad\qquad\qquad
& \t\delta \t\CA=\t\CB~,\cr
& \delta \t\CB= -2i \big(\sigma +i \epsdef \CL_V^{(a)} \big)\t\CA~, \qquad\qquad\qquad \qquad\quad\quad &\t\delta \t\CB=0~,\cr
& \delta \t\CC= -2i D_z \t\CA~,
\qquad\qquad\qquad \qquad\quad 
&\t\delta \t\CC=\t\CF~,\cr
& \delta \t\CF= -2i \big(\sigma +i \epsdef \CL_V^{(a)} \big)\t\CC +2i D_z  \t\CB + 2i \Lambda_z\t\CA
& \t\delta \t\CF=0~.
\eea
Using the  vector multiplet transformation rules \eqref{susyVector twisted i}-\eqref{susyVector twisted ii}, one can check that \eqref{susytranfoPhitwistBis}-\eqref{susytranfotPhitwistBis} realize the supersymmetry algebra 
\be\label{susy with gauge field i}
\delta^2=0~, \qquad \t\delta^2=0~, \qquad 
\{\delta,\t\delta\}=- 2i  \left(- \sigma+ i\epsdef \CL_V^{(a)} \right)~,
\ee
where $\sigma$ and $\CL_V^{(a)}$ act in the appropriate representation of the gauge group. 

We introduced the chiral and antichiral multiplets in complex coordinates to manifest the fact that their supersymmetry transformation rules are metric-independent. In concrete computations, however, it is useful to use the frame basis (see appendix \ref{app: conventions}). One translates between the coordinate and frame bases using the vielbein. For instance, $\CA^{\rm frame} = (e^z_1)^{r\over 2}\CA^{\rm coord}$ and $\CC^{\rm frame} = (e^z_1)^{r\over 2}e^{\bz}_{\b 1} \CC^{\rm coord}$. In the frame basis, the fields $\CA, \CB$ and $\CC, \CF$ have spin ${r\over 2}$ and ${r-2\over 2}$, respectively.

\subsubsection{Twisted chiral multiplet $\Omega$}
Another important short representation of the $\CN=(2,2)$ supersymmetry algebra is the twisted chiral multiplet $\Omega$. This multiplet has vanishing vector-like $R$-charge and vanishing central charge. Its components in $A$-twisted notation are
\be\label{Omega com}
\Omega=(\omega~, \, \CH_z~, \, \t\CH_\bz~, \, G)~,
\ee 
where $\omega$ and  $G$ are scalars. The supersymmetry  transformations of \eqref{Omega com} are
\bea\label{susy Omega}
&\delta\omega = -{2i \over \sqrt{g}} \epsdef V_z \t\CH_\bz~, \qquad 
&&  \t\delta \omega = -{2 i\over \sqrt{g}} \epsdef V_\bz \CH_z~,\cr
& \delta \CH_z = i \epsdef V_z G + 2 i \d_z \omega~, \qquad 
&&\t\delta \CH_z =0~,\cr
& \delta \t\CH_\bz = 0~, \qquad 
&&\t\delta \t\CH_\bz =-i \epsdef V_\bz G + 2 i \d_\bz \omega~,\cr
& \delta G= {4 i \over \sqrt{g}} \d_z \t\CH_\bz~, \qquad
&&\t\delta G = -{4 i \over \sqrt{g}} \d_\bz \CH_z~.
\eea
Similarly, the twisted antichiral multiplet has components
\be\label{t Omega com}
\t\Omega=(\omega~, \, \t h~, \, h~, \, \t G)~,
\ee 
where the four components are scalars, and  their supersymmetry transformations are
\bea\label{susy t Omega}
&\delta\t\omega = -2\t h~, \qquad 
&&  \t\delta \t\omega =2 h~,\cr
& \delta\t h= 0~, \qquad 
&&\t\delta \t h =\t G - {2 \over \sqrt{g}} \epsdef V_\bz \d_z \t\omega~,\cr
& \delta h = \t G + {2 \over \sqrt{g}} \epsdef V_z\d_\bz \t\omega~, \qquad 
&&\t\delta h=0~,\cr
& \delta \t G= {4 \over \sqrt{g}} \epsdef V_z \d_\bz  \t h~, \qquad
&&\t\delta \t G = {4  \over\sqrt{g}} \epsdef V_\bz \d_z  h~.
\eea
These multiplets realize the supersymmetry algebra  \eqref{susy algebra on fields} with $Z=0$. 
The comment of the previous subsection about coordinate versus frame basis  applies here as well. Note that the supersymmetry transformations do depend on the metric except when $\epsdef=0$.

\subsubsection{Twisted chiral multiplets from the vector multiplet}\label{subsec: twisted chiral from V}
Important examples of  twisted chiral multiplets are built from the vector multiplets $\CV$. More precisely, let us consider $U(1)_I$ an abelian factor in $\GG$, and denote by $\CV_I = \tr_I \CV$ the corresponding abelian vector multiplet.~\footnote{If $\GG=U(N)$, $\tr_I$ denotes the usual trace.} We can build the gauge-invariant twisted chiral multiplet
\be\label{SigmaTwistedChiral}
\Sigma_I = \left(\omega ~, \, \CH_1 ~,\, \t\CH_{\b 1} ~,\, G \right)^{\Sigma_I}=\; \left(\tr_I(\sigma)~, \,\tr_I(\Lambda_1)~, \, -\tr_I(\t\Lambda_{\b 1})~, \,   2i \tr_I(2i f_{1\b 1}) \right)~.
\ee
We also have the twisted antichiral multiplet
\be\label{tSigmaTwistedChiral}
\t\Sigma_I = \left(\t\omega~, \,\t h~,\, h~,\, \t G\right)^{\t\Sigma_I} =\; \left(\tr_I(\t \sigma)~, \, \tr_I(\t \lambda)~, \, -\tr_I (\lambda)~, \, -i \tr_I(D - 2i  f_{1\b 1})  \right)~.
\ee
More generally, we can build a twisted chiral multiplet with any gauge-invariant function of $\sigma$ as its lowest component, $\omega^\CO=\CO(\sigma)$.

Note that the $G$-term in \eqref{SigmaTwistedChiral} is slightly non-standard. The components \eqref{SigmaTwistedChiral} follow from our redefinition of $\sigma$ described in appendix \ref{app: conventions}, which is 
 natural 
in the presence of the $\epsdef$-deformation. One can also build another twisted chiral multiplet $\Sigma_I'$ from $\CV$,
\be\label{Sigmaprime}
\Sigma_I' =
\left(\tr_I(\sigma-\epsdef^2 V_1 V_{\b1} \t\sigma)~, \,\tr_I(\Lambda_1+i \epsdef V_1 \lambda)~, \, -\tr_I(\t\Lambda_{\b 1}- i \epsdef V_{\b 1} \t\lambda)~, \,  G^{\Sigma_I'} \right)~,
\ee
with
\be
 G^{\Sigma_I'}=  i \tr_I\left(D+2i f_{1\b 1}+ \t\sigma \CH + 2i \epsdef \left(V_1 D_{\b1}- V_{\b 1}D_1\right)\t\sigma\right)~,
\ee
where $\CH$ is the background supergravity field  \eqref{def CH}.
Importantly, $\Sigma_I$ and $\Sigma_I'$ differ by their $G$-term  even in the $\epsdef=0$ limit. The fact that there exist distinct choices of twisted chiral multiplets inside $\CV$ is a consequence of $S^2_\Omega$ only preserving two supercharges. (In flat space with four supercharges, only \eqref{Sigmaprime} with $\epsdef=0$ would be twisted chiral.)

\subsection{Supersymmetric Lagrangians}
One can easily construct supersymmetric actions on $S^2_\Omega$ \cite{Closset:2014pda},
\be
S= \int d^2x \sqrt{g}\, \SL~,
\ee
with $\delta\SL=\t\delta \SL=0$ up to a total derivative.
 Here we present the standard renormalizable actions and we study their $\delta$-, $\t \delta$-exactness properties.

\subsubsection{$D$-terms}
From the vector multiplet \eqref{V components}, one can build a gauge-invariant general multiplet of lowest component ${1\over 4 \e_0^2}\tr(\t\sigma\sigma)$. The corresponding $D$-term action reads
\bea\label{our YM Lag}
&\SL_{\t \Sigma \Sigma}&=& {1\over \e_0^2} \tr \Big(\half D_\mu \t\sigma  D^\mu\sigma  + \left(2 i f_{1\b 1}\right)^2   - 2 i f_{1\b 1}D \cr 
&&&\qquad\quad  +2i\t\Lambda_{\b 1}D_1 \lambda -2i \Lambda_1 D_{\b 1}\t\lambda - i \t\Lambda_{\b1} [\t\sigma, \Lambda_1]\Big)~.
\eea
Here $\e_0^2$ is the dimensionful Yang-Mills (YM) coupling.
Note that  \eqref{our YM Lag} is linear in $D$ and independent of $\epsdef$. This non-standard choice of supersymmetric Yang-Mills (SYM) term is on a par with our non-standard choice of $G^\Sigma$ in \eqref{SigmaTwistedChiral}. The Lagrangian \eqref{our YM Lag}  is also $\delta$- and $\t\delta$-exact (up to a total derivative):
\be\label{YM qexact}
\SL_{\t \Sigma \Sigma}= {1 \over \e_0^2}\delta\t\delta\, \tr\left(\t\sigma  f_{1\b 1}\right)~.
\ee
Equation \eq{our YM Lag}, however, is not a good starting point because it is degenerate. The more standard SYM Lagrangian can be obtained  by adding another $\delta\t\delta$-exact term to \eqref{YM qexact}:
\be\label{SYM Qex}
\SL_{\rm YM}=    {1 \over \e^2}\delta\t\delta\, \tr\left(\t\sigma  f_{1\b 1} -\half \t\lambda \lambda\right)~. 
\ee
This gives
\bea\label{YM standard}
&\SL_{\rm YM} &= &\,   {1\over \e_0^2} \tr \Big[\,\half D_\mu \t\sigma  D^\mu\sigma  + \half \left(2 i f_{1\b 1}\right)^2 -\half D^2 +{1\over 8}[\sigma, \t\sigma]^2  \cr
&&&\qquad\quad   +2i \t\Lambda_{\b 1}D_1 \lambda -2i \Lambda_1 D_{\b 1}\t\lambda
- i \t\Lambda_{\b 1}[\t\sigma,\Lambda_1]+i\t\lambda[\sigma,\lambda]+ \epsdef \t\lambda \CL_V^{(a)}\lambda \cr
&&&\qquad\quad - i \epsdef  (D-2 i f_{1\b 1})(V_1 D_{\b 1} - V_{\b 1} D_1) \t\sigma  -{i \epsdef\over 4} [\sigma, \t\sigma]\CL_V^{(a)}\t\sigma\cr
&&& \quad\qquad- 2\epsdef^2 V_1 V_{\b 1} D_1 \t\sigma D_{\b 1}\t\sigma\Big]~.
\eea
In particular, the first line of \eqref{YM standard} is the same as the bosonic part of the $\CN=(2,2)$ SYM Lagrangian in flat space.

The standard kinetic term for the chiral and antichiral multiplets $\Phi, \t\Phi$ coupled to $\CV$ reads:
\bea\label{Phi kinetic term}
&\SL_{\t\Phi\Phi} &=&  \;-4 \t\CA  D_1D_{\b 1}\CA -\t\CA \t\sigma\left(-\sigma + i \epsdef \CL_V^{(a)}\right)\CA  -\t\CF\CF + 2 i\t \CB D_1 \CC  - 2i \t \CC D_{\b1}\CB  \cr
&&& -{i\over 2}\t\CB\t\sigma\CB - 2i \t\CC \left(-\sigma + i \epsdef \CL_V^{(a)}\right)\CC 
+ i\t \CB\t\lambda\CA   +i \t\CA\lambda\CB +2i \t\CA\Lambda_1\CC\cr
&&& +2i \t\CC\t\Lambda_{\b1}\CA + \t\CA \left(D- 2i f_{1\b 1} +\half[\sigma, \t\sigma]- 2i \epsdef V_{\b1}D_1\t\sigma  \right)\CA~.
\eea
Here the vector multiplet fields are $\FR$-valued, and an overall trace over the gauge group is implicit. The components of $\Phi, \t\Phi$ are written in the frame basis. The Lagrangian \eqref{Phi kinetic term} is $\delta$-, $\t\delta$-exact:
\be\label{Phi Qex}
\SL_{\t\Phi\Phi}= \delta\t\delta\, \tr\left({i\over 2}\t\CA\t\sigma \CA +\t\CC\CC \right)~.
\ee

Another important $D$-term Lagrangian  is the ``improvement Lagrangian'' described in \cite{Closset:2014pda}. Let $f(\omega)$ be an arbitrary holomorphic function of $\omega^i$, the bottom components of some twisted chiral multiplets $\Omega^i$. The improvement Lagrangian on $S^2_\Omega$ is given by
\be\label{improv L}
\SL_{f} = -\half R f(\omega) -{i\over 2}\CH \left(G^i \d_i f(\omega) + 2 \CH^i_1 \t\CH^j_{\b1} \d_i \d_j f(\omega) \right)~,
\ee
which is marginal if $f(\omega)$ is dimensionless. (Any anti-holomorphic dependence drops out on $S^2_\Omega$.)

\subsubsection{Superpotential}\label{subsec: W}
Given a gauge-invariant holomorphic function $W(\CA)$ of the chiral multiplets $\Phi^i$, of $R$-charge $r=2$, one can write down the superpotential term
\be
\SL_{W} = \CF^i \d_i W +\left(\CB^i \CC^j + 2 i \epsdef V_1 \CC^i \CC^j \right)\d_i\d_j W
\ee
Note that $W=W(\CA^i)$ is a section of $\CK$---or a field of spin $1$, in the frame basis. Similarly, the  conjugate superpotential  $\t  W(\t\CA)$ leads to
\be
\SL_{\t W} = \t\CF^{ i} \d_i \t W -\left(\t\CC^i \t\CB^j - 2 i \epsdef V_{\b 1} \t\CC^i \t\CC^j \right)\d_i\d_j \t W~.
\ee
These Lagrangians are $\CF$- and $\t\CF$-terms, and they are therefore $\delta$-, $\t\delta$-exact due to \eqref{susytranfoPhitwistBis}, \eqref{susytranfotPhitwistBis}.

\subsubsection{Twisted superpotential}
Given a twisted chiral multiplet  \eqref{Omega com} and its conjugate \eqref{t Omega com}, one can build the $G$ and $\t G$-term Lagrangians
\be
\SL_G = G~, \qquad \SL_{\t G}= \t G + i \CH \t\omega~,
\ee
which are supersymmetric by virtue of \eqref{susy Omega}, \eqref{susy t Omega}. One can also see from \eqref{susy t Omega} that $\SL_{\t G}$ is $\delta$- and $\t\delta$-exact. 
Importantly, the $G$-term is {\it not} $\delta$- or $\t\delta$-exact. However, if $\epsdef\neq 0$, $\SL_G$ only fails to be exact at the fixed points of $V$. Using  \eqref{susy Omega}, one can show that
\be\label{SG almost Qexact}
S_G =\int d^2 \sqrt{g} \, G = {4\pi i \over \epsdef}\left(\omega_N -\omega_S\right) + \delta(\cdots)~.
\ee
Here $\omega_N$, $\omega_S$ denote $\omega$ inserted at the north and south pole, respectively.

Given a gauge-invariant holomorphic function $\CO(\sigma)= \h W(\sigma)$ and its conjugate $\t{\h W}(\t\sigma)$, we define the twisted superpotential terms
\be\label{W Lag in general}
\SL_{\h W} +   \SL_{ \t{\h W}}=\half G^{\h W}+\half\left( G^{\t{\h W} } + i\CH \t{\h W}\right)~.
\ee
In this case \eqref{SG almost Qexact} implies that
\be\label{S of Wh and susy}
S_{ \h W} =\int d^2 \sqrt{g} \,  \SL_{\h W}  = {2\pi i \over \epsdef}\left(\h W_N -\h W_S\right) + \delta(\cdots)~.
\ee
Of particular importance is the linear superpotential for an abelian factor $U(1)_I$ of the gauge group,
\be\label{linear_twisted_W}
\h W= \tau^I \tr_I(\sigma)~, \qquad\qquad \t{ \h W}= \t\tau^I \tr_I(\t\sigma)~,
\ee
where $\tr_I$ is defined like in section \ref{subsec: twisted chiral from V}. Here $\tau, \t\tau$ is the complexified Fayet-Iliopoulos coupling constant defined by
\be\label{def tau}
\tau^I = {\theta^I \over 2\pi} + i \xi^I~, \qquad \t\tau^I = -2 i \xi^I~.
\ee
The Lagrangian \eqref{W Lag in general} becomes
\be\label{FI Lag expanded}
\SL_{\rm FI} = i{\theta^I \over 2\pi} \tr_I\left(2 i f_{1\b 1}\right) - \xi^I \tr_I\left(D- \CH \t \sigma\right)~.
\ee
The non-standard choice of $\t\tau$ in \eqref{def tau} is a result of our choice of $\Sigma^I$, $\t\Sigma^I$ in \eqref{SigmaTwistedChiral}-\eqref{tSigmaTwistedChiral}.

\subsection{Supersymmetry equations}
Let us discuss the necessary and sufficient conditions for a particular configuration of  bosonic (dynamical or background) fields to preserve the two supersymmetries of $S^2_\Omega$. For a vector multiplet $\CV$, setting to zero the gaugini variations in \eqref{susyVector twisted i}, \eqref{susyVector twisted ii} gives:
\bea\label{SUSY eq for V nonAb}
&\CL_V^{(a)} \sigma=0~, \qquad
&&  D_1 \sigma + i \epsdef V_1 \left(2 i f_{1\b 1}\right)=0~,\cr
& \epsdef\CL_V^{(a)}\t\sigma + i [\sigma, \t\sigma]=0 ~,\qquad
&&D- 2 i f_{1\b 1} + i \epsdef \left(V_1 D_{\b1}- V_{\b1} D_1\right)\t\sigma=0~. 
\eea
For a pair of chiral and antichiral multiplets $\Phi, \t\Phi$ coupled to $\CV$, the supersymmetry equations correspond to setting the variations of the fermionic fields $\CB, \CC$ and $\t\CC, \t\CC$ to zero, in addition to \eqref{SUSY eq for V nonAb}:
\bea\label{SUSY Phi}
&\big(- \sigma +i \epsdef \CL_V^{(a)}\big)\CA=0~, &\qquad& D_\bz \CA=0~,   &\qquad& \CF=0~, \cr
&\big(\sigma +i \epsdef \CL_V^{(a)}\big)\t\CA=0~, &\qquad& D_z \t\CA=0~,  &\qquad& \t\CF=0~.
\eea
This implies, in particular, that a supersymmetric background for $\CA$ is a holomorphic section of the  vector bundle with connection $a_\mu$.

For a twisted chiral multiplet $\Omega$, the supersymmetry equations following from \eqref{susy Omega} are
\be\label{susy eq Omega}
\CL_V \omega =0~,\qquad   \epsdef G = -{4 i \over\sqrt{g}} \d_{|z|^2}\omega~,
\ee
while for the twisted antichiral multiplet $\t\Omega$ we have
\be
\epsdef \CL_V \t\omega =0~, \qquad  \t G = i \epsdef |z|^2 \d_{|z|^2}\t\omega~.
\ee
This applies in particular to the gauge-invariant twisted chiral multiplets built out of the gauge field. Consider the multiplet \eqref{SigmaTwistedChiral} for an abelian factor  $U(1)_I \subset \GG$. Assuming $\epsdef\neq 0$, it follows from \eqref{susy eq Omega} that 
\be\label{susy rel for flux}
\tr_I\left(2if_{1\b 1}\right) = -{2\over \epsdef \sqrt{g}} \d_{|z|^2}\, \tr_I(\sigma)~.
\ee
Let us denote by $k^I$  the quantized flux of the $U(1)_I$ gauge field through the sphere,
\be\label{def kI}
 {1\over 2\pi} \tr_I \int_{S^2_\Omega} {\rm d}a = {1\over 2\pi}  \int d^2 x \sqrt{g}\,  \tr_I\left(-2 i f_{1\b 1}\right) = k^I \in \Z~.
\ee
The supersymmetry relation \eqref{susy rel for flux}  implies that, for any supersymmetric configuration of $\tr_I\CV$, the flux \eqref{def kI} is related to the values of $\tr_I(\sigma)$ at the poles:
\be\label{rel flux to sigma}
k^I = -{1\over \epsdef}\tr_I\left(\sigma_N -\sigma_S\right)~.
\ee
This simple relation will play an important role in the following.

\section{GLSM and supersymmetric observables}\label{sec: GLSM and observ}
The theories of interest in this paper are $\CN=(2,2)$ supersymmetric GLSMs in two dimensions, consisting of the following ingredients:
\begin{itemize}
\item  A gauge group $\GG$ with Lie algebra ${\mathfrak g}$. The corresponding gauge field $a_\mu$ sits in a  ${\mathfrak g}$-valued vector multiplet $\CV$. 
\item Charged matter fields in chiral  and antichiral multiplets $\Phi_i, \t \Phi_i$, transforming in representations ${\mathfrak R}_i, \b{\mathfrak R}_i$ of 
 $\GG$, and with integer vector-like $R$-charges, $r_i \in \Z$.
We further assume that  
$\GG$ has  no decoupled factor, that is, for every element of the 
maximal torus $\GH$ of $\GG$  there is at least one charged chiral multiplet.
 The chiral multiplets can  also be  coupled to twisted masses whenever there is  a flavor symmetry (see section \ref{subsec: flavor sym} below).
\item A superpotential $W(\Phi_i)$, which must have $R$-charge $2$ in order to preserve the $R$-symmetry. Once we put the theory on $S^2_\Omega$ and use the $A$-twisted variables, the requirement that $R[W(\Phi)]=2$ is equivalent to   $W(\CA_i)$ being a holomorphic one-form, as explained in section \ref{subsec: W}.
 \item If $G$ contains some $U(1)$ factors, denoted $U(1)_I$, $I=1, \cdots, n$  with $n\leq \rk$, we consider the linear twisted superpotential \eqref{Wh linear}, 
\be\label{Wh linear bis}
\h W(\sigma) =  \tau^I \tr_I\left(\sigma\right)~,
\ee
 for $\sigma$ the $\Fg$-valued complex scalar in $\CV$. 
\end{itemize}

While the complexified FI couplings $\tau^I$  in \eqref{Wh linear bis} are classically marginal, they can run at one-loop.~\footnote{The $\beta$ functions of the holomorphic couplings $\tau^I$ are one-loop exact by a standard argument.} Their $\beta$ functions are given by 
\be\label{b0}
\beta^I \equiv \mu {d\tau^I\over d\mu} = -{b_0^I\over 2\pi i}~,\quad \qquad b_0^I = \sum_i \tr_{\mathfrak{R}_i} (t_I)~,
\ee
where the sum is over all the chiral multiplets $\Phi_i$, and $t_I \in i\mathfrak{h}$ is the generator for the subgroup $U(1)_I$. For instance, if $\GG=U(1)$ we have $b_0= \sum_i Q_i$, where the $Q_i$'s are the $U(1)$ charges of the chiral multiplets $\Phi_i$. Similarly, if we take $\GG=U(N)$ with $N_f$ fundamental and $N_a$ anti-fundamental chiral multiplets, we have $b_0= N_f- N_a$.

If $b_0^I\neq 0$, one  defines the RG-invariant scale $\Lambda_I$,
\be\label{def LambdaI}
\Lambda_I = \mu \,\exp\left({2\pi i \tau^I(\mu)\over b_0^I }\right)~.
\ee
This dynamically generated scale is very small when the  bare FI parameter $\xi^I$ is   very large and positive, $\xi^I \gg 0$.
If $b_0^I=0$, instead, the coupling $\tau^I$ is truly marginal and we   define the dimensionless parameter
\be\label{def qI}
q_I= e^{2\pi i \tau^I}~.
\ee
(Recall that $\tau^I \sim \tau^I +1$.) In practice, it is often convenient to use the parameters $q_I$ even when $b_0^I\neq 0$. The correct statements in term of the RG invariant quantities $(\Lambda_I)^{b_0^I}= \mu^{b_0^I} q_I (\mu)$ can be recovered by dimensional analysis.

\begin{table}[t]
\centering
\begin{tabular}{c|cccccccc}
 & $a_\mu$ & $\sigma$ &$\t\sigma$ & $\Lambda_1$ & $\lambda$ & $\t\Lambda_{\b 1}$ & $\t\lambda$ & $D$ \\
\hline
$R_A$ & $0$     & $2$ &       $-2$ &       $1$ & $-1$ & $1$ & $-1$ & $0$ 
\end{tabular}
\caption{$U(1)_A$ charges for the ($A$-twisted) fields in the vector multiplet.}
\label{tab:RAforV}
\end{table}
\begin{table}[t]
\centering
\begin{tabular}{c|cccccccc}
 & $\CA$ & $\CB$ &$\CC$ & $\CF$ & $\t\CA$ & $\t\CB$ & $\t\CC$ & $\t\CF$ \\
\hline
$R_A$ & $0$     & $1$ &       $-1$ &       $0$ & $0$ & $1$ & $-1$ & $0$ 
\end{tabular}
\caption{$U(1)_A$ charges for the ($A$-twisted) fields in the  chiral and antichiral multiplets.}
\label{tab:RAforPhi}
\end{table}

Classically, our GLSM with linear twisted superpotential \eqref{Wh linear} possesses an axial-like $R$-symmetry $U(1)_A$, whose charge we denote by $R_A$. The $R_A$-charges of the fields in the vector multiplet are given in Table \ref{tab:RAforV}. In particular, we have
\be
R_A[\sigma]=2~.
\ee
A twisted superpotential $\h W(\sigma)$ preserves $U(1)_A$ if and only if it has $R_A$-charge $2$, which is the case of \eqref{Wh linear bis}. The $U(1)_A$ charges for the components of the chiral and antichiral multiplets \eqref{compo Phi}, \eqref{compo tPhi} are given in Table \ref{tab:RAforPhi}. The supercharges themselves are charged under $U(1)_A$:
\be
[R_A, \CQ] =  \CQ~, \qquad [R_A, \t  \CQ]= \t  \CQ~.
\ee
Therefore the cohomology of $ \CQ, \t  \CQ$ is graded by $R_A$, 
 sometimes called the ghost number.

If $b_0^I \neq 0$ for at least one $U(1)_I$,  $U(1)_A$ is  gauge-anomalous at one-loop. The anomalous transformation of the path integral measure under $\varphi \rightarrow e^{i R_A[\varphi]\alpha} \varphi$ can be compensated by an anomalous shift of the $\theta$-angles:
\be\label{shift theta}
 \theta^I \rightarrow \theta^I + 2  \alpha b_0^I~.
\ee
The dynamically generated scales \eqref{def LambdaI} thus transform under $U(1)_A$ with $R_A[\Lambda_I]=2$. Gauge anomalies thus break the axial $R$-symmetry $U(1)_A$ to $\cap_I\, \Z_{2b_0^I}$ in the UV  (further breaking to the $\Z_2$ fermion number is expected to occur in the IR \cite{Witten:1993xi}).
Whenever $b^I_0 =0$ for all $U(1)_I$, the axial $R$-symmetry survives quantum-mechanically, and the theory is expected to flow to a non-trivial CFT in the infrared. By abuse of terminology, we  call this situation the conformal case. 

Upon coupling the GLSM to the curved-space background $S^2_\Omega$, $U(1)_A$ also suffers from a gravitational anomaly due to the twisted spins of the fermionic fields \cite{Morrison:1994fr}. In the GLSM under consideration, the anomalous transformation of the path integral measure $[\CD\varphi]$ under $U(1)_A$ is
\be\label{grav anomaly}
[\CD\varphi] \rightarrow e^{-2 i \,d_{\rm grav} \alpha }[\CD\varphi]~,\quad \qquad d_{\rm grav} = -{\dim}({\mathfrak g}) - \sum_i (r_i-1) {\dim}({\mathfrak R}_i)~.
\ee
In other words, a correlator  $\langle \CO \rangle$ has a $U(1)_A$ charge
\be
R_A[\langle \CO \rangle]=  R_A[\CO]-2\, d_{\rm grav}~.
\ee
When the GLSM is in a purely geometric phase, $d_{\rm grav}$ coincides with the complex dimension of the target space.~\footnote{This is true assuming that the theory flows to a non-linear sigma model (NLSM) on that target space $X_d$, such that all the NLSM chiral multiplets---which correspond to local coordinates on $X_d$---have vanishing $R$-charge. We have $d_{\rm grav}=d$ in the NLSM, and this must be equal to the GLSM anomaly by the 't Hooft anomaly matching condition.} 
For instance, for $\GG=U(N)$ with $N_f$ fundamentals of $R$-charge $r$, we have $d_{\rm grav}= -N^2 - (r-1)N_f N$. When $r=0$, the target space is the Grassmannian $G(N,N_f)$, which has complex dimension $N(N_f-N)$. 
When the GLSM flows to an interacting fixed point,  $3 d_{\rm grav}$ is the central charge of the IR $\cN=(2,2)$ SCFT \cite{Hori:2006dk}.

In addition to this 
anomaly, the $\Omega$-deformation parameter itself carries $R_A$-charge
\be
R_A[\epsdef]= 2~.
\ee
Therefore the $S^2_\Omega$  background with $\epsdef\neq 0$ breaks $U(1)_A$ explicitly \cite{Closset:2014pda}.

\subsection{Coupling to flavor symmetries}\label{subsec: flavor sym}
In general, the GLSM might enjoy a non-trivial flavor symmetry, that is, a non-$R$, continuous, global symmetry group acting on the chiral multiplets, which we denote by $\GF$. It is natural to turn on a background vector multiplet $\CV^F$ for $\GF$. To preserve supersymmetry on $S^2_\Omega$, the background vector multiplet must satisfy the conditions \eqref{SUSY eq for V nonAb}. When $\epsdef \neq 0$, this implies that we can turn on any $V$-invariant profile for $\sigma^F$, $\t \sigma^F$ in the Cartan  subgroup of $\GF$, with the accompanying background fields $2 i f_{1\b 1}^F$ and $D^F$ that satisfy  \eqref{SUSY eq for V nonAb}.  (When $\epsdef=0$, $\sigma, \t\sigma$ must be constant while any flux can be turned on independently.)

For simplicity, we will mostly restrict ourselves to the 
simpler case~\footnote{We will comment on how this assumption can be relaxed in the next section.} 
\be\label{simplified_background_flavor}
\sigma^F = m^F~, \qquad \t\sigma^F= \t m^F~, \qquad [m^F, \t m^F]=0~, \qquad 2 i f_{1\b 1}^F =D^F=0~,
\ee
with $m^F, \t m^F$ some constant background values for $\sigma^F, \t\sigma^F$. These parameters are called twisted masses.  
More precisely, if we have a family of chiral multiplets $\CR_{\rho^F}$ which realize some representation  ${\mathfrak R^F}$  of of the flavor Lie algebra,  with $\rho^F$ the weights of  ${\mathfrak R^F}$, then the holomorphic twisted mass of  $\CR_{\rho^F}$ is simply
\be\label{def mrho}
m_{\rho^F}= \rho^F(m^F)~.
\ee
In the following we will denote by $m_i$ the holomorphic mass $\eqref{def mrho}$ of a given chiral multiplet $\Phi_i$, with the understanding that $m_i=0$ if $\Phi_i$ is $\GF$-neutral.

\subsection{The Coulomb branch}\label{subsec: Coulomb branch}
Consider the Coulomb branch $\fM = \mathfrak{h}_\bC /W$ of the theory in flat space, corresponding to turning on constant expectation values for $\sigma, \t\sigma$. Here $ \mathfrak{h}$ is the Cartan subalgebra and $W$ is the Weyl group of $\mathbf{G}$. The two fields can be diagonalized simultaneously:~\footnote{We will often use $\sigma$ to denote both a Coulomb branch parameter and the actual field. This should cause no confusion.}
\be
\langle \sigma\rangle= {\rm diag}(\sigma_a)~, \qquad \langle\t\sigma\rangle={\rm diag}(\t \sigma_a)~, \qquad a=1, \cdots, {\rm rank}({\mathfrak g})~.
\ee
At a generic point on $\fM$, the gauge group is broken to its Cartan subgroup $\GH$,
\be
\GG\rightarrow \GH = \prod_{a=1}^{{\rm rank}(G)} U(1)_a~,
\ee
and all the fields are massive except for the $\GH$-valued vector multiplets $\CV_a$. The effective twisted superpotential on the Coulomb branch is \cite{Nekrasov:2009uh} 
\be\label{Weff MC}
\h W_{\rm eff}=    \h W+ \h W_{\rm mat} + \h W_{\rm vec}~, 
\ee
where the first term is the classical term \eqref{Wh linear bis},  the second term is the contribution from  the chiral multiplets $\Phi_i$ with twisted masses $m_i$,
\be
\h W_{\rm mat} = -{1\over 2\pi i} \sum_i \sum_{\rho_i\in \FR_i} (\rho_i(\sigma)+m_i) \left[\log{(\rho_i(\sigma)+m_i)} -1\right]~,
\ee
(there is a renormalization scale implicit in the logarithm), and the last term is the  contribution from the $W$-bosons, i.e. the vector multiplets for $\GG/\GH$:
\be
 \h W_{\rm vec} = {1\over 2\pi i}  \sum_{\alpha\in \Fg\setminus \Fh} \alpha(\sigma) \left[\log{ \alpha(\sigma)} -1\right]=  -\half \sum_{\alpha>0} \alpha(\sigma)~.
\ee
The only effect of this last contribution is to induce a subtle shift of the effective $\theta$-angles by a multiple of $\pi$. 
The vacuum equations coming from \eqref{Weff MC} are \cite{Witten:1993yc}
\be\label{vacuum equ}
e^{2\pi i \, \d_{\sigma_a} \h W_{\rm eff}}= 1~, \qquad\quad a=1, \cdots, \rm{dim}(\Fg)~.
\ee
In the case of massive theories, these equations have  come under great scrutiny in recent years in the context of the Bethe/gauge correspondence \cite{Nekrasov:2009uh}.
For future reference, let us also define the {\it effective $\tau$-couplings} on the Coulomb branch: 
\be\label{tau eff}
\tau_{\rm eff}^a \equiv \d_{\sigma_{a}}   \h W_{\rm eff} = \tau^a - \half \sum_{\alpha>0} \alpha^a -{1\over 2\pi i} 
 \sum_i \sum_{\rho_i\in \FR_i} \rho_i^a \log(\rho_i(\sigma)+m_i)
\ee
where $\tau^a \equiv \d_{\sigma_{a}}   \h W$. Finally, let us note that the flat space axial $R$-symmetry anomaly is seen on the Coulomb branch as a monodromy of \eqref{tau eff} under $\sigma\rightarrow e^{2i\alpha} \sigma$, which is compensated by the $\theta$-angle shift \eqref{shift theta}.

\subsection{Correlation functions on $S^2_\Omega$}
The interesting supersymmetric observables on $S^2_\Omega$ are the correlation functions of non-trivial supersymmetric operators, that is,  operators that are non-trivial in the cohomology of the supercharges $Q,  \t Q$. 
The only such local operators one can build out of the GLSM fields are functions of the complex scalar field $\sigma$. In the presence of the $\epsdef$-deformation, $\delta\sigma= \t\delta \sigma=0$ if and only if $V=0$. Therefore the operators must be  inserted at the north and south poles of $S^2_\Omega$,
\be\label{our susy ops}
 \CO^{(N)}(\sigma_N)~, \qquad   \CO^{(S)}(\sigma_S)~.
\ee
Here  $\CO^{(N)}, \CO^{(S)}$ denote two arbitrary gauge-invariant functions of $\sigma$.  The subscripts $N, S$ stand for the point of insertion at the north or south poles, $z=0, \infty$, respectively. 
We therefore consider the correlation functions
\be\label{our susy obs}
\left\langle \CO^{(N)}(\sigma_N) \CO^{(S)}(\sigma_S) \right\rangle~.
\ee
Note that  the operators \eqref{our susy ops} are not, in general, the only non-trivial supersymmetric local operators in the GLSM, but they are the only such operators one can build out of the elementary fields. More general supersymmetric operators, generally known as disorder operators, can be defined in term of singular boundary conditions in the path integral. In this work, we restrict our attention to the operators \eqref{our susy obs}.

If $\epsdef=0$, a gauge-invariant operator $\CO(\sigma)$ can be inserted anywhere on the sphere. Moreover, any correlation function
\be\label{Amode corr}
\left\langle \CO(\sigma)  \right\rangle_0 = \left\langle  \CO^{(1)} (\sigma(z_1,\bz_1))\cdots\CO^{(n)}(\sigma(z_n, \bz_n))   \right\rangle_0~, \qquad \CO=  \CO_1\cdots\CO_n~,
\ee
 is independent of the insertion points, since the derivatives $\d_z\sigma$, $\d_{\b z}\sigma$ are $\delta$- or $\t \delta$-exact. By a standard procedure \cite{Witten:1988ze}, one can construct the so-called descendant of the local operator $\CO=\CO(\sigma)$,~
\be\label{def descendant}
X_2(\CO) = -{i\over 4\pi} \int d^2 x \sqrt{g}\, G^{\CO}
= {1\over 2\pi} \int d^2 x \sqrt{g}\, \left(- 2 i f_{1\b 1}\, \d_\sigma \CO - i \Lambda_1 \t\Lambda_{\b 1}\, \d_\sigma^2 \CO  \right)~,
\ee
which is supersymmetric in the $A$-model.%
~\footnote{Strictly speaking, the last equality in \eqref{def descendant} only holds for $\GG= U(1)$, but $X_2$ takes this schematic form for any $\GG$.}
The normalization of \eqref{def descendant} has been chosen for future convenience. A general supersymmetric observable in the $A$-model includes such descendants.

The operator \eqref{def descendant} is also supersymmetric with $\epsdef\neq 0$. In that case, however, equation \eqref{SG almost Qexact} gives
\be
X_2(\CO) = { \CO(\sigma_N) -\CO(\sigma_S) \over\epsdef} + \cdots~,
\ee
where the ellipsis denotes a $\delta$-, $\t\delta$-exact term. Therefore, the insertion of any descendant $X_2$  into a supersymmetric correlation function on $S^2_\Omega$ is accounted for in \eqref{our susy obs}, being equivalent to the insertion of ${1\over \epsdef}\left(\CO(\sigma_N) -\CO(\sigma_S)\right)$.

\subsection{Parameter dependence and selection rules}
The correlation function \eqref{our susy obs} only depends on the supergravity background through the complex parameter $\epsdef$. For $\epsdef=0$, it is well-known that the theory is topological \cite{Witten:1988xj, Witten:1991zz}. In the general case, one can argue that any small deformations of the hermitian metric  preserving the Killing vector $K$ is $\delta$-, $\t\delta$-exact, similarly to the analysis of \cite{Closset:2013vra}.

The correlation function also depends on the complexified FI parameters $\tau^I$  through their exponentials $q_I$ defined in \eqref{def qI} (or through the RG-invariant scales \eqref{def LambdaI} if $\tau^I$ runs). This dependence is holomorphic, since the conjugate couplings $\t\tau^I$ in \eqref{def tau} are $\delta$-, $\t\delta$-exact. We can also have a dependence on the twisted masses $m^F$ for the flavor symmetry, and one can similarly argue that this dependence is holomorphic. In total, we therefore have a function
\be\label{our susy obs ii}
\left\langle \CO^{(N)}(\sigma_N) \CO^{(S)}(\sigma_S)  \right\rangle=  \; F\left(q_I~, m^F,  \epsdef\right)~,
\ee
which is locally holomorphic in all the parameters. 

On general grounds, \eqref{our susy obs ii} might suffer from ambiguities, corresponding to supersymmetric local terms one can add to the UV description. Allowed local terms of dimension two are severely restricted by supersymmetry (and by gauge and supergravity invariance). The parameters $q_I$ and $m^F$ are the lowest components of background twisted chiral multiplets of dimension $0$ and $1$, respectively, with all higher components set to zero. For a conformal $q_I$, the only allowed term is the improvement Lagrangian \eqref{improv L}, which gives $\SL_{f} = -\half R f(q)$ on the supersymmetric locus. Note that this term is topological due to the Gauss-Bonnet theorem:
\be
S_f= \int d^2 x\sqrt{g}\, \SL_{f}  =4\pi  f(q)~.
\ee
 For the twisted masses $m^F$, we can only turn on a linear twisted superpotential, which vanishes on the background \eqref{simplified_background_flavor}. 
For a non-conformal $\tau^I$, the dependence is through the scale $\Lambda_I$ and no local term is allowed either. Finally, $\epsdef$ can also be seen as the lowest component of a twisted chiral multiplet, of dimension $1$,  built out of the supergravity multiplet. No local term in $\epsdef$ is allowed by dimensional analysis. In conclusion, the only ambiguity in the definition of \eqref{our susy obs ii} is through an entire holomorphic function  of the conformal couplings $q^I$,
\be\label{ambiguity f(q)}
 F\left(q_I~, m^F,  \epsdef\right) \sim e^{-4\pi f(q)}  F\left(q_I~, m^F,  \epsdef\right)~.
\ee
Such local terms often appear as renormalization scheme ambiguities; see in particular \cite{Gerchkovitz:2014gta} for a discussion in a related context. On $S^2_\Omega$, however, all the one-loop determinants  that will appear in the localization computation  are actually finite, and $f(q)$ can be chosen in a scheme-independent way. We shall choose $f(q)=0$ in the following.
When the IR description  is in terms of a NLSM on a CY manifold $X_d$,  the ambiguity \eqref{ambiguity f(q)} corresponds to a K\"ahler transformation on the K\"ahler structure moduli space of $X_d$.

Dimensional analysis and the $U(1)_A$ axial $R$-symmetry lead to simple selection rules for \eqref{our susy obs ii}. 
Note that $q^I$ for a conformal $\tau^I$ has vanishing $R_A$-charge, while $\Lambda_I$ for a non-conformal coupling, the twisted masses $m^F$ and $\epsdef$ all have $R_A=2$.
Without loss of generality, consider the insertion of operators of definite $R_A$-charge in \eqref{our susy obs ii},
\be
R_A[\CO^{(N)}]= r_A(N)~, \qquad R_A[\CO^{(S)}]= r_A(S)~, \qquad r_A(\CO)= r_A(N)+r_A(S)~.
\ee
Taking into account the gravitational anomaly \eqref{grav anomaly}  of $U(1)_A$, we have
\be
R_A[ F\left(q_I~, m^F,  \epsdef\right)]= r_A(\CO) - 2 d_{\rm grav}~.
\ee
Consider first the case of a conformal GLSM flowing to a non-singular CFT, that is, $b_0^I=0$ $\forall I$ and $m^F=0$. Then we must have 
\be\label{selec rule conf}
 F\left(q_I~,  0, \epsdef\right) =  (\epsdef)^{\half r_A(\CO) - d_{\rm grav}}\; F_c(q_I)~,
\ee
with the condition $r_A(\CO)\geq  2 d_{\rm grav}$ because the answer should be smooth in the $\epsdef\rightarrow 0$ limit. In  particular, we recover the usual ghost number selection rule $r_A(\CO)=  2 d_{\rm grav}$ for $\epsdef=0$.
More generally, we have 
\be\label{selec rule}
F\left(q_I~, \Lambda_{I'}~, m^F,  \epsdef\right)\sim  (\epsdef)^j \left(\prod_{I'} \Lambda_{I'}^{ b_0^{I'}k'_{I'} }\right) (m^F)^l  \, F_c(q_I)~,
\ee
where $I$ and $I'$ denote conformal and non-conformal gauge couplings, respectively, and $j+  b_0^{I'}k'_{I'} + l =  \half r_A(\CO) - d_{\rm grav}$ with $j\geq 0$. This is of course schematic. Interestingly, we can have negative powers of the twisted mass $m^F$. (A slightly finer selection rule can be obtained by realizing that the  $m^F$ term should appear as a singlet of the flavor group $\GF$, which can disallow some values of $l$.)

\section{Localizing the GLSM on the Coulomb branch}\label{section: der1}
In this section, we outline the derivation of the Coulomb branch formula \eqref{formula intro 1}-\eqref{formula intro 2}. Some of the more technical steps are presented in section \ref{section: derivation} and in appendix. We also discuss the specialization \eqref{formula Amodel intro} to the $A$-twisted GLSM (the $\epsdef=0$ limit).

\subsection{The Coulomb branch formula}\label{subsec: summary}
Our main result is that  the correlation function \eqref{our susy obs} can be computed exactly by a sum of Jeffrey-Kirwan (JK) residues  \cite{JK1995, 1999math......3178B,2004InMat.158..453S}:
\bea
&\vev{\cO^{(N)}(\sigma_N) \cO^{(S)}(\sigma_S)}
=\cr
&\qquad={(-1)^{N_*} \ov |W|} \sum_{k \in \Gamma_{\mathbf{G}^\vee}} q^k
\sum_{\hs_* \in \tfM^{k}_\text{sing}}
\underset{\hs = \hs_*}{\text{JK-Res}} \left[ \mathbf{Q}(\hs_*),  \xi_\eff^\text{UV}  \right]
\mathbf{I}_k (\cO^{(N)},\cO^{(S)})~,
\label{main formula}
\eea
of the differential form
\be
\mathbf{I}_k (\cO^{(N)},\cO^{(S)}) =
\cZ^\oneloop_k (\hs) \cO^{(N)} \left( \hs -{\epsdef k \ov 2} \right)
\cO^{(S)}\left(\hs +{\epsdef k \ov 2} \right)
d \hs_1 \wedge \cdots \wedge d \hs_\rk~,
\label{density}
\ee
on  $\tfM \cong \bC^\rk$, in each topological sector $k$.

Let us explain \eq{main formula} in more detail.  We denote by $k \in \Gamma_{\mathbf{G}^\vee} \subset i\mathfrak{h}$ the magnetic fluxes that label the topological sectors \cite{Goddard:1976qe}, where $\mathfrak{h}$ is the Cartan subalgebra of the gauge algebra $\mathfrak{g}$. The lattice $\Gamma_{\mathbf{G}^\vee} \cong \bZ^\rk$ is the integral lattice of magnetic fluxes, which can be obtained from $\Gamma_\mathbf{G}$, the weight lattice of electric charges of $\mathbf{G}$ within the vector space $i\mathfrak{h}^*$, by \cite{Englert:1976ng,Kapustin:2005py}
\be
\Gamma_{\mathbf{G}^\vee} = \setcond{k}{\rho(k) \in \bZ ~~ \forall \rho \in \Gamma_\mathbf{G}} \,,
\ee
where $\rho(k)$ is given by the canonical pairing of the dual vector spaces, which we elaborate on shortly.  Let us also introduce the notation $\vec k \in \bZ^n$ to denote the fluxes in the free part $U(1)^n$ of the center of  $\mathbf{G}$. We have
\be
q^k \equiv \exp(2 \pi i \sum_{I=1}^n (\vec \tau)_I (\vec k)_I) =  \exp(2 \pi i \tau(k))~.
\label{vqvK0}
\ee
The complex FI parameter $\vec \tau \in \bC^n$ lies in the central sub-algebra $\mathfrak{c}_\bC^* \subset \mathfrak{h}_\bC^* \subset \mathfrak{g}_\bC^*$ of the dual of the Lie algebra $\mathfrak{g}$, but is also an element of $\mathfrak{h}_\bC^*$ by the embedding of the center into the Cartan subgroup $\mathbf{H}$---thus the equality of \eq{vqvK0}. The pairing $\tau(k)$ is the canonical pairing between elements of $\mathfrak{h}_\bC^*$ and $\mathfrak{h}_\bC$. In particular, for elements $V \in \mathfrak{h}^*_\bC$ and $W \in \mathfrak{h}_\bC$ of the vector spaces over $\bC$, we write $V(W) \equiv \sum_{a=1}^\rk V^a W_a$, 
where the label ``$a$" is used to index the basis $t_a$ of $i\mathfrak{h}$, as well as elements of the dual basis $t_a^*$ of $i\mathfrak{h}^*$. 
The parameter $\hs=(\hs_a)$ is the complex coordinate on $\tfM = \mathfrak{h}_\bC \equiv \bC^\rk$, which is the cover of the ``Coulomb branch moduli space" $\fM = \tfM / W$, where the quotient is taken with respect to the Weyl group $W= {\rm Weyl}(\mathbf{G})$. The ``one-loop'' factor $\cZ^\oneloop_{k} (\hs;\epsdef)$ is the product of all the one-loop determinants of charged fields on the Coulomb branch:
\be
\cZ^\oneloop_{k} (\hs;\epsdef) =   Z^\text{vector}_{k} (\hs;\epsdef) \prod_i Z^{\Phi_i}_{k} (\hs;\epsdef) \,.
\ee
For each chiral multiplet $\Phi_i$, we have
\be\label{Zphi introsec}
Z^{\Phi_i}_{k} (\hs;\epsdef)=  \prod_{\rho_i \in \mathfrak{R}_i }\left( 
\epsdef^{r_i-\rho_i(k)-1} {\Gamma({\rho_i(\hs) +m_i^F \ov \epsdef} + {r_i- \rho_i(k) \ov 2})
\ov \Gamma({\rho_i(\hs) +m_i^F \ov \epsdef} - {r_i- \rho_i(k) \ov 2}+1) } \right) \,,
\ee
 where $\rho_i$ denote the weights of the representation $\mathfrak{R}_i$ of $\mathbf{G}$ in which $\Phi_i$ transforms, 
whereas $m_i^F$ and $r_i$ are the twisted mass and the $R$-charge of $\Phi_i$. 
The result \eqref{Zphi introsec} has a $R$-charge-dependent sign ambiguity, which we could not determine. This leads to the sign ambiguity $(-1)^{N_*}$ in \eqref{main formula}. We propose a prescription to fix  this sign in subsection \ref{subsec: oneloop det} below.
The vector multiplet factor $ Z^\text{vector}_{k} (\hs;\epsdef)$ receives contribution from all the $W$-bosons and their superpartners $\CV^{(\alpha)}$, indexed by the roots $\alpha$ of the gauge group. These $W$-boson multiplets contribute  like an adjoint chiral multiplet of $R$-charge $2$.

It is convenient to collect together the labels $(i,\rho_i)$ for the component of weight $\rho_i$ of the $i$-th chiral multiplet $\Phi_i$ and the label $\alpha$ for the $W$-boson multiplet $\cV^{(\alpha)}$ into a collective label $\cI$  for all the components of the charged fields in the theory. In this notation, 
\bea\label{one_loop_collective}
& \cZ^\oneloop_{k}(\hs;\epsdef) = \prod_\cI Z_{k}^\cI(\hs;\epsdef)~,\cr
&Z_{k}^\cI(\hs;\epsdef)
=  \epsdef^{r_\cI-Q_\cI(k)-1} {\Gamma({Q_\cI (\hs) +m^F_\cI \ov \epsdef} + {r_\cI- Q_\cI(k) \ov 2})
\ov \Gamma({Q_\cI( \hs) +m^F_\cI \ov \epsdef} - {r_\cI- Q_\cI( k) \ov 2}+1) }
\,,
\eea
where $Q_\cI \in i\mathfrak{h}^*$ is the charge of the field component $\cI$ of the theory, and $r_\cI$ its $R$-charge. That is,  $Q_\cI$ equals $\rho_i$ for the component of labels $(i,\rho_i)$ of a chiral multiplet $\Phi_i$, and it equals $\alpha$ for a $W$-boson multiplet  $\cV^{(\alpha)}$ associated to the root $\alpha$, and similarly for $r_\cI$.

In each flux sector labelled by $k$, the singular locus of the one-loop determinant $\cZ_k^\oneloop$ arises from codimension-one poles  located on hyperplanes in $\tfM$, and the intersections thereof in higher codimension. 
Using the collective label $\cI$, these hyperplanes are given by:
\be
H^{n}_\cI =
\Big \{\, \hs ~:~
Q_\cI ( \hs )
= -m^F_\cI-\epsdef \left( n+{r_\cI-Q_\cI (k ) \ov 2} \right)  \, \Big \}~,
 ~~~ n \in [0,-r_\cI+Q_\cI( k)]_\text{int} ~.
\label{hyperplanes}
\ee
We  use the notation $[A,B]_\text{int}$ to denote the set of integers
\be
[A,B]_\text{int} \equiv
\{ \, n ~:~
n \in \bZ
\quad \text{and} \quad
A \leq m \leq B \, \} \,,
\ee
which is empty when $A>B$. 
These singular hyperplanes are due to all the components of the  charged chiral and vector multiplets. However,  any hyperplane $H_\alpha^n$ coming from a $W$-boson multiplet $\cV^{(\alpha)}$ is actually non-singular, because of additional zeros in the determinant of the oppositely charged  multiplet $\cV^{(-\alpha)}$.%
~\footnote{We do not introduce additional indices that label only the hyperplane singularities coming from modes of the chiral fields, since we lose nothing by treating the hyperplanes $H_\alpha^{n}$ as codimension-one poles of $\cZ_k^\oneloop(\hs;\epsdef)$ with residue zero.} 
In \eq{main formula}, we denote by $\tfM^{k}_\text{sing}$ the collection of complex-codimension-$\rk$ singularities of $\cZ_{k}^\oneloop (\hs;\epsdef)$, which come from the intersection of $s\geq \rk$ hyperplanes $H^{n_1}_{\cI_1}, \cdots, H^{n_s}_{\cI_s}$. Finally, $\mathbf{Q}(\hs_*)$   denotes the collection of charges $Q_{\cI_1},\cdots,Q_{\cI_s}$  determining the orientations of the singular hyperplanes which intersect at $\hs_*$.

The   Coulomb branch formula \eq{main formula} gives the correlation function $\vev{\cO^{(N)}(\sigma_N) \cO^{(S)}(\sigma_S)}$ as a sum of Jeffrey-Kirwan (JK) residues of the codimension-$\rk$ poles of the meromorphic $\rk$-form \eq{density} on $\tfM$. To  define the JK residue, one needs to specify an additional vector $\eta \in i\mathfrak{h}^*$. In particular, for non-degenerate codimension-$\rk$ poles, where exactly $\rk$ hyperplanes are intersecting, we have
\be
\underset{\hs = \hs_*}{\text{JK-Res}} \left[(Q_i),\,  \eta  \right]
{d\hs_1 \wedge \cdots \wedge d \hs_\rk
\ov Q_1(\hs) \cdots Q_\rk (\hs)} =
\begin{cases}
{1 \ov |\det(Q_i)|}
&\text{if}~\eta \in \text{Cone}(Q_i) \\
0
&\text{if}~\eta \notin \text{Cone}(Q_i) \,,
\end{cases}
\ee
by definition, where $(Q_i)= (Q_1,\cdots,Q_\rk)$, and $\text{Cone}(Q_i)$ is the cone in $i\mathfrak{h}^*$ generated by the $Q_i$'s. We  give the complete definition of the JK residue in subsection \ref{subsec: JK} below.   In   \eq{main formula}, the vector $\eta$ must be set to $\xi_\eff^\text{UV} \in i \mathfrak{h}^*$, which is defined as follows. Let us first note that the coupling $\t\tau$ in \eqref{def tau} is $Q$-exact. It could thus be taken arbitrary as
\be\label{def txi}
\t\tau = -{2i\over \e^2}\t\xi~,
\ee
\emph{independent} of the physical complexified FI parameter $\tau$. Here $\e^2$ is a dimensionless parameter which will appear in the supersymmetric localization  computation, and $\t\xi  \in i \mathfrak{h}^*$ is  finite and otherwise arbitrary.
We define:
\be
\xi_\eff^\text{UV}
=\t\xi + {1 \ov 2 \pi}b_0 \lim_{R \ra \infty} \log R
\label{xieff}
\ee
with
\be\label{def bo in hstar}
b_0 \equiv \sum_\cI Q_\cI   \,.
\ee
Note that this definition of $b_0 \in i \mathfrak{h}^*$ is equivalent to the one given by equation \eq{b0}. For theories with conformal IR fixed points, $b_0$ vanishes and $\xi_\eff^\text{UV} =\t \xi$. When $b_0 \neq 0$, the vector ${\xi_\eff^\text{UV}} \in i \mathfrak{h}^*$ is defined to lie within the cone $\text{Cone}(Q_1,\cdots,Q_p)$ spanned by $Q_1, \cdots, Q_p$ if there exists an $R_0 > 1$ such that
\be
\t\xi + {1 \ov 2 \pi} b_0 \log R
\in \text{Cone}(Q_1,\cdots,Q_p)
\qquad \text{for all} \quad R \geq R_0 \,.
\ee
Thus there exists a large enough $R$ that can be used to define $\xi_\eff^\text{UV}$ for all practical purposes. The value of $\t\xi$ is arbitrary, but we shall always choose it to be parallel and pointing in the same direction to the physical FI parameter $\xi$ which appears in $\tau$. This is because, even though the result \eqref{main formula} is formally true for any choice of $\t\xi$, the $q$ series in \eq{main formula} might not converge otherwise. Discussion on this issue is presented at the end of section \ref{ss: general}.

This concludes  the executive summary of our main result \eq{main formula}. For the remainder of the section, we explain how we arrive at this result and we present some important background material. More technical aspects of the derivation are presented in section \ref{section: derivation}.

\subsection{Localization on the Coulomb branch}
The exact computation of the correlation function \eqref{our susy obs} in \eq{main formula} is possible because of supersymmetry. By a standard argument---see e.g. \cite{Witten:1991zz},  we expect the path integral to {\it localize} on the subspace  $\CM_{\rm susy}$ of supersymmetric configurations $\{\varphi\}$ such that $\delta \varphi=\t\delta \varphi =0$, where $\varphi$ runs over all the fields in the path integral. In the present case, $\CM_{\rm susy}$ is given by the solutions to the supersymmetry equations \eqref{SUSY eq for V nonAb}-\eqref{SUSY Phi} for the bosonic fields, together with their fermionic counterparts. However, $\CM_{\rm susy}$ is still too large and complicated to be of much use. In particular, it includes arbitrary profiles of $\sigma = \sigma(|z|^2)$.  Another potential issue is that there are  fermionic zero modes   on $S^2_\Omega$, which must be treated carefully before localizing to $\CM_{\rm susy}$.

Both of these issues can be addressed by introducing a {\it localizing action}, $S_{\rm loc}$. This is a $\delta$-, $\t\delta$-exact action for all the fields in the theory, which plays two roles. First of all,  it further restricts $\CM_{\rm susy}$ to a more manageable subset given by its supersymmetric saddles. Secondly, it introduces a convenient kinetic term for the fluctuations around the localization locus. 
For the GLSM, one can consider  two distinct localizing actions, leading to two different {\it localization schemes}  \cite{Benini:2012ui}. By construction,  the final answer is independent of such schemes, but the explicit formula resulting from one or the other localization procedure can be more or less wieldy.  In this section, we consider the ``Coulomb branch'' localization, which leads to the simplest answer. The so-called ``Higgs branch'' localization is discussed in section \ref{section: Higgs branch loc}.

Let us consider the localizing action
\be\label{Sloc coulomb}
\SL_{\rm loc} = {1\over \e^2}\left(\SL_{YM} -\t\xi(D- 2 i f_{1\b 1} -\t\sigma \CH)  \right) + {1\over \g^2}\SL_{\t\Phi\Phi}~.
\ee
Consider first the limit $\e^2\rightarrow 0$, so that we localize the vector multiplet using the  the YM action \eqref{YM standard}:
\be\label{YM standard bis}
\SL_{\rm YM} =    {1\over \e_0^2} \tr \Big[\,\half D_\mu \t\sigma  D^\mu\sigma  + \half \left(2 i f_{1\b 1}\right)^2 -\half D^2 +{1\over 8}[\sigma, \t\sigma]^2 + \cdots \Big]~,
\ee
where the ellipsis denotes the fermion contributions plus some terms of higher order in $\epsdef$.~\footnote{The YM coupling $\e^2_0$ is a mass scale that we keep fixed while we send the dimensionless coupling $\e^2$ to zero.} The gauge-fixing of this action is considered in appendix \ref{App: det}. For future convenience,  we also introduced a $\delta$, $\t\delta$-exact coupling $\t\xi$ in \eqref{Sloc coulomb}, which we defined in \eqref{def txi}.

 To perform the path integral over the vector multiplet, one should specify reality conditions for the bosonic fields.  We choose
\be\label{reality condition}
\sigma^\dagger= \t\sigma~, \qquad (2i f_{1\b 1})^\dagger = 2 i f_{1\b1}~.
\ee
In particular, we take the gauge field to be real. The proper contour for $D$ is rather more complicated, as we shall discuss.
Let us denote by $\CM_{\rm susy}^{\R}$ the intersection of $\CM_{\rm susy}$ with \eqref{reality condition}. On $\CM_{\rm susy}^{\R}$, we must have
\be\label{cond on MsusyR}
D= 2i f_{1\b 1}\left(1 - 2 |\epsdef|^2 V_1 V_{\b 1}\right)~, \qquad [\sigma, \t\sigma]=0~.
\ee
 The ``Coulomb branch'' localization locus is the intersection of $\CM_{\rm susy}^{\R}$ with the saddles of \eqref{YM standard bis}. Na\"ively, one would integrate out the auxiliary field $D$ to find $D= - i \epsdef \left(V_1 D_{\b1}- V_{\b1} D_1\right)\t\sigma$. Plugging into the supersymmetry equations \eqref{SUSY eq for V nonAb}, this would imply the supersymmetric saddle
\be\label{naive_saddle}
\sigma= {\rm constant}~, \quad\qquad 2i f_{1\b 1} =0~.
\ee
However, this cannot be the whole story. For consistency, the path integral should include a  sum over all allowed $G$-bundles.%
~\footnote{Although sometimes a restricted sum over topological sectors can be consistent \cite{Pantev:2005rh, Pantev:2005wj, Pantev:2005zs, Hellerman:2006zs, Seiberg:2010qd}, we take the simplest approach of summing over all such sectors, weighted by the $\theta^I$-angles.}
 This implies, at the very least, that some saddles have non-vanishing fluxes \eqref{def kI} for the $U(1)_I$ factors in $\GG$.

To accommodate the proper sum over topological sectors, we should consider more general field configurations in $\CM_{\rm susy}^{\R}$, and worry about the auxiliary field $D$ a posteriori.
Thanks to \eqref{cond on MsusyR}, we can use the gauge freedom to diagonalize $\s$ and $\t\s$ simultaneously on $\CM_{\rm susy}^{\R}$. The localization locus is the ``Coulomb branch'', 
\be\label{sigma vev loc}
\sigma= {\rm diag }(\sigma_a)~, \qquad\quad \t\sigma=  {\rm diag }(\t\sigma_a)~, \qquad\quad a= 1, \cdots, \rk~,
\ee
for some $\sigma_a= \sigma_a(|z|^2)$ to be determined.
A generic expectation value \eqref{sigma vev loc}  breaks $\GG$ to its Cartan subgroup,
\be
\GH = \prod_{a=1}^\rk U(1)_a~. 
\ee
Upon choosing the diagonal gauge for $\sigma$, the path-integral generalization of the Weyl integral formula  leads to a sum over all $\GH$-bundles  \cite{Blau:1994rk, Blau:1995rs}. (It is in keeping with a Coulomb branch intuition that we should be able to  deal with an abelianized theory throughout.) Such bundles are the line bundles $\oplus_a L_a$  characterized by the first Chern classes
\be
c_1(L_a) = {1\over 2\pi}\int_{S^2} (da)_a \equiv  k_a~.
\ee
The fluxes $k_a$ are GNO quantized \cite{Goddard:1976qe,Englert:1976ng,Blau:1994rk,Kapustin:2005py} and lie in a discrete subspace $k = (k_a) \in \Gamma_{\GG^\vee}$ of $i\mathfrak{h}$. In any  $k$-flux sector, supersymmetry imposes the relations
\be\label{constraint on k sigma a}
k_a = {(\sigma_a)_S -(\sigma_a)_N\over \epsdef}~,
\ee
which can be derived like \eqref{rel flux to sigma}.

To gain some intuition, it is useful to minimize 
\be\label{kin sigma toy}
 \half \int d^2 x \sqrt{g}\, \d_\mu \t\sigma_a \d^\mu\sigma_a
\ee
on $\CM_{\rm susy}^{\R}$, with the constraint \eqref{constraint on k sigma a}. One finds:
\be\label{sigma saddle toy model}
\sigma_a =  
\begin{cases} 
\h\sigma_a - \epsdef{k_a\over 2}  &\qquad {\rm if}\quad z=0~, \\
 \h\sigma_a  &\qquad {\rm if}\quad z\neq 0, \infty~, \\
 \h\sigma_a + \epsdef{k_a\over 2}  &\qquad {\rm if}\quad z=\infty~,
\end{cases}
\ee
with $\h\sigma_a \in \C$ a constant. This obviously satisfies \eqref{constraint on k sigma a}.  By supersymmetry, the profile of the field strength $(2 i f_{1\b 1})_a$ is determined precisely in terms of $\sigma_a$: 
\be\label{susy rel for flux bis}
\left(2if_{1\b 1}\right)_a = -{2\over \epsdef \sqrt{g}} \d_{|z|^2}\, \sigma_a~.
\ee
Plugging \eqref{sigma saddle} into \eqref{susy rel for flux bis}, we see that the flux is given by Dirac $\delta$-functions at the poles,
\be\label{f saddles}
\left(- 2if_{1\b 1}\right)_a  = {2\pi \over  \sqrt{g} }{k_a\over 2}\left(\delta_N + \delta_S\right)~.
\ee
The field configurations \eqref{sigma saddle toy model}-\eqref{f saddles}  are thus equal to \eqref{naive_saddle} away from the  the fixed points of the  Killing vector $V$, and they thus  solve the equations of motions that follow from \eqref{YM standard bis} away from the poles. Their singular behavior at the poles is itself rather mild; in particular,  these ``singular saddles'' have  vanishing action by construction (since the field configuration is supersymmetric and the SYM action is $\delta$-exact).

More generally, let us {\it conjecture} that there exists a proper ``saddle'' of the localizing action \eqref{Sloc coulomb}---that is, a vector multiplet configuration which minimizes the real part of the action, in a given topological sector. (By supersymmetry, this minimum is in fact a configuration of zero action.)  Such a saddle always has a bosonic zero mode, which shifts $\sigma$ by an arbitrary constant $\h\sigma$. Since we also know that  \eqref{constraint on k sigma a} holds by supersymmetry, this determines the value of $\sigma$ at the north and south poles:
\be\label{sigma saddle}
(\sigma_a)_N = \h\sigma_a - \epsdef{k_a\over 2}~, \qquad\quad
(\sigma_a)_S = \h\sigma_a + \epsdef{k_a\over 2}~, 
\ee
where the constant mode $\h\sigma$ is here  defined as the average, $\h\sigma= \half(\sigma_N+ \sigma_S)$.  We further assume that the saddle $\sigma$ is unique up to the shift by $\h\sigma$, in any given flux sector. This is all we need for everything that follows.
Ultimately, we regard the final answer we obtain and the agreement with the Higgs branch localization scheme of section \ref{section: Higgs branch loc} as the most convincing arguments in support of our conjecture.

Let us now comment on the integration contour of the auxiliary field $D$. The na\"ive contour obtained by analytic continuation from flat space takes $D$ to be purely imaginary. Instead, we should take 
\be\label{def hD}
D_a= D_a^{\rm susy} + i \h D_a~,
\ee
where $D_a^{\rm susy}$ is the supersymmetric saddle solution, which is related to the gauge field flux by \eqref{cond on MsusyR}.  That is, the $D$ contour is taken to pass through the supersymmetric saddle (at $\h D=0$). The contour for $\h D$ itself will be discussed more thoroughly in the following. Let us just note that it should go to $\pm \infty$ on the real axis to provide a damping factor in the path integral, and that it should be such that it does not introduce any tachyonic instabilities for the chiral multiplets to which it couples. 

Finally, let us remark that we have implicitly considered a limit where $\e^2\rightarrow 0$ before taking $\g^2\rightarrow 0$, so that we can discuss the vector multiplet saddles and worry about the matter contribution later. In fact, this procedure turns out to be incorrect \cite{Benini:2013nda, Benini:2013xpa}. At some special values of $\h\sigma$, extra massless modes from the matter sector can appear in the $\g^2\rightarrow 0$ limit, leading to singularities which invalidate the localization argument. The proper treatment of these singularities shall be discussed at length in section \ref{section: derivation}.

\subsection{Gaugino zero-modes}\label{ss:gaugino}
We denote $\fM \cong \tfM / W$ the Coulomb branch, and $\tfM = \bC^\rk$ its cover, spanned by $\h\sigma_a, \t{\h\sigma}_a$ in \eqref{sigma saddle}, for any $k$.  Here, $W$ is the Weyl group of $\mathbf{G}$.
It is clear from  \eqref{YM standard} that the scalar gauginos $\lambda_a, \t\lambda_a$ have constant zero modes on $\fM$.
 (The  gauginos $\Lambda_1, \t\Lambda_{\b 1}$, on the other hand, have no zero modes  on the sphere.)  The vector multiplet path integral then reduces to an integral over the constant modes $\h\sigma_a, \t{\h\sigma}_a, \lambda_a, \t\lambda_a$  in each flux sector. However, the integral over $\fM$ has singularities at points where some of the chiral multiplet scalars become massless. To regulate them, it is convenient to keep  some constant modes $\h D_a$ of the auxiliary fields $D_a$  as an intermediate step \cite{Benini:2013nda, Benini:2013xpa}. The field $\h D$ was defined in \eqref{def hD}, and in the rest of this paper $\h D$ will simply denote its constant mode.
It is clear from \eqref{Phi kinetic term}  that this gives an extra mass squared $\rho_i(\h D)$ to  the chiral multiplet scalars.

The consideration of $\h D$ is  especially useful because  the constant modes
\be\label{zero mode multiplet}
\CV_a^{(0)} = \left(\h\sigma_a~,\, \t{\h\sigma}_a~,\,  \lambda_a~,\,  \t\lambda_a~,\, \h D_a\right)~
\ee
transform non-trivially under a residual supersymmetry
\bea\label{residue susy}
&\delta\h \sigma_a=0~, \qquad
\delta \t{\h\sigma}_a = -2 \t\lambda_a~, \qquad \delta\lambda_a =- \h D_a~, \qquad \delta\t\lambda_a=0~, \qquad \delta \h D_a= 0~, \cr
&\t\delta\h \sigma_a=0~, \qquad \t\delta \t{\h\sigma}_a = -2 \lambda_a~, \qquad \t\delta\lambda_a =0~, \qquad\quad \t\delta\t\lambda_a= \h D_a~, \qquad \t\delta \h D_a= 0~,
\eea
which follows from \eqref{susyVector twisted i}-\eqref{susyVector twisted ii}.
The full path integral can be written as
\be\label{path int inter i}
{1 \ov |W|}\int_{\tfM} \sum_k \prod_a d^2 \h\sigma_a  \int_\Gamma \prod_a d\h D_a \int \prod_a (d\lambda_a d\t\lambda_a)\; \CZ_k(\h\sigma, \t{\h\sigma}, \lambda, \t\lambda, \h D)~,
\ee
where $\CZ_k$ is the result of the path integration over all  modes other than the zero-mode multiplets \eqref{zero mode multiplet}, including the matter fields in chiral multiplets.
Note that the equation \eq{path int inter i} can be obtained by first taking the integral over $\tfM$ and identifying the integration variables $\hs_a$, $\t \hs_a$, $\lambda_a$, $\t \lambda_a$ and $\hD_a$ up to actions of the Weyl group. The integrand of \eq{path int inter i}, however, is invariant under the Weyl group, and hence the integral $\int_\fM$ may be replaced by ${1 \ov |W|} \int_\tfM$.
The supersymmetry \eqref{residue susy} implies that
\be\label{susy eq 0}
 \delta \CZ_k = \left(- 2 \t\lambda_a {\d\over \d  \t{\h\sigma}_a} - \h D_a  {\d\over \d  \lambda_a} \right) \CZ_k=0~,
\ee
and similarly for $\t\delta \CZ_k=0$.
In section \ref{ss: general}, we show that the path integral \eqref{path int inter i} can be written as
\be\label{path int inter ii}
{1 \ov |W|}\sum_k \int_{\tfM}\prod_a d^2 \h\sigma_a  \int_\Gamma \prod_a d\h D_a \; \det_{b c}(h^{bc})\; \CZ_k(\h\sigma, \t{\h\sigma},0, 0 , \h D)
\ee
upon integrating out the gaugino zero modes, thanks to supersymmetry. Here, $h^{ab}$ is a symmetric tensor that satisfies
\be\label{rel on MC i}
 \b\d \CZ_k(\h\sigma, \t{\h\sigma},0, 0 , \h D) = \half \h D_a\,  h^{ab} \, \CZ_k(\h\sigma, \t{\h\sigma},0, 0 , \h D)\, d\t{\h\sigma}_b
\ee
and the relation
\be\label{rel on MC ii}
\b \d^a h ^{bc} - \b \d^b h ^{ac}=0~. 
\ee
As we explain in  detail in section \ref{section: derivation}, closely following  similar computations  in \cite{Benini:2013nda, Benini:2013xpa, Hori:2014tda}, the integration cycle $\Gamma$ for the $\h D$ integration is determined by requiring that the modes of the chiral multiplet scalars not be tachyonic.

Note that, on  the localization locus, $\h D=0$ and the integrand $\cZ_k$ in \eqref{path int inter ii} is holomorphic in $\hs$ due to \eqref{susy eq 0}:
\be\label{cZ holo}
 \CZ_k(\h\sigma, \t{\h\sigma},0, 0 , 0) =  \CZ_k(\h\sigma)~.
\ee


\subsection{Classical action}
On the supersymmetric locus $\CM_{\rm susy}$, all $\delta$- or $\t\delta$-exact actions vanish. The only classical contribution to \eqref{path int inter ii} thus  comes from the twisted superpotential $\h W(\sigma)$. Due to \eqref{S of Wh and susy}, we have
\be\label{S classical gen}
e^{-S_{\rm cl}(k)} =e^{{2\pi i \over \epsdef}\left(\h W\left(\h \sigma + {k\over 2}\epsdef\right) -\h W\left(\h \sigma - {k\over 2}\epsdef\right)\right)}~,
\ee
for a generic superpotential $\h W$. In our case, we have \eqref{Wh linear bis} and thus
\be\label{S classical}
e^{-S_{\rm cl}(k)} = \prod_{I=1}^n q_I^{\tr_I(k)} \equiv q^k ~,
\ee
with $q_I$ defined in \eqref{def qI}. Note that \eqref{S classical} is independent of $\epsdef$, and topological.

\subsection{One-loop determinants}\label{subsec: oneloop det}
Here we introduce the fluctuation determinants for all the massive fields around the localization locus, which enter into \eqref{cZ holo}.
Their computation is relegated to appendix \ref{App: det}.
Let us define the function 
\be
{\mathbf Z}^{({\mathbf r})}(x; \epsdef) = \epsdef^{{\mathbf r}-1} \, {\Gamma({x\over \epsdef} +{{\mathbf r}\over 2})\over \Gamma({x\over \epsdef} +{2- {\mathbf r}\over 2})} = ~\epsdef^{{\mathbf r}-1} \,\left( {x\over \epsdef} +{2- {\mathbf r}\over 2} \right)_{{\mathbf r}-1} ~,
\ee
where  $x\in \C$, ${\mathbf r}\in \Z$, and $(y)_n$ denotes the extension of the Pochhammer symbol to $n\in\bZ$. Since ${\mathbf r}$ is integer, the function can be written as a finite product: 
\be\label{Zr explicit}
{\mathbf Z}^{({\mathbf r})}(x; \epsdef) = \left\{ 
\begin{array}{ll}   \prod_{m=-{{\mathbf r}\over 2}+1}^{{{\mathbf r}\over 2}-1} (x+ \epsdef m)  &\qquad {\rm if}\quad {\mathbf r}>1~, \\
 1 &\qquad {\rm if}\quad  {\mathbf r}=1~, \\
   \prod_{m= -{|{\mathbf r}|\over 2}}^{{|{\mathbf r}|\over 2}} (x + \epsdef m)^{-1}&\qquad {\rm if}\quad {\mathbf r}<1~.
\end{array} \right.
\ee
On a supersymmetric saddle with flux $k$, one finds that the one-loop determinant from a chiral multiplet $\Phi$ of $R$-charge $r$, gauge charges $Q^a$  under $\GH$, and twisted mass $m^F$ is
\be\label{Zphi oneloop}
Z^\Phi_k ={\mathbf Z}^{(r-Q(k))}\left(Q(\h\sigma) + m^F; \epsdef\right)~,
\ee
where $\hs=\half (\s_N+\s_S)$.
Note that $Z_k^\Phi$ has $Q(k)-r +1$ simples poles if $Q(k)-r\geq 0$, and no poles otherwise. The poles corresponds to zero-modes of the bottom component $\CA$ of the chiral multiplet, which are holomorphic sections of $\CO(Q(k)-r)$. 
We will discuss those modes in more detail in section \ref{section: Higgs branch loc}. Similarly, the zeros of \eqref{Zphi oneloop} for $Q(k)-r < -1$ correspond to zero-modes of the  fermionic   field $\CC$ in the chiral multiplet.

There is a sign ambiguity in the determination of \eqref{Zphi oneloop}. Following a similar discussion in \cite{Morrison:1994fr}, we propose to assign $+1$ to chiral multiplets of $R$-charge $0$ and $-1$ to chiral multiplets of $R$-charge $2$. This introduces the overall factor $(-1)^{N_*}$ in \eqref{main formula}, where $N_*$ is the number of field components of $R$-charge $2$ in the GLSM. This prescription is consistent with all the examples worked out in sections \ref{sec: Examples 1} and \ref{sec:_Amodel_examples}. It would be interesting to derive it, and to generalize it to any integer $R$-charge.

The total contribution from the matter sector is
\be\label{Zk matter}
Z_k^{\rm matter}(\h \sigma; \epsdef) = \prod_i \prod_{\rho_i \in\FR_i}  {\mathbf Z}^{(r_i-\rho_i(k))}\left(\rho_i(\h\sigma) + m^F_i; \epsdef\right)~.
\ee
So far, we have assumed for simplicity that the background gauge field for the flavor symmetry $\GF$ has no flux, see \eqref{simplified_background_flavor}. This assumption can be relaxed in the obvious way, treating the background vector multiplet on the same footing as a dynamical vector multiplet: the only effect is to shift $r_i-\rho_i(k)\to r_i-\rho_i(k)-\rho^F_i(k^F)$ in \eqref{Zk matter} and to change the Dirac quantization condition for the vector-like $R$-charge to $r_i-\rho^F_i(k^F)\in\bZ$. 
One can then gauge a subgroup of the flavor symmetry $\GF$ by summing over the associated fluxes $k^F$ and integrating over the associated complex scalars $m^F$.

As argued in appendix \ref{App: det},  a $W$-boson multiplet $\CV^{(\alpha)}$ for the simple root $\alpha$  has the same one-loop determinant as a chiral multiplet of $R$-charge $r=2$ and $\GH$-charges $\alpha^a$. The massive fluctuations along $\GH$ itself are completely paired between bosons and fermions, leading to a trivial contribution. Thus, the total contribution for the vector multiplet reads: 
 \bea\label{Zk vector}
&Z_k^{\rm vector}(\h\sigma; \epsdef)  &=&\; \prod_{\alpha\in \Fg\setminus \Fh} {\mathbf Z}^{(2-\alpha(k))}(\alpha(\hs); \epsdef) \cr
&&=& \; (-1)^{\sum_{\alpha>0}(\alpha(k) +1)}\;  \Delta\left(\hs + {k\over 2}\epsdef\right) \Delta\left(\hs - {k\over 2}\epsdef\right)~,
\eea
where $\alpha>0$ denotes the positive roots, and  we introduced the  function
\be
\Delta(x)\equiv \prod_{\alpha>0} \alpha(x)~,
\ee
which is  the Vandermonde determinant of $\GG$. Note that there is no sign ambiguity in this case, since the roots come in pairs $\alpha, -\alpha$.

The one-loop determinant of all the fields around the localization locus with flux $k$ is then given by
\be
\cZ^\oneloop_k (\hs;\epsdef) = Z^\text{vector}_k(\hs;\epsdef) Z_k^\text{matter}(\hs;\epsdef) \,.
\ee
As explained in subsection \ref{subsec: summary}, it is convenient to collect all the labels of the components $(i,\rho_i)$ of the chiral multiplets $\Phi_i$ and $\alpha$ of the $W$-boson multiplets $\cV^{(\alpha)}$ into a collective label $\cI$. Taking $Q_\cI$ to be the charge of the component, $m^F_\cI$ to be its twisted mass and $r_\cI$ to be its $R$-charge, \emph{i.e.}
\be
(Q_\cI, m^F_\cI, r_\cI) =
\begin{cases}
(\rho_i, m^F_i, r_i ) &\text{when}~~\cI=(i,\rho_i) \\
(\alpha, 0, 2 ) &\text{when}~~\cI=\alpha
\end{cases} \,,
\ee
we can simply write:
\be
\cZ^\oneloop_k (\hs;\epsdef) = \prod_\cI Z_k^\cI (\hs;\epsdef)
= \prod_\cI  \mathbf{Z}^{(r_\cI - Q_\cI (k))} (Q_\cI(\hs)+m^F_\cI;\epsdef) \,.
\ee

The analytic structure of $\cZ^\oneloop_k$ plays an important role in the computation of correlators. The singular loci of $\cZ^\oneloop_k$ can be specified by the oriented hyperplanes \eqref{hyperplanes}.
All singularities of $\cZ^\oneloop_k$ lie at hyperplanes and at their intersections. In particular, any codimension-$p$ singularity lies at the intersection of $p$ or more hyperplanes. When the charges $Q_\cI$ of the hyperplanes intersecting at the singularity lie within a half-space of $i\mathfrak{h}^*$, the singularity is said to be projective. If a codimension-$p$ singularity comes from $p$ linearly-independent hyperplanes intersecting, the singularity is said to be non-degenerate.

\subsection{The  Jeffrey-Kirwan residue}\label{subsec:Coulomb_branch_formula}
From the previous discussion, upon integrating out all the constant modes of the theory save for  $\hs$, we expect  the correlator $\left\langle  \CO^{(N)}(\s_N) \CO^{(S)}(\s_S)\right\rangle$ to be of the form:
\be
\sum_k q^k \oint \left( \prod_a d \hs_a \right) \cZ_k^\oneloop(\hs;\epsdef)\CO^{(N)}\left(\h\sigma-\epsdef {k\over 2}\right)  \CO^{(S)}\left(\h\sigma+\epsdef {k\over 2}\right) \,.
\ee
The contour of this integral is, for now,  an unspecified $\rk$-(real)-dimensional cycle within $\tfM$. Note that the operator insertions at the north and south poles depend on $\h\sigma_a \mp {k_a\over 2}\epsdef$, since they are evaluated on the saddle \eqref{sigma saddle}. From the analytic structure of $\cZ_k^\oneloop$, it is thus natural to expect the correlators to be a sum of residues of codimension-$\rk$ poles of the integrand,
\be\label{formula integrand}
\mathbf{I}_k\left(\CO^{(N)}\CO^{(S)}\right)
=\cZ_k^\oneloop(\hs;\epsdef)
\CO^{(N)}\left(\h\sigma-\epsdef {k\over 2}\right)
\CO^{(S)}\left(\h\sigma+\epsdef {k\over 2}\right)
 d\h\sigma_1\wedge \cdots \wedge d\h\sigma_\rk~.
\ee
 We prove in section \ref{section: derivation} that the final answer for the correlation function is in fact given by
\be\label{formula roughly}
\left\langle  \CO^{(N)}(\sigma_N) \CO^{(S)}(\sigma_S)\right\rangle ={(-1)^{N_*} \over |W|}\sum_{k \in \Gamma_{\mathbf{G}^\vee}} \, q^k \,  {\rm JK{\text-}Res}\left[\xi_{\rm eff}^{\rm UV}\right] \mathbf{I}_k\left(\CO^{(N)}\CO^{(S)}\right)~\,.
\ee
The symbol  ${\rm JK{\text-}Res}\left[\xi_{\rm eff}^{\rm UV}\right] \mathbf{I}_k$ in  \eqref{formula roughly}  is a short-hand for a sum over the Jeffrey-Kirwan residues at all the codimension-$\rk$ ``poles'' of $\mathbf{I}_k$:
\be\label{JK residues expanded}
{\rm JK{\text-}Res}\left[ \xi_{\rm eff}^{\rm UV}\right] \mathbf{I}_k= \sum_{\hs_* \in \tfM^\text{sing}_k}{\rm JK{\text-}Res_{\,\h\sigma=\h\sigma_*} }\left[{\bf Q}(\hs_*), \xi_{\rm eff}^{\rm UV}\right] \mathbf{I}_k \left(\CO^{(N)}\CO^{(S)}\right)~.
\ee
Recall that $\tfM^\text{sing}_k$ denotes the location of all the codimension-$\rk$ poles of $\cZ^\oneloop_k$, and $\mathbf{Q}(\hs_*)$ denotes the set of all charges $Q_\cI$ of hyperplanes $H^\cI_n$ crossing through $\hs_*$. We shall explain the JK residue prescription in the following subsection. 

The sum over fluxes $k$ in \eqref{formula roughly} is weighted by the classical factor \eqref{S classical}. 
In examples, it is useful to formally generalize it to
\be\label{class contrib}
e^{-S_{\rm cl}(k)} = \prod_{a=1}^\rk q_a^{k_a} = \exp(2\pi i \tau(k))~,
\ee
where the complexified FI parameter $\tau$ is now taken to be a generic element of $\mathfrak{h}^*_\bC$. In order for this term to be gauge invariant, $\tau$ must be restricted to the subspace $\mathfrak{c}^*_\bC \subset \mathfrak{h}^*_\bC$ spanned by the generators of the center of $\mathbf{G}$. For instance, for $\GG= U(N)$ we should take $q_a= q$, $a=1, \cdots, N$. Nevertheless,  it is often convenient and sometimes necessary to keep $\tau$ to be general until the end of a computation. We follow \cite{Halverson:2013eua} and refer to the auxiliary theory with generic $\tau \in \mathfrak{h}^*_\bC$ as the \emph{Cartan theory} associated to the non-abelian GLSM. 
In cases in  which some of the FI parameters run, one should replace $q_a$ in \eqref{S classical} by  the RG invariant scale \eqref{def LambdaI},
\be
q_a = \Lambda^{b_0^a}~, \qquad {\rm if} \quad b_0^a \neq 0~.
\ee
Note that we have implicitly set the RG scale $\mu$ to $1$ throughout.

The formula \eqref{formula roughly}-\eqref{formula integrand} is the more precise form of the  Coulomb branch formula  \eqref{formula intro 1}-\eqref{formula intro 2} promised in the introduction. Incidentally, one can easily see that the anomalous $U(1)_A$ transformation of the path integral in \eqref{grav anomaly} and the $R_A$-gauge anomaly are explicitly realized by \eqref{formula roughly}-\eqref{formula integrand}, using the 1-loop determinants \eqref{one_loop_collective}.

\subsection{GLSM chambers and JK residues}\label{subsec: JK}
Any set of $\rk$ distinct charges $(Q_{\cI_1}, \cdots, Q_{\cI_\rk})$ defines a cone in $i\Fh^\ast$, denoted by
\be\label{def C(Q)}
\text{Cone}(Q_{\cI_1}, \cdots, Q_{\cI_\rk})~.
\ee
The union of all such cones spans a subspace $\CK \subseteq i\Fh^\ast$. $\CK$ can be subdivided in minimal cones ({\it chambers}) of maximal dimension $\rk$, that  meet on codimension-one {\it walls}. The {\it phases} \cite{Witten:1993yc, Morrison:1994fr} of the associated Cartan theory are determined by the chamber which the FI parameter $(\xi^a)$ belongs to.
The phase diagram of the Cartan theory refines the phase diagram of the non-abelian GLSM \cite{Hori:2006dk}, that is obtained by the projection $(\xi^a)\mapsto(\xi^I)$ from $i\mathfrak{h}^*$ 
to the physical FI parameter space $i\mathfrak{c}^*$.

The effective FI  parameters on the Coulomb branch  were defined in \eqref{tau eff}. 
We denote by $\xi_{\rm eff}^{\rm UV}$ the effective FI parameters on the Coulomb branch {\it at infinity} on $\fM$, that is, the effective coupling in the limit $|\sigma| \rightarrow \infty$:
\be
\xi_\eff^\text{UV} = \t\xi + {1 \ov 2 \pi} b_0 \lim_{R \ra \infty} \log R \,,
\ee
with $b_0 \in i\mathfrak{c}^* \subset i\mathfrak{h}^*$ as in \eqref{def bo in hstar}.
The chamber that $\xi_{\rm eff}^{\rm UV}$ lies in determines the phase of the Cartan theory. More rigorously, $\xi_\eff^\text{UV}$ is defined to be in a chamber $\mathfrak{C}$ of $\CK$, or for that matter any $\rk$-dimensional cone $\mathfrak{C} \subset i\mathfrak{h}^*$, if
\be\label{xi rigorous}
\exists ~R_0 >1 \quad \text{such that for all} \quad
R \geq R_0, \qquad
\t\xi + {1 \ov 2 \pi} b_0 \log R \in \mathfrak{C} \,.
\ee
When $b_0 \neq 0$, the finite piece $\t\xi$ is irrelevant unless $b_0$ lies at the boundary of multiple chambers. That is, when $b_0$ lies safely within a chamber $\mathfrak{C}_0$ of $\CK$, the Cartan theory has only one phase $\mathfrak{C}_0$. When $b_0$ lies at the boundary of multiple chambers $\mathfrak{C}_0 , \cdots, \mathfrak{C}_p$, the Cartan theory can be taken to be in any one of these chambers by turning on a finite $\t\xi$. When $b_0 =0$, i.e., when the theory flows to a conformal fixed point in the infrared, $\xi_\eff^\text{UV}  = \t\xi$ can span the entire $i\mathfrak{h}^*$-space and can be in any chamber of $\CK$. As we discussed in subsection \ref{subsec: summary}, we always choose to align $\t\xi$ with the physical FI parameter $\xi$.

Let us now define the JK residue at a singular point, or codimension-$\rk$ singularity, $\h\sigma_*$ of $\cZ^\oneloop_k (\hs;\epsdef)$. Recall that all such singularities come from intersections of $\rk$ or more hyperplanes $H_{\cI_1}^{n_1}, \cdots, H_{\cI_s}^{n_s}$.~%
\footnote{Recall that the hyperplane $H_\cI^n$ is a singular locus of $\cZ^\oneloop_k$ only when $\cI$ is an index for a component of the chiral field, i.e., $\cI = (i,\rho_i)$. The hyperplanes $H_\alpha^n$ with vector labels can be thought of as codimension-one poles of $\cZ_k^\oneloop(\hs;\epsdef)$ with residue zero.}
Let us assume that $\h\sigma_*=0$, so that the singular hyperplanes are given by the equations
\be\label{ns hyper}
Q_{\cI_1}(\hs)=0~, \quad \cdots, \quad Q_{\cI_{s}}(\hs)=0~.
\ee
This can be achieved by shifting $\hs$ appropriately. We have defined
\be\label{bfQs}
{\bf Q} (\hs_*)= \{ Q_{\cI_1}, \cdots,  Q_{\cI_{s}} \}~, \qquad\quad s\geq \rk~,
\ee
the set of $\GH$-charges defining the hyperplanes \eqref{ns hyper}. We further assume  that this arrangement of hyperplanes is {\it projective}---that is, the vectors ${\bf Q} (\hs_*)$ are contained in a half-space of $i\Fh^\ast$. This important technical assumption is satisfied in ``most'' cases of interest. We comment on the non-projective case below.

Let us denote by $R_{{\bf Q} (\hs_*)}$ the ring of rational holomorphic $\rk$-forms, with poles  on the hyperplane arrangement \eqref{ns hyper}.
Let us also define $S_{{\bf Q} (\hs_*)} \subset R_{{\bf Q} (\hs_*)}$  the linear span of 
\be
\omega_S=  \prod_{Q_j \in Q_S}{1\over Q_j(\sigma)} \; d\h\sigma_1 \wedge \cdots\wedge d\h\sigma_\rk~, 
\ee  
where $Q_S$ denotes any subset of $\rk$ distinct charges in ${\bf Q}(\hs_*)$. (There are thus ${s \choose \rk}$ distinct $Q_S$.)
There also exists a natural projection \cite{1999math......3178B}
\be\label{proj R to S}
\pi : R_{{\bf Q} (\hs_*)}\rightarrow S_{{\bf Q} (\hs_*)}~,
\ee
whose exact definition we shall not need. 
 The JK residue on $S_{{\bf Q} (\hs_*)}$ is defined by
\be\label{def JK}
{\rm JK{\text-}Res_{\,\h\sigma=0} }\left[{\bf Q} (\hs_*), \eta \right]\, \omega_S =  
 \left\{ 
\begin{array}{ll} {1\over |\det(Q_S)| }  &\qquad {\rm if}\quad \eta\in \text{Cone}(Q_S)~, \\
 0 &\qquad {\rm if}\quad \eta\notin \text{Cone}(Q_S)~,
\end{array} \right.
\ee
in terms of a vector $\eta \in \Fh^\ast$.  More generally,  the JK residue of any holomorphic $\rk$-form  in $R_{{\bf Q} (\hs_*)}$ is defined as the composition of \eqref{proj R to S} with \eqref{def JK}. This definition, along with equation \eq{xi rigorous} is enough to compute the correlation functions with \eq{formula roughly}.

In general, there are  ${s \choose \rk}$ homologically distinct $\rk$-cycles that one can define on the complement of the hyperplane arrangement \eqref{ns hyper}, and the definition \eqref{def JK} determines a choice of cycle for any $\eta \in \Fh^\ast$ (or rather, for any chamber).  It has been proven that   \eqref{def JK} always leads to a consistent choice of contour \cite{1999math......3178B, 2004InMat.158..453S}.

Finally, let us comment on the case of a {\it non-projective} hyperplane arrangement. In that case, one can often turn on twisted masses to split the non-projective singular point into a sum of projective singularities.  The occurrence of a non-projective singularity typically signals the presence of a non-normalizable vacuum, for instance in the case of GLSMs for non-compact geometries. To be more precise, the existence of a non-projective singularity implies that there is a point in $\tfM$ where $s > \rk$ hyperplanes $H^{\cI_1}_{n_1},\cdots,H^{\cI_s}_{n_s}$ collide, where
\be
c_1 Q_{\cI_1} + \cdots + c_s Q_{\cI_p}
= c_1 \rho_{i_1} + \cdots + c_s \rho_{i_s} = 0\,,
\ee
for some positive integers $c_p$. We have made the components of the involved chiral fields explicit. In order for the hyperplanes to have poles at this point, it must be the case that $Q_{\cI_p} (k) \geq r_{\cI_p}$ for all $r_{\cI_p}$. This implies the existence of the gauge invariant chiral operator
\be
\cO \equiv \sum_{w \in W} w \left( \prod_{p=1}^s (\Phi_{i_p,\rho_{i_p}})^{c_p} \right) \,,
\ee
with $R$-charge
\be
r_\cO = \sum_p c_p r_{i_p} \leq \sum_p c_p \rho_{i_p} (k) = 0
\ee
which is allowed to take a vacuum expectation value by the $D$-term equations.

Given the choice of $\eta = \xi_\eff^\text{UV}$, only a subset of the flux sectors $k \in \Gamma_{\mathbf{G}^\vee}$ contribute to the computation of the correlation function. In particular, let us define the set of $\rk$-tuples of components
\be
RCS(\eta) \equiv
\setcond{S}{S=\{{\cI_1},\cdots,{\cI_\rk}\},~
\eta \in \text{Cone}(Q_{\cI_1},\cdots,Q_{\cI_\rk})}
\ee
(the set of ``relevant component sets") that contribute to the JK residue for the vector $\eta \in i\mathfrak{h}^*$. The $k$-flux sectors that contribute to the correlators for the choice of $\xi_\eff^\text{UV}$ are given by $\Gamma_\text{flux} ({\xi_\eff^\text{UV}}) \subset \Gamma_{\mathbf{G}^\vee}$ for
\be\label{chamber for k}
\Gamma_\text{flux} (\eta) \equiv
\Gamma_{\mathbf{G}^\vee} \cap \left( \bigcup_{S \in RCS(\eta)}
\setcond{k}{Q_{\cI}(k)-r_{\cI} \geq 0~~ \text{for all}~~\cI \in S} \right) \,.
\ee
Note that $\Gamma_\text{flux} (\eta)$ only depends on the chamber of $\CK$ in which $\eta$ lies, and not on $\eta$ itself. This is true also of the JK residue.

\subsection{$A$-model correlation functions}
The $\epsdef =0$ limit of the Coulomb branch formula \eqref{formula roughly} computes the  $A$-model correlation function \eqref{Amode corr}. We find:
\be\label{formula Amodel i}
\left\langle  \CO(\sigma) \right\rangle_0 ={(-1)^{N_*} \over |W|}\sum_{k}\, q^k \,  {\rm JK{\text-}Res}\left[\xi_{\rm eff}^{\rm UV}\right]   \cZ_k^\oneloop(\h\sigma;0)\, \CO\left(\h\sigma\right)\,   d\h\sigma_1\wedge \cdots \wedge d\h\sigma_\rk  ~,
\ee
with
\be\label{formula Amodel integrand}
\cZ_k^\oneloop(\h\sigma;0)=  (-1)^{\sum_{\alpha>0}(\alpha(k) +1)} \prod_{\alpha>0} \alpha(\h\sigma)^2 \, \prod_i \prod_{\rho_i \in\FR_i} \left(\rho_i(\h\sigma)+m_i^F\right)^{r_i - 1 - \rho_i(k)}~.
\ee
This Coulomb branch formula can be favorably compared to many known results for the $A$-twisted GLSM. In particular, in the abelian case $\GG=U(1)^n$, the exact correlation functions were first obtained by Morrison and Plesser using toric geometry \cite{Morrison:1994fr}. The JK residue formula \eqref{formula Amodel i} for such theories was first obtained in \cite{2004InMat.158..453S} from a more mathematical perspective. Both references focussed on the case of a complete intersection $X$ in a compact toric variety $V$, which is realized by an abelian GLSM with chiral multiplets of $R$-charge $0$ (corresponding to $V$) and some chiral multiplets of $R$-charge $2$ (corresponding to the superpotential terms which restrict $V$ to $X$). We comment on such models using the Coulomb branch formula in section \ref{section: Q Cohomology}.

One can also write \eqref{formula Amodel i} as
\be\label{formula Amodel ii}
\left\langle  \CO(\sigma) \right\rangle_0 ={(-1)^{N_*}\over |W|}\sum_{k}\,  {\rm JK{\text-}Res}\left[\xi_{\rm eff}^{\rm UV}\right]     e^{2\pi i \tau_{\rm eff}(k)}  \cZ_0^\oneloop(\h\sigma;0)\, \CO\left(\h\sigma\right)\,   d\h\sigma_1\wedge \cdots \wedge d\h\sigma_\rk  ~,
\ee
where the classical and one-loop contributions have been recombined into the Coulomb branch effective couplings $\tau_{\rm eff}^a(\h\sigma)$ defined in \eqref{tau eff}, with $\tau_{\rm eff}(k)= \sum_a \tau_{\rm eff}^a(\h\sigma) k_a$, and 
\be
\cZ_0^\oneloop(\h\sigma;0) =  (-1)^{\half {\rm dim}(\Fg/\Fh)} \prod_{\alpha>0} \alpha(\h\sigma)^2 \, \prod_i \prod_{\rho_i \in\FR_i} \left(\rho_i(\h\sigma)+m_i^F\right)^{r_i - 1}~.
\ee

Let us assume that $\xi_\eff^\text{UV}$ lies within a chamber $\mathfrak{C}$ of $\CK$ such that $\Gamma_\text{flux}(\xi_\eff^\text{UV})$, as defined in equation \eq{chamber for k}, is entirely contained within a discrete cone $\Lambda \subset \Gamma_{\mathbf{G}^\vee}$ that satisfies the following properties:
\begin{itemize}
\item $\Lambda$ is given by
\be
\Lambda = \setcond{k}{k=\sum_A {n_A \kappa^A } + r^{(0)}, \quad
n_A \in \bZ_{\geq 0}}
\ee
for some $r^{(0)} \in \Gamma_{\mathbf{G}^\vee}$, where $\kappa^1,\cdots,\kappa^\rk \in i\mathfrak{h}$ is a basis of $\Gamma_{\mathbf{G}^\vee}$.
\item $\Lambda \cap \Gamma_\text{flux}(\eta) = \emptyset$ for $\eta \notin \mathfrak{C}$.
\end{itemize}
The second assumption implies that for $k \in \Lambda$, all the poles of $\cZ_k^\oneloop(\h\sigma;0)$ are counted in the JK residue and thus
\bea
&{\rm JK{\text-}Res}\left[\xi_{\rm eff}^{\rm UV}\right]
\cZ_k^\oneloop(\h\sigma;0)\, \CO\left(\h\sigma\right)
d\sigma_1\wedge \cdots \wedge d\h\sigma_\rk \\
&\qquad\qquad
= \oint_{\p \tfM}
\cZ_k^\oneloop(\h\sigma;0)\, \CO\left(\h\sigma\right)
d\sigma_1\wedge \cdots \wedge d\h\sigma_\rk \,.
\eea
where $\p \tfM$ is the $\rk$-torus at infinity. In particular, the contour is the same for all $k \in \Lambda$, which lets us sum over all $k \in \Lambda$ to arrive at
\be\label{formula Amodel iii}
\left\langle  \CO(\sigma) \right\rangle_0 =
{(-1)^{N_*} \over |W|}\oint_{\p \tfM}
\left( \prod_{a=1}^\rk {d\h\sigma_a \over 2\pi i} \right)
{e^{2 \pi i r^{(0)}_a  \d_{\h\sigma_a} \h W_{\rm eff}}
\over \prod_{A=1}^\rk ( 1 - e^{2\pi i \kappa^A_a \d_{\h\sigma_a} \h W_{\rm eff}})}
\cZ_0^{\oneloop}(\h\sigma;0)\,  \CO\left(\h\sigma\right)~,
\ee
with $\h W_{\rm eff}$ as defined in section \ref{subsec: Coulomb branch}. The sum over the repeated indices $a$ in the formula is understood.  This  instanton-resummed expression makes it obvious that the $A$-twisted correlations functions are singular on the ``singular locus'' defined as the set of values of $q_a$ for which some of the vacuum equations
\be\label{vacuum equ ii}
 e^{2\pi i \kappa^A_a \d_{\h\sigma_a}\h W_{\rm eff}(\hs; q)}=1
\ee
are trivially satisfied (for any $\hs$). (Note that \eqref{vacuum equ ii} is the proper form of \eqref{vacuum equ} in general, with $\kappa^A_a=\delta^A_a$  for unitary gauge groups.)

When the theory is fully massive, i.e., when the theory has a finite number of distinct ground states and a mass gap, the critical points of $\h W_{\rm eff}$ become {\it isolated}, and the integral simply picks the iterated poles at the critical points. Let us denote:
\be
P= \left \{ \h\sigma_P \,\big|\,  e^{2\pi i \kappa^A_a \d_{\h\sigma_a} \h W_{\rm eff}(\h\sigma_B)}=1
~~\text{for all}~A=1,\cdots,\rk~\right\}~.
\ee
The formula \eqref{formula Amodel iii}  becomes
\be\label{Coulomb sum}
\left\langle  \CO(\sigma) \right\rangle_0 ={(-1)^{N_*}\over |W|} {1\over (-2\pi i)^\rk} \sum_{\h\sigma_P \in P}  { \cZ_0^\oneloop (\h\sigma_P;0)\,  \CO\left(\h\sigma_P\right)\over H(\h\sigma_P)}~,
\ee 
with $H(\h\sigma) = \det_{AB}\left(\kappa^A_a \kappa^B_b \d_{\h\sigma_a} \d_{\h\sigma_b} \h W_{\rm eff}\right)$ the Hessian of $ \h W_{\rm eff}$.  The exact same formula
has been recently derived in \cite{Nekrasov:2014xaa}, using different methods. In the abelian case and in the special case when all the chiral fields have vanishing $R$-charge, the formula \eqref{Coulomb sum} was also found earlier in \cite{Melnikov:2005tk}.
It is also a  natural generalization of the formula of \cite{Vafa:1990mu} for $A$-twisted Landau-Ginzburg models of twisted chiral multiplets.~\footnote{More precisely, \cite{Vafa:1990mu} considered $B$-twisted LG models of chiral multiplets, which is the same thing by the  $\Z_2$ mirror automorphism of the $\CN=(2,2)$ supersymmetry algebra.}

\section{Derivation of the  Coulomb branch formula}\label{section: derivation}

We derive the formula for the partition function in the Coulomb branch in this section. The techniques used in the derivation are equivalent to those used in arriving at the elliptic genus \cite{Benini:2013nda,Benini:2013xpa} or the index of supersymmetric quantum mechanics \cite{Hori:2014tda}. Our situation has more similarity with the latter case, as the integral over the Coulomb branch parameter is taken over a non-compact space. Let us first summarize the overall picture of getting at the partition function by replicating the arguments of \cite{Benini:2013nda,Benini:2013xpa,Hori:2014tda}.

As explained before, the partition function, upon naive localization on the Coulomb branch, can be written as the holomorphic integral
\be
\ZS = \sum_{k} q^k \oint \left(\prod_a d \hs_a \right)
\cZ^\oneloop_k (\hs) \,,
\label{naive}
\ee
where the contour of integration is an unspecified real dimension-$r$ cycle in $\tfM$. In this section, and only in this section, we use
\be
r = \rk
\ee
for sake of brevity. We also make the $\epsdef$-dependence of the one-loop determinant implicit, i.e.,
\be
\cZ^\oneloop_k(\hs) \equiv
\cZ^\oneloop_k(\hs;\epsdef)
\ee
to shorten equations.
 
We follow \cite{Benini:2013nda,Benini:2013xpa,Hori:2014tda} to evaluate the partition function. Decomposing the action of the gauge theory into
\be\label{Coulomb Lagrangian for 5}
\mathscr{L} = {1 \ov \e^2} \mathscr{L}_{YM}+
{1 \ov \g^2} \mathscr{L}_{\t \Phi \Phi} +
\mathscr{L}_{\t {\h W}} +
\mathscr{L}_{\h W} \,,
\ee
the path integral of the theory can be formally written as
\be
\ZS=
{1 \ov |W|}\int_\tfM \prod_a (d \hs_a d \t \hs_a) F_{\e, \g} (\hs,\t \hs) \,.
\ee
While the twisted superpotential ${\h W}$ is given by that of equation \eqref{linear_twisted_W}, the $\delta$ and $\t \delta$-exact term $\mathscr{L}_{\t {\h W}}$ is set to be
\be
\mathscr{L}_{\t {\h W}} = -{\t\xi \ov \e^2} (D - 2i f_{1 \b 1} -\t \sigma \cH) \,,
\ee
as explained in the previous section. The $\e \ra 0$ limit of the partition function then can be obtained by the integral
\be
\ZS=
{1 \ov |W|} \sum_{k} q^k
\lim_{\substack{\epsilon \ra 0 \\ \e \ra 0}} \int_{\tfM \setminus \De}
\prod_a (d \hs_a d \t \hs_a) F^{k}_{\e, 0} (\hs,\t \hs)
\label{overall}
\ee
in a double scaling limit of $\epsilon$ and $\e$, which we soon explain. $F_{\e,0}^k$ is obtained via an integral over the gauginos $\lambda_a$, $\t \lambda_a$ and the auxiliary fields $\hD_a$ along the Coulomb branch localization locus. $\Delta_\eps$ is a union of tubular neighborhoods of the codimension-one poles of $\cZ_{k}(\hs)$, i.e., the hyperplanes $H^\cI_n$, of size $\eps$. The parameter $\eps$ is to be distinguished from the omega deformation parameter $\epsdef$. The tubular neighborhoods are carved out precisely around points of the ``Coulomb branch" where massless modes of the chiral fields may develop. The radius $\eps$ is taken to scale with respect to $\e$ so that $\eps < \e^{M_*+1}$ as $\e$ is taken to zero, where $M_*$ is the maximal number of modes that become massless at any of the poles.~\footnote{As pointed out in \cite{Benini:2013nda,Benini:2013xpa}, it must be assumed that at any point $\hs_*$, the charges of the field components whose modes become massless at $\hs_*$ lie within a half plane of $i\mathfrak{h}^*$, i.e., the configuration of hyperplanes intersecting at $\hs_*$ must be projective. Throughout this section, we assume this is always the case, and refer to this property as ``projectivity." Furthermore, a theory for which at every $\hs_*$ there are at most $r$ singular hyperplanes intersecting is said to be ``non-degenerate" \cite{Benini:2013xpa}. When there is a non-projective hyperplane configuration responsible for a pole of the integrand, it can be made projective by formally giving generic twisted masses to all the components to the charged fields involved---the path integral of the original theory can be obtained by evaluating the partition function (or operator expectation values) in this ``resolved" theory and taking the twisted masses to the initial values.} 

In this limit, the integral \eq{overall} becomes the contour integral \eq{naive} along a particular middle-dimensional cycle in $\tfM$, whose value is summarized by the formula:
\be
\ZS = {(-8\pi^2)^r \ov |W|} \sum_{k} q^k
\sum_{\hs_* \in \tfM^{k}_\text{sing}}
\underset{\hs = \hs_*}{\text{JK-Res}} \left[ \mathbf{Q}(\hs_*),  \xi_\eff^\text{UV} \right] \cZ^\oneloop_{k} (\hs) \,.
\ee
In this section, we omit the factor $d\hs_1 \wedge \cdots \wedge d \hs_r$ in this expression. The result for the physical correlation functions follows straightforwardly, once the contour of integration is understood for the partition function:
\bea\label{main equation repeated in 5}
&\vev{\cO^{(N)}(\s_N)\cO^{(S)}(\s_S)}
=\\
&
{(-1)^{N_*} \ov |W|} \sum_{k} q^k \!\!\!
\sum_{\hs_* \in \tfM^{k}_\text{sing}}\!\!\!
\underset{\hs = \hs_*} {\text{JK-Res}}
\left[ \mathbf{Q}(\hs_*),  \xi_\eff^\text{UV}  \right]  \!~
\cZ^\oneloop_k (\hs)
\cO^{(N)}\! \left( \hs -{\epsdef k \ov 2} \right)
\cO^{(S)}\! \left(\hs +{\epsdef k \ov 2} \right) 
 \,.
\eea
Here we have fixed the overall normalization constant $(-1)^{N_*}$ of the physical correlators. There are two sets of data going into the overall normalization. An overall proportionality constant $(-8\pi^2)^{-r}$ is introduced, coming from the normalization of the Coulomb branch coordinates. The coefficient has been fixed empirically to match geometric computations. The sign $(-1)^{N_*}$ originates from the sign ambiguity of the one-loop determinants of the chiral fields, as explained in the previous section. Our prescription for $N_*$ is explained in section \ref{subsec: oneloop det}.

To arrive at this result, we follow the tradition of \cite{Benini:2013nda,Benini:2013xpa,Hori:2014tda} and examine the rank-one path integral in detail to gain insight. We subsequently generalize to the theories with gauge groups with higher rank. The main outcome of this section is that the methods of  \cite{Benini:2013nda,Benini:2013xpa,Hori:2014tda} generalize with minimal modification to the problem at hand. We apply these methods to arrive at the final answer.

\subsection{$U(1)$ theories} \label{section:rank one}

Let us write the path integral of the rank-one theory in the form
\be
\ZS=
\int d \hs d \t\hs d \lambda d\t\lambda d \h D
\cZ_{\e,\g} (\hs, \t\hs, \lambda, \t\lambda, \h D) \,,
\label{integral}
\ee
where the integral is taken with respect to constant modes $\hs$, $\t\hs$, $\lambda$, $\t\lambda$ and the auxiliary field $\h D$ on the Coulomb branch localization locus. As explained before, these constant modes form a supersymmetric multiplet for a supersymmetric integration measure $\cZ_{\e,\g}$. While for $\e, \g \neq 0$ we do not know the exact value of $\cZ_{\e,\g}$, supersymmetry implies that for any supersymmetric function $\cZ$ of the constant-mode supermultiplet,
\be
\hat D \p_\lambda \p_{\t\lambda} \cZ|_{\lambda=\t \lambda =0} =
2 \p_{\t\hs} \cZ|_{\lambda=\t \lambda =0} \,.
\ee
Hence, upon integrating the gaugino zero modes, the integral \eq{integral} must take the form
\be
\ZS=
{2 \ov |W|} \int_{\idom} d \hs d \b\hs  \p_{\t \hs}
\int d \h D {1 \ov \h D} \cZ_{\e,\g} (\hs, \b\hs, \h D) \,,
\label{der}
\ee
where we have denoted
\be
\cZ_{\e,\g} (\hs, \b\hs, \h D) \equiv \cZ_{\e,\g} (\hs, \b\hs,0,0, \h D) \,.
\ee
Note that $\tfM = \bC$ in the rank-one case. We have imposed the reality condition \eq{reality condition} on the integration contour to take $\t \hs = \b \hs$.

Let us now examine what the contour for $\hD$ should be. To do so, let us begin with taking the limit%
~\footnote{We have set the dimensionful Yang-Mills coupling to $\sqrt{{\rm \vol}(S^2)}$ for simplicity. As can be seen in latter parts of this section, this choice is not important, as the classical contribution of the Yang-Mills Lagrangian to the action does not affect the final answer.}
\be
\lim_{\substack{\e \ra 0 \\ \g \ra 0}}
\cZ_{\e,\g} (\hs, \b\hs, \h D)
= \sum_{k \in \bZ} q^k
\lim_{\e \ra 0} \left( e^{-{\hD^2 \ov 2\e^2}+{4\pi i \t \xi \ov \e^2} \hat D }
\cZ_{k} (\hs,\b \hs, \hat D) \right) \,.
\label{limit}
\ee
$\cZ_k(\hs,\b \hs,\hD)$ is given by the product of one-loop determinants of the components of the charged fields of the theory in the flux-$k$ sector:
\be
\cZ_k(\hs,\b \hs,\hD) = \prod_\cI Z_k^\cI (\hs, \b \hs,\hD) \,.
\ee
The computation of these determinants is presented in appendix \ref{App: det}. It is useful to split the contribution from the component $\cI$ of the charged field into two pieces 
\bea
Z^{\cI}_{{k}} (\hs,\b \hs, \hD) =
Z^{(0)}_{r_\cI - Q_\cI k} (\Sigma,\b \Sigma,D) \cdot
Z^\pos_{r_\cI-Q_\cI k} (\Sigma,\b \Sigma,D)
|_{\Sigma=Q_\cI \hs + m^F_\cI,~D = Q_\cI \hD} \,,
\eea
where
\be
Z^{(0)}_\mathbf{r} (\Sigma,\b \Sigma,D) = 
\begin{cases}
\prod_{m=-\mathbf{r}/2+1}^{\mathbf{r}/2-1}
(\Sigma + \epsdef m) &\text{if} \quad \mathbf{r}>1 \,, \\
1 &\text{if} \quad \mathbf{r}=1 \,, \\
\prod_{m=-|\mathbf{r}|/2}^{|\mathbf{r}|/2} {\b \Sigma \ov
\b \Sigma (\Sigma + \epsdef m) + i D}
&\text{if} \quad \mathbf{r}<1 \,,
\end{cases}
\label{Z0}
\ee
and 
\be
Z^{\pos}_{\mathbf{r}}(\Sigma,\b \Sigma,D)
= \prod_{\substack{|m| \leq j \\[2pt] j > j_0 (\mathbf{r})}}
{\b \Sigma ( \Sigma + \epsdef m) + j(j+1)
- {\mathbf{r} \ov 2} ({\mathbf{r}\ov 2}-1)
\ov 
i D+\b \Sigma (\Sigma + \epsdef m) + j(j+1)
- {\mathbf{r} \ov 2} ({\mathbf{r}\ov 2}-1)} \,,
\label{Zpos}
\ee
where
\be
j_0 (\mathbf{r}) = {|\mathbf{r}-1 |\ov 2} - {1 \ov 2} \,.
\ee
We have split up the chiral fields into individual components with charge $Q_\cI$. When $\hD$ is set to zero in $\cZ_k (\hs,\b \hs, \hD)$, we recover the holomorphic one-loop integral $\cZ_k (\hs)$:
\be\label{cZk}
\cZ_k (\hs,\b \hs, \hD=0) = \cZ_k (\hs)
\ee
since $Z^\cI_k (\hs, \b \hs, 0) = Z^\cI_k (\hs)$.

\begin{figure}[!t]
\centering\includegraphics[width=12cm]{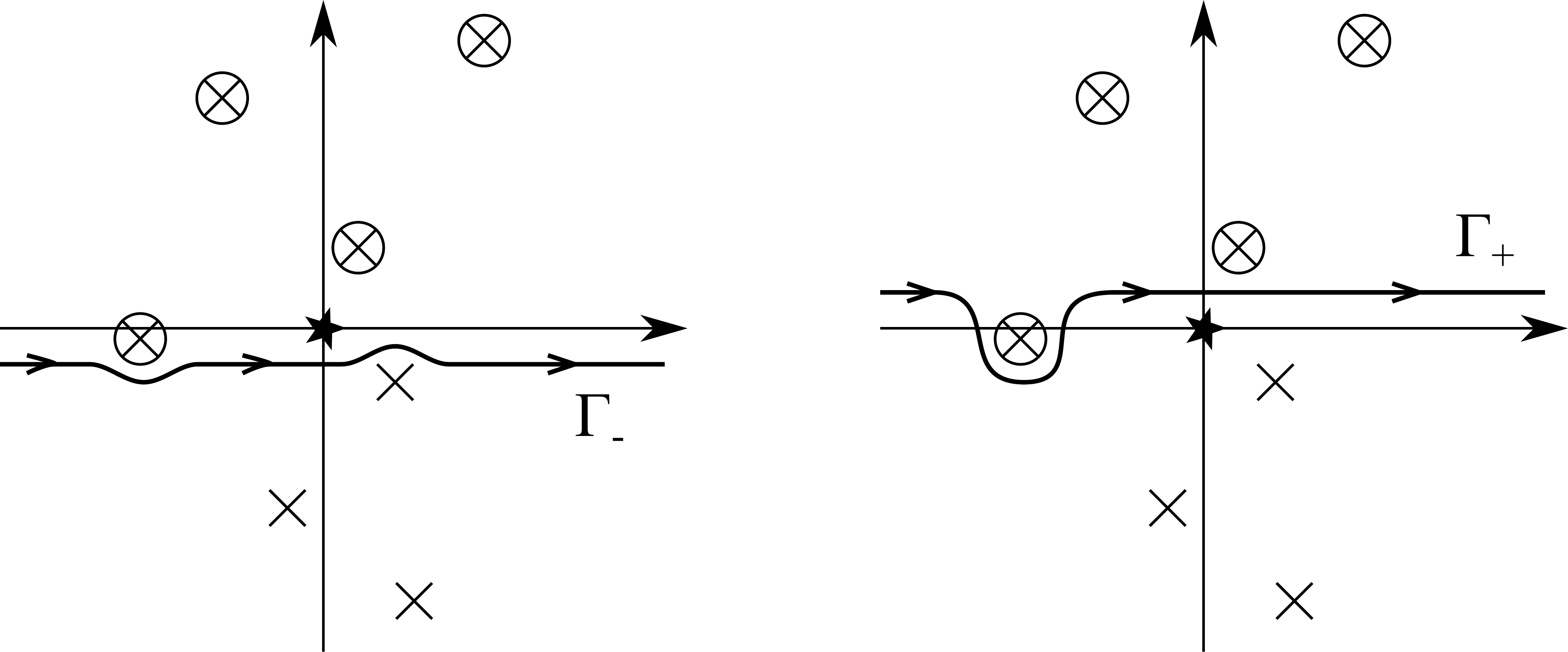}
\caption{\small Two different choices of contours $\Gamma_-$ and $\Gamma_+$ for the $\hD$-integral depicted on the $\hD$-plane. The poles of the integrand coming from positively charged fields are marked by $\otimes$, while the those coming from negatively charged fields are marked $\times$. The pole at the origin $\hD = 0$ is marked by a $\star$.}
\label{f:gammapm}
\end{figure}

From the denominator of $Z^{\cI}_{k}$, we can see that the contour of integration for $\hD$ is rather subtle. Examining the eigenmodes of the chiral fields of effective $R$-charge 
\be
\mathbf{r}_{\cI} \equiv r_\cI - Q_\cI k
\ee
around a constant background field $\hs$, $\b \hs$ and $\hD$,
we find that there exist complex bosonic eigenmodes $\phi_{j,m}$ of the chiral fields $\Phi_i$ with eigenvalues
\be
\Delta^{j,m}_\text{bos} = i Q_\cI \hD+ 
\overline{(Q_\cI \hs + m^F_\cI)} (Q_\cI \hs + m^F_\cI + \epsdef m)+ j(j+1)
- {\mathbf{r}_{\cI} \ov 2} ({\mathbf{r}_{\cI}\ov 2}-1) \,,
\ee
for $j > j_0$ and $|m| \leq j$ in general, and
\be
\Delta^{j_0,m}_\text{bos} = i Q_\cI \hD+
\overline{(Q_\cI \hs + m^F_\cI)}  (Q_\cI \hs + m^F_\cI + \epsdef m) \,, \quad
|m| \leq j_0
\ee
when $\mathbf{r}_{\cI} < 1$, that are uncanceled by fermionic ones in the flux sector $k$. Now this implies that when $\g$ is very small, the action of the theory contains the term:
\be
-{1 \ov \g^2} \Delta_\text{bos}^{j,m} \phi_{j,m} \phi_{j,m}^* \,.
\ee
When the real part of $\Delta_\text{bos}^{j,m}$ is negative, the Gaussian integral is unstable, and the saddle point approximation is unreliable. In more physical terms, the limit $\g \ra 0$ is the limit where perturbation theory is exact, but the perturbation theory is not well-defined when ${\rm Re} \, \Delta_\text{bos}^{j,m}$ is negative for some $j$, $m$. Hence the $\hat D$ integration contour must be such that
\be
Q_\cI {\rm Im} \hD \leq 
{\rm Re} \left[ \overline{(Q_\cI \hs + m^F_\cI)}  (Q_\cI \hs + m^F_\cI+ \epsdef m) + j(j+1)
- {\mathbf{r}_{\cI} \ov 2} ({\mathbf{r}_{\cI}\ov 2}-1) \right] \,,
\ee
for all allowed values of $j$ and $m$. More precisely, the contour $\Gamma$ of $\hD$ must satisfy the following conditions:
\begin{itemize}
\item $\Gamma$ asymptotes from $-\infty$ to real $+\infty$.
\item All poles of $Z^{\cI}_k(\hs,\b\hs,\hD)$ with positive charge $Q_\cI$ (``positive poles") lie above $\Gamma$.
\item All poles of $Z^{\cI}_k(\hs,\b\hs,\hD)$ with negative charge $Q_\cI$ (``negative poles") lie below $\Gamma$.
\end{itemize}
The latter two conditions can be stated more ``covariantly" as the following:
\begin{itemize}
\item All poles of $Z^{\cI}_k(\hs,\b\hs,\hD)$ must lie above $\Gamma$ in the $Q_\cI \hD$ plane.
\end{itemize}
Since the integration measure of \eq{der} in the localization limit is (locally) holomorphic with respect to $\hD$, the integral is completely determined if the asymptotics of $\Gamma$ and its position with respect to all the poles are specified. Hence there is one more crucial choice in completely specifying the choice for $\Gamma$. We see from \eq{der} that the fermion integral yields an additional pole of the integrand at $\hD =0$. $\Gamma$ can be chosen to be on either side of this pole. Hence we define two contours $\Gamma_+$ and $\Gamma_-$ so that
\begin{itemize}
\item The pole $\hD =0$ lies below $\Gamma_+$.
\item The pole $\hD =0$ lies above $\Gamma_-$.
\end{itemize}
The contours $\Gamma_\pm$ are depicted in figure \ref{f:gammapm}.

\begin{figure}[!t]
\centering\includegraphics[width=6cm]{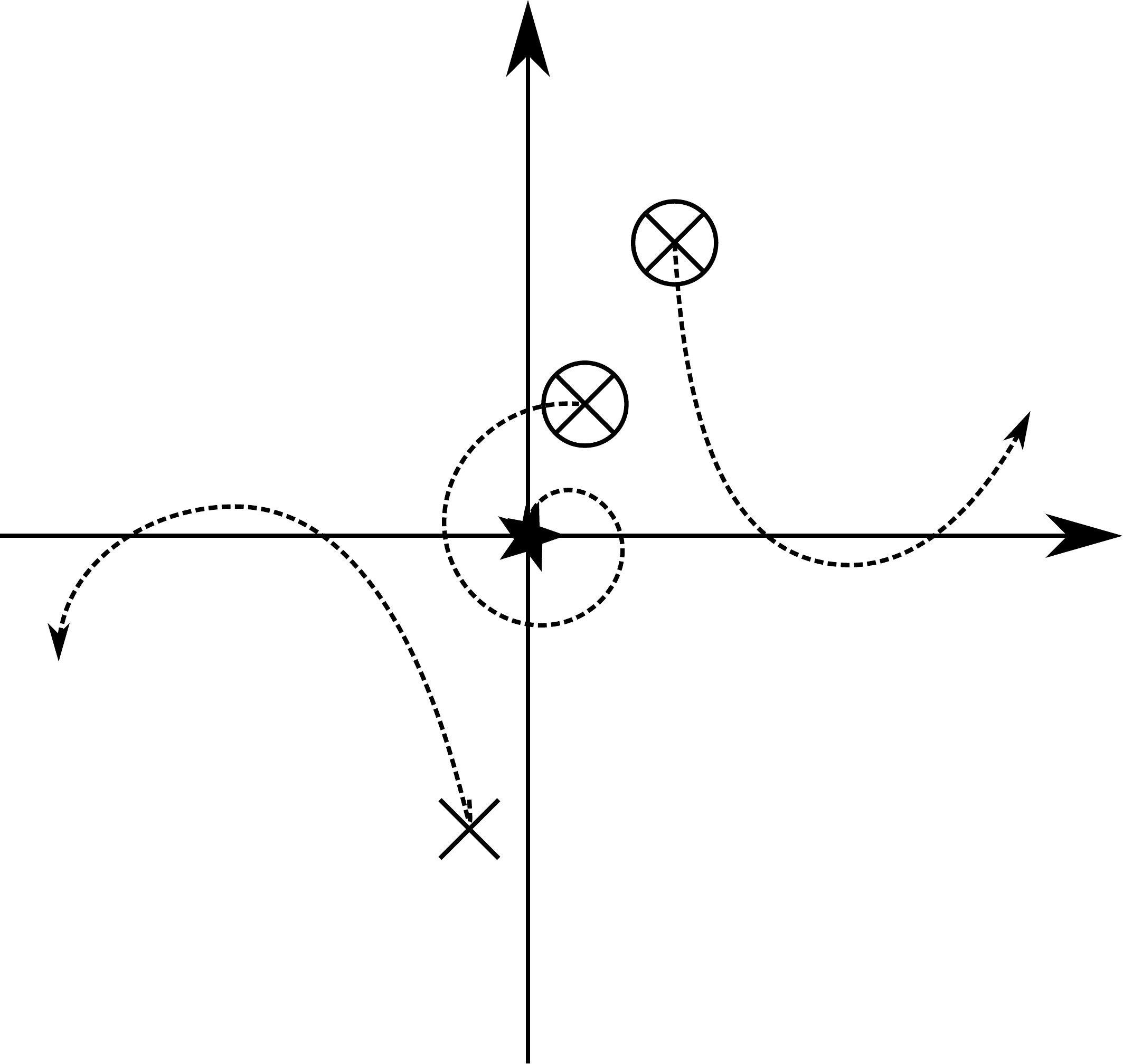}
\caption{\small Schematic depiction of behavior of poles of $\cZ_k (\hs,\b \hs, \hD)$ in the $\hD$-plane with respect to variation of $\hs$.}
\label{f:poles_basic}
\end{figure}

The poles of $\cZ_k$ behave more erratically than the poles of the integrands studied in \cite{Benini:2013nda,Benini:2013xpa,Hori:2014tda}. For the elliptic genus and the Witten index, poles of $\cZ_k$ coming from positively / negatively charged fields with respect to $\hD$ stayed strictly in the upper / lower-half plane of $\hD$, respectively. In our case, some poles cross the real $\hD$-axis, as depicted in figure \ref{f:poles_basic}.

\begin{figure}[!t]
\centering\includegraphics[width=10cm]{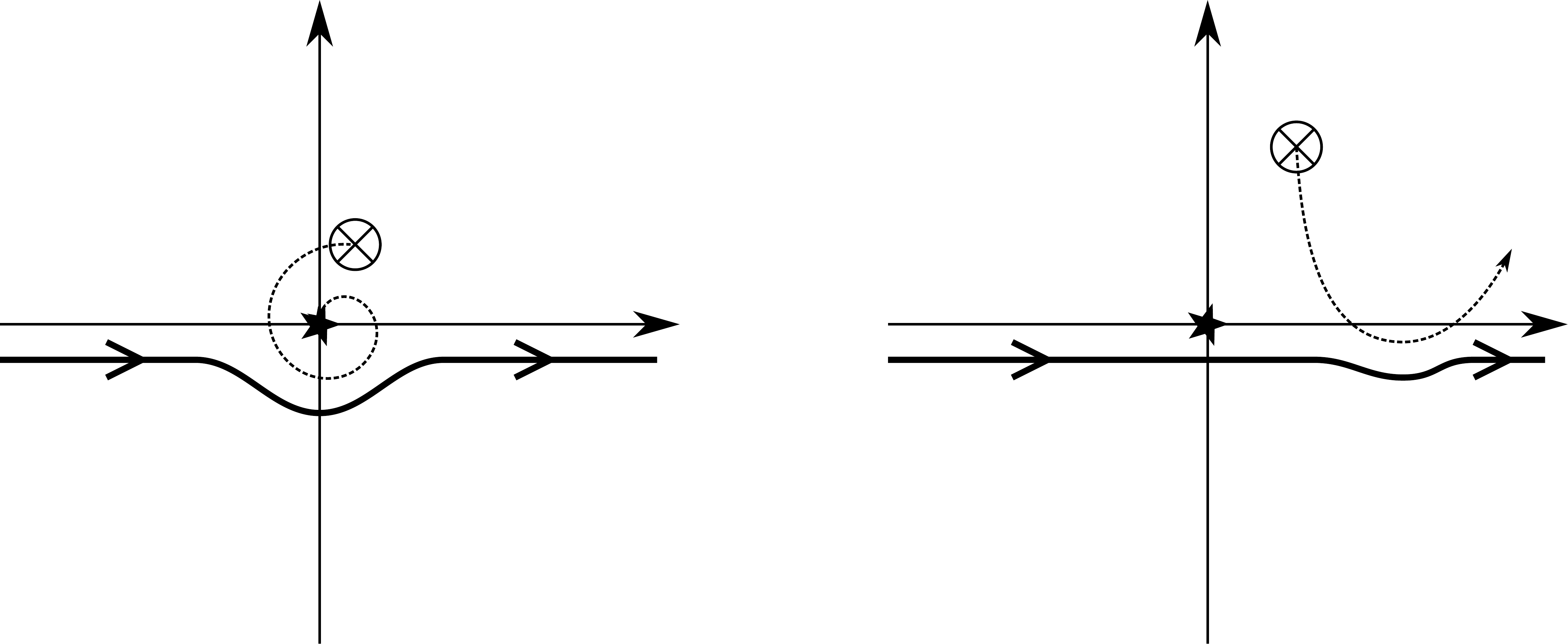}
\caption{\small Deformation of the contour $\Gamma_-$ in the $\hD$-plane as a positive pole swoops below the real axis. When negative poles are far enough away (which can always be made the case by taking small $\epsdef$ due to the assumption of projectivity), the contour $\Gamma_-$ can be taken to be parallel along the real axis by taking the imaginary part of $\Gamma_-$ to be $-i \delta$ for $\delta > K |\epsdef|^2$ for an order-one constant $K$.}
\label{f:poles_upper}
\end{figure}

\begin{figure}[!t]
\centering\includegraphics[width=15cm]{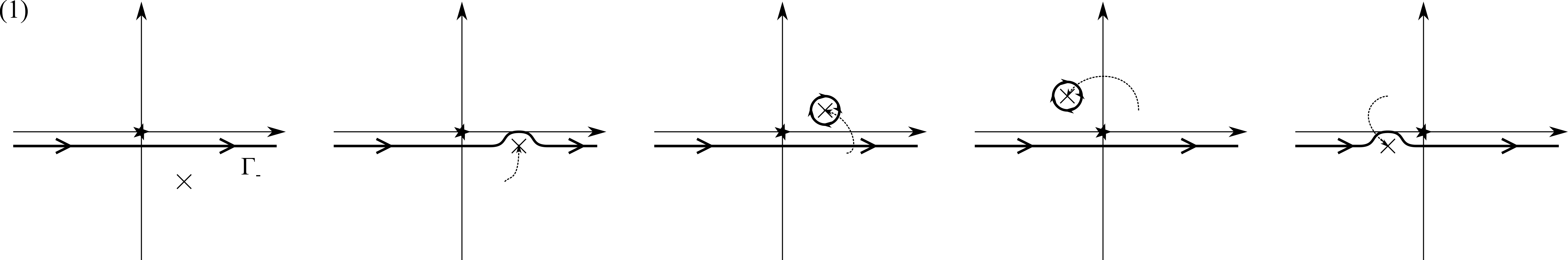} \\[10pt]
\includegraphics[width=15cm]{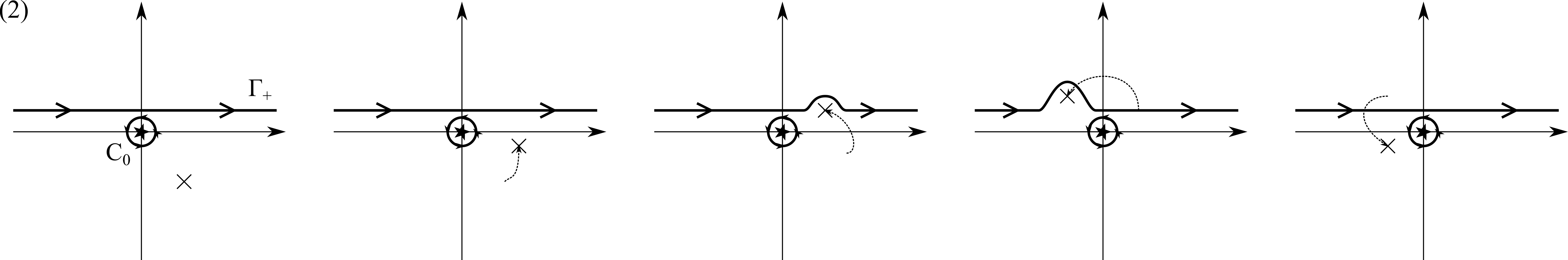}
\caption{\small Two homologically equivalent ways of deforming of the contour $\Gamma_-$ in the $\hD$-plane as a negative pole circles around the origin. One can either split a small portion of the contour and move it around with the pole, or pass it through the origin to obtain $\Gamma_+ + C_0$, where $C_0$ is a tight contour around the origin. When all positive poles are far enough away, the contour $\Gamma_+$ can be made parallel to the real axis, with imaginary part $i \delta$ with $\delta > K |\epsdef|^2$ for a constant $K$ of order-one.}
\label{f:poles_lower}
\end{figure}

The contours $\Gamma_-$ and $\Gamma_+$ are nevertheless well defined for generic values of $\hs$---they are not, only for the following real codimension-two loci:
\begin{itemize}
\item $\Gamma_+$ is ill-defined when a positive pole collides with the origin.
\item $\Gamma_-$ is ill-defined when a negative pole collides with the origin.
\end{itemize}
Upon such collisions, a positive or negative pole unavoidably crosses the contour, leading to a possible singularity of the $\hD$-integral, as we see shortly. One may worry about positive or negative poles swooping to the other side of the real $\hD$ axis from their half-plane, but this can be dealt with. In particular, at small $\epsdef$, the only poles that cross the real $\hD$ axis are poles $\hD_* (\hs,\b \hs)$ of the $Z^{(0)}$ factor of the one-loop determinants of the charged fields. In particular, this only happens when $\hD_* (\hs,\b\hs) \sim \cO (\epsdef)$, and the imaginary part of the location of the pole is of order $\cO(\epsdef^2)$. Hence these poles can be kept from crossing the contour of integration by a small deformation of this order. The path integral for macroscopic values of $\epsdef$ can be obtained from the ones with small $\epsdef$ by analytic continuation. Additional complications ensue as certain poles may circle around the origin, but this can also be accommodated. The appropriate maneuvers for $\Gamma_-$ are depicted in figures \ref{f:poles_upper} and \ref{f:poles_lower}. The final worry might be that a positive pole may coincide with a negative pole at some value of $\hs$. Such a situation can be avoided by taking $\epsdef$ to be very small, computing the path integral, and then analytically continuing for macroscopic values of $\epsdef$. Throughout this section, we thus work with the convenient assumption that $\epsdef$ is very small. We discuss the contours $\Gamma_-$ and $\Gamma_+$ in further detail at the end of this subsection.

It is worth emphasizing that the only poles $\hD_* (\hs,\b \hs)$ that collide with the origin and can make the integrand
\be
F^k_{\e,0} = \p_{\b \hs}
\int_\Gamma d \h D {1 \ov \h D}
e^{-{\hD^2 \ov 2\e^2}+{4\pi i \t \xi \ov \e^2} \hat D }
\cZ_{k} (\hs,\b \hs, \hat D)
\label{Dint}
\ee
of the $\hs$-integral singular, are the poles of the $Z^{(0)}$ piece of the one-loop determinants. We have made the contour-dependence of $F_{\e,0}$ implicit. For macroscopic values of $\epsdef$ the poles of $Z^\pos$ can collide with the origin, but their effects are benign, and does not give rise to any singularities. In fact, when $\epsdef$ is taken to be very small compared to other masses of the theory, the imaginary part of the poles of the $Z^\pos$ piece has magnitude of order $\cO(1)$---thus for small enough $\epsdef$, the poles of $Z^\pos$ never even come near the real axis of $\hD$. The integral \eq{Dint} at macroscopic values of $\epsdef$ can be obtained via analytic continuation from small values of $\epsdef$.

The poles of the $Z^{(0)}$ piece of the one-loop determinant of charged field component $\cI$, as can be seen from equation \eq{Z0}, collide with the origin $\hD =0$ when
\be
Q_\cI \hs = - m_\cI^F - \epsdef \left( n+{\mathbf{r}_{\cI} \ov 2} \right)    \quad
\text{for $n \in [0,-\mathbf{r}_{\cI}]_\text{int}$}\,,
\quad \text{or} \quad
\overline{Q_\cI \hs + m_\cI^F} = 0 \,.
\ee
It is simple to see that the singularity is absent when $Q_\cI \hs + m_\cI^F = 0$ unless $\mathbf{r}_{\cI}$ is even, due to the factor $\overline{(Q_\cI \hs + m_\cI^F)}^{~\mathbf{r}_{\cI}+1}$ present in the numerator of $Z^{(0)}_{\mathbf{r}_{\cI}}$. We hence find that the values of $\hs$ at which $F^k_{\e,0}$ can become singular narrows down to precisely the poles of $\cZ_k (\hs)= \cZ_k (\hs,\b \hs, \hD =0)$, as expected. Recall that $\De$ is defined to be a tubular neighborhood of these points. For future purposes, it is useful to split the poles of $\cZ_k (\hs)$, or rather, the potentially singular loci of $F^k_{\e,0}$, into two groups---the ``positive" loci $H_+$ at which the positive poles of $\cZ_k (\hs, \b \hs, \hD)$ with respect to $\hD$ collide with $\hD =0$ and the ``negative" loci $H_-$ at which the negative poles of $\cZ_k (\hs, \b \hs, \hD)$ do. Let $\Dek^+$ / $\Dek^-$ denote the tubular neighborhoods around the positive / negative loci respectively.

Let us now, for definiteness, take the contour of integration to be $\Gamma_-$ and evaluate the path integral \eq{der} in the localization limit where $\e, \g \ra 0$. The integral is given by
\bea
\ZS&=
{2 \ov |W|}
\sum_{k \in \bZ} q^k 
\lim_{\substack{\e \ra 0\\ \eps \ra 0}}
\int_{\tfM \setminus \Dek} d \hs d \b\hs  F^k_{\e,0} \\
&=
{2 \ov |W|} 
\sum_{k \in \bZ} q^k 
\lim_{\substack{\e \ra 0\\ \eps \ra 0}}
\int_{\tfM \setminus \Dek} d \hs d \b\hs  \p_{\b \hs}
\int_{\Gamma_-} d \h D {1 \ov \h D}
e^{-{\hD^2 \ov 2\e^2}+{4\pi i \t \xi \ov \e^2}\hat D }
\cZ_{k} (\hs,\b \hs, \hat D) \,.
\label{rank one 1}
\eea
The path integral is a total derivative in $\b \hs$, and hence it is important to understand the boundaries of $\tfM \setminus \Dek$:
\be
\p \tfM \setminus \Dek = \p \tfM - \p \Dek^+ - \p \Dek^- \,.
\ee
The boundary of $\Dek$ consists of counter-clockwise contours encircling the potentially singular loci of $F^k_{\e,0}$. Meanwhile, the boundary of $\tfM = \bC$ can be obtained by first constructing a large circular counter-clockwise contour of radius $R_\tfM$, which encircles all the singular loci, and taking the limit $R_\tfM \ra \infty$.

\begin{figure}[!t]
\centering\includegraphics[width=15cm]{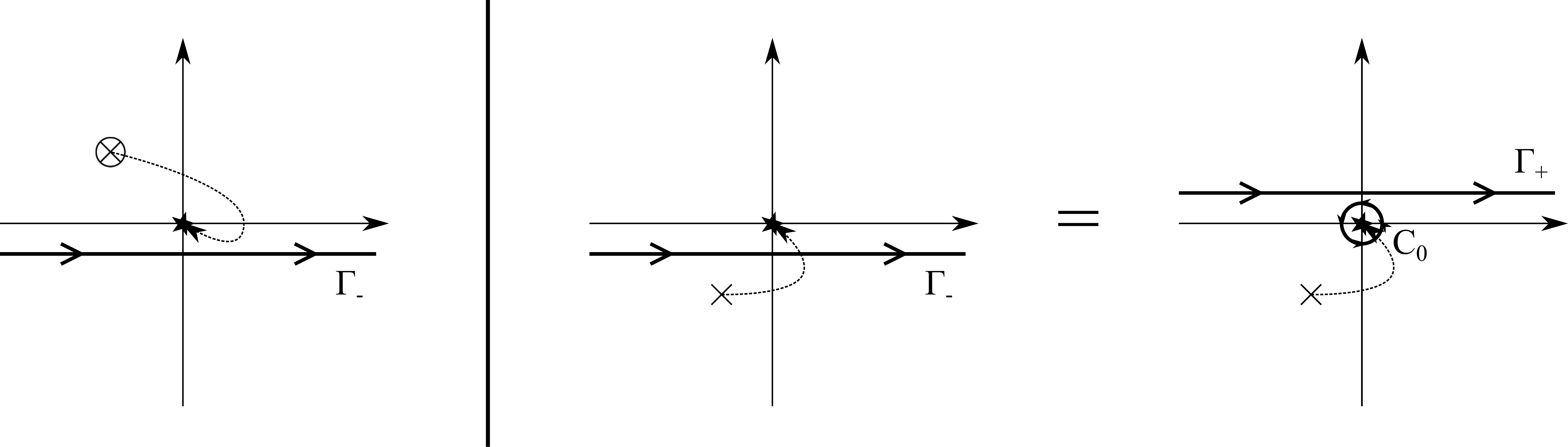}
\caption{\small The behavior of poles of $\cZ(\hs,\b\hs,\hD)$ in the $\hD$-plane as $\hs$ varies inside $\Dek$. The left panel depicts a positive pole colliding with the origin as $\hs$ is taken to a singular point within $\Dek^+$. The $\Gamma_-$-contour integral is smooth and bounded. The right panel depicts a negative pole colliding with the origin as $\hs$ is taken to a singular locus within $\Dek^-$. The $\Gamma_-$-contour integral is singular---the singular element can be isolated as the contour integral around $C_0$.}
\label{f:poles_crossing}
\end{figure}

Let us first examine the integral
\be
\mathbf{Z}_k^{\eps,+} =
- \int_{\p \Dek^+} d \hs 
\int_{\Gamma_-} d \h D {1 \ov \h D}
e^{-{\hD^2 \ov 2\e^2}+{4\pi i \t \xi \ov \e^2} \hat D }
\cZ_{k} (\hs,\b \hs, \hat D) \,.
\ee
The $\hD$-integral is bounded and well-defined within $\Dek^+$. This is because the pole of $\cZ_{k} (\hs,\b \hs, \hD)$ with respect to $\hD$ that collides with $\hD=0$ does not cross $\Gamma_-$, as depicted in the left panel of figure \ref{f:poles_crossing}, thereby making the integrand of the $\Gamma_-$ integral to be smooth across the integration contour. We therefore see that
\be
\lim_{\substack{\e \ra 0 \\ \eps \ra 0}} \mathbf{Z}_k^{\eps,+} = 0 \,.
\ee

The integral
\be
\mathbf{Z}_k^{\eps,-} =
- \int_{\p \Dek^-} d \hs 
\int_{\Gamma_-} d \h D {1 \ov \h D}
e^{-{\hD^2 \ov 2\e^2}+{4\pi i \t \xi \ov \e^2} \hat D }
\cZ_{k} (\hs,\b \hs, \hat D)
\ee
is more interesting. Since a pole of $\cZ_k  (\hs,\b \hs, \hat D)$ necessarily crosses the contour $\Gamma_-$ to collide with the pole of the integrand at $\hD =0$, the integral along $\Gamma_-$ is expected to be singular at some point within $\Dek^-$. An efficient way to deal with this singularity, as pointed out in  \cite{Benini:2013nda,Benini:2013xpa,Hori:2014tda}, is to deform the contour $\Gamma_- = \Gamma_+ + C_0$, where $C_0$ is a small contour encircling the origin in the counter-clockwise direction. The $\hD$ integral over $\Gamma_+$, as argued before, is smooth and bounded as a function of $\hs$ and $\b \hs$ in $\Dek^-$ as $\e$ is taken to be very small. The contour integral around $C_0$ can be evaluated explicitly, due to \eq{cZk}. We thus arrive at
\bea
\lim_{\substack{\e \ra 0 \\ \eps \ra 0}} \mathbf{Z}_k^{\eps,-} &=
- \lim_{\substack{\e \ra 0 \\ \eps \ra 0}} 
\int_{\p \Dek^-} d \hs 
\int_{C_0} d \h D {1 \ov \h D}
e^{-{\hD^2 \ov 2\e^2}+{4\pi i \t \xi \ov \e^2} \hat D }
\cZ_{k} (\hs,\b \hs, \hat D) \\[3pt]
&=-2 \pi i\int_{\p \Dek^-} d \hs  \cZ^\oneloop_{k} (\hs) \\
&=4 \pi^2 \sum_{\hs_{*} \in H_-} \res_{\hs=\hs_{*}}\cZ^\oneloop_{k} (\hs) \,.
\eea

Finally, let us examine the integral along the boundary at infinity:
\be
\mathbf{Z}_k^\infty = \int_{\p \tfM} d \hs 
\int_{\Gamma_-} d \h D {1 \ov \h D}
e^{-{\hD^2 \ov 2\e^2}+{4\pi i \t \xi \ov \e^2} \hat D }
\cZ_{k} (\hs,\b \hs, \hat D) \,.
\label{Zinf}
\ee
To evaluate this integral, one must understand how to control the behavior of the parameter $\t\xi$ as the coupling $\e$ is taken to infinity. We choose to keep $\t\xi$ constant. This is called the ``Higgs scaling limit" in \cite{Hori:2014tda}.%
~\footnote{Note that our $\t \xi$ can be identified with $\e^2 \zeta$ of \cite{Hori:2014tda}, where $\zeta$ is the FI parameter of the one-dimensional ``gauge" theories studied therein.}
This is because we wish to compute correlators of the IR theories obtained by flowing from various Higgs phases of the GLSM. To make the discussion more concrete, let us first restrict to values of $\t \xi$ that lie in a chamber of a ``geometric phase." We then see that the classical ``Higgs phase" field configurations of the localization Lagrangian \eq{Coulomb Lagrangian for 5} engineers the IR geometry whose K\"ahler structure is given by $\t\xi$. Hence in order for the Coulomb branch path integral to compute correlators which eventually are to be identified with those of the non-linear sigma model of an IR geometry, this value must be kept macroscopic as $\e$ is taken to be small. An equivalent argument holds for all phases of the GLSM.

\begin{figure}[!t]
\centering\includegraphics[width=13cm]{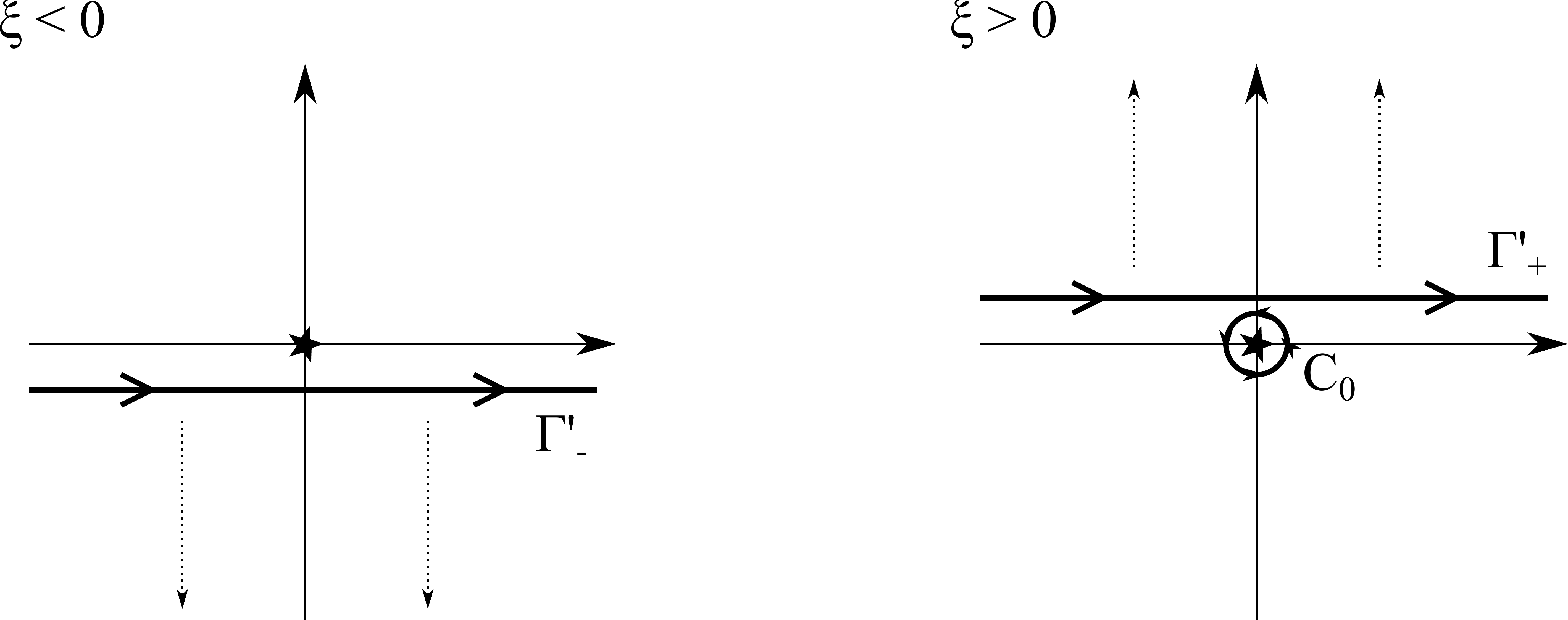}
\caption{\small The $\hD '$ integration contour in the $\e \ra 0$ limit. When $\xi < 0$, the contour $\Gamma_-'$ can be deformed away in the lower half plane. When $\xi>0$, the contour must be deformed away to the upper half plane, thus forcing the integral to pick up the pole at the origin.}
\label{f:poles_crossing_infinity}
\end{figure}

We therefore arrive at
\be
\mathbf{Z}_k^\infty = \int_{\p \tfM} d \hs 
\int_{\Gamma_-'} d \h D' {1 \ov \h D'}
e^{-{\e^2 {\hD'}{}^2 \ov 2}+4\pi i \t \xi \hat D' }
\cZ_{k} (\hs,\b \hs, \e^2 \hat D') \,.
\label{hat D prime}
\ee
by the reparametrization $\hat D = \e^2 \hat D'$. The contour of the integrand of \eq{hat D prime} is depicted in figure \ref{f:poles_crossing_infinity}. Before going further, we must acknowledge an order of limits issue here. This integral depends on whether the boundary of $\p \tfM$ is taken to infinity faster than $\e$ is taken to zero. We assume that $R$ must grow much faster, as the domain of integration should in principle span the entire $\tfM$ regardless of the coupling. To be more precise, take the $|\hs| \ra \infty$ and $\e \ra 0$ limit so that
\be\label{def R}
R \equiv {|\hs|}^{\e^2} \ra \infty \,.
\ee
In other words, $\log |\hs|$ is taken to grow faster than ${1 \ov \e^2}$. The behavior of the determinant $Z_k^\cI(\hs,\b\hs,\hD)$ has been studied in detail in appendix \ref{apsub: large sigma}. There, we provide strong evidence that
\be
\lim_{\substack{\e \ra 0 \\ R \ra \infty}}
\cZ^\cI_k (\hs,\b\hs,\e^2 \hD') =
\lim_{\substack{\e \ra 0 \\ R \ra \infty}}
Z_k^{\cI} (\hs)
e^{ 2 i (1+ \epsdef' \alpha_{\epsdef'})(\log R)  Q_\cI \hD' } \,.
\label{large R}
\ee
Here, $\epsdef'$ is defined to be
\be
\epsdef' = {{\b \hs} \ov |\hs| }\epsdef \,,
\ee
while $\alpha_{\epsdef'}$ is a function of $\epsdef'$ that is smooth and bounded in a small enough neighborhood of $\epsdef'=0$. This limit comes from estimating the $\zeta$-function regularization of the determinants $Z^\pos$ defined in equation \eq{Zpos}. We thus find that
\be\label{with alpha}
\lim_{\substack{\e \ra 0 \\ R \ra \infty}}
e^{-{\e^2 {\hD'}{}^2 \ov 2}+4\pi i \t \xi \hat D' }
\cZ_{k} (\hs,\b \hs, \e^2 \hat D') =
\lim_{\substack{\e \ra 0 \\ R \ra \infty}}
e^{4\pi i \left[\t\xi+{1 \ov 2\pi} b_0 (1+ \epsdef' \alpha_{\epsdef'}) \log R \right] \hD'}\cZ_k^\oneloop (\hs) \,.
\ee
Recall that $b_0 = \sum_\cI Q_\cI$. We can absorb the real part of the factor $(1+ \epsdef' \alpha_{\epsdef'})$ in front of $\log R$ in the equation \eq{with alpha} into $R$. Then, defining
\be
{\xi_\eff^\text{UV}} = \t\xi + {1 \ov 2 \pi}b_0 \log R\,,
\ee
we arrive at the integral
\be
\lim_{\e \ra 0}\mathbf{Z}_k^\infty =
\lim_{R \ra \infty}
\int_{\p \tfM} d \hs 
\int_{\Gamma_-'} d \h D' {1 \ov \h D'}
e^{4\pi i {(\xi_\eff^\text{UV} + {i \ov 2\pi} b_0 \cO(|\epsdef|) \log R)} \hat D' }
\cZ_k^{(0)} (\hs,\b \hs,0)  \,,
\ee
where we have indicated that the exponent develops a real part parametrically smaller than $\xi_\eff^\text{UV}$. In the presence of this real part, i.e., when $b_0 \neq 0$, the contour $\Gamma_-'$ needs to be slightly bent downward in the $\hD$-plane by an angle of order $\cO(|\epsdef|)$ to exhibit desirable asymptotic behavior. This does not alter the position of the contour with respect to any of the poles of the integrand. Now when ${\xi_\eff^\text{UV}} < 0$, the contour can be deformed away in the lower half of the complex $\hD$-plane. On the other hand, when ${\xi_\eff^\text{UV}} >0$, the contour must be deformed away in the upper half plane. In this process, the contour $C_0$ around the origin is picked up. In conclusion, we arrive at
\be
\mathbf{Z}_k^\infty = 
2 \pi i \Theta(\xi_\eff^\text{UV}) \int_{\p \tfM} d \hs  \cZ^{(0)}_k (\hs,\b \hs,0)
=  2 \pi i \Theta(\xi_\eff^\text{UV}) \int_{\p \tfM} d \hs 
\cZ^\oneloop_k (\hs)\,,
\ee
where $\Theta(\xi)$ is defined to be
\be
\Theta(\xi) =
\begin{cases}
1 & \xi >0 \,, \\
0 & \xi < 0 \,.
\end{cases}
\ee

Since $\cZ_k (\hs)$ is a holomorphic function on $\idom$, it follows that
\be
2\pi i \int_{\p \tfM} d \hs 
\cZ_k (\hs) = - 4 \pi^2 \sum_{\hs_* \in H_+ \cup H_-}
\res_{\hs = \hs_*} \cZ^\oneloop_k (\hs) \,.
\ee
Putting everything together, we arrive at the result:
\bea
\ZS &=
{2 \ov |W|} \sum_{k \in \bZ} q^k
\lim_{\substack{\e \ra 0 \\ \eps \ra 0}}
\left(
\mathbf{Z}_k^{\eps,-} +
\mathbf{Z}_k^{\eps,+} +
\mathbf{Z}_k^{\infty}  \right) \\
&= {- 8 \pi^2 \ov |W|} \sum_{k \in \bZ} q^k
\left(
\Theta(\xi_\eff^\text{UV}) \sum_{\hs_* \in H_+}
\res_{\hs = \hs_*} \cZ^\oneloop_k (\hs)
-\Theta(-\xi_\eff^\text{UV}) \sum_{\hs_* \in H_-}
\res_{\hs = \hs_*} \cZ^\oneloop_k (\hs)
\right) \,.
\eea

In doing the integral \eq{rank one 1}, we could have chosen the $\hD$-contour to be $\Gamma_+$:
\bea
\ZS'&=
{2 \ov |W|} 
\sum_{k \in \bZ} q^k 
\lim_{\substack{\e \ra 0\\ \eps \ra 0}}
\int_{\tfM \setminus \Dek} d \hs d \b\hs  \p_{\b \hs}
\int_{\Gamma_+} d \h D {1 \ov \h D}
e^{-{\hD^2 \ov 2\e^2}+{4\pi i \t \xi \ov \e^2} \hat D }
\cZ_{k} (\hs,\b \hs, \hat D) \,.
\label{rank one 2}
\eea
This would not have made a difference in the final result. Since $\Gamma_+-\Gamma_- = -C_0$,
\bea
\int_{\Gamma_+-\Gamma_-} d \h D {1 \ov \h D}
e^{-{\hD^2 \ov 2\e^2}+{4\pi i \t \xi \ov \e^2} \hat D }
\cZ_{k} (\hs,\b \hs, \hat D)=-\int_{C_0} d \h D {1 \ov \h D}
= -2 \pi i \cZ^\oneloop_k (\sigma)
\eea
and thus
\be
\ZS' - \ZS = - {4 \pi i \ov |W|} 
\sum_{k \in \bZ} q^k 
\lim_{\substack{\e \ra 0\\ \eps \ra 0}}
\int_{\tfM \setminus \Dek} d \hs d \b\hs  \p_{\b \hs} \cZ^\oneloop_k (\sigma) = 0\,,
\ee
since $\cZ^\oneloop_k (\sigma)$ is smooth and holomorphic on $\tfM \setminus \Delta_{\eps,k}$.

Let us conclude this section by commenting on the geometry of the contour of integration for $\hD$, to prepare ourselves for moving on to gauge theories with higher rank. In the rank-one case, the topology of the $\hD$-plane and contours is simple enough to visualize, so that the consistency of the prescription for the contours $\Gamma_\pm$ is manifest. When the $\hD$-space is multi-dimensional, it appears rather tricky to keep track of the topology of the contour. In order for us to apply the machinery of \cite{Benini:2013xpa,Hori:2014tda}, we need to describe the contours as hyperplanes. For the rest of the section, we demonstrate that this can be done, as we take $\epsdef$ to be
\be
\eps \ll \epsdef \ll \mu_0 \leq 1 \,,
\label{scales}
\ee
where $\mu_0$ is the minimum of all other mass scales present in the theory:
\be
\mu_0 = \min (1, \mu_\text{min})
\ee
where $\mu_{min}$ is the lightest twisted mass. Following the usual strategy, the contour $\Gamma$ for the $\hD$-integral is defined via hyperplanes as we take $\epsdef$ to be very small compared to $\mu_0$. The path integral for macroscopic $\epsdef$ can be obtained by analytically continuing the result for small $\epsdef$.

\begin{figure}[!t]
\centering\includegraphics[width=15cm]{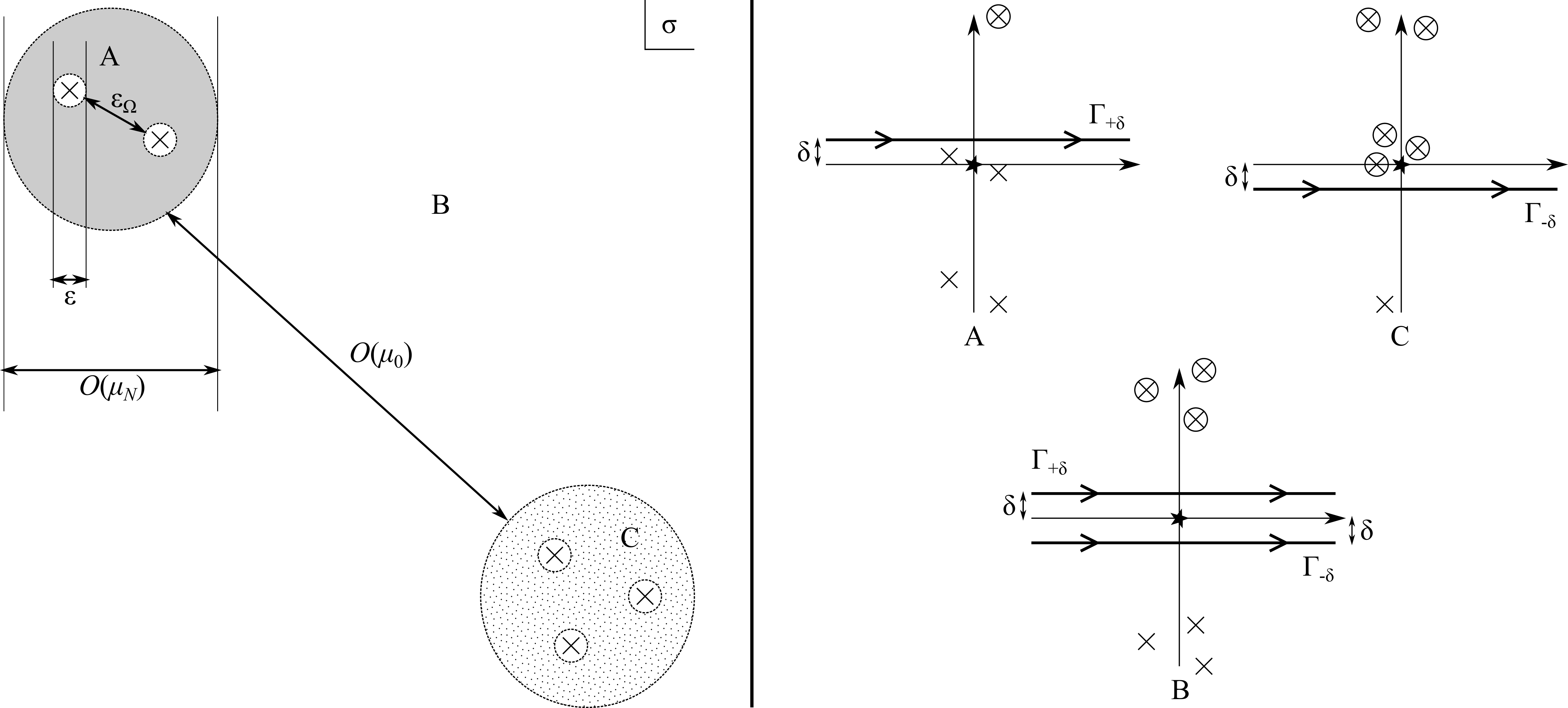}
\caption{\small Distribution of poles of $\cZ_k (\hs,\b \hs,\hD)$ in the $\hD$ plane as $\hs$ is varied. $A$ / $C$ is an $\cO(\mu_N)$-neighborhood of poles of $Z_k^\cI(\hs)$ with $Q_\cI>0$ / $Q_\cI <0$, respectively. The poles of $Z_k^\cI(\hs)$ themselves, within $A$ and $C$ are separated by a distance of order $\epsdef$. In $B$, the positive poles lie above $\Gamma_{+\delta}$ and the negative poles lie below $\Gamma_{-\delta}$, where $\delta$ is of order $\cO(|\epsdef|)$---there are no poles of $\cZ_k (\hs,\b\hs,\hD)$ between the two contours, as they are all away at a distance $\cO(\mu_N^2)$ from the real axis. In $A$, while some negative poles cross the real $\hD$-axis, $\Gamma_{+\delta}$ still can be used to divide the positive and negative poles. This is because the negative poles that cross $\hD=0$ still lie within a range $\cO(\epsdef^2)$ of the real line, while the positive poles are far away from the real line at distances of $\cO(\mu_N^2)$. In $C$, $\Gamma_{-\delta}$ divides the positive and negative poles.}
\label{f:sigma}
\end{figure}

Recall that the poles of $Z^\cI_k (\hs)$ lie at the ``hyperplanes" \eq{hyperplanes}:
\be
\hs = \hs^\cI_{*,n} \equiv \left[- m_\cI^F-\epsdef \left( n+{\mathbf{r}_{\cI} \ov 2} \right) \right]/ Q_\cI,\quad
n \in [0,-\mathbf{r}_{\cI}]_\text{int} \,.
\ee
Assuming the separation of scales \eq{scales}, all the poles of $Z^\cI (\hs, \b \hs, \hD)$ with respect to $\hD_\cI$ lie above the contour
\be
{\rm Im} \hD_\cI = - \mathbf{r}_{\cI}^2 |\epsdef|^2  
\ee
in the $\hD_\cI$-plane where we have defined
\be
\hD_\cI \equiv Q_\cI \hD \,.
\ee
Meanwhile, all poles of $Z^\cI (\hs, \b \hs, \hD)$ (with respect to $\hD_\cI$) lie above the contour
\be
{\rm Im} \hD_\cI = x^2 |\epsdef|^2  
\ee
in the $\hD_\cI$-plane when
\be
|Q_\cI \hs - Q_\cI \hs^\cI_{*,n}| > (x+\mathbf{r}_{\cI}) |\epsdef|
\ee
for all poles $\hs^\cI_{*,n}$ of $Z^\cI_k (\hs)$. The poles of $Z^\cI_k (\hs)$ all lie close to each other at a distance of $\cO(\epsdef)$.

Let us take a small connected open set $N_{\cI}$ that contains the domain
\be
\{ \, \hs \, :\, |Q_\cI \hs - Q_\cI \hs^\cI_{*,n}| \leq \mu_N^2 \,  \} \,,
\ee
for
\be
\epsdef \ll \mu_N^2 \ll \mu_N \ll \mu_0 \leq 1 \,.
\ee
$N_{\cI}$ is of size $\cO(\mu_N) \ll \mu_0$ in the $\hD$-plane. Outside of $\bigcup_\cI N_\cI$, the positive poles of $Z_k (\hs,\b \hs,\hD)$ with respect to $\hD$ lie above
\be
\Gamma_{+\delta} = \setcond{\hD}{{\rm Im\,}\hD = \delta}
\ee
and negative poles lie below
\be
\Gamma_{-\delta}=\setcond{\hD}{{\rm Im\,}\hD=-\delta}
\ee
for any $\delta$ of $\cO(|\epsdef|)$. Meanwhile, inside $N_\cI$ with $Q_\cI >0$, positive poles may swoop down below the real $\hD$-axis, but do not go below $\Gamma_{-\delta}$. Due to the separation of scales, all negative poles of $Z_k (\hs,\b \hs,\hD)$ stay far below $\Gamma_{-\delta}$ within such $N_\cI$. Likewise, inside $N_\cI$ with $Q_\cI<0$, negative poles may go above the real axis of $\hD$, but do not go over $\Gamma_{+\delta}$. All positive poles of $Z_k (\hs,\b \hs,\hD)$ are far above $\Gamma_{+\delta}$ within such $N_\cI$. The situation is depicted in figure \ref{f:sigma}.

The upshot is for any point in $\idom$, there exists a real line of the complex $\hD$-plane that lies below all the negative poles, and that lies above all the positive poles. One can use these hyperplanes to represent the topological prescription of the contour. Let us denote
\be
N_+ = \bigcup_{\cI, Q_\cI >0} N_\cI, \quad
N_- = \bigcup_{\cI, Q_\cI <0} N_\cI \,,
\ee
which consist of regions of size $\cO(\mu_N)$. $\Gamma_\pm$, which we have defined topologically can be redefined in the following way:
\begin{itemize}
\item $\Gamma_+$ is given by $\Gamma_{+\delta}$ in $\tfM \setminus (\De \cup N_+)$, and by $(\Gamma_{-\delta} - C_0)$ in $N_+ \setminus \De$.
\item $\Gamma_-$ is given by $\Gamma_{-\delta}$ in $\tfM \setminus (\De \cup N_-)$, and by $(\Gamma_{+\delta} + C_0)$ in $N_- \setminus \De$.
\end{itemize}
An important point to emphasize is that this is true as long as $\delta$ is of order $\cO(\epsdef)$. This provides a nice handle on the $\hD$-contours that proves useful as we move on to computing the path integral in the general case.

\subsection{The general case} \label{ss: general}

Let us now prove the formula \eq{main formula} for gauge theories with general gauge groups. The result follows from the analysis of \cite{Benini:2013xpa,Hori:2014tda} straightforwardly---in this section, we explain why that analysis 
can be extended to our case. To be more precise, we show that:
\begin{enumerate}[(1)]
\item The partition function can be written in the form
\be
\ZS=
{1 \ov |W|} \sum_{k} q^k
\lim_{\substack{\epsilon \ra 0 \\ \e \ra 0}}
\int_{\Gamma \ltimes \tfM \setminus \De} \mu
\ee
with
\be
\mu = {1 \ov r!} \cdot
e^{-{\hD^2 \ov 2\e^2} +{4\pi i \t\xi \ov \e^2} \hD} \cdot
\cZ_{k} (\hs,\b\hs,\hD) d^r \hs
\wedge (\nu(d \hD))^{\wedge r}
\label{mu}
\ee
where $\nu$ is defined such that
\be
\nu( V ) =   h^{ab} d \b \hs_a \wedge  V_b \,.
\label{nu}
\ee
for any form $V$ valued in $\mathfrak{h}_\bC$. $h^{ab}$ is given by equation \eq{habdef} and satisfies the properties \eq{hab}. $\cZ_{k}(\hs,\b\hs,\hD)$ satisfies
\be
\cZ_{k}(\hs,\b\hs,\hD=0) = \cZ^\oneloop_{k}(\hs) \,.
\ee
\item The $\hD$-contour of integration is equivalent to that of \cite{Benini:2013xpa,Hori:2014tda}.
\item The asymptotics of $\mu$ are such that when the covector $\eta$ \cite{Benini:2013xpa,Hori:2014tda} that defines the $\hD$ contour is aligned with
\be
\xi_\eff^\text{UV} = \t\xi + {1 \ov 2\pi} b_0 \lim_{R \ra \infty} \log R
\ee
as defined in section \ref{subsec: summary}, the contribution from the cells adjacent to the boundary of $\tfM$ at infinity vanishes.~\footnote{As is evident from various examples presented further into the paper, the parameter $\t\xi$, when restricted to the dual of the center of the lie algebra $i \mathfrak{c}^* \subset i \mathfrak{h}^*$ may not be generic enough. In this case, $\t\xi$ is to be deformed within $i \mathfrak{h}^*$ slightly to a generic value, and then taken to its initial value by analytic continuation.}
\end{enumerate}
These conditions are enough to arrive at \eq{main formula}.

In this section, we choose to follow the notation of \cite{Benini:2013xpa,Hori:2014tda}, without much alteration. Since our result follows rather straightforwardly from the methods of \cite{Benini:2013xpa,Hori:2014tda}, we do not rehash all the steps taken in those papers with the level of rigor of their presentation. Instead, we simply review their notation and results relevant in the present setting and explain why they can be re-run for our computation.

Condition (1) follows from supersymmetry. Repeating previous arguments, the partition function is obtained by the integral~\footnote{Note that the definition of $\cZ_{k}$ in this section differs from that of section \ref{ss:gaugino} by an exponential factor only dependent on $\hD$. 
This does not affect the supersymmetry relations or properties of $h^{bc}$, as $\hD_a$ are invariant under supersymmetry transformations, as can be seen from equation \eq{susy new}.}
\be
\ZS=
{1 \ov |W|} \sum_{k} q^k
\lim_{\substack{\epsilon \ra 0 \\ \e \ra 0}}
\int_{\Gamma \ltimes \tfM \setminus \De}
\prod_a (d\hs_a d \b \hs_a d \hD_a d\lambda_a d \t \lambda_a)
e^{-{\hD^2 \ov 2\e^2} + {4\pi i \t \xi \ov \e^2}( \hD)}
\cZ_{k} (\hs, \b \hs, \lambda, \t \lambda, \hD) \,.
\ee
Here, $\hD^2$ is understood to be the inner-product of an element of $\mathfrak{h}_\bC$ defined by the Lie algebra, while $\t\xi (\hD)$ is the pairing between elements of $\mathfrak{h}_\bC^*$ and $\mathfrak{h}_\bC$. Note that $\t\xi$ in this term is not to be taken as the $n$-dimensional vector in the free subalgebra of $\mathfrak{g}$, but rather as an element of $\mathfrak{h}^*_\bC$ with $r$ components. Recall that $\cZ_{k} (\hs, \b \hs, \lambda, \t \lambda, \hD)$, which we denote $\cZ_{k}$  for sake of brevity, satisfies the supersymmetry equations
\be
\left( -2 \t \lambda_a {\p \ov \p \b \hs_a} - \hD_a {\p \ov \p \lambda_a} \right) \cZ_{k} =
\left( -2 \lambda_a {\p \ov \p \b \hs_a} + \hD_a {\p \ov \p \t \lambda_a} \right) \cZ_{k} =0 \,.
\label{susy new}
\ee
Let us assume that there exists a symmetric tensor $h^{ab}$ satisfying
\be
\p_{\b \hs_a} \cZ_{k} (\hs, \b \hs, \hD)
= {1 \ov 2} \hD_b h^{ba} \cZ_{k} (\hs, \b \hs, \hD), \quad
\p_{\b \hs_c} h^{ab}=\p_{\b \hs_a} h^{cb} \,.
\label{hab}
\ee
Then we find that
\be
\cZ_{k} (\hs, \b \hs, \hD) \exp(h^{bc} \t \lambda_b \lambda_c)
\ee
is a solution to \eq{susy new}. This implies that
\be
\cZ_{k} (\hs, \b \hs, \lambda, \t \lambda, \hD)
= \cZ_{k} (\hs, \b \hs, \hD) \exp(h^{bc} \t \lambda_b \lambda_c)
+\cC(\hs, \b \hs, \lambda, \t \lambda, \hD)
\ee
where $\cC$ is a solution of \eq{susy new} whose bottom component is zero:
\be
\cC_0 \equiv \cC |_{\lambda = \t \lambda = 0}  =0\,.
\ee

Let us now show that the top component $\cC |_{\t \lambda_1 \lambda_1 \cdots \t \lambda_r \lambda_r}$ of $\cC$ vanishes. The $\prod_a (\t \lambda_a \lambda_a)$ component of $\cC$ can be written as 
\be
\left(\prod_a \hD_a \right) \cC|_{\t \lambda_1 \lambda_1 \cdots \t \lambda_r \lambda_r}
={1 \ov r!} \left[ \left(\prod_b  \hD_a {\p^2 \ov \p \lambda_b \p \t \lambda_a} \right)
\cC \right]_{\lambda=\t \lambda=0} \,.
\ee
Meanwhile, the supersymmetry relation implies that
\be
\cD_b \cC =
2\cA_b \cC \,,
\ee
where we have defined the operators 
\be
\cD_b \equiv \hD_a {\p^2 \ov \p \lambda_b \p \t \lambda_a}, \quad
\cA_b \equiv {\p \ov \p \b \hs_b} - \lambda_a {\p^2 \ov \p \b \hs_a \p \lambda_b} \,.
\ee
The commutation relation between these operators are given by
\be
[\cD_a,\cD_b]=[\cA_a,\cA_b] =0\,, \qquad
[\cD_a,\cA_b] = -2 {\p \ov \p \b \hs_a} \cD_b \,.
\ee
We therefore see that 
\be
{1 \ov r!} \left(\prod_{b=1}^r \cD_b \right) \cC =
{2 \ov r!} \left(\prod_{b=1}^{r-1} \cD_b \right) \cA_r \cC = \cdots
\ee
can be ultimately written as a linear combination of terms of the form
\be
{\p \ov \p \b \hs_{a_1}} \cdots {\p \ov \p \b \hs_{a_p}}
 \cA_{b_1} \cdots \cA_{b_{r-p}} \cC \,.
\ee
The lowest components of these terms vanish, since they must be multiples of derivatives of $\cC_0$, which is zero. We therefore arrive at
\be
\cC|_{\t \lambda_1 \lambda_1 \cdots \t \lambda_r \lambda_r} = 0, \qquad
\text{when}\quad \cC|_{\lambda=\t \lambda=0} = 0\,.
\ee
Hence we find that the highest component of $\cZ_{k} (\hs, \b \hs, \lambda, \t \lambda, \hD)$ is given by
\be
\cZ_{k}
|_{\t \lambda_1 \lambda_1 \cdots \t \lambda_r \lambda_r}  =
\cZ_{k} (\hs, \b \hs, \hD) \exp(h^{bc} \t \lambda_b \lambda_c)
|_{\t \lambda_1 \lambda_1 \cdots \t \lambda_r \lambda_r}
= \det_{bc} (h^{bc}) \cZ_{k}  (\hs, \b \hs, \hD)\,.
\ee
It follows that 
\bea
\ZS&=
{1 \ov |W|} \sum_{k} q^k
\lim_{\substack{\epsilon \ra 0 \\ \e \ra 0}}
\int_{\Gamma \ltimes \tfM \setminus \De}
\prod_a (d\hs_a d \b \hs_a d \hD_a)
e^{-{\hD^2 \ov 2\e^2} + {4\pi i \t \xi \ov \e^2} (\hD)}
\det_{bc} (h^{bc}) \, \cZ_{k} (\hs, \b \hs, \hD) \\
&={1 \ov |W|} \sum_{k} q^k
\lim_{\substack{\epsilon \ra 0 \\ \e \ra 0}}
\int_{\Gamma \ltimes \tfM \setminus \De} \mu \,.
\label{muint}
\eea
The function $\cZ_{k}(\hs,\b \hs,\hD)$ is obtained by multiplying all the one-loop determinants of charged field components in the theory labeled by $\cI$ when the background zero modes $\hs_a, \b \hs_a$ and $\hD_a$ are turned on:
\be
\cZ_{k}(\hs,\b \hs,\hD) = \prod_\cI Z_{k}^\cI (\hs, \b \hs,\hD) \,.
\label{cZcomp}
\ee
Recall that the determinants of a component of a charged vector field is equivalent to that of a charged chiral field with $R$-charge $2$. The one-loop determinant of a charged field component $\cI$, whose computation is presented in appendix \ref{App: det}, is given by 
\bea
Z^{\cI}_{{k}} (\hs,\b \hs, \hD) =
Z^{(0)}_{r_\cI - Q_\cI (k)} (\Sigma,\b \Sigma,D) \cdot
Z^\pos_{r_\cI-Q_\cI (k)} (\Sigma,\b \Sigma,D)
|_{\Sigma=Q_\cI (\hs) + m_\cI^F,~D = Q_\cI (\hD)} \,,
\eea
where $Z^{(0)}_\mathbf{r}$ and $Z^\text{pos}_\mathbf{r}$ have been defined in equations \eq{Z0} and \eq{Zpos}, respectively. It follows that
\be
\cZ_{k}(\hs,\b\hs,\hD=0) = \prod_\cI Z^\cI_{k} (\hs) = \cZ^\oneloop_{k} (\hs) \,,
\label{cZzero}
\ee
as claimed. We find that the symmetric tensor $h^{ab}$ given by
\be
h^{ab}
= 2i \sum_{\cI} Q_a^\cI Q_b^\cI
F_{r_\cI-Q_\cI (k)} (\Sigma,\b \Sigma,D)|_{\Sigma=Q_\cI(\hs)+m_\cI^F,D=Q_\cI(\hD)} \,,
\label{habdef}
\ee
satisfies the conditions \eq{hab}. Here, $F_\mathbf{r} (\Sigma,\b\Sigma,D)$ is defined to be
\bea
F_\mathbf{r} (\Sigma,\b\Sigma,D) \equiv &~
\Theta(\mathbf{r} <1) \sum_{m=-|\mathbf{r}|/2}^{|\mathbf{r}|/2}
{1 \ov \b \Sigma (iD+ \b \Sigma(\Sigma+\epsdef m))} \\
&\qquad + \sum_{\substack{|m|<j \\ j > j_0(\mathbf{r})}}
{\Sigma+\epsdef m \ov \Delta_{j,m,\mathbf{r}}(\Sigma,\b \Sigma) 
\left(iD+ \Delta_{j,m,\mathbf{r}}(\Sigma,\b \Sigma) \right)}  \,,
\eea
with
\be
\Delta_{j,m,\mathbf{r}}(\Sigma,\b \Sigma) =
\b \Sigma(\Sigma+\epsdef m)+ j(j+1)
-{\mathbf{r}\ov 2} \left( {\mathbf{r} \ov 2} -1 \right) \,.
\ee
In particular, it can be explicitly verified that
\be
\p_{\b\hs_c}h^{ab}=\p_{\b\hs_a}h^{cb}
= 2i \sum_{\cI} Q_a^\cI Q_b^\cI Q_c^\cI
\p_{\b \Sigma}F_{r_\cI-Q_\cI (k)} (\Sigma,\b \Sigma,D)
|_{\Sigma=Q_\cI(\hs)+m_\cI^F,D=Q_\cI(\hD)} \,.
\ee
Note that the poles of $h^{ab}$ with respect to $\hD$ are a subset of the poles of $\cZ_{k} (\hs,\b\hs,\hD)$. Now defining
\be
\mu_{Q_1 \cdots Q_s} \equiv {(-2)^s \ov (r-s)!}
e^{-{\hD^2 \ov 2\e^2} +{4\pi i \t \xi \ov \e^2} \hD} \cdot
\cZ_{k} (\hs,\b\hs,\hD) d^r \hs
\wedge (\nu(d \hD))^{\wedge {(r-s)}} \wedge {d Q_1(D) \ov Q_1 (D)} \wedge
\cdots \wedge {d Q_s(D) \ov Q_s(D)} \,,
\label{muQ}
\ee
we find that
\be
d \mu_{Q_0 \cdots Q_s} = \sum_{l=0}^s (-1)^{s-l} \mu_{Q_0 \cdots \widehat{Q}_l \cdots Q_s} \,,
\label{dmu}
\ee
where $\widehat{~}$ implies omission. Note that for generic values of $\hs$, the only singular loci of $\mu_{Q_1 \cdots Q_s}$ with respect to $\hD$ when $\hD$ is real are given by the hyperplanes
\be
Q_l (\hD) =0 \,.
\ee

The inclusion of operators in the path integral does not alter these relations. Recall that the operator insertions $\cO(\sigma)$ at the north / south pole of the sphere corresponds to inserting a polynomial $\cO(\hs \pm {\epsdef k \ov 2})$ of $\hs_a$ in the integral. Since this polynomial only depends on $\hs_a$, this does not affect any of the relations the integration measure $\mu$ satisfies.

Now taking an appropriate cell-decomposition of the integration domain $\idom$, we find that the integral \eq{muint} can be written as \cite{Benini:2013xpa,Hori:2014tda}:
\be
\int_{\Gamma \ltimes \idom} \mu = -\sum_{p=1}^r ~\sum_{(\cI_1,n_1) < \cdots < (\cI_p,n_p)}~
\int_{\Gamma \times S_{(\cI_1,n_1)\cdots(\cI_p,n_p)}} \mu_{Q_{\cI_1} \cdots Q_{\cI_p}} \,,
\label{stokes}
\ee
due to \eq{dmu}. Some notation needs to be explained. Denoting the $\eps$-neighborhood of a hyperplane $H$ as $\Delta_\eps (H)$, we define
\be
S_{(\cI,n)} \equiv \p \Delta_\eps \cap \p \Delta_\eps(H^n_{\cI}) \,.
\ee
We also define the ``contour at infinity"
\be
S_\infty \equiv - \overline{(\p \tfM) \setminus \Delta_\eps}
\ee
where $\p \tfM$ is understood as the infinite radius limit of a $(2r-1)$-dimensional sphere enclosing a $2r$-dimensional ball inside $\tfM$. Let us denote the set of indices $(\cI,n)$ by $\mathscr{I}_S$, and $\mathscr{I}=\mathscr{I}_S \cup \{ \infty \}$. We order the indices so that $(\cI,n) < \infty$ for all $(\cI,n)$ and
\be
(\cI,n) < (\cI',n') \qquad
\text{if}\quad \cI < \cI' \quad
\text{or} \quad \cI = \cI'~\text{and}~n<n' \,.
\ee
Then $S_{(\cI_1,n_1)\cdots(\cI_p,n_p)}$ is defined as
\be
S_{(\cI_1,n_1)\cdots(\cI_p,n_p)} = \bigcap_{s=1}^p S_{(\cI_s,n_s)} \,.
\ee
Hence in order to compute the integral, one must define the $\hD$-contour $\Gamma$ for $\mu_{Q_{\cI_1} \cdots Q_{\cI_p}}$ when $\hs$ take values on $S_{(\cI_1,n_1)\cdots(\cI_p,n_p)}$. The vector $Q_\infty$, for now, is taken to be arbitrary.

Now let us show that condition (2) holds true. In particular, we construct the $\hD$ contour $\Gamma$ fibering over the domain of integration for $\hs$ and $\b\hs$, i.e., $\idom$, and find that it is equivalent to the $\hD$-contour used in \cite{Benini:2013xpa,Hori:2014tda}. What we show is that the manipulations shifting the $\hD$-contours carried out in those references can be carried out equivalently in our case as well.
Before doing so, let us extend upon the discussion at the end of the previous section. Due to the rather erratic behavior of poles of the integrand with respect to $\hD$, it is rather difficult to describe the geometry of the contour $\Gamma$ for macroscopic $\epsdef$. Upon assuming the separation of scales \eq{scales}, we gain enough control over the behavior of poles to describe the contours using hyperplanes, as demonstrated at the end of section \ref{section:rank one}. Following the process presented there, we can show for all charged field components $\cI$ that there exists a tube
\be
\cN_\cI = N_\cI \times {\rm Ker} Q_\cI \,,
\ee
such that
\begin{itemize}
\item $N_\cI$ is a disc of radius $\cO(\mu_N)$ encircling all the hyperplanes $H^\cI_n$, where
\be
\epsdef \ll \mu_N^2 \ll \mu_N \ll \mu_0
\ee
\item When $\hs \notin \cN_\cI$, all poles of $Z^\cI_{k} (\hs,\b\hs,\hD)$ with respect to $Q_\cI (\hD)$ lie above the contour
\be 
\Gamma_{Q_\cI,\delta} \equiv
\setcond{Q_\cI(\hD)}{\imag Q_\cI(\hD) = \delta}
\ee
on the $Q_\cI(\hD)$-plane if $\delta$ is a positive number of $\cO(|\epsdef|)$.
\item Even when $\hs \in \cN_\cI$, all poles of $Z^\cI_{k} (\hs,\b\hs,\hD)$ with respect to $Q_\cI (\hD)$ lie above the contour
\be 
\Gamma_{Q_\cI,-\delta} =
\setcond{Q_\cI(\hD)}{\imag Q_\cI(\hD) = - \delta}
\ee
on the $Q_\cI(\hD)$-plane if $\delta$ is a positive number of $\cO(|\epsdef|)$.
\end{itemize}
By assumption of projectivity, we can make $\epsdef$ and, accordingly, $\mu_N$ sufficiently small so that for any set of indices
\be
I=\{ \cI_1, \cdots, \cI_p\} \,,
\ee
the intersection
\be
\bigcap_{\cI \in I} \cN_\cI = \emptyset
\ee
when $Q_{\cI_1}, \cdots, Q_{\cI_p}$ do not lie within a half-plane of $i\mathfrak{h}^*$. Note that
\be
S_{(\cI,n)} \subset \cN_\cI \,.
\ee

Let us denote
\be
\cN_\text{tot} = \bigcup_\cI \cN_\cI \,.
\ee
We see that for values of $\hs \in \tfM \setminus \cN_\text{tot}$ the contour
\be
\Gamma = \setcond{\hD}{\imag \hD = \delta}
\label{generic}
\ee
for an $r$-dimensional vector $\delta$ of magnitude of $\cO(|\epsdef|)$ divides all the poles of the theory correctly, and thus defines a valid contour of integration for the integration density $\mu_{Q_{\cI_1} \cdots Q_{\cI_p}}$ for any $\cI_1,\cdots,\cI_p$. More precisely, the poles of $\cZ^\cI_{k} (\hs,\b\hs,\hD)$ lie above the projection of $\Gamma$ to the $Q_\cI (\hD)$-plane for all $\cI$. The construction of $\cN_\cI$ along with projectivity enables us to make a stronger statement. Let $\hs$ be a point in $\idom$ and let
\be
I(\hs) \equiv
\setcond{\cI}{\hs \in \cN_\cI} \,.
\ee
Then, $\Gamma$ of equation \eq{generic} is a valid contour of integration for $\hD$ at $\hs$---i.e., divides up the poles correctly---if $\delta$ is a vector of magnitude $\cO(|\epsdef|)$ and
\be
Q_\cI(\delta) < 0\,, \qquad
\text{for all} \quad \cI \in I(\hs) \,.
\ee
At this point we can define the $\hD$-integral of any $\mu_{Q_{\cI_1} \cdots Q_{\cI_p}}$ by specifying a contour $\Gamma$ with respect to $\delta$ on $\tfM \setminus \cN_\text{tot}$, computing the $\hD$-integral there, and analytically continuing to regions within $\cN_\cI$. This provides an existence proof for the appropriate ``deformation" of $\Gamma$ into all points in $\idom$. In order to evaluate \eq{stokes}, however, we need to provide more details about the $\hD$-integration contours on $S_{(\cI_1,n_1)\cdots(\cI_p,n_p)}$, which lie inside $\bigcap_{s=1}^p \cN_{\cI_s}$.

We now have the necessary tools for consistently assigning contours $\Gamma$ for the integration \eq{stokes}. The main difference of our situation compared to \cite{Benini:2013xpa,Hori:2014tda} is that in order for the contour of integration to be defined as a linear combination of products of hyperplanes and tori, the contour $\Gamma$ must be defined via the shifted $\hD$-contours, even before considering the integral \eq{stokes}, for points $\hs$ that lie inside any $\cN_\cI$. As we show shortly, the ``shifting'' is equivalent to that introduced in those works, as long as the shifts $\delta$ are taken to be of size $\cO(|\epsdef|)$. Note that $\mu_{Q_{\cI_1} \cdots Q_{\cI_p}}$ vanishes unless the charges $Q_{\cI_1}, \cdots, Q_{\cI_p}$ are linearly independent. Also, in evaluating equation \eq{stokes} we only need to specify the integration contour $\Gamma$ for the integral of $\mu_{Q_{\cI_1} \cdots Q_{\cI_p}}$ within $\bigcap_{s=1}^p \cN_{\cI_s}$.

Let us construct a $\hD$-contour $\Gamma$ for some covector $\eta \in i\mathfrak{h}^*$. Here $\eta$ is a book-keeping device that specifies the topology of the integration contour. In the previous section, $i\mathfrak{h}^*$ was one-dimensional, and $\eta>0$ / $\eta<0$ for the choice of contour $\Gamma_+$ / $\Gamma_-$ there. We assume that $\eta$ is generic so that it cannot be written as a sum of less than $r$ vectors $Q_\cI$. Given such a $\eta$, and a set of indices $\{ \cI_1, \cdots, \cI_p, {\cJ}_1, \cdots, {\cJ}_q \}$ for which the charges
\be
Q_{\cI_1}, \cdots, Q_{\cI_p}, Q_{{\cJ}_1}, \cdots, Q_{{\cJ}_q} 
\ee
are linearly independent, we define the vector $\delta_{\cI_{1} \cdots \cI_p \cancel{\cJ}_1 \cdots \cancel{\cJ}_q}$ that satisfies the following conditions:
\begin{itemize}
\item It is orthogonal to $Q_{\cI_s}$:
\be
Q_{\cI_1} (\delta_{\cI_{1} \cdots \cI_p \cancel{\cJ}_1 \cdots \cancel{\cJ}_q}) = \cdots
= Q_{\cI_p} (\delta_{\cI_{1} \cdots \cI_p \cancel{\cJ}_1 \cdots \cancel{\cJ}_q}) = 0 \,.
\ee
\item For any $\cJ$ such that
\be
\cN_\cJ \cap \left(\bigcap_{s=1}^q \cN_{\cJ_s}\right) \neq \emptyset \,,
\ee
the following holds:
\be
Q_{{\cJ}} (\delta_{\cI_{1} \cdots \cI_p \cancel{\cJ}_1 \cdots \cancel{\cJ}_q}) < 0 \,.
\ee
\item Its magnitude is of order $\cO(|\epsdef|)$, i.e., for any $\cJ$,
\be
Q_{{\cJ}} (\delta_{\cI_{1} \cdots \cI_p \cancel{\cJ}_1 \cdots \cancel{\cJ}_q}) = 0 \,,
\qquad \text{or} \qquad
Q_{{\cJ}} (\delta_{\cI_{1} \cdots \cI_p \cancel{\cJ}_1 \cdots \cancel{\cJ}_q}) \sim \cO(|\epsdef|)\,. 
\ee
\item $\delta_{\cI_1 \cdots \cI_p}$, with no slashed indices, satisfy the condition
\be
\eta(\delta)>0, \qquad
\eta(\delta_{\cI_1 \cdots \cI_p})>0 \,.
\ee
\end{itemize}
Using these vectors, we define the contours
\be
\Gamma_{\cI_{1} \cdots \cI_p \cancel{\cJ}_1 \cdots \cancel{\cJ}_q} =
\setcond{\hD \in \mathfrak{h}_\bC}{\imag\hD=
\delta_{\cI_{1} \cdots \cI_p \cancel{\cJ}_1 \cdots \cancel{\cJ}_q},~
Q_{\cI_s} (\hD)= 0~\text{for all $s$}} \times \ell_{\cI_{1} \cdots \cI_p}
\ee
where $\ell_{\cI_{1} \cdots \cI_p}$ is a small $p$-torus encircling $\bigcap_{s=1}^p \{\, Q_{\cI_s}(\hD) = 0\,\}$. As before,
\be
\Gamma = \setcond{\hD}{\imag \hD = \delta} \,.
\ee
We see that $\Gamma_{\cI_{1} \cdots \cI_p \cancel{\cJ}_1 \cdots \cancel{\cJ}_q}$ has the topology $\bR^{r-p}\times T^{p}$.

Now let us consider a point $\hs$ on $\idom$. Let
\be
\cI'_1, \cdots, \cI'_{p'} \in I(\hs) \,.
\ee
Then, for any $\cI_1, \cdots, \cI_p, \cJ_1, \cdots, \cJ_q$ such that
\be
\{ \cI'_1, \cdots, \cI'_{p'} \}
\subset
\{ \cI_1, \cdots, \cI_p, \cJ_1, \cdots, \cJ_q \}
\subset I(\hs) \,,
\ee
with linearly independent
\be
Q_{\cI_1}, \cdots, Q_{\cI_p}, Q_{{\cJ}_1}, \cdots, Q_{{\cJ}_q}  \,,
\ee
the contour
\be
\Gamma_{\cI_{1} \cdots \cI_p \cancel{\cJ}_1 \cdots \cancel{\cJ}_q}
\ee
divides the poles of $\mu_{Q_{\cI'_1} \cdots Q_{\cI'_{p'}}}$ correctly at $\hs$, by construction. Hence it is a valid contour of integration for $\mu_{Q_{\cI'_1} \cdots Q_{\cI'_{p'}}}$ at the point $\hs$.

Meanwhile, in $\hs \in \tfM \setminus \cN_\text{tot}$, the contours for the integral of $\mu_{Q_{\cK_1} \cdots Q_{\cK_t}}$ 
satisfy the relation
\begin{align}
\Gamma_{\cI_{1} \cdots \cI_p}
&\cong
\Gamma_{\cI_{1} \cdots \cI_p \cancel{\cJ}_1 \cdots \cancel{\cJ}_q}+
\Theta(Q_{\cJ_1} (\delta_{\cI_1, \cdots, \cI_p}))
\Gamma_{\cI_{1} \cdots \cI_p \cJ_1 \cancel{\cJ}_2  \cdots \cancel{\cJ}_q}+
\cdots \nonumber \\
&+ \prod_{t=1}^k \Theta(Q_{\cJ_s} (\delta_{\cI_1, \cdots, \cI_p}))
\Gamma_{\cI_{1} \cdots \cI_p \cJ_{s_1} \cdots \cJ_{s_k}
\cancel{\cJ}_{s_1'} \cdots \cancel{\cJ}_{s_{q-k}'}} + \cdots \label{decomp}
\\
&+ \prod_{s=2}^p \Theta(Q_{\cJ_s} (\delta_{\cI_1, \cdots, \cI_p}))
\Gamma_{\cI_{1} \cdots \cI_p \cancel{\cJ}_1 \cJ_2  \cdots {\cJ}_q}
+\prod_{s=1}^p \Theta(Q_{\cJ_s} (\delta_{\cI_1, \cdots, \cI_p}))
\Gamma_{\cI_{1} \cdots \cI_p {\cJ}_1 \cJ_2  \cdots {\cJ}_q}  \nonumber
\end{align}
for
\be
\{ \cK_s \} \subset \{ \cI_s \} \cup \{ \cJ_s \} \,.
\ee
The right-hand side of \eq{decomp} is a sum over $2^q$ terms where each index $\cJ_s$ appears either slashed or not. The congruence \eq{decomp} is at the level of homology on the ``punctured" complex $\hD$-space where hyperplanes of poles of $\mu_{Q_{\cK_1} \cdots Q_{\cK_t}}$ with respect to $\hD$ are removed. Hence, any contour, once split up in the manner \eq{decomp} can be continued to a point $\hs$ within
\be
\hs \in \bigcap_{\cI \in \{ \cI_s \} \cup \{ \cJ_s \}} \cN_{\cI} \,.
\ee

An important property of the contour $\Gamma_{\cI_{1} \cdots \cI_p \cJ_{s_1} \cdots \cJ_{s_k}
\cancel{\cJ}_{s_1'} \cdots \cancel{\cJ}_{s_{q-k}'}} $ in 
\eq{decomp} is that if
\be
\cK_s \in \{ {\cJ}_{s_l'} \} \,,
\label{slashed}
\ee
\emph{i.e.}, if any of the indices $\cK_s$ show up as a slashed index in the contour, then
\be
\lim_{\substack{\e \ra 0 \\ \eps \ra 0}}
\int_{\Gamma_{\cI_{1} \cdots \cI_p \cJ_{s_1} \cdots \cJ_{s_k}
\cancel{\cJ}_{s_1'} \cdots \cancel{\cJ}_{s_{q-k}'}}
\times S_{(\cK_1,n_1)\cdots(\cK_t,n_t)}} \mu_{Q_{\cK_1} \cdots Q_{\cK_t}} =0\,.
\label{vanishing}
\ee
This is because equation \eq{slashed} implies that the contour of \eq{vanishing} for the $\hD$ integral is such that $Q_{\cK_t} (\hD) <0$ and $Q_{\cK_t} (\hD) \sim \cO(|\epsdef|)$ on the contour. This means that all the poles of $\mu_{Q_{\cK_1} \cdots Q_{\cK_t}}$ in the $Q_{\cK_t} (\hD)$-plane, including the one at the origin, lie above the contour, and no poles cross the contour as $\eps$ is taken to zero. Hence the integrand vanishes in the limit $\eps \ra 0$.

We can now describe the integration contour $\Gamma$ for $\mu_{Q_{\cK_1} \cdots Q_{\cK_t}}$ for points $\hs$ inside $\bigcap_{s=1}^t \cN_s$. This in particular implies that $\cK_s \in I(\hs)$ for all $s$. Therefore, from previous discussions, the contour $\Gamma$ can be continued to such a point $\hs$ by the decomposition:
\bea
\Gamma
&=
\Gamma_{\cancel{\cK}_1 \cdots \cancel{\cK}_t}+
\Theta(Q_{\cK_1} (\delta))
\Gamma_{\cK_1 \cancel{\cK}_2  \cdots \cancel{\cK}_t}+
\cdots \\
&\qquad+ \prod_{s=2}^t \Theta(Q_{\cK_s} (\delta))
\Gamma_{\cancel{\cK}_1 \cK_2  \cdots {\cK}_t}
+\prod_{s=1}^t \Theta(Q_{\cJ_s} (\delta))
\Gamma_{{\cK}_1 \cK_2  \cdots {\cK}_t} \,.
\eea
This is precisely the contour used in \cite{Benini:2013xpa,Hori:2014tda} to evaluate the integrals of $\mu_{Q_{\cK_1} \cdots Q_{\cK_t}}$. Taking \eq{stokes}, and using \eq{vanishing} we arrive at the expression
\be
\int_{\Gamma \ltimes \idom} \mu = -\sum_{p=1}^r ~\sum_{(\cI_1,n_1) < \cdots < (\cI_p,n_p)}~
\prod_{s=1}^p \Theta(Q_{\cI_s} (\delta))
\int_{\Gamma_{\cI_1\cdots \cI_p} \times S_{(\cI_1,n_1)\cdots(\cI_p,n_p)}}
\!\!\!\!\!\!\!\!\!\!\!\!\!\!\!\!\!\!
\mu_{Q_{\cI_1} \cdots Q_{\cI_p}} \,.
\ee

Equation \eq{dmu}, the decomposition rule \eq{decomp}, and the vanishing rule \eq{vanishing} enable us to replicate the manipulations of \cite{Benini:2013xpa,Hori:2014tda} further to arrive at
\begin{align}
&\int_{\Gamma \ltimes \idom} \mu =\\
&= (-1)^r \sum_{(\cI_1,n_1) < \cdots < (\cI_r,n_r)< \infty}
\left(
\prod_{\cJ \in \{ \cI_1, \cdots, \cI_r \}}
\Theta(Q_\cJ (\delta_{\cI_1 \cdots \widehat{\cJ} \cdots \cI_r}))
\right) P_{(\cI_1,n_1) \cdots (\cI_r,n_r) } + \mathfrak{B} \,, \nonumber
\end{align}
where $P_{(\cI_1,n_1) \cdots (\cI_r,n_r)}$ is defined by
\be
P_{(\cI_1,n_1) \cdots (\cI_r,n_r)} \equiv
\lim_{\substack{\e \ra 0 \\ \eps \ra 0}}
\int_{\Gamma_{\cI_1\cdots \cI_r} \times S_{(\cI_1,n_1)\cdots(\cI_r,n_r)}}
\!\!\!\!\!\!\!\!\!\!\!\!\!\!\!\!\!\!
\mu_{Q_{\cI_1} \cdots Q_{\cI_r}}
\ee 
and $\mathfrak{B}$ is the ``boundary term" which we soon discuss. As noted in \cite{Benini:2013xpa},
\be
\Theta_{\{ \cI_1, \cdots, \cI_r \},\eta}
\equiv \prod_{\cJ \in \{ \cI_1, \cdots, \cI_r \}}
\Theta(Q_\cJ (\delta_{\cI_1 \cdots \widehat{\cJ} \cdots \cI_r})) =
\begin{cases}
1 &\text{if}~\eta \in \text{Cone} (Q_{\cI_1},\cdots,Q_{\cI_r})\\
0 &\text{otherwise.}
\end{cases}
\ee
Meanwhile, by definition, $\Gamma_{\cI_1 \cdots \cI_r}$ is an $r$-torus surrounding the point $\bigcap_s \{ Q_{\cI_s} (\hD) = 0 \}$ in complex $\hD$-space. Recalling the definition of $\mu_{Q_{\cI_1} \cdots Q_{\cI_r}}$ \eq{muQ}, and the relation \eq{cZzero}, i.e.,
\be
\cZ_{k}(\hs,\b\hs,\hD=0) = \cZ^\oneloop_{k}(\hs) \,,
\ee
we arrive at
\be
\int_{\Gamma_{\cI_1\cdots \cI_r} \times S_{(\cI_1,n_1)\cdots(\cI_r,n_r)}}
\!\!\!\!\!\!\!\!\!\!\!\!\!\!
\mu_{Q_{\cI_1} \cdots Q_{\cI_r}}
= (-4\pi i)^r \lim_{\substack{\e \ra 0 \\ \eps \ra 0}}
\int_{S_{(\cI_1,n_1)\cdots(\cI_r,n_r)}}
\!\!\!\!\!\!\!\!\!\!\!\!\!\!\!\!\!\!
\cZ^\oneloop_{k} (\hs)
d^r \hs
\ee
when $\cI_s \neq \infty$. Therefore we find that
\bea
\ZS &= {(4 \pi i)^r \ov |W|} \sum_{k} q^k
\sum_{\substack{\mathfrak{P}=\{ \cI_1, \cdots, \cI_r \} \cancel{\ni} \infty}} 
\Theta_{\mathfrak{P},\eta}
\lim_{\substack{\e \ra 0 \\ \eps \ra 0}}
\int_{S_{(\cI_1,n_1)\cdots(\cI_r,n_r)}}
\!\!\!\!\!\!\!\!\!\!\!\!\!\!\!\!\!\!
\cZ^\oneloop_{k} (\hs)
d^r \hs  + \mathfrak{B} \,.
\label{penultimate}
\eea
It has been shown in \cite{Benini:2013xpa} that the first term in the previous line can be identified with the Jeffrey-Kirwan residue:
\be
\sum_{\substack{\mathfrak{P}=\{ \cI_1, \cdots, \cI_r \} \cancel{\ni} \infty}} 
\Theta_{\mathfrak{P},\eta}
\lim_{\substack{\e \ra 0 \\ \eps \ra 0}}
\int_{S_{(\cI_1,n_1)\cdots(\cI_r,n_r)}}
\!\!\!\!\!\!\!\!\!\!\!\!\!\!\!\!\!\!
\cZ_{k} (\hs)
d^r \hs
= (2\pi i)^r \sum_{\hs_* \in \tfM^{k}_\text{sing}}
\underset{\hs = \hs_*}{\text{JK-Res}} [\mathbf{Q}(\hs_*),\eta]\, \cZ^\oneloop_{k} (\hs) \,.
\label{JK eta}
\ee
$\tfM^{k}_\text{sing}$ are the collection of codimension-$r$ singular points of $\cZ^\oneloop_{k} (\hs)$ on $\tfM$. We use $\mathbf{Q}(\hs_*)$ to denote the charge of the singular hyperplanes $Q_{\cI_1},\cdots,Q_{\cI_s}$ $(s \geq r)$ colliding at $\hs_*$. Note that when $S_{(\cI_1,n_1)\cdots(\cI_r,n_r)}$ is surrounding a non-degenerate codimension-$r$ pole $\hs_*$ of $\cZ^\oneloop_{k} (\hs)$, we get the iterated residue:
\be
\int_{S_{(\cI_1,n_1)\cdots(\cI_r,n_r)}}
\!\!\!\!\!\!\!\!\!\!\!\!\!\!\!\!\!\!
\cZ^\oneloop_{k} (\hs)
d^r \hs
= (2 \pi i)^r \, \underset{\hs = \hs_*}{\text{Res}} \,\cZ^\oneloop_{k} (\hs) \,.
\ee

To arrive at the final result, let us show that condition (3) holds. As explained in \cite{Hori:2014tda}, the boundary term $\mathfrak{B}$ is given by a linear combination of integrals of the form
\be
\lim_{\substack{\e \ra 0 }}
\int_{\Gamma_{\cI_1\cdots \cI_{q-1}} \times S_{(\cI_1,n_1)\cdots(\cI_{q-1},n_{q-1}) \infty}}
\!\!\!\!\!\!\!\!\!\!\!\!\!\!\!\!\!\!
\mu_{Q_{\cI_1} \cdots Q_{\cI_q-1} Q_\infty}
\label{infinity terms}
\ee
We take 
\be
Q_\infty = \xi_\eff^\text{UV} \quad \text{and} \quad
\eta = \xi_\eff^\text{UV} = \t\xi+{1 \ov 2\pi} b_0 \lim_{R \ra \infty} \log R \,.
\label{choice}
\ee
The precise definition of the $R \ra \infty$ limit is elaborated upon in sections \ref{subsec: summary} and \ref{subsec: JK}. As before, upon taking the limit $\e \ra 0$, we also take the boundary radius to grow faster than $\exp({{1 \ov \e^2}})$ so that
\be
R = |\hs|^{\e^2} \ra \infty \,.
\ee
Now when restricted to the contour $S_{(\cI_1,n_1)\cdots(\cI_{q-1},n_{q-1}) \infty}$, we find that
\bea
&\lim_{\substack{\e \ra 0 \\ R \ra \infty}}
e^{-{\e^2 {\hD'}{}^2 \ov 2}+4\pi i \t \xi (\hat D') }
\cZ_{k} (\hs,\b \hs, \e^2 \hat D') = \\
&\quad =\lim_{\substack{\e \ra 0 \\ R \ra \infty}}
e^{4\pi i \left[\t \xi+{1 \ov 2\pi} Q_{\cI_1 ,\cdots,\cI_{q-1}}
(1 + \epsdef' \alpha_\epsdef') \log R \right](\hD')}\cZ_k^\oneloop (\hs)
\label{cell at infinity}
\eea
where
\be
Q_{\cI_1 ,\cdots,\cI_{q-1}} = \sum_{Q_\cJ \notin P(Q_{\cI_1},\cdots,Q_{\cI_{q-1}})} Q_\cJ \,.
\ee
$P(Q_{\cI_1},\cdots,Q_{\cI_{q-1}})$ is the plane in $i \mathfrak{h}^*$ spanned by $Q_{\cI_1},\cdots,Q_{\cI_{q-1}}$. This is because if $Q_\cJ \in P(Q_{\cI_1},\cdots,Q_{\cI_{q-1}})$,
\be
|\overline{(Q_\cJ(\hs) + m^F_\cI )}(Q_\cJ(\hs) + m^F_\cI  + \epsdef)| \leq \cO(\mu_0) \,,
\ee
where $\mu_0$ is the macroscopic scale set by the masses of the theory. We thus find, for such indices $\cJ$ that
\be
\lim_{\substack{\e \ra 0 \\ |\hs| \ra \infty}} Z^\cJ_k (\hs,\b\hs,\e^2 \hD) = Z^\cJ_k (\hs) \,.
\ee
Meanwhile, for $Q_\cJ \notin P(Q_{\cI_1},\cdots,Q_{\cI_{q-1}})$
\be
|\overline{(Q_\cJ(\hs) + m^F_\cI )}(Q_\cJ(\hs) + m^F_\cI  + \epsdef) | \sim \cO(|\hs|) \,,
\ee
and thus
\be
\lim_{\substack{\e \ra 0 \\ R \ra \infty}} Z^\cJ_k (\hs,\b\hs,\e^2 \hD) =
\lim_{\substack{\e \ra 0 \\ R \ra \infty}} e^{2i (1 + \epsdef' \alpha_\epsdef')(\log R) Q_\cJ   \hD }Z^\cJ_k (\hs) \,,
\ee
from the behavior of the $\zeta$-regulated piece $Z^\pos$ of the determinant explained in appendix \ref{apsub: large sigma}. We hence arrive at the result \eq{cell at infinity}.

From \eq{cell at infinity}, the $\hD$-integral of the terms \eq{infinity terms} in $\mathfrak{B}$ is of the form
\bea
\lim_{\substack{\e \ra 0 \\ R \ra \infty}}
\int_{\Gamma_{\cI_1 \cdots \cI_{q-1}}}
&f( \cdots,\xi_\eff^{UV} (\e^2 \hD'))
\wedge {d \xi_\eff^{UV} (\hD') \ov \xi_\eff^{UV} (\hD')} \\
&\exp\left[ 
4 \pi i \left( \t \xi + {1 \ov 2 \pi} Q_{\cI_1 \cdots \cI_{q-1}}
(1 + \epsdef' \alpha_{\epsdef'})\log R \right)(\hD')  \right] \,,
\label{asymptotic}
\eea
where $f$ is a form that does not affect the asymptotics of the $\xi_\eff^{UV}(\hD)$-integral. Let us denote
\be
\xi' \equiv \lim_{\substack{R \ra \infty}}
\left( \t \xi + {1 \ov 2 \pi}Q_{\cI_1 \cdots \cI_{q-1}} \log R \right) \,.
\ee
By definition of the contours $\Gamma_{\cI_1 \cdots \cI_{q-1}}$, the integration contour for the variable $\xi_\eff^\text{UV} (\hD)$ lies within the plane
\be
Q_{\cI_1} (\hD) = \cdots = 
Q_{\cI_{q-1}} (\hD) = 0
\label{gamma plane}
\ee
in $\mathfrak{h}_\bC$. We therefore find that on $\Gamma_{\cI_1 \cdots \cI_{q-1}}$,
\be
\xi_\eff^\text{UV} (\hD) = \xi'(\hD) \,,
\ee
since
\be
Q_{\cJ} (\delta_{\cI_1 \cdots \cI_{q-1}}) (\hD)= 0
\qquad \text{for} ~
Q_{\cJ} \in P(Q_{\cI_1}, \cdots Q_{\cI_{q-1}}) 
\quad \text{when}~
\hD \in \Gamma_{\cI_1 \cdots \cI_{q-1}}\,.
\ee
Thus the integral \eq{asymptotic} may be rewritten as
\bea\label{asymptotic prime}
\lim_{\substack{\e \ra 0 \\ R \ra \infty}}
\int_{\Gamma_{\cI_1 \cdots \cI_{q-1}}}
&f( \cdots,\xi' (\e^2 \hD'))
\wedge {d \xi' (\hD') \ov \xi' (\hD')} \\
& \quad
\cdot \exp\left[ 4 \pi i \left( \xi' + {i \ov 2\pi} b_0 \cO(|\epsdef|) \log R \right) (\hD')  \right] \,.
\eea
As before, we have absorbed the real part of $(1+\epsdef'\alpha_{\epsdef'})$ in to the definition of $R$, and indicated the existence of a parametrically small real part to the exponent in \eq{asymptotic prime}. Meanwhile, $\Gamma_{\cI_1 \cdots \cI_{q-1}}$ is situated within the plane \eq{gamma plane} so that
\bea
\imag \xi'(\hD) = \imag \xi_\eff^\text{UV}(\hD)
= \xi_\eff^\text{UV} (\delta_{\cI_1 \cdots \cI_{q-1}})
> 0 \,,
\eea
and thus the contour of integration is above the pole $\xi'(\hD')=0$ and the $\xi'(\hD')$-integral and can be deformed away to positive infinity in the $\xi'(\hD)$ plane.%
~\footnote{When $b_0 \neq 0$, the definition of $\Gamma_{\cI_1 \cdots \cI_{q-1}}$ must be such that its projection to the $\xi'(\hD)$ plane is bent slightly upwards with angle $\cO(|\epsdef|)$. This slight modification to the definition of $\Gamma_{\cI_1 \cdots \cI_{q-1}}$ does not affect its position with respect to any of the poles of the integrand.}
Therefore the integral at boundaries of $\idom$ touching $\p \tfM$ vanish, given the choice \eq{choice} of $\eta$ and $Q_\infty$:
\be
\lim_{\substack{\e \ra 0 \\ \eps \ra 0}}
\int_{\Gamma_{\cI_1\cdots \cI_{q-1} \infty} \times S_{(\cI_1,n_1)\cdots(\cI_{q-1},n_{q-1}),\infty}}
\!\!\!\!\!\!\!\!\!\!\!\!\!\!\!\!\!\!\!\!\!\!\!\!\!\!\!\!\!\!\!\!\!\!\!\!
\mu_{Q_{\cI_1} \cdots Q_{\cI_{q-1}}(Q_\infty=\xi_\eff^\text{UV})} =0 \,.
\label{infinity vanishing}
\ee

As demonstrated in the previous section, the integral \eq{stokes} does not depend on the choice of $\eta$ or $Q_\infty$ \cite{Benini:2013xpa,Hori:2014tda}. Thus setting $\eta =Q_\infty=\xi_\eff^\text{UV}$ and combining \eq{penultimate} with \eq{JK eta} and \eq{infinity vanishing}, we arrive at:
\bea\label{ZS for higher rank}
\ZS =
{(-8 \pi^2)^r \ov |W|} \sum_{k} q^k
\sum_{\hs_* \in \tfM^{k}_\text{sing}}
\underset{\hs = \hs_*}{\text{JK-Res}} [\mathbf{Q}(\hs_*),\xi_\eff^\text{UV}]\, \cZ^\oneloop_{k} (\hs) \,.
\eea
The physical correlators straightforwardly follow, leading to the main result \eq{main formula}.

The derivation given in this section does not rely much on the details we have assumed about the localization saddles parametrized by the Coulomb branch coordinates $\hs$ and $\t\hs$, and the determinants $Z^\cI_k(\hs,\t\hs,\hD)$. In fact, much of the effort in this section has been geared toward proving that $Z^\cI_k(\hs,\t\hs,\hD)$ satisfy certain desirable properties. The properties needed to arrive at the formula \eq{ZS for higher rank} are given by the following:
\begin{itemize}
\item $\hs$ and $\t\hs$ are independent coordinates on the moduli space of Coulomb branch saddles.
\item $Z^\cI_k(\hs,\t\hs,\hD)$ factors into the function $Z^{(0)}$ defined in equation \eq{Z0}, and a function $Z^\pos$ whose poles with respect to $\hD$ safely lie in the upper-half of the complex $\hD$ plane for any $\hs$ and $\t \hs$ for small enough $\epsdef$.
\item The asymptotics of $Z^\pos$.
\end{itemize}

Let us end this section with a remark on the JK vector $\xi_\eff^\text{UV}$. Note that while $\t\xi$ is not the physical FI parameter, as explained below equation \eq{Zinf}, it acts as a FI parameter for the localizing Lagrangian \eq{Coulomb Lagrangian for 5}. More precisely, the fictitious UV FI parameter $\xi_\eff^\text{UV}$ determines the ``phase" of the Higgs branch of the localizing Lagrangian. We thus expect that the main formula \eq{main formula} for the correlators to be a convergent series in $q$ only when $\xi$ and $\xi_\eff^\text{UV}$ are aligned, i.e., when the phase of the localizing Lagrangian coincides with the phase of the physical theory. Even when the physical FI parameter $\xi$ is taken so that these phases do not match, equation \eq{main formula} still should produce a formally correct series expansion of the correlators, although the series is not expected to be convergent for such values of $\xi$.


\section{Quantum cohomology and recursion relations in $\epsdef$} \label{section: Q Cohomology}

In this section, we investigate relations that expectation values of operators must satisfy from a general perspective, while the correlators of specific theories are computed explicitly in the following sections. In particular, when the $\epsdef$ parameter is turned on, the correlators of the theory satisfy certain recursion relations that can be used to efficiently compute them.
Such relations are more easily derivable in abelian theories---for each gauge group factor $U(1)_a$ of an abelian theory, we find that
\begin{align}
\label{recursion}
&\left\langle{\cO^{(N)}(\sigma_N) \cO^{(S)}(\sigma_S)
\prod_{i, \, Q_i^a > 0} \, \prod_{l=0}^{Q_i^a-1}
\left[ Q_i( \sigma_N) +m^F_i
+\epsdef \left( {r_i \ov 2} + l \right) \right]} \right\rangle \\
&~ = q_a \cdot \left\langle{
\cO^{(N)}(\sigma_N\!-\epsdef\delta(a)) \cO^{(S)}(\sigma_S)
\prod_{i, \, Q_i^a < 0} \, \prod_{l=0}^{|Q_i^a|-1}
\left[ Q_i(\sigma_N) +m^F_i
+\epsdef \left( {r_i \ov 2} + l \right) \right]
} \right\rangle \,, \nonumber
\end{align}
assuming that all the singularities of $\cZ^\oneloop_k$ are projective.%
~\footnote{These relations can be violated, given that there exist non-projective singularities of $\cZ^\oneloop_k$ in the theory. Explicit examples are presented in the following sections.}
We have defined the vector $\delta(a) \in i\mathfrak{h}$ by
\be
\delta(a)_b = \delta_{ab} \,.
\ee
As before, the index $i$ labels the chiral fields of the theory. The product on the left-hand side of this equation is over the chiral fields $i$ with positive charge under $U(1)_a$, while the product over $i$ on the right-hand side is over chiral fields that have negative charge. 
For sake of brevity, we focus on recursion relations for operator insertions at the north pole, with the south pole insertion $\cO^{(S)}(\sigma_S)$ in \eqref{recursion} as a spectator (relations for operators at the south pole are obtained by replacing $N\leftrightarrow S$ and $\epsdef \leftrightarrow - \epsdef$). It is immediate to see that the identity \eq{recursion} reduces to the quantum cohomology ring relations in the limit $\epsdef \ra 0$, where we obtain the $A$-twisted theory.

The relations \eq{recursion} can be summarized efficiently by considering the generating function
\be
F(z) \equiv \left\langle e ^{z^b \sigma_b}|_N \right\rangle
\ee
for the operator insertions at the north pole. Equation \eq{recursion} can be straightforwardly translated as a set of differential equations for $F(z)$:
\bea\label{PF equation from recursion}
&\prod_{i, \, Q_i^a > 0} \, \prod_{l=0}^{Q_i^a-1}
\left[ Q_i^b \p_{z^b} +m^F_i
+\epsdef \left( {r_i \ov 2} + l \right) \right] F(z) \\
&= q_a e^{-z^a\epsdef}\prod_{i, \, Q_i^a < 0} \, \prod_{l=0}^{|Q_i^a|-1}
\left[ Q_i^b \p_{z^b} +m^F_i
+\epsdef \left( {r_i \ov 2} + l \right) \right] F(z) \,.
\eea

We investigate abelian GLSMs in section \ref{ss: Abelian} and present a proof for the formula \eq{recursion}.  This relation follows from the structure of the function $\cZ^\oneloop_k$. We verify the quantum restriction formula of \cite{Morrison:1994fr,Hori:2000kt}, and derive its $\epsdef$-deformed version in a similar fashion. An elegant presentation of the recursion relations for non-abelian theories is still lacking, although expectation values of gauge invariant operators of the theory can be computed using the associated Cartan theory. In section \ref{ss: UN}, we restrict ourselves to discussing how to verify the quantum cohomology relations of correlators of the $A$-twisted variables for $U(N_c)$ theories with fundamental and anti-fundamental matter.

\subsection{Abelian GLSMs} \label{ss: Abelian}

Let us consider abelian GLSMs with $\mathbf{G}=U(1)^\rk$ and matter $\Phi_i$ carrying charge $Q^a_i$ under the gauge group $U(1)_a$. We normalize the gauge fields so that the charges, and hence the fluxes, are quantized to be integers. Let us now derive some relations between vacuum expectation values of operators that follow from the properties of the one-loop determinant $\cZ^\oneloop_k(\hs;\epsdef)$. We focus, for sake of simplicity, on the case when operators are inserted at the north pole of the sphere. Insertions on the south pole can be treated in a similar fashion.

For abelian theories, the one-loop determinant may be written as
\be
\cZ^\oneloop_k(\hs;\epsdef)
= \prod_i \left(
\epsdef^{r_i - Q_i (k)-1}
{\Gamma({Q_i(\hs)+m_i^F \ov \epsdef}+{r_i-Q_i(k) \ov 2}) \ov
\Gamma({Q_i(\hs)+m_i^F \ov \epsdef}-{r_i-Q_i(k) \ov 2}+1)}
\right)
\ee
One may explicitly verify the identity
\bea\label{one loop rel}
&\cZ_k^\oneloop(\hs;\epsdef)
\prod_{i, \, Q_i^a > 0} \, \prod_{l=0}^{Q_i^a-1}
\left[ Q_i^b \left(  \hs_b - {\epsdef k_b \ov 2}\right)+m^F_i
+\epsdef \left( {r_i \ov 2} + l \right) \right] \\
&=\cZ_{k-\delta(a)}^\oneloop(\hs+{\epsdef \delta(a)\ov 2};\epsdef) \\
&\quad \cdot
\prod_{i, \, Q_i^a < 0} \, \prod_{l=0}^{|Q_i^a|-1}
\left[ Q_i^b \left(  \hs_b+\epsdef{\delta_{ab} \ov 2} - {\epsdef (k_a-\delta_{ab}) \ov 2}\right)+m^F_i
+\epsdef \left( {r_i \ov 2} + l \right)\right]
\eea
for each $a$, where we have defined the vector $\delta(a) \in i\mathfrak{h}$ by
\be
\delta(a)_b = \delta_{ab} \,.
\ee
Now the JK residue
\bea\label{recursion lhs}
\sum_k q^k \, &
\text{JK-Res} \left[\xi_\eff^\text{UV}\right]
\cZ_k^\oneloop(\hs;\epsdef)
\cO^{(N)}(\hs-\epsdef{k \ov 2})\cO^{(S)}(\hs+\epsdef{k \ov 2}) \\
&\qquad\cdot \prod_{i, \, Q_i^a > 0} \, \prod_{l=0}^{Q_i^a-1}
\left[ Q_i^b \left(  \hs_b - {\epsdef k_b \ov 2}\right)+m^F_i
+\epsdef \left( {r_i \ov 2} + l \right) \right]
\eea
yields the vacuum expectation value
\be
\left\langle{\cO^{(N)}(\sigma_N)\cO^{(S)}(\sigma_S)
\prod_{i, \, Q_i^a > 0} \, \prod_{l=0}^{Q_i^a-1}
\left[ Q_i( \sigma_N) +m^F_i
+\epsdef \left( {r_i \ov 2} + l \right) \right]} \right\rangle \,. 
\ee
We choose to neglect the volume form $d \hs_1 \wedge \cdots \wedge d \hs_\rk$ in the residue formulae such as \eq{recursion lhs} to avoid clutter. Meanwhile, the sum
\bea\label{recursion rhs}
\sum_k q^k &
\text{JK-Res} \left[\xi_\eff^\text{UV}\right]
\cZ_{k-\delta(a)}^\oneloop(\hs+{\epsdef \delta(a)\ov 2};\epsdef)
\cO^{(N)}(\hs-\epsdef{k \ov 2})\cO^{(S)}(\hs+\epsdef{k \ov 2})\\
&\quad \cdot
\prod_{i, \, Q_i^a < 0} \, \prod_{l=0}^{|Q_i^a|-1}
\left[ Q_i^b \left(  \hs_b+\epsdef{\delta_{ab} \ov 2} - {\epsdef (k_a-\delta_{ab}) \ov 2}\right)+m^F_i
+\epsdef \left( {r_i \ov 2} + l \right)\right]
\eea
can be massaged into
\bea
q_a \sum_k q^k 
\text{JK-Res} \left[\xi_\eff^\text{UV}\right]
& \cZ_{k}^\oneloop(\hs;\epsdef) \cO^{(N)}(\hs-\epsdef{k \ov 2}-\epsdef \delta(a))
\cO^{(S)}(\hs+\epsdef{k \ov 2})\\
&\quad \cdot
\prod_{i, \, Q_i^a < 0} \, \prod_{l=0}^{|Q_i^a|-1}
\left[ Q_i^b \left(  \hs_b - {\epsdef k_a \ov 2}\right)+m^F_i
+\epsdef \left( {r_i \ov 2} + l \right)\right]
\eea
by shifting $\hs \ra \hs-{1 \ov 2}\epsdef \delta(a)$ and $k \ra k+\delta(a)$. This expression is equal to
\be
q_a \cdot \left\langle{
\cO^{(N)}(\sigma_N\!-\epsdef \delta(a))
\cO^{(S)}(\sigma_S)
\prod_{i, \, Q_i^a < 0} \, \prod_{l=0}^{|Q_i^a|-1}
\left[ Q_i(\sigma_N) +m^F_i
+\epsdef \left( {r_i \ov 2} + l \right) \right]
} \right\rangle \,.
\ee
We thus arrive at the desired relation: 
\begin{align}
\label{recursion 2}
&\left\langle{\cO^{(N)}(\sigma_N) \cO^{(S)}(\sigma_S)
\prod_{i, \, Q_i^a > 0} \, \prod_{l=0}^{Q_i^a-1}
\left[ Q_i( \sigma_N) +m^F_i
+\epsdef \left( {r_i \ov 2} + l \right) \right]} \right\rangle \\
&~ = q_a \cdot \left\langle{
\cO^{(N)}(\sigma_N\!-\epsdef\delta(a)) \cO^{(S)}(\sigma_S)
\prod_{i, \, Q_i^a < 0} \, \prod_{l=0}^{|Q_i^a|-1}
\left[ Q_i(\sigma_N) +m^F_i
+\epsdef \left( {r_i \ov 2} + l \right) \right]
} \right\rangle \,. \nonumber
\end{align}
Using these relations, the correlation functions involving operators of degree
\be
\max\left( \sum_{i,Q_{i}^a>0} Q_{i}^a\,, \,\sum_{i,Q_{i}^a<0} |Q_{i}^a| \right)
\ee
in $\sigma$ inserted at the north pole of the sphere can be written in terms of correlators with operators of lower degree. In special cases, the relation \eq{recursion 2} may simplify further, by replacing $\cO^{(N)}(\sigma_N)$ by $\cO^{(N)}(\sigma_N)/f(\sigma_N)$ for some polynomial $f$, given that $f(\sigma_N)$ divides
\be
\prod_{i, \, Q_i^a > 0} \, \prod_{l=0}^{Q_i^a-1}
\left[ Q_i( \sigma_N) +m^F_i
+\epsdef \left( {r_i \ov 2} + l \right) \right]
\ee
and $f(\sigma_N-\epsdef \delta(a))$ divides
\be
\prod_{i, \, Q_i^a < 0} \, \prod_{l=0}^{|Q_i^a|-1}
\left[ Q_i(\sigma_N) +m^F_i
+\epsdef \left( {r_i \ov 2} + l \right) \right] \,.
\ee
This applies, for example, to the quintic GLSM studied in section \ref{subsec:_quintic}.

Now in deriving \eq{recursion 2} we have assumed that the operation $\text{JK-Res} \left[\xi_\eff^\text{UV}\right]$ is well-defined. Thus, the formula \eq{recursion 2} may fail in the presence of non-projective singularities of the differential forms of \eq{recursion lhs} or \eq{recursion rhs}. In those cases, the non-projective singularities must be tamed by introducing additional twisted masses to the theory and expectation values must be computed by first doing the computation in the deformed theory and by gradually turning off the twisted masses. The expectation values with vanishing twisted masses obtained this way are not guaranteed to satisfy the recursion relations \eq{recursion 2}.

Let us conclude the discussion of abelian theories by showing that the quantum restriction formula of \cite{Morrison:1994fr,Hori:2000kt} for abelian GLSMs is naturally realized as a property of the one-loop determinant $\cZ^\oneloop_k$. The quantum restriction formula is relevant to computing quantum correlators of complete intersections in compact toric varieties. Let us review the presentation of the formula following \cite{Hori:2000kt}. A compact toric variety $X$ is engineered by an abelian GLSM with gauge group $\mathbf{G}=U(1)^\rk$ with chiral fields $\Phi_i$ whose charges $Q_i^a$ lie within a half-plane of $i\mathfrak{h}^*$. We would like to understand the GLSM that engineers the non-linear sigma model of a complete intersection manifold $M$ defined by the equations
\be
G_\beta (\Phi_i) =0 \,, \quad \beta=1,\cdots,\ell
\ee
in the toric variety in the infra-red limit.  The $G_\beta (\Phi_i)$ are charge $d_\beta^a$-operators. We now introduce the fields $P_\beta$ with charge $-d_{\beta}^a$ to the theory, and add the superpotential
\be
W = \sum_{\beta=1}^\ell P_\beta G_\beta(\Phi_i) \,.
\ee
In this theory, which we denote $\mathscr{T}_M$, the $R$-charges of the $\Phi_i$ fields are taken to be $0$, while those of the $P$-fields are taken to be $2$. This theory engineers the desired manifold $M$ in the IR in the $\xi>0$ phase  \cite{Witten:1993yc}.
Meanwhile, we can also consider the theory $\mathscr{T}_V$, that engineers the \emph{non-compact} toric variety defined by the fields $\Phi_i$ and $P_\beta$ together. In this theory, all the $R$-charges of the fields are set to zero. The quantum restriction formula relates the expectation values of $A$-twisted operators of theory $\mathscr{T}_M$ and theory $\mathscr{T}_V$ (with vanishing twisted masses) by
\be\label{qrf}
\vev{\cO(\sigma)}_{\mathscr{T}_M,0}
=(-1)^\ell \left\langle
\cO(\sigma) \prod_{\beta=1}^\ell d_\beta(\sigma)^2
\right\rangle_{\mathscr{T}_V,0} \,.
\ee
The subscript ``$0$" in equation \eq{qrf} signifies that the expectation value is taken in the $A$-twisted theory.

The relation \eq{qrf} is a natural consequence of the properties of $\cZ^\oneloop_k (\hs;\epsdef)$. Let us denote the one-loop determinant in flux sector $k$ of the theory $\mathscr{T}_M$ and $\mathscr{T}_V$ as $\cZ^\oneloop_{M,k} (\hs;\epsdef)$ and $\cZ^\oneloop_{V,k} (\hs;\epsdef)$, respectively. It then follows that 
\be
\cZ^\oneloop_{M,k} (\hs;\epsdef) =
\cZ^\oneloop_{V,k} (\hs;\epsdef)
\prod_{\beta=1}^\ell \left(d_\beta(\sigma)^2-\epsdef{d_\beta(k)^2 \ov 4} \right) \,.
\ee
We thus arrive at the deformed quantum restriction formula:
\be\label{deformed qrf}
\vev{\cO(\sigma_N)\cO(\sigma_S)}_{\mathscr{T}_M}=
(-1)^\ell \left\langle
\cO(\sigma_N)\cO(\sigma_S)
\prod_\beta
\left(d_\beta(\sigma_N)d_\beta(\sigma_S) \right)
\right\rangle_{\mathscr{T}_V} \,,
\ee
where the sign $(-1)^\ell$ follows from the prescription of section \ref{subsec: oneloop det}. This equation reduces to the relation \eq{qrf} in the $A$-twisted theory.

\subsection{Non-abelian GLSMs} \label{ss: UN}

Let us now discuss the quantum cohomology of theories with non-abelian gauge symmetry. While the correlators of gauge invariant operators of the theory can be efficiently computed using the recursion relations \eq{recursion 2} by utilizing the associated Cartan theory, these relations have not yet been written in an elegant way comparable to the quantum cohomology relations that can be found, for example, in \cite{Givental:1993nc,Witten:1993xi,Astashkevich:1993ks,Nekrasov:2009uh}.

We can check in certain examples that the correlation functions computed using localization techniques satisfy the required quantum cohomology relations. The representative example is a $U(N_c)$ theory with $N_f$ fundamental and $N_a$ antifundamental chiral fields. Let us also make the technical assumption that the $R$-charges of the fields are favorable, so that equation \eq{formula Amodel iii} is applicable. Turning on generic twisted masses $-m^F_1,\cdots,-m^F_{N_f}$ for the fundamentals and $\t m^F_1,\cdots, \t m^F_{N_a}$ for the antifundamentals, the $A$-twisted correlators of the theory can be written as
\bea\label{UN-A}
\vev{\cO(\sigma)}_0 = \mathscr{N} \oint_{ \p \tfM}
&\left( \prod_{a=1}^{N_c} {d \hs_a \ov 2\pi i}
\prod_{i=1}^{N_f} (\hs_a -m_i^F)^{r_i-1}
\prod_{\t i=1}^{N_a} (-\hs_a +\t m_{\t i}^F)^{\t r_{\t i}-1} \right) \\
&\cdot \prod_{a<b} (\hs_a - \hs_b)^2
\cdot
{e^{2\pi i r^a_0 \p_{\hs_a} \h W_\eff} \ov
\prod_a (1- e^{2 \pi i \p_{\hs_a} \h W_\eff})}
\cdot
 \cO(\hs)\,.
\eea
for some integers $r_0^a$. We can be cavalier about the overall normalization constant $\mathscr{N}$ for the purposes of this section, as we are interested in verifying the quantum cohomology relations.

Now let us insert any quantum cohomology relation of the form
\be
f(\sigma)=0
\ee
in the expectation value, where $f$ can be written as a Weyl-invariant polynomial of the variables $\sigma_a$. For generic twisted masses, the quantum cohomology relations are the relations that the isolated solutions $(\sigma_a)$ to
\be\label{UN solutions}
e^{2 \pi i \p_{\sigma_a} \h W_\eff}=1, \qquad
\sigma_a \neq \sigma_b \quad
\text{for} \quad
a \neq b
\ee
satisfy \cite{Nekrasov:2009uh}. That is, $f(\sigma_0)=0$ for any solution $\sigma_0$ of \eq{UN solutions}. Assuming that generic twisted masses are turned on, the integral \eq{UN-A} is given by the sum of residues of the integrand located precisely at the solutions of equation \eq{UN solutions}. Note that the non-degeneracy condition of the solutions is enforced by the Vandermonde determinant in the integral \eq{UN-A}. Thus we find that
\be
\vev{f(\sigma) \cO(\sigma)}_0 = 0
\ee
for any quantum cohomology relation $f$. This proof is expected to extend to a large class of examples, including quiver gauge theories. It would be interesting to lift some of the simplifying assumptions we have made in this section and see if the quantum cohomology relations  still  can be derived.

\section{Examples: Correlators with $\Omega$-deformation}\label{sec: Examples 1}

In this section, we apply the Coulomb branch localization formula \eqref{formula roughly} to some abelian gauge theories and we compute correlators with all the insertions at the north pole. (This is for simplicity of presentation. Considering insertions at both the north and south poles is straightforward.)  These correlation functions satisfy the finite difference equations \eqref{recursion}, that may be used to compute them recursively.  We also study generating functions of north pole correlators, which satisfy differential equations of Picard-Fuchs type \cite{Candelas:1990rm} due to the finite difference equations \eqref{recursion}.

\subsection{The abelian Higgs model}\label{subsec:_AHM}

The abelian Higgs model is a $U(1)$ gauge theory with a single chiral multiplet $\Phi$ of charge $Q$, which we assume to be positive in the following with no loss of generality. The effective FI parameter in the UV is $\xi_{\rm eff}^{\rm UV}=+\infty$, forcing a non-vanishing VEV for $\Phi$. The vacuum moduli space consists of $Q$ points related by a residual $\bZ_Q$ gauge symmetry. In the UV geometric phase where $\Phi$ takes VEV, the JK residue ${\rm JK{\text-}Res}\left[\xi_{\rm eff}^{\rm UV}\right]$ is a sum of residues at the poles of the 1-loop determinant of $\Phi$,
\begin{equation}
 Z_k^{\Phi}(\hs;\epsdef) = 
\begin{cases}
\prod\limits_{p=0}^{Qk} (Q\hs +(p-Q\frac{k}{2})\epsdef)^{-1}  & Q k\geq 0\\
1 & Qk=-1 \\
\prod\limits_{p=0}^{-Qk-2} (Q\hs + (1 + Q\frac{k}{2} + p)\epsdef) & Qk\leq -2
\end{cases} ~.\label{1-loop_AHM}
\end{equation}
$Z_k^\Phi$ has poles only for $k\geq 0$ if $Q>0$, therefore the sum over $k$ in \eqref{formula roughly} reduces to $k\geq 0$. This is the first instance of a general phenomenon that we explained in section \ref{subsec: JK} and that we will encounter repeatedly: the summation over $k$ effectively reduces to the closure \eqref{chamber for k} of the cone dual to the cone in FI parameter space that defines the UV phase of the GLSM, up to finite shifts due to the R-charge.  This general structure arises naturally in the approach of \cite{Morrison:1994fr} to the sum over instantons and it is satisfying to see it appear also in our formalism. In the case of the abelian Higgs model, the GLSM is in the phase $\xi>0$ in the UV: correspondingly, the sum over $k$ reduces to $k \geq 0$.

Inserting twisted chiral operators only at the north pole and shifting the integration variable $\hs\to\hs+\frac{k}{2}\epsdef$ (in other words, the new $\hs$ is $\hs_N$), the localization formula \eqref{formula roughly} becomes 
\be\label{AHM_Coulomb}
\langle  \sigma_N^n \rangle = \sum_{k=0}^\infty \, q^k  \sum_{\ell=0}^{Qk}  \res_{\hs=-\frac{\ell}{Q}\epsdef}  \frac{\hs^n}{\prod\limits_{p=0}^{Qk} (Q \hs + p \epsdef)} = -\sum_{k=0}^\infty \, q^k \res_{\hs=\infty} \frac{\hs^n}{\prod\limits_{p=0}^{Qk} (Q \hs + p \epsdef)}~.
\ee
From the latter expression in \eqref{AHM_Coulomb} it is straightforward to obtain the first few expectation values
\begin{equation}\label{correlators_AHM}
\begin{split}
\langle 1 \rangle = \frac{1}{Q}~, \qquad \langle  \sigma_N^n \rangle =0\quad (n=1,\dots,Q-1)~, \qquad \langle  \sigma_N^Q \rangle =\frac{1}{Q^{Q+1}} q~.
\end{split}
\end{equation}
Higher correlators can be computed from the first expression in \eqref{AHM_Coulomb}, which yields
\begin{equation}\label{correlators2_AHM}
\begin{split}
\langle  \sigma^n_N \rangle =  \frac{(-\epsdef)^n}{Q^{n+1}} \sum_{k=0}^\infty (q \epsdef^{-Q})^k \sum_{l=0}^{Qk} \frac{(-1)^l l^n}{(Qk-l)!l!}~,
\end{split}
\end{equation}
or alternatively using the identity \eqref{recursion} 
\be\label{AHM_finite_diff}
\langle \cO(\s_N)
\prod_{p=0}^{Q-1} (Q \s_N+p \epsdef)\rangle = q \langle \cO(\s_N-\epsdef)\rangle~,
\ee
which follows directly from \eqref{AHM_Coulomb}. \eqref{AHM_finite_diff} is the $\epsdef$-deformed version of the twisted chiral ring (or quantum cohomology) relation $(Q\s)^Q = q$, and allows to compute recursively the higher correlators $\langle  \sigma^n_N \rangle$ with $n\ge Q$ from the lower ones.%
~\footnote{A similar formula holds for insertions at the south pole, with $\s_N\to \s_S$ and $\epsdef\to -\epsdef$. More generally, one can consider insertions at the north and south pole. Then operators inserted at one pole are spectators in the recursion relation for operators inserted at the other pole.}

The finite difference equation \eqref{AHM_finite_diff} can be recast into a differential equation in $z$ for the generating function 
\be\label{generating_fn_AHM}
F(z) = \langle e^{z \s_N}\rangle = \sum_{n=0}^\infty \frac{z^n}{n!}  \langle \s^n_N\rangle
\ee
of twisted chiral correlators inserted at the north pole:
\be\label{differential_AHM}
\left[\prod_{p=0}^{Q-1} (Q \d_z +p \epsdef) - q e^{-z\epsdef} \right] F(z) = 0~.
\ee
Changing variable at non-vanishing $\epsdef$ from $z$ to 
\be
\mathbf{q}_z=  q e^{-z\epsdef} (-\epsdef)^{-Q}~,
\ee
the differential equation \eqref{differential_AHM} can be written in the standard Picard-Fuchs form
\be\label{PF_AHM}
\left[\prod_{p=0}^{Q-1} (Q \Theta -p) - \mathbf{q}_z \right] F(z) = 0~,
\ee
where $\Theta\equiv \mathbf{q}_z\frac{\d}{\d\mathbf{q}_z}$. 
In this case the generating function $F(z)$ has the simple closed form
\be
F(z)=\frac{1}{Q^2} \sum _{m=1}^Q \exp\left(2\pi i \left(\h{W}_m(\mathbf{q}_z) - \h{W}_m(\mathbf{q}_0) \right)\right)
\ee
in terms of the on-shell twisted effective superpotential in the $m$-th vacuum \be
\h{W}_m(q)=\frac{1}{2\pi i} q^{1/Q} e^\frac{2 \pi  i m}{Q}
~, \qquad m=1,\dots,Q~.
\ee
Here $\{\exp (2\pi i \h{W}_m(\mathbf{q}_z))\}_{m=1}^Q$ is a basis of solutions of \eqref{differential_AHM} or \eqref{PF_AHM}, and the coefficients are fixed by the initial condition  $F(z)=\frac{1}{Q}+\cO(z^Q)$ around $z=0$, according to \eqref{correlators_AHM}. 

In the $A$-model limit $\epsdef\to 0$, the generating function reduces to
\be
F(z)_{\epsdef=0}=\langle e^{z\s}\rangle_{0} =\frac{1}{Q^2} \sum _{m=1}^Q \exp\left( \frac{1}{Q}  e^\frac{2 \pi  i m}{Q} q^{1/Q} z\right)~,
\ee
while the correlators \eqref{AHM_Coulomb} are most easily computed from 
\be\label{AHM_Coulomb_Amodel}
\begin{split}
\langle  \sigma^n \rangle_{0} &= \sum_{k=0}^\infty \, q^k \res_{\hs=0} \frac{\hs^n}{(Q \hs)^{Qk+1}}= \begin{cases}
\frac{1}{Q}\left(\frac{q^{1/Q}}{Q} \right)^n  & n\in Q \bZ\\
0  & n \notin Q \bZ
\end{cases}~.
\end{split}
\ee
or equivalently by resumming the instanton series:
\be\label{AHM_Coulomb_Amodel2}
\begin{split}
\langle  \sigma^n \rangle_{0} &= -\sum_{k=0}^\infty \, q^k \res_{\hs=\infty} \frac{\hs^n}{(Q \hs)^{Qk+1}}= -\res_{\hs=\infty} \frac{(Q\hs)^{Q-1}\hs^n}{(Q \hs)^Q- q}=\\
&= \sum_{m=1}^Q \,\,\res_{\hs=\frac{1}{Q}e^{\frac{2\pi i m}{Q}} q^\frac{1}{Q}}\, \frac{(Q\hs)^{Q-1}\hs^n}{(Q \hs)^Q- q}= \begin{cases}
\frac{1}{Q}\left(\frac{q^{1/Q}}{Q} \right)^n  & n\in Q \bZ\\
0  & n \notin Q \bZ
\end{cases}~.
\end{split}
\ee

\subsection{$\mathbb{CP}^{N_f-1}$}\label{subsec:_CP}

The gauged linear sigma model for the projective space $\mathbb{CP}^{N_f-1}$ is a $U(1)$ gauge theory with $N_f$ chiral multiplets $X_i$ of charge $1$ \cite{Witten:1993yc}. We will consider $N_f\geq 2$ in the following. The effective FI parameter $\xi_{\rm eff}\to +\infty$ in the UV, where the vacuum moduli space is precisely $\mathbb{CP}^{N_f-1}$ with homogeneous coordinates $X_i$. If twisted masses $m_i$ are introduced for the matter fields, with $\sum_{i=1}^{N_f} m_i=0$, the vacuum moduli space reduces to the $N_f$ fixed points of the toric $U(1)^{{N_f}-1}$ action (the maximal torus of the $SU({N_f})$ flavor symmetry). 

In the UV geometric phase where $X_i$ take VEV, the JK residue is a sum of residues at the poles of the 1-loop determinants of $X_i$,
\begin{equation} 
 Z_k^{X_i}(\hs,m_i;\epsdef) =  
\begin{cases}
\prod\limits_{p=0}^{k} (\hs -m_i +(p-\frac{k}{2})\epsdef)^{-1}  & k\geq 0\\
1 & k=-1 \\
\prod\limits_{p=0}^{-k-2} (\hs - m_i + (1 + \frac{k}{2} + p)\epsdef) & k\leq -2
\end{cases} ~.\label{1-loop_CP}
\end{equation}
Due to the pole structure of \eqref{1-loop_CP}, the sum over $k$ reduces to the closure $k\geq 0$ of the cone dual to $\xi>0$. 

Inserting twisted chiral operators only at the north pole and shifting the integration variable $\hs\to\hs+\frac{k}{2}\epsdef$ as before, the localization formula \eqref{formula roughly} becomes 
\be\label{Coulomb_CP}
\begin{split}
\langle  \sigma^n_{N} \rangle &= \sum_{k=0}^\infty \, q^k  \sum_{i=1}^{N_f} \sum_{l_i=0}^{k} \, \res_{\hs=m_i-l_i\epsdef} \, \frac{\hs^n}{\prod\limits_{j=1}^{N_f} \prod\limits_{l_j=0}^{k} (\hs -m_j+ l_j \epsdef)} =\\
& = -\sum_{k=0}^\infty \, q^k  \res_{\hs=\infty}  \frac{\hs^n}{\prod\limits_{j=1}^{N_f} \prod\limits_{l_j=0}^{k} (\hs -m_j+ l_j \epsdef)}~.
\end{split}
\ee
From the latter expression in \eqref{Coulomb_CP} we easily obtain the lowest expectation values
\begin{equation}\label{correlators_CP}
\begin{split}
\langle  \sigma^n_N \rangle =0\quad (n\le {N_f}-2)~, \qquad \langle  \sigma^{{N_f}-1}_N \rangle =1~, \qquad \langle  \sigma^{N_f}_N \rangle =\sum_{i=1}^{N_f} m_i=0~.
\end{split}
\end{equation}
Higher correlators can be computed from the first expression in \eqref{Coulomb_CP}, which yields
\be\label{correlators2_CP}
\begin{split}
\langle  \sigma^n_N \rangle =   \sum_{k=0}^\infty (q \epsdef^{-1})^k \sum_{i=1}^{N_f} \sum_{l_i=0}^k \frac{(-1)^{l_i} (m_i-l_i \epsdef)^n}{ (k-l_i)!l_i! \prod\limits_{j=1}^{N_f} \prod\limits_{l_j=0}^k (m_i -m_j -(l_i-l_j) \epsdef)^{1-\delta_{ij}}}~,
\end{split}
\ee
or alternatively using the identity \eqref{recursion}, that in this case reads
\be\label{CP_finite_diff}
\langle \cO(\s_N)
\prod_{j=1}^{N_f} (\s_N-m_j)\rangle = q \langle \cO(\s_N-\epsdef)\rangle~
\ee
and that allows to compute the correlators $\langle  \sigma^n_N \rangle$ with $n\ge {N_f}$ recursively  from  lower ones. More explicitly, writing the characteristic polynomial
\be\label{char_poly}
\prod_{j=1}^{N_f}(\s-m_j) = \s^{N_f}+\sum_{k=2}^{N_f} s_k(m) \s^{{N_f}-k}
\ee
in terms of the symmetric polynomials of the twisted masses
\be\label{sym_poly}
s_k(m) = (-1)^k \sum_{i_1<\dots<i_k} m_{i_1}\dots m_{i_k} ~, \qquad k=2,\dots,{N_f}~,
\ee
the finite difference equation \eqref{CP_finite_diff} determines iteratively
\be\label{recur_CP}
\langle \s^{r+{N_f}}_N \rangle = -\sum_{k=2}^{N_f} s_k(m) \langle \s^{r+{N_f}-k}_N \rangle
+ q \sum_{h=0}^r \begin{pmatrix} r\\ h\end{pmatrix} (-\epsdef)^h \langle \s^{r-h}_N \rangle ~.
\ee

The difference equation \eqref{CP_finite_diff} can again be recast into a differential equation for the generating function $F(z) = \langle e^{z \s}|_N\rangle $ of twisted chiral correlators inserted at the north pole:
\be\label{differential_CP}
\left[\prod_{j=1}^{N_f} (\d_z -m_j) - q e^{-z\epsdef} \right] F(z) = 0~.
\ee
Changing variable at non-vanishing $\epsdef$ from $z$ to 
\be
\mathbf{q}_z=  q e^{-z\epsdef} (-\epsdef)^{-{N_f}}~,
\ee
the differential equation \eqref{differential_CP} takes the Picard-Fuchs form
\be\label{PF_CP}
\left[\prod_{j=1}^{N_f} \left(\Theta + \frac{m_j}{\epsdef}\right) - \mathbf{q}_z \right] F(z) = 0~,
\ee
where $\Theta\equiv \mathbf{q}_z\frac{\d}{\d\mathbf{q}_z}$. The general solution of this Picard-Fuchs equation is given in terms of generalized hypergeometric functions as
\be\label{PF_sol_CP}
\sum_{j=1}^{N_f} c_j (-{\bf q}_z)^{-\frac{m_j}{\epsdef}} {}_0 F_{{N_f}-1}\left(\left\{1+\frac{m_i-m_j}{\epsdef}  \right\}_{\substack{i=1\\ i\neq j}}^{N_f}  \Big| {\bf q}_z \right)
\ee
but it seems difficult to determine the coefficients $c_j$ as functions of the twisted masses and $\epsdef$ such that $F(z)=\frac{z^{{N_f}-1}}{({N_f}-1)!} + \cO(z^{N_f})$ in accord with \eqref{correlators_CP}. We leave this problem to future work. 

In the $A$-model limit $\epsdef\to 0$, correlators are given by the formula
\be\label{nonvanishing_corr_CP_Amodel_mass}
\langle  \s^n \rangle_{0} = \sum_{k=0}^\infty q^k \sum_{j=1}^{N_f} \res_{\hs=m_j} \frac{\hs^n}{\prod_{i=1}^{N_f} (\hs -m_i)^{k+1}} = - \res_{\hs=\infty} \frac{\hs^n}{\prod_{i=1}^{N_f} (\hs -m_i) - q} 
\ee
and can be computed recursively from \eqref{correlators_CP} using the twisted chiral ring operator relation $\prod_{i=1}^{N_f}(\s -m_i) = q$, which implies 
\be\label{recur_CP_Amodel}
\langle \s^{r+{N_f}} \rangle_{0} = -\sum_{k=2}^{N_f} s_k(m) \langle \s^{r+{N_f}-k} \rangle_{0}
+ q \langle \s^r \rangle_{0} ~.
\ee

If the twisted masses $m_i$ vanish, the only non-vanishing correlators
are 
\be\label{nonvanishing_corr_CP_Amodel_nomass}
\langle  \sigma^{{N_f}(h+1)-1} \rangle_{0} = 
 q^h ~, \qquad h=0,1,2,\dots,
\ee
and the generating function is 
\be
F(z)_{\epsdef=0}= \frac{z^{{N_f}-1}}{({N_f}-1)!} ~{}_0 F_{{N_f}-1}\left(\left\{1+\frac{j}{{N_f}}  \right\}_{j=1}^{{N_f}-1}  \Big| q\frac{z^{N_f}}{{N_f}^{N_f}} \right)~.
\ee

\subsection{The quintic}\label{subsec:_quintic}

The quintic Calabi-Yau threefold can be engineered using a $U(1)$ GLSM with 5 fields $X_i$ of gauge charges $Q=1$ and R-charges $r=0$, and one field $P$ of charge $Q=-5$ and $r=2$, subject to a superpotential $W=P F(X)$, with $F(X)$ a homogeneous quintic polynomial in $X_i$ \cite{Witten:1993yc}. The GLSM flows to a nontrivial CFT, and correspondingly the FI parameter $\xi$ is marginal, therefore $\xi_{\rm eff}^{\rm UV}=\xi$. The detailed form of the Coulomb branch localization formula \eqref{formula roughly} is sensitive to the phase of the GLSM, even though the final result is independent of the phase, being analytic in $q$. Before discussing the two phases of the GLSM, corresponding to positive or negative FI parameter $\xi$, we list for future reference the 1-loop determinants of the matter fields:
\begin{align}
 Z_k^{X_i}(\hs;\epsdef) &=  
\begin{cases}
\prod\limits_{p=0}^k (\hs +(p-\frac{k}{2})\epsdef)^{-1}  & k\geq 0\\
1 & k=-1 \\
\prod\limits_{p=0}^{-k-2} (\hs + (1 + \frac{k}{2} + p)\epsdef) & k\leq -2
\end{cases} \qquad (i=1,\dots,5) \label{1-loop_X_quintic}\\
 Z_k^{P}(\hs;\epsdef) &= 
\begin{cases}
\prod\limits_{j=0}^{5k} (-5 \hs +(\frac{5}{2}k- j)\epsdef)  & k\geq 0\\
\prod\limits_{j=0}^{-5k-2}(-5 \hs +(\frac{5}{2}k+ 1+j)\epsdef)^{-1} & k\leq -1~.
\end{cases} \label{1-loop_P_quintic}
\end{align}

\subsubsection{Geometric phase}\label{subsec:_quintic_geom}

For $\xi>0$ the GLSM is in the geometric phase: the positively charged fields $X_i$ are forced to take VEV by the $D$-term equation, whereas the vev of $P$ vanishes by the $F$-terms provided $F(X)$ is generic. Modding out by the $U(1)$ gauge symmetry, $X_i$ are homogeneous coordinates of $\mathbb{CP}^4$. Finally, the $F$-term equation $F(X)=0$ associated to $P$ cuts out the  quintic hypersurface in $\mathbb{CP}^4$. 

In the geometric phase $\xi>0$ where $X_i$ take VEV, the JK residue ${\rm JK{\text-}Res}\left[\xi 
\right]$ is a sum of residues at the poles of the 1-loop determinants of $X_i$.
We see from the poles of \eqref{1-loop_X_quintic} that the sum over $k$ reduces to the dual cone $k\geq 0$. Inserting twisted chiral operators only at the north pole and shifting the integration variable $\s\to\s+\frac{k}{2}\epsdef$, the localization formula \eqref{formula roughly} becomes 
\be\label{Coulomb_quintic}
\begin{split}
\langle  \sigma^n_N \rangle &= - \sum_{k=0}^\infty \, q^k  \sum_{l=0}^{k}  \res_{\hs=-l\epsdef} \frac{\prod\limits_{j=0}^{5k} (-5\hs -j \epsdef) }{\prod\limits_{p=0}^{k} (\hs +p \epsdef)^5} \hs^n  =\\
& = \sum_{k=0}^\infty \, q^k   \res_{\hs=\infty} \frac{\prod\limits_{j=0}^{5k} (-5\hs -j \epsdef) }{\prod\limits_{p=0}^{k} (\hs +p \epsdef)^5} \hs^n = \epsdef^{n-3} \sum_{k=0}^\infty \, q^k   \res_{z=\infty} \frac{\prod\limits_{j=0}^{5k} (-5z -j) }{\prod\limits_{p=0}^{k} (z +p )^5} z^n~,
\end{split}
\ee
where we inserted an overall minus sign because of the field $P$ of $R$-charge $2$, as explained in section \ref{subsec: oneloop det}. 
 By explicit computation, we find the correlators
\be\label{correlators_quintic}
\begin{split}
\langle  \sigma^n_N \rangle &= 0  \qquad\qquad (n=0,1,2)\\
\langle  \sigma^3_N \rangle &= \frac{5}{1+5^5 q}\\
\langle  \sigma^4_N \rangle &= \epsdef \frac{2\cdot 5^6 q}{(1+5^5 q)^2}\\
\langle  \sigma^5_N \rangle &= \epsdef^2 \frac{5^5 q(-17 + 13\cdot 5^5 q)}{(1+5^5 q)^3}\\
\langle  \sigma^6_N \rangle &= \epsdef^3 \frac{2 \cdot 5^5 q(13 - 125000 q + 68359375 q^2)}{(1+5^5 q)^4}\\
& \vdots
\end{split}
\ee
In the $A$-model limit $\epsdef=0$, the only non-vanishing correlator is the Yukawa coupling $\langle  \sigma^3 \rangle_0 = \frac{5}{1+5^5 q}$ computed using the GLSM description in \cite{Morrison:1994fr} and originally in \cite{Candelas:1990rm}. The higher correlators, which become non-trivial with the Omega deformation, can again be computed recursively using the identity \eqref{recursion}, which here reads 
\be\label{finite_diff_quintic1}
\langle \cO(\s_N)\s^5_N\rangle = q \langle \cO(\s_N-\epsdef)  \prod_{j=1}^5 (-5\s_N+j \epsdef) \rangle~.
\ee
In fact, since division by $\s_N$ does not introduce extra poles, we can safely substitute $\cO(\s_N)\to \cO(\s_N)/\s_N$ to obtain the more general identity
\be\label{finite_diff_quintic2}
\langle \cO(\s_N)\s^4_N\rangle = -5q \langle \cO(\s_N-\epsdef)  \prod_{j=1}^4 (-5\s_N+j \epsdef) \rangle~,
\ee
which allows to compute the higher correlators $\langle \s^{3+j}_N\rangle$ recursively from $\langle \s^{3}_N\rangle=\langle \s^3\rangle_{0}$ and the vanishing lower correlators.

The difference equation \eqref{finite_diff_quintic2} translates into a differential equation for the generating function $F(z) = \langle e^{z \s_N}\rangle $ of twisted chiral correlators inserted at the north pole:
\be\label{differential_quintic}
\left[\d_z^4 +5 q e^{-z\epsdef} \prod_{j=1}^4(5 \d_z-j\epsdef) \right] F(z) = 0~.
\ee
Changing variable at non-vanishing $\epsdef$ from $z$ to 
\be
\mathbf{q}_z=  - q e^{-z\epsdef}~,
\ee
the differential equation \eqref{differential_quintic} becomes the celebrated Picard-Fuchs equation 
\be\label{PF_quintic}
\Theta^4 F = 5 \mathbf{q}_z \prod_{j=1}^4(5 \Theta+j)F~,
\ee
for the mirror of the quintic \cite{Candelas:1990rm}.

\subsubsection{Landau-Ginzburg phase}\label{subsec:_quintic_LG}

For $\xi<0$, the charge $-5$ field $P$ is forced to take VEV by the $D$-term equation, whereas the charge $1$ fields $X_i$ vanish by the $F$-term equations. The low energy physics is described by a Landau-Ginzburg orbifold for the $5$ $X_i$ fields, with a homogeneous quintic superpotential $W(X)=\langle P \rangle F(X)$ and a residual $\bZ_5$ gauge symmetry. For this reason this non-geometric phase is called Landau-Ginzburg phase.

In the $\xi<0$ phase, ${\rm JK{\text-}Res}\left[\xi\right]$ is a sum of residues at the poles of the 1-loop determinant of $P$. We see from \eqref{1-loop_P_quintic} that the sum over $k$ reduces to the shifted dual cone $-5k-2\ge 0$, hence $k\le -1$. Inserting twisted chiral operators only at the north pole and shifting the integration variable $\s\to\s+\frac{k}{2}\epsdef$, the localization formula \eqref{formula roughly} becomes  
\be\label{Coulomb_quintic_LG}
\begin{split}
\langle  \sigma^n_N \rangle &= \sum_{k=-\infty}^{-1} \, q^k  \sum_{l=0}^{-5k-2}  \res_{\hs=\frac{1}{5}(l+1)\epsdef} \frac{\prod\limits_{p=0}^{-k-2} (\hs+(p+k+1)\epsdef)^5}{\prod\limits_{j=0}^{-5 k-2} (-5\hs + (j+1) \epsdef)}
 \hs^n  =\\
& = - \sum_{k=-\infty}^{-1} \, q^k  \res_{\hs=\infty} \frac{\prod\limits_{p=0}^{-k-2} (\hs+(p+k+1)\epsdef)^5}{\prod\limits_{j=0}^{-5 k-2} (-5\hs + (j+1) \epsdef)}
 \hs^n ~,
\end{split}
\ee
where we adopted the short-hand notation that $\prod_{p=0}^{-k-2} (\dots) = 1$ for $k=-1$. We have checked explicitly at low $n$ that the correlators \eqref{Coulomb_quintic_LG} computed in the LG phase match those \eqref{Coulomb_quintic} computed in the geometric phase, giving \eqref{correlators_quintic}. Note that while correlators computed in the geometric phase $\xi>0$ are given by a Taylor series in $q$, correlators computed in the LG phase $\xi<0$ are given by a Taylor series in $q^{-1}$. 
Showing that the two computations agree requires resumming the Taylor series to an analytic function as in \eqref{correlators_quintic}.
In fact, it is easy to see that the correlators $\langle \s_N^n \rangle$ computed in the two phases are equal for any $n$. This follows from the equality for $n=0,1,2,3$ together with the fact that the identity \eqref{finite_diff_quintic2} holds for correlators computed in any of the two phases.

\subsection{The resolved $\mathbb{WCP}^4_{1,1,2,2,2}$}

\begin{table}[t]
\centering
\begin{tabular}{c|cc  ccc c| c}
&  $X_1$ & $X_2$ &$Y_1$ &$Y_2$ & $Y_3$ &$Z$ & FI \\
\hline
$U(1)_1$ & $0$ &$0$ & $1$ &$1$ & $1$& $1$ &$\xi_1$ \\ 
$U(1)_2$ & $1$ & $1$ & $0$& $0$ & $0$& $-2$  & $\xi_2$ \\ \hline
$m^F$ & $-m_1^X$& $-m_2^X$& $-m_1^Y$& $-m_2^Y$& $-m_3^Y$& $0$ &
\end{tabular}
\caption{$U(1)^2$ charges and twisted masses of the chiral multiplets in the GLSM for the (resolved)  projective space $\mathbb{WCP}^4_{1,1,2,2,2}$.}
\label{tab:MPexample}
\end{table}

\begin{figure}[t]
\centering\includegraphics[width=5cm]{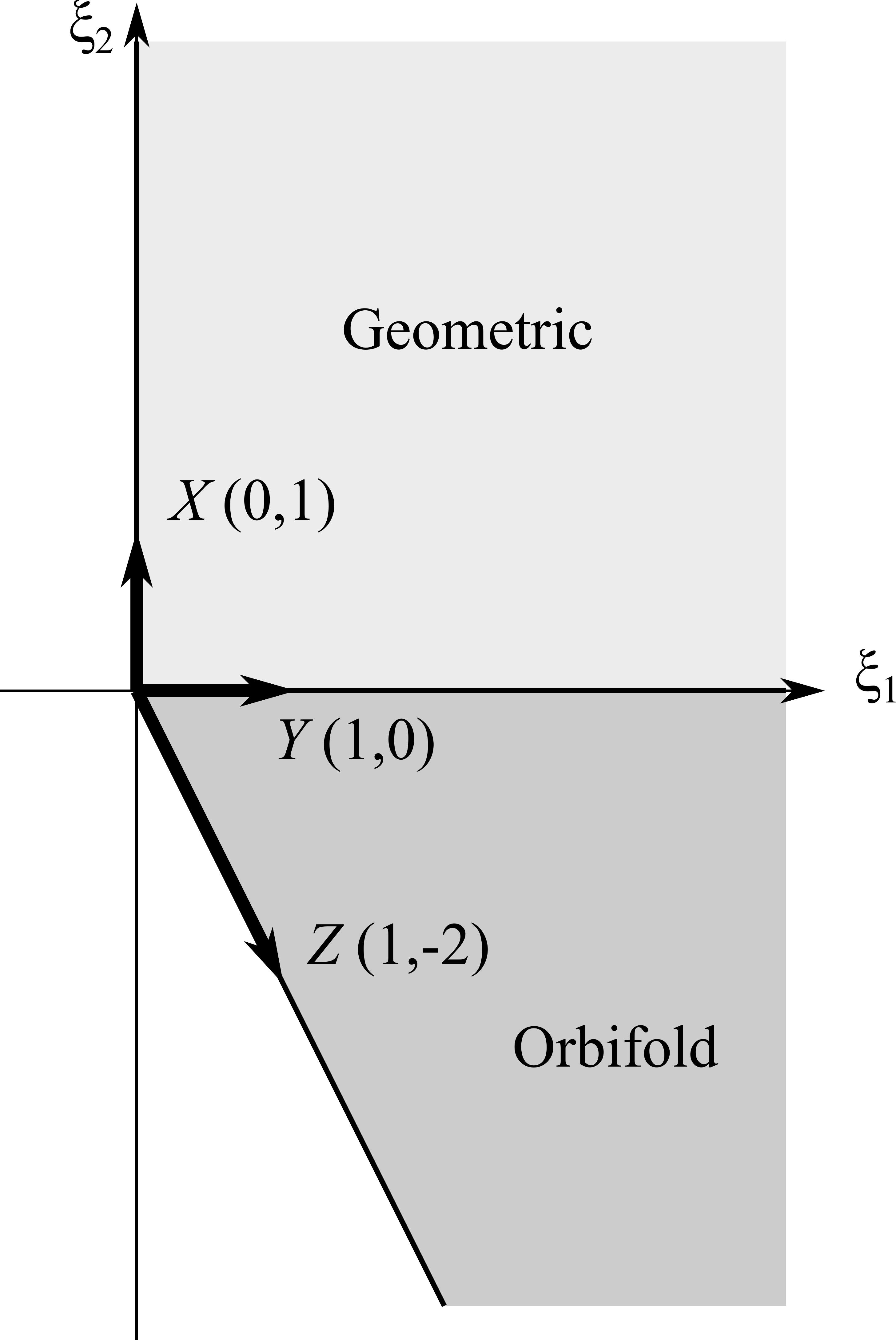}
\caption{\small  The two phases of the (resolved) $\mathbb{WCP}^4_{1,1,2,2,2}$ toric variety.}
\label{f:WCPphase}
\end{figure}

The last example of this section is a $U(1)^2$ GLSM with six chiral fields: $X_i$ ($i=1,2$), $Y_j$ ($j=1,2,3$), and $Z$.  All the chiral multiplets have vanishing vanishing $R$-charge, and their gauge charges and twisted masses are given in Table \ref{tab:MPexample}.  This abelian GLSM was studied in detail in \cite{Morrison:1994fr}.  For $\xi_1>1$, $\xi_2>0$, it engineers a toric variety of dimension $4$, obtained by blowing up the curve of $\Z_2$ singularities inside $\mathbb{WCP}^4_{1,1,2,2,2}$ (the size of the blown-up curve is given by $\xi_2$). 
For $2\xi_1+  \xi_2 >0$ and $\xi_2<0$, we obtain the unresolved space instead. These so-called geometric and orbifold phases are  depicted in Figure \ref{f:WCPphase}, in FI parameter space. We have $b_0^1=4 >0$ and $b_0^2=0$, so that $\xi_{\rm eff}^{\rm UV}= (+ \infty, \xi_2)$.

The flavor symmetry group is $SU(2)\times SU(3)\times U(1)$. For simplicity, we only turn on twisted masses in the $SU(2)\times SU(3)$ subgroup,  which we denote by $m_i^X$ and $m_j^Y$ (see  Table \ref{tab:MPexample}). It is also convenient to define the Casimirs $u_n(m^X) = \sum_{i=1}^2 (m_i^X)^n$, $u_n(m^Y) =\sum_{j=1}^3 (m_j^Y)^n$, with $u_1(m^X)= u_1(m^Y)=0$. Let us introduce the notation:
\be\label{def Fm1m2}
F_{m_1, m_2}\equiv \langle \sigma_1^{m_1} \sigma_2^{m_2}\rangle_N~.
\ee
The subscript $N$ denotes a north pole insertion.
 By the  selection rule \eqref{selec rule}, we have
\be\label{selec rule WP model}
F_{m_1, m_2}  \sim (\epsdef)^j \Lambda_1^{4 k_1} (m^F)^l \, F_c(q_2)~, \qquad m_1+ m_2 = 4 k_1 + 4 + j +l~,
\ee
with $j \geq 0$. Here $\Lambda_1^4=q_1$, setting the energy scale $\mu=1$ as in the rest of the article. Because the target space is compact, the massless limit is nonsingular and we must also have $l \geq 0$.
The one-loop determinants of the matter fields are:
\begin{align}
 Z_k^{X_i}(\hs;\epsdef) &= 
\begin{cases}
\prod\limits_{p=0}^{k_2} (\hs_2 - m_i^X  +(p-\frac{k_2}{2})\epsdef)^{-1}~,  & k_2\geq 0~,\\
1~, & k_2=-1~, \\
\prod\limits_{p=0}^{-k_2-2} (\hs_2 - m_i^X + (1 + \frac{k_2}{2} + p)\epsdef)~, & k_2\leq -2~,
\end{cases} \qquad (i=1,2) \label{1-loop_X_WP} \\
 Z_k^{Y_j}(\hs;\epsdef) &= 
\begin{cases}
\prod\limits_{p'=0}^{k_1} (\hs_1 - m_j^Y  +(p'-\frac{k_1}{2})\epsdef)^{-1}~,  & k_1\geq 0~,\\
1~, & k_1=-1~, \\
\prod\limits_{p'=0}^{-k_1-2} (\hs_1 - m_j^Y + (1 + \frac{k_1}{2} + p')\epsdef)~, & k_1\leq -2~,
\end{cases} \quad ~(j=1,2,3) \label{1-loop_Y_WP} \\
 Z_k^{Z}(\hs;\epsdef) &= 
\begin{cases}
\prod\limits_{p''=0}^{ k_1 -2k_2} (\hs_1 - 2 \hs_2 +(p''-\frac{k_1}{2} + k_2 )\epsdef)^{-1}~,  & k_1- 2 k_2 \geq 0~,\\
1~,  & k_1-2 k_2=-1 ~,\\
\prod\limits_{p''=0}^{-k_1+2 k_2-2} (\hs_1 - 2 \hs_2 +(1+\frac{k_1}{2} - k_2 + p'')\epsdef)~,  & k_1- 2 k_2 \leq -2 ~.
\end{cases} \label{1-loop_Z_WP}
\end{align}
The singular hyperplanes  from  \eqref{1-loop_X_WP}-\eqref{1-loop_Z_WP} are:
\bea
& H_{X}^{i, p_X} =\left\{\hs_2 - m_i^X  +(p_X-\frac{k_2}{2})\epsdef =0 \right\}~, \qquad && p_X = 0, \cdots, k_2~,\cr
& H_{Y}^{j, p_Y} =\left\{  \hs_1 - m_j^Y  +(p_Y-\frac{k_1}{2})\epsdef =0 \right\}~, \qquad  && p_Y = 0, \cdots, k_1~, \cr
& H_{Z}^{p_Z} =\left\{  \hs_1 - 2 \hs_2 +(p_Z-\frac{k_1}{2} + k_2 )\epsdef =0 \right\}~, && p_Z = 0, \cdots, k_1- 2 k_2~.
\eea
The elementary singularities that contribute to the JK residues  are
\be
\omega_{Q_X Q_Y} = { d\hs_1 \wedge d\hs_2\over \hs_1\hs_2}~, \quad 
\omega_{Q_X Q_Z} = { d\hs_1 \wedge d\hs_2\over \hs_2(\hs_1 - 2\hs_2)}~, \quad 
\omega_{Q_Y Q_Z} = { d\hs_1 \wedge d\hs_2\over \hs_1(\hs_1 - 2\hs_2)}~, 
\ee
from the intersections $H_X \cap H_Y$, $H_X \cap H_Z$, and $H_Y \cap H_Z$, respectively. From the definition \eqref{def JK}, one  finds:
\bea\label{JK res WP expl}
&{\rm JK{\text-}Res_{\,\h\sigma=0} }\left[ \xi_{\rm\eff}^{\rm UV} \in {\rm Geometric}\right]\, f(\hs) d\hs_1 \wedge d\hs_2  = \, \res_{\hs_1 =0}\,\res_{\hs_2=0} f(\hs) ~,\cr
&{\rm JK{\text-}Res_{\,\h\sigma=0} }\left[ \xi_{\rm\eff}^{\rm UV} \in {\rm Orbifold}\right]\,f(\hs) d\hs_1 \wedge d\hs_2  = \, \res_{\hs_2 =0}\,\res_{\hs_1= 2\hs_2} f(\hs)~.
\eea
Note that  the order of  the residues in \eqref{JK res WP expl} is crucial. Each residue is taken with the remaining integration variable generic and away from the hyperplanes. The JK residue at any of the singularities in $\tfM$ is given by \eqref{JK res WP expl} after a translation. The singularities are always projective. For generic twisted masses, they are also all regular.

\subsubsection{The geometric phase with $m^F\neq 0$}
In the geometric phase,  the only fluxes that contribute to the JK residue are in the window $k_1\geq 0$, $k_2\geq 0$, and we have
\be\label{Fm1m2 expl i}
F_{m_1, m_2}= \sum_{k_1, k_2=0}^{\infty} q_1^{k_1} q_2^{k_2}  U_{k_1, k_2, m_1, m_2}~.  
\ee 
Consider first the case of generic twisted masses $m_i^X, m_j^Y$. Each factor $U_{k_1, k_2, m_1, m_2}$ splits into contributions from $H_X\cap H_Y$ and $H_X \cap H_Z$:
\be\label{Fm1m2 expl ii}
 U_{k_1, k_2, m_1, m_2}=  U_{k_1, k_2, m_1, m_2}^{XY}+  U_{k_1, k_2, m_1, m_2}^{XZ}~.
\ee
The $H_{Y}\cap H_{Z}$ singularities do not contribute to the JK residue in the geometric phase, in line with the absence of vacua where $Y$ and $Z$ take VEV while $X$ vanishes. 

The contribution from $H_X\cap H_Y$  reads:
\bea
 &U_{k_1, k_2, m_1, m_2}^{XY}=\cr
&\;\sum_{i=1}^2 \sum_{j=1}^3 \sum_{p_X=0}^{k_2}\sum_{p_Y=0}^{k_1}  \res_{\hs_1 =0}\,\res_{\hs_2=0} \,  (\mathbf{I}_{XY})_{k_1, k_2}^{p_X, p_Y, i, j}\, (\hs_1 + m_j^Y - p_Y\epsdef)^{m_1}(\hs_2 + m_i^X - p_X \epsdef)^{m_2}~,
\eea
where we shifted the integration variables $\hs_a$ in each summand so that the singularity is always at $\hs_a=0$. Here we defined
\bea\label{I XY contr}
 &  (\mathbf{I}_{XY})_{k_1, k_2}^{p_X, p_Y, i, j}  = \prod_{i'=0}^2\prod_{p=0}^{k_2}{1\ov \hs_2 -(m_{i'}^X - m_i^X)+(p-p_X)\epsdef} \, \cr
&\quad\times \prod_{j'=1}^3 \prod_{p'=0}^{k_1} {1\over \hs_1 -(m_{j'}^Y -m_j^Y) + (p'-p_Y)\epsdef}\cr
&\quad \times  \begin{cases}
\prod\limits_{p''=0}^{ k_1 -2k_2} (\hs_1 - 2 \hs_2+ m_j^Y - 2 m_i^X +(p''-p_Y+ 2 p_X )\epsdef)^{-1}\\
1\\
\prod\limits_{p''=0}^{-k_1+2 k_2-2} (\hs_1 - 2 \hs_2  + m_j^Y - 2 m_i^X+(p''- p_Y + 2 p_X + k_1 - 2 k_2+1)\epsdef)~,
\end{cases} 
\eea
where the three cases in the third line are like in \eqref{1-loop_Z_WP}. The contribution from $H_X\cap H_Z$ reads
\bea
 &U_{k_1, k_2, m_1, m_2}^{XZ}=\cr
&\;\sum_{i=1}^2  \sum_{p_X=0}^{k_2}\sum_{p_Z=0}^{k_1-2 k_2}  \res_{\hs_1 =0}\,\res_{\hs_2=0} \,  (\mathbf{I}_{XZ})_{k_1, k_2}^{p_X, p_Z, i}\, (\hs_1 + m_i^X -( p_Z+ 2 p_X)\epsdef)^{m_1}(\hs_2 + m_i^X - p_X \epsdef)^{m_2}~,
\eea
with $(\mathbf{I}_{XZ})_{k_1, k_2}^{p_X, p_Z, i}$ an expression similar to \eqref{I XY contr}. 

From the  expression \eqref{Fm1m2 expl i}-\eqref{Fm1m2 expl ii}, one can compute the correlation functions $F_{m_1, m_2}$ explicitly, at least at low order in $m_1, m_2$ and $q_2$. Let us define $L=m_1+ m_2$,  the ``level'' of $F_{m_1, m_2}$. The first non-trivial correlation functions occur at level $4$:
\bea\label{resummed first correlators}
& F_{m_1, m_2}=0~, \quad && \forall m_1, m_2~:\; m_1+m_2\leq  3~,\cr
& F_{4,0}= 2~, \quad 
&& F_{3,1}= 1~, \quad\cr
& F_{2,2}= -{2 q_2\ov 1- 4 q_2}~, \quad 
&& F_{1,3}= { q_2(1+ 4 q_2)\ov (1- 4 q_2)^2}~, \qquad 
& F_{0,4}= -{2 q_2^2(3 + 4 q_2)\ov (1- 4 q_2)^3}~.
\eea
These exact resummed expressions are more easily obtained in the $m^F=0$, $\epsdef=0$ limit of the Coulomb branch formula, which we discuss below.

\subsubsection{Recursion relations}
The explicit expression  \eqref{Fm1m2 expl i}-\eqref{Fm1m2 expl ii} is unwieldy but perfectly general.%
~\footnote{That is, given that the twisted masses are generic. For special values, some singularities from  $H_X\cap H_Y$ and  $H_X\cap H_Z$ can coincide, and one must be careful not to over-count.}
  Fortunately, the recursion relations \eqref{recursion} discussed in the previous section allow us to compute \eqref{def Fm1m2} recursively. In our model,  \eqref{recursion} reads
\bea\label{recursion WP model}
&\langle \sigma_1^{l_1}\sigma_2^{l_2} (\s_1 - 2 \s_2)\prod_{j=1}^3 (\s_1 + m_j^Y)\rangle_N 
= q_1 \langle (\sigma_1-\epsdef)^{l_1}\sigma_2^{l_2}\rangle_N~,\cr
& \langle \sigma_1^{l_1}\sigma_2^{l_2}\prod_{i=1}^3 (\s_2 + m_i^X)\rangle_N 
=q_2 \langle  (\s_1- 2\s_2)(\s_1- 2 \s_2 +\epsdef)  \sigma_1^{l_1}(\sigma_2-\epsdef)^{l_2} \rangle_N~,
\eea
with $l_1, l_2\geq 0$. it is easy to check that \eqref{resummed first correlators} indeed satisfies these relations.
 At any level $L=m_1+ m_2$, \eqref{recursion WP model} provides $2 L-4$ linear relations between the level $L$ and lower level correlators . (We have $L-3$ relations from the first line of  \eqref{recursion WP model}, and $L-1$ relations from the second one.) This gives an overdetermined set of recursion relations for the $L\geq 5$ correlators (at each level, there are $2 L-4$ equations for $L+1$ undetermined correlators). Therefore, any correlation function can be determined recursively with the initial data \eqref{resummed first correlators}. This  is easily implemented on a computer. At level $L=5$, one finds:
\bea\nn
&F_{5,0}= F_{4,1}= F_{3,2}=0~,\quad
&&F_{2,3}= {3 q_2 \epsdef\over (1- 4 q_2)^2}~, \cr
&F_{1,4}= -{2 q_2 (1+ 12 q_2)\epsdef \ov (1- 4 q_2)^3}~, \quad
&&F_{0,5}= {30 q_2^2 (1+ 4 q_2)\epsdef\ov (1- 4 q_2)^4}~.
\eea
These correlators are proportional to $\epsdef$,  in agreement with \eqref{selec rule WP model}. 
At level $L=6$, we find:
\footnotesize
\bea\nn
& F_{6,0}= 8 u_2(m^X) + 2 u_2(m^Y)= 2 F_{5,1}~, \quad\qquad F_{4,2} =2 u_2(m^X) - { q_2 u_2(m^Y)\ov 1-4q_2}~,\cr
&F_{3,3}=u_2(m^X)+   { q_2(1+4q_2) u_2(m^Y)\ov (1-4 q_2)^2}~,\cr
& F_{2,4}= - {4 q_2 (1-2 q_2)  u_2(m^X)\ov (1- 4 q_2)^2} - {2 q_2^2 (3+ 4 q_2) u_2(m^Y) + 2 q_2 (2 + 7 q_2)\epsdef^2\ov (1- 4 q_2)^3}~,\cr
& F_{1,5} = {2 q_2(1+ 6 q_2^2 - 8 q_2^3)u_2(m^X)\ov (1- 4 q_2)^3} +{q_2^2(1 + 24 q_2 + 16 q_2^2)u_2(m^Y)  +q_2 (3  + 95 q_2 + 140 q_2^2)\epsdef^2 \ov (1- 4 q_2)^4} ~,\cr
& F_{0,6}  = -{2 q_2^2(9+ 16 q_2 -16 q_2^2) u_2(m^X)\ov (1- 4 q_2)^4} -{ 2 q_2^3(5 +40 q_2+ 16 q_2^2) + 2 q_2^2(51 + 515 q_2 + 420 q_2^2)\epsdef^2 \ov (1-4 q_2)^5}~.
\eea
\normalsize
At this order, we have $\epsdef^2$ terms and the first appearance of the mass terms, as expected. One can continue this process level by level indefinitely. The dependence on $ q_1=\Lambda_1^4$ kicks in at level $8$.

\subsubsection{Geometric and orbifold phases in the $\epsdef=0$ and $m^F=0$ limit}
In the $\epsdef=0$ limit, we have some considerable simplifications. In order to compare to   \cite{Morrison:1994fr}, we also take $m_i^X= m_j^Y=0$. 
Consider first the geometric phase.  We have \eqref{Fm1m2 expl i} with:
\be
U_{k_1, k_2, m_1, m_2} =  \res_{\hs_1 =0}\,\res_{\hs_2=0} \,   {\hs_1^{m_1} \hs_2^{m_2}\ov  \sigma_2^{2(k_2+1)} \sigma_1^{3(k_1+1)} (\sigma_1-2\sigma_2)^{k_1-2k_2+1}}~.
\ee
This expression is  very easy to evaluate analytically, to obtain:~\footnote{Here the binomial coefficient $\mat{m \cr n}$ is understood to be equal to zero if $n<0$.}
\be
U_{k_1, k_2, m_1, m_2}=  (-2)^{2k_2-m_2+1} \mat{-k_1+2k_2-1\cr 2k_2+1-m_2}\, \delta_{m_1+m_2, 4 k_1+ 4}~.
\ee
This matches precisely the results of   \cite{Morrison:1994fr}. Summing  the $q_2$ series for $m_1+m_2=4$, we obtain \eqref{resummed first correlators}.
In the orbifold phase, instead, we have the expansion 
\be
F_{m_1, m_2} = \sum_{k_1 \geq 0, \, k_1 - 2 k_2\geq 0}  q_1^{k_1} q_2^{k_2} U^{\rm orb}_{k_1, k_2, m_1, m_2}~,
\ee
where the sum is over the fluxes in the dual cone to ${\rm Cone}(Q_Y, Q_Z)$.
The JK residue \eqref{JK res WP expl} gives
\bea
& U^{\rm orb}_{k_1, k_2, m_1, m_2}&=&\;  \res_{\hs_2 =0}\,\res_{\hs_1= 2\hs_2} {\hs_1^{m_1} \hs_2^{m_2}\ov  \sigma_2^{2(k_2+1)} \sigma_1^{3(k_1+1)} (\sigma_1-2\sigma_2)^{k_1-2k_2+1}}~,\cr
&&=&\;  2^{2 k_2-4k_1 + m_1 -3}  \mat{- 3 k_1 - 3 + m_1 \cr k_1 - 2 k_2} \, \delta_{m_1+m_2, 4 k_1+ 4}~.
\eea
The correlation functions agree as analytic functions across the phases, as expected. For instance, upon resumming 
\be
F_{m_1, 4- m_1} =  \sum_{k_2 \leq 0} q^{k_2} U^{\rm orb}_{0, k_2, m_1, 4-m_1}~,
\ee
in the orbifold phase, we recover \eqref{resummed first correlators}. 


\section{Examples: Correlators of $A$-twisted GLSMs}\label{sec:_Amodel_examples}

In this section we switch off the Omega-deformation parameter $\epsdef$ and compute correlators in some interesting examples of $A$-twisted gauged linear sigma models. 
The correlation functions no longer depend on the location of the operators on the sphere and thus the subscripts denoting the insertion points are dropped.~\footnote{In this section, we thus utilize the variables $N$ and $S$ to denote numerical parameters of the theory. These variables should not be confused with subscripts indicating the insertion locus of operators.}

In the abelian case, we consider a GLSM for a non-compact orbifold  studied in \cite{Melnikov:2005tk}. 
The authors of \cite{Melnikov:2005tk} found some puzzling violations of the quantum cohomology relations in that model. In our formalism, this is expected  due to the presence of non-projective singularities in the  Coulomb branch integrand. Physically, this signals a singular behavior of some correlation functions as the twisted masses are sent to zero. 

In the non-abelian case, we restrict for simplicity to unitary gauge groups, although our formalism applies generally. We compute twisted chiral correlators involving Casimir invariants of the form $\Tr(\sigma^j)$, where $\sigma$ is an adjoint matrix of a unitary gauge group factor. For correlators involving only dimension-one Casimirs ($j=1$), which lead to Yukawa couplings  in the Calabi-Yau models, we successfully compare our formulas with available results obtained using Picard-Fuchs equations and mirror symmetry \cite{Batyrev:1998kx}.
Our formalism also allows us to compute correlators involving higher Casimir invariants, which, to our knowledge, could not be computed with previous  techniques.

\subsection{$\bC^3/\bZ_{(2N+1)(2,2,1)}$}

Here we consider the GLSM  for the non-compact orbifold $\bC^3/\bZ_{(2N+1)(2,2,1)}$ studied in \cite{Melnikov:2005hq}. We will compare genus zero topological correlators computed in our formalism with the results of \cite{Melnikov:2005hq,Melnikov:2005tk}, fixing an ambiguity for certain constant correlators that violate the quantum cohomology relations of the theory with vanishing twisted masses.

The GLSM in question has $U(1)^2$ gauge group and $5$ chiral multiplets of zero $R$-charge and gauge charges as in Table \ref{tab:Melnikov-Plesser}. We consider $N>2$, so that the axial $R$-symmetry has a mixed anomaly with the $U(1)_1$ gauge group: consequently the effective FI parameters in the UV are $\xi_{\rm eff}^{\rm UV}=(-\infty, \xi_2)$.

\begin{table}[t]
\centering
\begin{tabular}{c|c c c c c | c}
 & $X_1$ & $X_2$ & $Y$ & $Z$ & $W$ & FI \\
\hline
$U(1)_1$ & $1$ & $1$ & $1$ & $-N$ & $-1$ & $\xi_1$ \\
$U(1)_2$ & $0$ & $0$ & $1$ & $1$ & $-2$ & $\xi_2$ 
\end{tabular}
\caption{Gauge charges of matter fields and FI parameters in the $\bC^3/\bZ_{(2N+1)(2,2,1)}$ GLSM.}
\label{tab:Melnikov-Plesser}
\end{table}

\begin{figure}[!t]
\centering\includegraphics[width=7cm]{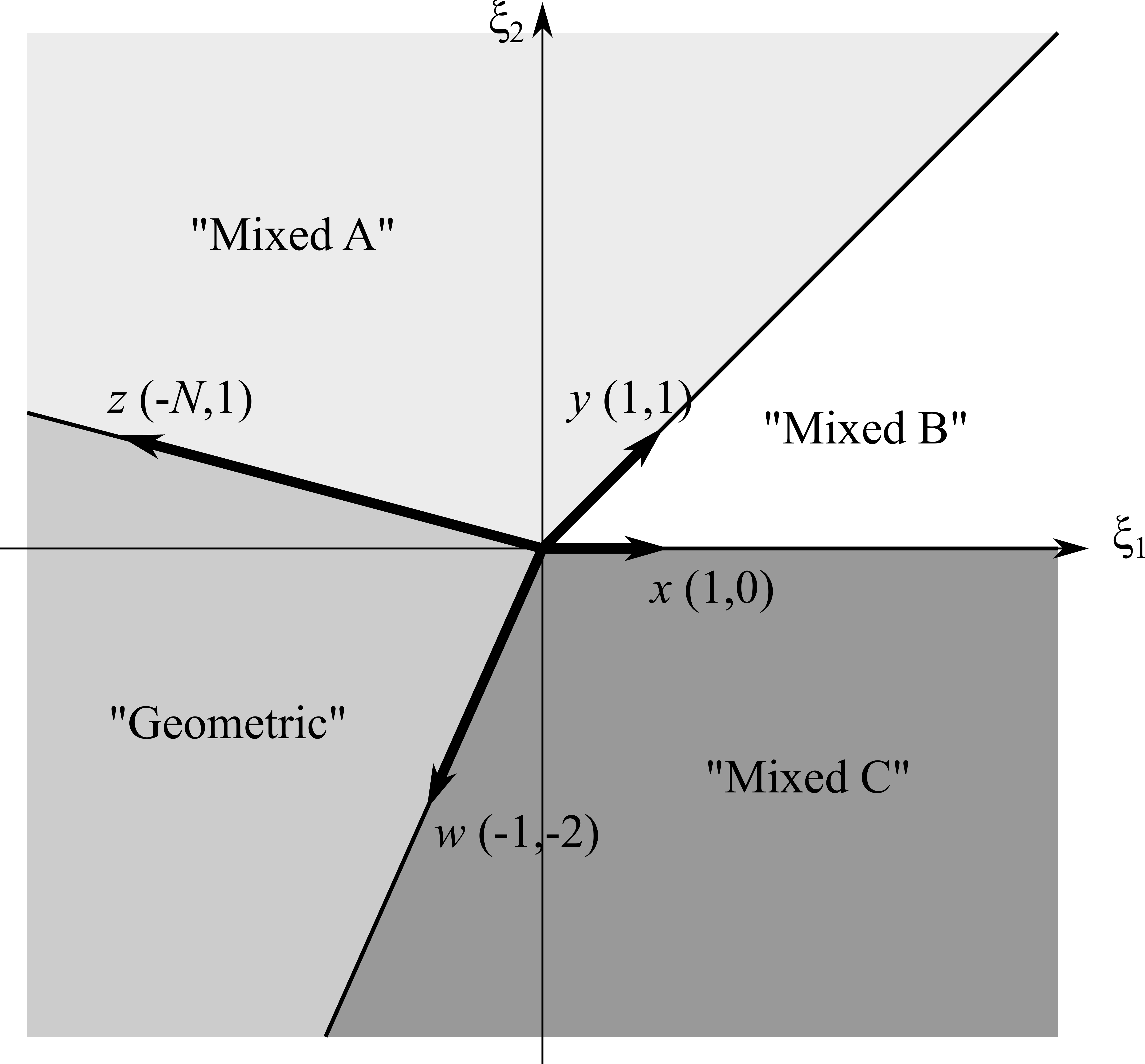}
\caption{\small The classical phase diagram of the $\bC^3/\bZ_{(2N+1)(2,2,1)}$ GLSM. }
\label{f:phases_MelPl}
\end{figure}

The classical phase diagram of the model is shown in Fig. \ref{f:phases_MelPl} \cite{Melnikov:2005tk}. Recall that in our formalism we are only interested in the phases that are probed by the GLSM in the UV, once quantum corrections to the FI parameters are included. Since $\xi_{\rm eff}^{\rm UV}=(-\infty, \xi_2)$, 
 the GLSM only probes the so called geometric phase $-2\xi_1+\xi_2>0$, $-\xi_1-N \xi_2>0$ in the UV, out of the four classical phases. In the geometric phase the fields $Z$ and $W$ are forced to acquire VEVs by the $D$-term equations, breaking the gauge group to $\bZ_{2N+1}$ and leaving a $\bC^3/\bZ_{(2N+1)(2,2,1)}$ orbifold, with $\bC^3$ parametrized by $X_1$, $X_2$ and $Y$. 

Let $\s_1$ and $\s_2$ be the complex scalars in the vector multiplets associated to the gauge groups $U(1)_1$ and $U(1)_2$. We wish to compute the correlators 
\be\label{correlators_MelPl}
F_{a,b}=\langle \s_1^a \s_2^b\rangle_0  
\ee
of this GLSM for vanishing twisted masses and compare to the results of \cite{Melnikov:2005tk,Melnikov:2005hq}. The selection rule for the anomalous axial $R$-symmetry implies that the topological correlator $F_{a,b}$ vanishes unless $a+b=3+(2-N) k_1$, where $k_1$ is the flux of the $U(1)_1$ gauge group. Therefore $F_{3+(2-N)n-b,b}\propto q_1^n$.

In defining and computing the Jeffrey-Kirwan residue, we face the technical complication that the arrangement of hyperplanes meeting at $\s_1=\s_2=0$ is not projective. We remedy this as in \cite{Benini:2013xpa,Park:2012nn} by turning on a small common twisted mass $m$ for the matter fields. This splits the multiple intersection at the origin into $6$ simple intersections of pairs of hyperplanes at separate points, which are projective arrangements. We will define  ${\rm JK{\text-}Res}\left[\xi_{\rm eff}^{\rm UV} \right]$ in the presence of the twisted mass $m$, which we take to zero at the end of the computation. 
With $m$ turned on, the 1-loop determinants of the matter fields are
\be\label{1loopdet-MelPl}
\begin{split}
Z_k^{X_i} &= (\hs_1+m)^{-(k_1+1)} ~\,\qquad \qquad \qquad\qquad Z_k^{Y} = (\hs_1+\hs_2+m)^{-(k_1+k_2+1)} \\
Z_k^{Z} &= (-N\hs_1+\hs_2+m)^{-(-N k_1+k_2+1)} \qquad Z_k^{W} = (-\hs_1-2\hs_2+m)^{-(-k_1-2k_2+1)} ~.
\end{split} 
\ee

The Jeffrey-Kirwan residue in the UV geometric phase is a residue at the intersection of the hyperplanes associated to $Z$ and $W$, namely the point $(\hs_1,\hs_2)=(\frac{3}{2N+1}m, \frac{N-1}{2N+1}m)$. It takes the form 
\be\label{JK_regul-MelPl}
{\rm JK{\text-}Res}\left[\xi_{\rm eff}^{\rm UV} \right] f(\hs_1,\hs_2,m) d\hs_1 \wedge d\hs_2 
= \res_{\hs_1=\frac{3}{2N+1}m} ~\res_{\hs_2=\frac{1}{2}(-\hs_1+m)} ~f(\hs_1,\hs_2,m)~.
\ee
 Since the residue only picks the poles of $Z_k^Z$ and $Z_k^W$, the summation over $k$ is effectively reduced to the dual cone
$-Nk_1+k_2\geq 0$, $ -k_1-2k_2\geq 0$.  
The localization formula for the topological correlators in the vanishing twisted mass limit is therefore 
\be\label{Coulomb_MelPl}
F_{a,b} =  \lim_{m\to 0} \sum_{k\in \bZ^2 }
q_1^{k_1} q_2^{k_2} \res_{\hs_1=\frac{3}{2N+1}m} ~\res_{\hs_2=\frac{1}{2}(-\hs_1+m)}  (\hs_1^a \hs_2^b Z_k^{X_1}Z_k^{X_2}Z_k^{Y}Z_k^{Z}Z_k^{W} ) ~.
\ee
Using \eqref{Coulomb_MelPl}, we can compute all the correlators of non-negative axial $R$-charge, $R_A=2(a+b-3)\geq 0$. The  correlators of negative $R_A$-charge diverge like $m^{a+b-3}$. Instead, correlators of non-negative $R_A$-charge have a finite $m\to 0$ limit. As we now explain, correlators of positive axial $R_A$-charge behave differently from correlators of zero axial $R_A$-charge.

To understand the difference, let us consider a na\"ive version of the localization formula with vanishing twisted masses, where the order of the $m\to 0$ limit and the residue in \eqref{Coulomb_MelPl} is reversed. When applied to 
$F_{3+(2-N)n-b,b}$ with $n\leq 0$, this na\"ive formula reproduces  the result of \cite{Melnikov:2005tk}.%
~\footnote{In particular, the contour integral formula for correlators in the geometric phase presented  in \cite{Melnikov:2005tk} can be obtained by summing the instantons in the na\"ive version of our formula \eqref{Coulomb_MelPl}. } 
For $n<0$ we find that our formula \eqref{Coulomb_MelPl} gives the same result: limit and residue commute.%
~\footnote{The statement extends to all $F_{a,b}$ correlators of positive axial $R$-charge ($a+b>3$). Those which violate the $m=0$ selection rule for the axial $R$-symmetry vanish in the $m\to 0$ limit.} 
It was also argued in \cite{Melnikov:2005tk} that their naive formula is incorrect for the $n=0$ correlators $F_{3-b,b}$. Note that the formula follows from $n<0$ correlators and quantum cohomology relations at vanishing twisted masses. Instead, the $F_{3-b,b}$ correlators must be quantum cohomology violating constants (with respect to $q_1$ and $q_2$), that could not be determined using their methods. Applying our localization formula \eqref{Coulomb_MelPl} to the correlators in question, we indeed obtain the constants:
\be\label{qc_viol_constants_of_MelPl}
F_{3-b,b} = \frac{3^{2-b}}{4} \frac{(N-1)^b}{(N+1) (N+2)^2 (2 N+1)}~, \qquad b=0,1,2,3.
\ee
We are led to conclude that the $m\to 0$ limit and the residue do not commute even for the finite and $R_A$-neutral correlators $F_{3-b,b}$. 

Let us emphasize that the proper quantum cohomology relations do hold in the massive theory, in agreement with the discussion of  section \ref{section: Q Cohomology}. The violation of the na\"ive relations in the massless theory are a symptom of the fact that the theory is singular (albeit in a mild fashion: only a finite number of correlators diverge) when we send the twisted mass to zero.

\subsection{Calabi-Yau complete intersections in Grassmannians}\label{subsec:HoriTong}

\begin{table}[t]
\centering

\begin{tabular}{c|c c| c}
&  $\Phi_i^a$ & $P^\alpha$ & FI \\
\hline
$U(N)$ & $\mathbf{N}$ & $\det^{-Q_\alpha}$ & $\xi$ \\ \hline
$U(1)_{R}$ & $0$ & $2$ & 
\end{tabular}

\caption{$U(N)$ gauge representations and vector R-charges 
of the chiral multiplets in the GLSM for complete intersections in Grassmannians.}
\label{tab:HoriTong}
\end{table}

In \cite{Hori:2006dk}, Hori and Tong studied a $U(N)$ gauge theory with $N_f$ chiral multiplets $\Phi_i$ in the fundamental representation and $S$ chiral multiplets $P^\alpha$ transforming in the $\det^{-Q_\alpha}$ representation of the gauge group, with $Q_\alpha>0$ (see Table \ref{tab:HoriTong}). We will focus on the case $N_f=\sum_\alpha Q_\alpha$, so that the axial $R$-symmetry is anomaly-free and the vacuum moduli space is a noncompact Calabi-Yau manifold of complex dimension $(N_f-N)N+S$. For $\xi>0$ the vacuum moduli space is the total space of $\oplus_\alpha \cO(-Q_\alpha) \rightarrow G(N,N_f)$. To obtain a compact Calabi-Yau, the superpotential 
\be\label{W_HoriTong}
W = \sum_{\alpha=1}^S P^\alpha G_\alpha(B)~,
\ee
is introduced, where $G_\alpha$ are generic degree $Q_\alpha$ polynomials in the baryons 
\be\label{baryons}
B_{i_1\dots i_N} = \epsilon_{a_1\dots a_N} \Phi^{a_1}_{i_1}\dots \Phi^{a_N}_{i_N} ~.
\ee
The vector $R$-symmetry assigns charge $2$ to $P_\alpha$ and $0$ to $\Phi_i$. 

For $\xi> 0$, the fundamentals acquire VEV and the $F$-term equations are solved by $P^\alpha=0$ and $G_\alpha(B)=0$. The GLSM is in a geometric phase: the low energy theory is a nonlinear sigma model on a compact Calabi-Yau $X_{Q_1,\dots,Q_S} \subset G(N,N_f)$, which is the complete intersection of the hypersurfaces $G_\alpha(B)=0$ in the Grassmannian $G(N,N_f)$. 
The compact Calabi-Yau has complex dimension $N(N_f-N)-S$. Excluding abelian examples which give hypersurfaces in projective spaces, there are six threefolds in this class, which were studied from a geometric viewpoint in \cite{Batyrev:1998kx} and from a physical viewpoint in \cite{Hori:2006dk}. We will list them and compute their topological correlators below.  In the $\xi<0$ phase, $P^\alpha$ acquire VEV while the fundamentals vanish, leaving a residual $PSU(N)$ gauge group.

Let us now focus on the geometric phase $\xi>0$. As we will explain better in section \ref{subsec:_Cartan_HT}, there is no need to use the associated Cartan theory to define the localization formula in this case. The reason is that the non-abelian gauge group is completely broken in this phase, and the instanton sums are absolutely convergent. Hence the topological correlators in the geometric phase are simply given by
\be\label{correlators_Hori_Tong}
\begin{split}
\langle \CO(\sigma) \rangle_0 &= (-1)^{\frac{N(N-1)}{2}+S} \frac{1}{N!} \sum_{k_a=0}^\infty ((-1)^{N-1} q)^{\sum_{a=1}^N k_a}  
\\
&  \oint_{(\hs_a=0)} \prod_{a=1}^N \frac{d\hs_a}{2\pi i} \, 
\prod_{1\leq a<b\leq N}(\hs_a-\hs_b)^2 ~
\frac{\prod\limits_{\alpha=1}^S(-Q_\alpha \sum\limits_{a=1}^N \hs_a)^{1+Q_\alpha \sum_a k_a}}{\prod\limits_{a=1}^N \hs_a^{N_f(k_a+1)}}
~\CO(\hs)~,
\end{split}
\ee
where we inserted the sign $(-1)^S$ due to the $S$ fields of $R$-charge $2$, to obtain positive intersection numbers.

In the CY case $N_f=\sum_{\alpha=1}^S Q_\alpha$, the selection rule for the axial $R$-symmetry implies that the correlator vanishes unless $\CO(\sigma)$ is a homogeneous polynomial of degree equal to the complex dimension $d=N(N_f-N)-S$ of the CY. This result is also easily derived from the previous formula. 

The instanton series in \eqref{correlators_Hori_Tong} can be resummed to give 
\be\label{correlators_Hori_Tong_resummed}
\begin{split}
\langle \CO(\sigma) \rangle_0 
&= (-1)^{\frac{N(N-1)}{2}}  \frac{1}{N!}\oint \prod_{a=1}^N \frac{d\hs_a}{2\pi i} \, \prod_{1\leq a<b\leq N}(\hs_a-\hs_b)^2~ \cdot \\ 
& \cdot  \frac{\prod\limits_{\alpha=1}^S(Q_\alpha \sum\limits_{a=1}^N \hs_a)}{\prod\limits_{a=1}^N \left[ \hs_a^{N_f}+(-1)^N q \prod\limits_{\alpha=1}^S(-Q_\alpha \sum\limits_{b=1}^N \hs_b)^{Q_\alpha}   \right]} ~ \CO(\hs) ~,
\end{split}
\ee
where the contour integral is now around poles in $\hs_a$ proportional to positive powers of $q$. The contour picks all the zeros of the denominator, namely all the solutions of the vacuum equations $e^{2\pi i\d_{\hs_a} \t W_{\rm eff}(\hs)}=1$. (We will be more precise about the details at the end of the section.) 
This expression manifests the connection with the twisted chiral ring relations and the quantum Coulomb branch vacua of the theory. Note in particular that the correlators diverge along the \emph{singular locus} in moduli space where a noncompact Coulomb branch arises \cite{Morrison:1994fr,Hori:2006dk,Jockers:2012zr}, because then the poles in $\sigma$ (\ie{ the quantum Coulomb vacua}) are no longer isolated. We will indeed see that the correlators are rational functions in $q$ with poles along the singular locus $\Delta(q)=0$. 

We now list the twisted chiral correlators for the Hori-Tong GLSM that engineer $CY_3$ complete intersections in Grassmannians, and compare our results to the Yukawa couplings obtained in \cite{Batyrev:1998kx} using mirror symmetry. In the following we use the notation 
\be\label{Casimir}
u_j(\sigma)\equiv \Tr(\sigma^j)
\ee
for the Casimir invariants of $\sigma$. 

\begin{itemize}
\item $X_4\subset G(2,4)$: the correlators are
\be\label{correlators_X4_G24}
\langle u_1(\sigma)^3 \rangle = \frac{8}{1-2^{10}q}~, \qquad \langle u_2(\sigma)u_1(\sigma) \rangle = 0~.
\ee
The denominator shows that the singular locus is $q=2^{-10}$. 
\item $X_{3,1^2}\subset G(2,5)$: the correlators are
\be
\begin{split}
\langle u_1(\sigma)^3 \rangle &= \frac{15}{1+11 \left(3^3 q\right)-\left(3^3 q\right)^2} \\ \langle u_2(\sigma)u_1(\sigma) \rangle &= \frac{3(1-54q)}{1+11 \left(3^3 q\right)-\left(3^3 q\right)^2}~.
\end{split}
\ee
$\langle u_1(\sigma)^3 \rangle$ agrees with the Yukawa coupling computed in \cite{Batyrev:1998kx}, with $q=-z$.
\item $X_{2^2,1}\subset G(2,5)$: the correlators are
\be
\begin{split}
\langle u_1(\sigma)^3 \rangle &= \frac{20}{1+11 \left(2^4 q\right)-\left(2^4 q\right)^2} \\ \langle u_2(\sigma)u_1(\sigma) \rangle &= \frac{4(1-32q)}{1+11 \left(2^4 q\right)-\left(2^4 q\right)^2}~.
\end{split}
\ee
$\langle u_1(\sigma)^3 \rangle$ agrees with the Yukawa coupling computed in \cite{Batyrev:1998kx}, with $q=-z$. 
\item $X_{2,1^4}\subset G(2,6)$: the correlators are
\be
\begin{split}
\langle u_1(\sigma)^3 \rangle &= \frac{28}{\left(1+2^2 q\right) \left(1-27 \left(2^2 q\right)\right)}\\ 
\langle u_2(\sigma)u_1(\sigma) \rangle &=\frac{8(1+18q)}{\left(1+2^2 q\right) \left(1-27 \left(2^2 q\right)\right)}~.
\end{split}
\ee
$\langle u_1(\sigma)^3 \rangle$ agrees with the Yukawa coupling computed in \cite{Batyrev:1998kx}, with $q=z$. 
\item $X_{1^7}\subset G(2,7)$: the correlators are 
\be
\begin{split}
\langle u_1(\sigma)^3 \rangle &= \frac{14 (3+q)}{1+57 q -289 q^2  -q^3} \\ 
\langle u_2(\sigma)u_1(\sigma) \rangle &=\frac{14 (1-9q)}{1+57 q -289 q^2  -q^3}~.
\end{split}
\ee
$\langle u_1(\sigma)^3 \rangle$ agrees with the Yukawa coupling computed in \cite{Batyrev:1998kx}, with $q=-z$. This is the celebrated R\o dland Calabi-Yau \cite{Rodland}.
\item $X_{1^6}\subset G(3,6)$: the correlators are 
\be
\begin{split}
\langle u_1(\sigma)^3 \rangle &= \frac{42}{(1-q) (1-64 q)} \\ 
\langle u_2(\sigma)u_1(\sigma) \rangle &=0 \\
\langle u_3(\sigma) \rangle &= -\frac{6(1-8q)}{(1-q) (1-64 q)} 
\end{split}
\ee
$\langle u_1(\sigma)^3 \rangle$ agrees  (up to a typo) with the Yukawa coupling computed in \cite{Batyrev:1998kx}, with $q=z$. 

\end{itemize}

\subsubsection{The associated Cartan theory and the $\xi<0$ phase}\label{subsec:_Cartan_HT}

As we have explained, the proper way to deal with a non-abelian GLSM, which has fewer FI parameters than the rank of the gauge group, is to consider the associated Cartan theory \cite{Halverson:2013eua} (see also \cite{Jockers:2012dk}). This procedure is necessary to discuss phases of the non-abelian GLSM where the gauge group is not Higgsed to a finite group, as is the case in the $\xi<0$ phase of the models of Hori and Tong.

The Cartan theory associated to the non-abelian $U(N)$ GLSM is a GLSM with gauge group the maximal torus $U(1)^N$, with the same chiral matter as in the $U(N)$ theory plus extra chiral multiplets of vector $R$-charge $2$ associated to the $W$-bosons of $U(N)$. As we have explained, the chiral multiplets originating from $W$-bosons do not contribute poles to the 1-loop determinants, because of a cancellation between opposite roots $\pm \alpha$. They are spectators in the following analysis of the phase diagram, the pole structure of the integrand and the JK residue.

The advantage of the Cartan theory is to have FI parameters $(\xi_1,\dots,\xi_N)$ belonging to the dual of the Cartan subalgebra of $U(N)$. The presence of as many FI parameters as the rank of the gauge group ensures that in the interior of each chamber in FI space that defines a phase, the gauge group is Higgsed completely (up to a finite group) and the instanton sums are absolutely convergent. We will therefore formulate the localization formula in the associated Cartan model with FI parameters $\vec{\xi}=(\xi_1,\dots,\xi_N)$ in the dual of the Cartan subalgebra and correspondingly instanton factors $\vec{q}=(q_1,\dots,q_N)$, and take the physical limit $\vec{q}=(q_1,\dots,q_N)\to q (1,\dots,1)$ at the end of the computation: 
\be\label{correlators_Hori_Tong_Cartan}
\begin{split}
\langle \CO(\sigma) \rangle_0 &= \lim_{\substack{\vec{\xi}\to \xi(1,\dots,1)\\ \vec{q}\to q (1,\dots,1)}} ~ (-1)^{\frac{N(N-1)}{2}+S} \frac{1}{N!} \sum_{\vec{k}\in \bZ^N} \prod_{a=1}^N ((-1)^{N-1} q_a)^{k_a} 
\\
& {\rm JK{\text-}Res}\left[\vec{\xi}\right]
\,d^N \hs ~ 
\prod_{1\leq a<b\leq N}(\hs_a-\hs_b)^2 ~
\frac{\prod\limits_{\alpha=1}^S(-Q_\alpha \sum\limits_{a=1}^N \hs_a)^{1+Q_\alpha \sum_a k_a}}{\prod\limits_{a=1}^N \hs_a^{N_f(k_a+1)}}
~\CO(\hs)~.
\end{split}
\ee

To specify the JK residue, we need to discuss the phase structure of the Cartan theory. The charge vectors of the chiral multiplets that originate from the fundamentals $\Phi^a$ and the determinant fields $P^\alpha$ define $N+1$ rays generating $N+1$ chambers in FI space $\bR^N$. Therefore the Cartan theory has $N+1$ phases, which we now discuss. 

If $\xi_a>0$ for all $a=1,\dots,N$, then all the fields $\Phi^a$, $a=1,\dots,N$, take VEV. The JK residue is an iterated residue at the poles of the 1-loop determinants of fields $\Phi^a$, namely $\hs_a=0$ for all $a$: 
\be\label{JK_regul-HoriTong_firstphase}
\begin{split}
&{\rm JK{\text-}Res}\left[\vec{\xi} \right] f(\hs) d^N \hs 
= \, \res_{\hs_N=0}\, \dots \,\res_{\hs_1=0} ~f(\hs) ~.
\end{split}
\ee
The instanton sum reduces to the dual cone $k_a\geq 0$ for all $a=1,\dots,N$, therefore it is a Taylor series in $q_1, q_2, \dots, q_N$. We call this phase the $0$-th phase.

If $-\xi_{\bar a}>0$ and $\xi_a-\xi_{\bar a}>0$ for all $a=1,\dots,N$ different from ${\bar a}$, then 
the fields $P^\alpha$ and all the fields $\Phi^a$ except for $\Phi^{\bar a}$ take VEV. There is a total of $N$ phases of this kind, depending on the choice of ${\bar a}$. The JK residue is a residue at the poles of the corresponding 1-loop determinants, given by 
\be\label{JK_regul-HoriTong_secondphase}
\begin{split}
&{\rm JK{\text-}Res}\left[\vec{\xi} \right] f(\hs) d^N \hs 
= - \,\res_{\hs_N=0}\, \dots \,\res_{\hs_{\bar{a}+1}=0}\, \res_{\hs_{\bar{a}-1}=0}\dots \,\res_{\hs_1=0}~ \res_{\hs_{\bar{a}}=-\sum_{\substack{a=1\\ a\neq \bar{a}}}^N\hs_a } ~f(\hs) ~.
\end{split}
\ee
Taking into account that $R[P^\alpha]=2$, the instanton sum reduces to the shifted dual cone defined by $-\sum_{b=1}^N k_b\geq 1$ and $k_a\geq 0$ for all $a\neq{\bar a}$: it is therefore a Taylor series in $1/q_{\bar a}$ and in $q_a/q_{\bar a}$ for all $a\neq {\bar a}$. 
We call this phase the $\bar{a}$-th phase.

Note that thanks to the $R$-charge $2$ of the determinant fields $P^\alpha$, for any dual cone in $\vec{k}$-space only $N$ types of 1-loop determinants out of $N+1$ can simultaneously have poles, reflecting the previous phase structure. The corresponding arrangement of hyperplanes is therefore projective in each phase.

Let us now discuss the physical limit $\vec{q}=(q_1,\dots,q_N)\to q (1,\dots,1)$. If $|q|<1$, that is if we are in the geometric phase $\xi>0$, the FI parameter $\xi(1,\dots,1)$ is in the interior of the $0$-th phase described above. Correspondingly, the instanton sum is absolutely convergent. This explains why we did not need to use the Cartan theory to discuss the geometric phase of the Hori-Tong GLSM.

If instead $|q|>1$, that is if we are in the phase $\xi<0$, the FI parameter $\xi(1,\dots,1)$ lies along a generator of the cone of the Cartan theory, the common boundary of phases $1$, $2$, \dots, $N$. At this boundary only $P^\alpha$ and none of the $\Phi^a$ fields acquire VEV. In the non-abelian theory, $U(N)$ is only broken to $PSU(N)$. In the Cartan theory, $N-1$ out of the $N$ $U(1)$ gauge factors are not broken. Correspondingly, $N-1$ out of $N$ instanton sums in the Hori-Tong GLSM are at the radius of convergence. Moving to the Cartan theory and perturbing the FI parameter $(\xi_1, \dots,\xi_N)$ away from $\xi(1,\dots,1)$ so that it enters one of the $N$ phases above, the instanton series falls within its radius of convergence and can be safely resummed. We can finally take the physical limit $\vec{q}=(q_1,\dots,q_N)\to q (1,\dots,1)$, which is non-singular. 

\begin{figure}[!t]
\centering\includegraphics[width=7cm]{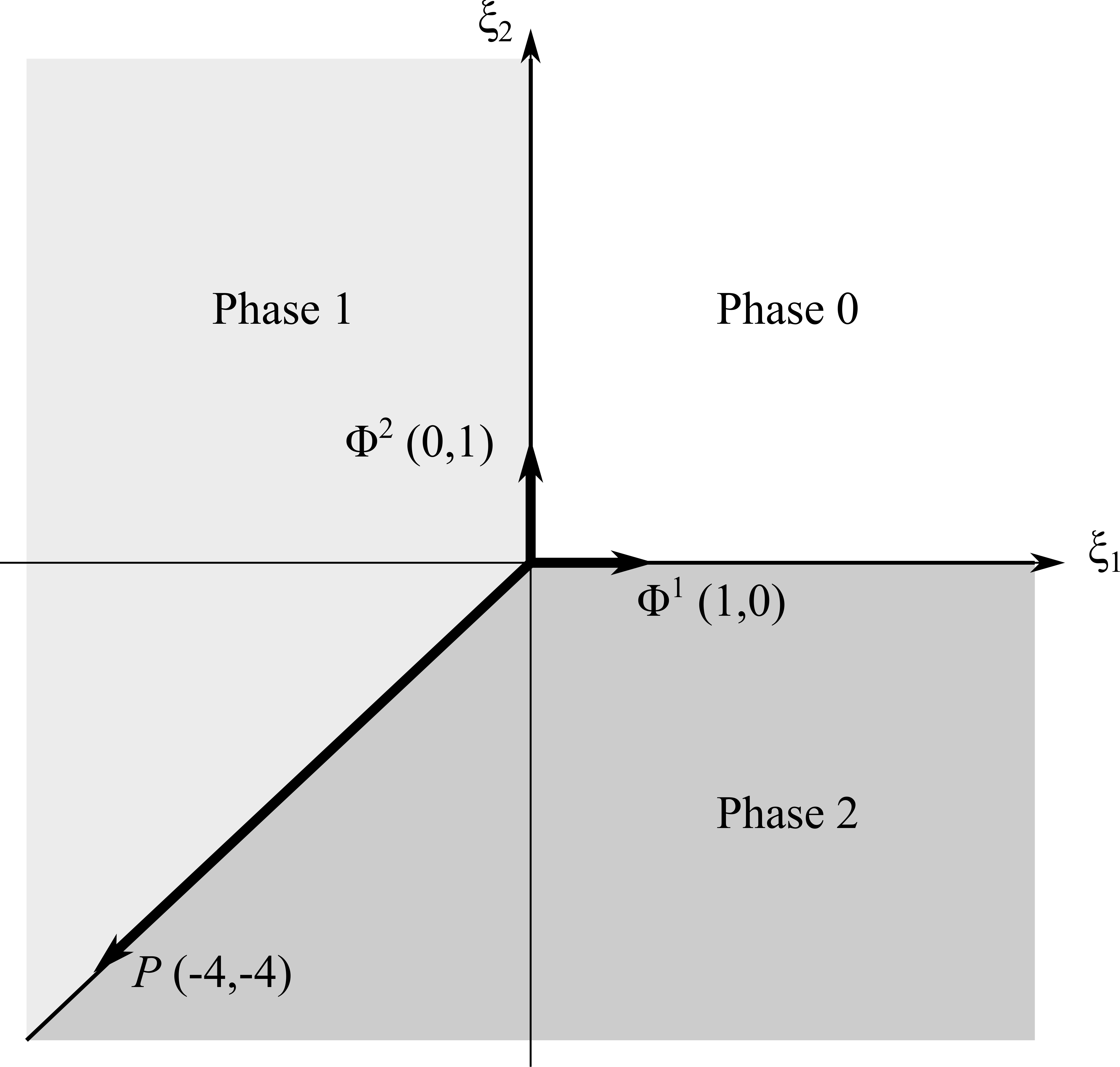}
\caption{\small The phase diagram of the Cartan theory associated to the $X_4\subset G(2,4)$ GLSM. }
\label{f:phases_HT}
\end{figure}

Let us exemplify this discussion in the case of $X_4\subset G(2,4)$, which is based on a $U(2)$ gauge theory. The phase diagram of the associated Cartan theory is shown in Figure \ref{f:phases_HT}. The correlators \eqref{correlators_Hori_Tong_Cartan} are given by a Taylor series in $q_1, q_2$ in the $0^{\rm th}$ phase $\xi_1>0,\xi_2>0$, by a Taylor series in $q_1^{-1}, q_2 q_1^{-1}$ in the $1^{\rm st}$ phase $-\xi_1>0, -\xi_1+\xi_2>0$, and by a Taylor series in  $q_2^{-1}, q_1 q_2^{-1}$ in the $2^{\rm nd}$ phase $-\xi_2>0, \xi_1-\xi_2>0$. 

Resumming the Taylor series in any of the three phases, the correlators of Casimir invariants in the Cartan theory (the arguments of the limit in \eqref{correlators_Hori_Tong_Cartan}) are found to be
\be\label{correlators_cartan}
\begin{split}
\langle u_1(\sigma)^3 \rangle_{\rm Cartan} &= \frac{8 (1+1536 (q_1+q_2)-720896 (q_1^2+q_2^2)-2752512 q_1 q_2)}{\Delta(q_1,q_2)} \\
\langle u_2(\sigma)u_1(\sigma) \rangle_{\rm Cartan} &= \frac{2097152 (q_1 - q_2)^2 (-3 + 256 (q_1 + q_2))}{\Delta(q_1,q_2)}
~,
\end{split}
\ee
where 
\be
\Delta(q_1,q_2)= \mathrm{Resultant}_x \left(1 + 256 q_1 (1 + x)^4 , x^4 + 256 q_2 (1 + x)^4 \right)~
\ee
is a quartic polynomial whose vanishing gives the singular locus of the Cartan theory. The independence of the correlators on the phase is due to their analyticity in $q_1, q_2$. In the physical limit $(q_1,q_2)\to q(1,1)$,  \eqref{correlators_cartan} reduce to the correlators \eqref{correlators_X4_G24} that we computed previously in the geometric phase of the non-abelian theory.

\subsubsection{Resumming the instantons and phase independence of correlators}

The correlators \eqref{correlators_Hori_Tong_Cartan} can be argued to be independent of the phase of the associated Cartan theory by resumming the instanton series as follows. We first rewrite
\be\label{instanton_sum}
\mathbf{I}(\hs,q)\equiv \sum_{\vec{k}} \prod_{a=1}^N \left[(-1)^{N-1} q_a \frac{\prod\limits_{\alpha=1}^S(-Q_\alpha \sum\limits_{a=1}^N \hs_a)^{Q_\alpha}}{\hs_a^{N_f}} \right]^{k_a} = \sum_{\vec{k}} e^{2\pi i\sum_{a=1}^N  k_a \d_a \h W_{\rm eff}(\hs)}
\ee
where $\h {W}_{\rm eff}(\s)$ is the effective twisted superpotential, with
\be\label{Weff_HoriTong}
\begin{split}
2\pi  i\, \h W_{\rm eff}(\hs) &= 2\pi i \sum_{a=1}^N \tau_a \hs_a -\pi i \sum_{a=1}^N (N+1-2a) \hs_a + \\
& - N_f \sum_{a=1}^N \hs_a(\log \hs_a-1) - \sum_{\alpha=1}^S (-Q_\alpha \sum_{a=1}^N \hs_a) (\log(-Q_\alpha \sum_{a=1}^N \hs_a)-1)~.
\end{split}
\ee

In the $0$-th phase where $\xi_a>0$ for all $a=1,\dots,N$, the instanton sum is over $k_a\geq 0$ for all $a=1,\dots,N$. Performing the summation, \eqref{instanton_sum} becomes 
\be\label{instanton_sum_first_phase}
\mathbf{I}_0(\hs,q) = \prod_{a=1}^N \frac{1}{1-e^{2\pi i \d_a \h W_{\rm eff}(\hs)}}~.
\ee

In the $\bar{a}$-th phase where $-\xi_{\bar a}>0$ and $\xi_a-\xi_{\bar a}>0$ for all $a=1,\dots,N$ different from ${\bar a}$, the instanton sum is over $\sum_{b=1}^N k_b \leq -1$ and $k_a\geq 0$ for all $a\neq{\bar a}$. Performing the summation, \eqref{instanton_sum} becomes 
\be\label{instanton_sum_second_phase}
\begin{split}
\mathbf{I}_{\bar{a}}(\hs,q) &= \frac{e^{-2\pi i\d_{\bar{a}} \h W_{\rm eff}(\hs)}}{1-e^{-2\pi i\d_{\bar{a}} \h W_{\rm eff}(\hs)}} \prod_{\substack{a=1\\ a\neq \bar{a}}}^N \frac{1}{1-e^{2\pi i(\d_a-\d_{\bar{a}}) \h W_{\rm eff}(\hs)}} =\\
&= - \frac{1}{1-e^{2\pi i\d_{\bar{a}} \h W_{\rm eff}(\hs)}} \prod_{\substack{a=1\\ a\neq \bar{a}}}^N \frac{1}{1-e^{2\pi i(\d_a-\d_{\bar{a}}) \h W_{\rm eff}(\hs)}} 
\end{split}
\ee

After resummation, the JK residues \eqref{JK_regul-HoriTong_firstphase} and \eqref{JK_regul-HoriTong_secondphase} become iterated residues at the poles of the previous expressions, which are the quantum Coulomb branch vacua that solve $\setcond{\hs}{ e^{2\pi i \d_a \h W_{\rm eff}(\hs)}=1~\forall a=1,\dots,N}$.%
~\footnote{We are glossing over a subtlety here: 
when quantum Coulomb vacua exist, they are not isolated but instead form a non-compact one-dimensional Coulomb branch \cite{Hori:2006dk}. This happens when $q$ is at the singular locus. We will be more precise on the definition of the residue in the next subsection.}
 Even though \eqref{instanton_sum_first_phase} and \eqref{instanton_sum_second_phase} are different, they coincide at their poles, \emph{i.e.} on-shell in the twisted chiral ring.~\footnote{The relative minus sign between \eqref{instanton_sum_second_phase} and \eqref{instanton_sum_first_phase} is compensated by the relative minus sign between \eqref{JK_regul-HoriTong_firstphase} and \eqref{JK_regul-HoriTong_secondphase} when computing the residue.}
This shows the equality of the correlators across all phases.

\subsubsection{Resumming the instantons and the general residue formula}

In this section we elaborate on \eqref{correlators_Hori_Tong_resummed} and the previous discussion, and present a simple residue formula that allows to compute the topological correlators in all Hori-Tong models. We consider the physical limit $q_a=q$ for all $a=1,\dots,N$ to simplify some of the following formulas. The generalization to the associated Cartan theory is straightforward, though no longer necessary after resumming the instanton series.

We have seen that after resumming the instantons the quantum Coulomb branch vacuum equations naturally appear in the integrand. (We will refer for definiteness to formula \eqref{instanton_sum_second_phase} and take $\bar{a}=N$ with no loss of generality.) It is important to note that the quantum Coulomb branch vacua, when they exist, are not isolated \cite{Hori:2006dk}: given a solution $(\hs_1,\dots,\hs_N)$ to the vacuum equations, the rescaled $\lambda (\hs_1,\dots,\hs_N)$ is also a solution for any $\lambda\in \bC$. This is a general property of $R_A$-anomaly-free GLSM, that flow to non-trivial fixed points: the quantum Coulomb branch vacuum equations are invariant under complexified $U(1)_A$ transformations, that is a common rescaling of all the $\t\sigma$ variables. In the case at hand, the vacuum equations are therefore $N$ equations for $N-1$ variables, which only have solutions if $q$ is at the singular locus $\Delta(q)=0$ where the equations become dependent. In that case the rescaling mode parametrizes a non-compact Coulomb branch and the CFT is singular.

Let us then change variables from $(\hs_1,\dots,\hs_N)$ to $(x_1,\dots,x_{N-1},y)$, where 
\be\label{new_variables}
x_a = \frac{\hs_a}{\hs_N} \quad (a=1,\dots,N-1)~, \qquad y = \prod_{a=1}^N \hs_a^{1/N}~,
\ee
so that the rescaling $(\hs_1,\dots,\hs_N)\mapsto\lambda (\hs_1,\dots,\hs_N)$ translates to $y\mapsto \lambda y$ with $x_a$ fixed. 

In these new variables, using the notation $x_N=1$ and keeping $\hs_N=y \prod_{a=1}^{N-1} x_a^{1/N}$ temporarily to shorten some formulas, we obtain
\begin{align}\label{instanton_sum_HT}
\mathbf{I}_{N}(x,q) &= - \frac{1}{1-(-1)^{N-N_f-1}q \prod\limits_{\alpha=1}^S Q_\alpha^{Q_\alpha} (\sum\limits_{a=1}^N x_a)^{N_f} } \prod_{a=1}^{N-1} \frac{1}{1-x_a^{-N_f}} ~,\\
\label{Z00_HT}
\cZ^\oneloop_0 (x,\hs_N;0) &= (-1)^{\frac{N(N-1)}{2}} \prod_{1\leq a<b\leq N} (x_a-x_b)^2 \cdot \prod_{\alpha=1}^S Q_\alpha \cdot \frac{(\sum_{a=1}^N x_a)^S}{\prod_{a=1}^{N-1} x_a^{N_f}} \hs_N^{-N-d}~,
\end{align}
where $d=N(N_f-N)-S$ is the complex dimension of the compact Calabi-Yau. Finally, the dimension $d$ inserted operator is 
\be\label{CO_HT}
\cO(\hs) = \hs_N^d \cO(x)
\ee
and the integration measure is
\be\label{measure_HT}
d^N \hs = \hs_N^N \frac{dy}{y} \prod_{a=1}^{N-1} d x_a~.
\ee

Collecting all these ingredients together, we see that the factors of $\hs_N$ cancel out. 
Next, we see that the integral over the non-compact mode $y$ decouples, giving $\oint \frac{dy}{2\pi i y}=1$.
This is because the quantum Coulomb branch vacuum equations do not involve the rescaling mode $y$, as is visible from  \eqref{instanton_sum_HT}. The integral over $y$ simply imposes the $U(1)_A$ selection rule. We are left with $N-1$ contour integrals over $x_a$, $a=1,\dots,N-1$. The contour encircles the poles that solve the independent vacuum equations $x_a^{N_f}=1$ for all $a=1,\dots,N-1$, which are given by 
\be\label{poles_HT}
x_a = \omega_{N_f}^{m_a}~, \qquad m_a=0,\dots,N_f-1
\ee
for all $a=1,\dots,N-1$, where $\omega_{N_f}=e^{2\pi i/N_f}$ denotes an $N_f$-th root of unity. In summary, the correlators 
of the Hori-Tong GLSMs are given by the residue formula 
\be\label{residue_formula_HT}
\begin{split}
\langle \cO(\s)\rangle_0 &= \prod_{\alpha=1}^S Q_\alpha\cdot \frac{(-1)^{\frac{N(N-1)}{2}}}{N!}    \sum_{(m_1,\dots,m_{N-1})\in \bZ_{N_f}^{N-1}} \res_{x_{N-1}=\omega_{N_f}^{m_{N-1}}}\dots \res_{x_{1}=\omega_{N_f}^{m_{1}}}\\
& 
\qquad \qquad \qquad \frac{\left[\prod\limits_{1\leq a<b\leq N} (x_a-x_b)^2 \right] \cdot (\sum\limits_{a=1}^N x_a)^S \cdot \cO(x)}{\left[\prod\limits_{a=1}^{N-1} (x_a^{N_f} -1)\right]\left[1+(-1)^{N-N_f}q \prod\limits_{\alpha=1}^S Q_\alpha^{Q_\alpha} (\sum\limits_{a=1}^{N} x_a)^{N_f} \right]} ~,
\end{split}
\ee
where again $x_N=1$ is understood. 

Formula \eqref{residue_formula_HT} immediately reproduces the correlators computed using formula \eqref{correlators_Hori_Tong} and presented 
in section \ref{subsec:HoriTong} for Calabi-Yau threefolds. It can also be easily applied to GLSMs that engineer higher-dimensional Calabi-Yau manifolds.

The residue formula \eqref{residue_formula_HT} for twisted chiral correlators is closely related to the analysis of the quantum Coulomb branch carried out in \cite{Hori:2006dk} using the effective twisted superpotential that arises from integrating out all massive fields at a generic Coulomb branch vacuum.%
~\footnote{The integers $n_a$ introduced in \cite{Hori:2006dk} are related to ours by  $(n_1,\dots,n_{N-1},n_N)=(m_1,\dots,m_{N-1},0)$.} The authors of \cite{Hori:2006dk} therefore considered all the quantum Coulomb branch vacua around which all matter fields and $W$-bosons are massive, that is
$\hs_a\neq 0$ for all $a$, $\sum_{a=1}^N \hs_a \neq 0$, and $\hs_a\neq \hs_b$ for all $a\neq b$. These conditions are implemented automatically in \eqref{residue_formula_HT}: there are no poles at $x_a=0,\infty$, whereas poles such that $\sum_{a=1}^N x_a = 0$ or $x_a = x_b$ for some $a\neq b$ have vanishing residue thanks to the numerator.

\subsection{The Gulliksen-Neg\aa rd $CY_3$}

\begin{table}[t]
\centering
\begin{tabular}{c|ccc|c}
 & $\Phi_\alpha$ & $P^i$ & $X_i$ & FI \\ \hline
$U(1)_\sigma$ & $+1$ & $-1$ & $0$ & $\xi_0$ \\
$U(2)_\Sigma$ & $\mathbf{1}$ & $\mathbf{2}$ & $\overline{\mathbf{2}}$ & $\xi$ \\ \hline
$U(1)_{R}$ & $0$ & $0$ & $2$ & 
\end{tabular}
\caption{Gauge representations and vector $R$-charges of the chiral multiplets in the PAX GLSM for the determinantal Gulliksen-Neg\aa rd $CY_3$.}
\label{tab:JKLMR_GN}
\end{table}

The computations of the previous subsection can be repeated for the PAX/PAXY gauged linear sigma models introduced in \cite{Jockers:2012zr} to describe determinantal Calabi-Yau varieties. PAX and PAXY models are related by the duality of \cite{Hori:2011pd}. In this subsection we compute topological correlators in the PAX model of the Gulliksen-Neg\aa rd  Calabi-Yau threefold \cite{Gulliksen}. 

The matter content of the PAX model for the Gulliksen-Neg\aa rd  $CY_3$ of \cite{Jockers:2012zr} is listed in Table \ref{tab:JKLMR_GN}. The gauge group is $U(1)\times U(2)$, with complex scalars $\s$ and $\Sigma$ in the adjoint representation of $U(1)$ and $U(2)$ respectively. There are $8$ chiral multiplets $\Phi_\alpha$ of charge $+1$ under $U(1)$, $4$ chiral multiplets $P^i$ in the bifundamental representation of $U(2)\times U(1)$, and $4$ chiral multiplets $X_i$ in the antifundamental representation of $U(2)$, subject to a superpotential
\be\label{W_PAX_GN}
W=\tr (P A(\Phi) X)~,
\ee
where $A(\Phi)=\sum_{\alpha=1}^8 A^\alpha \Phi_\alpha$ and $A^\alpha$ are $8$ constant $4\times 4$ matrices. 

For simplicity we will work in phase I of the GLSM of \cite{Jockers:2012zr}, corresponding to the cone $\xi_0+2\xi > 0$, $\xi> 0$ in FI space. Here $\xi_0$ and $\xi$ are FI parameters for the $U(1)$ and $U(2)$ gauge groups respectively. In this phase the fields $\Phi$ and $P_a$ acquire VEV, whereas $X^a$ do not ($a=1,2$ is a $U(2)$ gauge index). In the associated Cartan theory, with FI parameters $(\xi_1,\xi_2)$ for the Cartan of $U(2)$, the phase where $\Phi$ and $P_a$ acquire VEV is given by the chamber $\xi_0+\xi_1+\xi_2>0$, $\xi_1>0$, $\xi_2>0$. Phase I of the non-abelian GLSM is obtained when $\xi_1=\xi_2=\xi>0$ and lies in the interior of this cone. Therefore the instanton sum over the closure of the dual cone $k_0\geq 0$, $k_1-k_0\geq 0$, $k_2-k_0\geq 0$ is convergent even for physical values of the instanton expansion factor for $U(2)$, $q_1=q_2=q$, to which we restrict in the following.

The topological correlators are given by
\be\label{correlators_PAX_GN}
\begin{split}
\langle \CO(\sigma,\Sigma) \rangle_0 &= - \sum_{k_0=0}^\infty \sum_{k_1=k_0}^\infty
\sum_{k_2=k_0}^\infty
q_0^{k_0} (-q)^{k_1+k_2} \res_{\hs=0} \res_{\hS_2=\s} \res_{\hS_1=\hs} \frac{1}{2} 
(\hS_1-\hS_2)^2 \cdot \\
& \qquad \cdot \hs^{-8(k_0+1)} \prod_{a=1}^2 \left[(-\hS_a)^{4(k_a+1)} (-\hs+\hS_a)^{-4(-k_0+k_a+1)}\right]
\,\CO(\hs,\hS)~.
\end{split}
\ee

To compare with the notation of \cite{Jockers:2012dk}, we introduce $z=q_0q^2$ and $w=-q$, so that the above formula expresses the correlator as a Taylor series in $z$ and $w$. By the selection rule for the axial $R_A$-symmetry, only correlators cubic in $\sigma$, $\Sigma$ do not vanish. 

We can obtain a simple alternative formula for the correlators by resumming the instantons as in the previous section. Changing variables to $x_1=\hS_1/\hs$, $x_2=\hS_2/\hs$, $y=(\hS_1\hS_2\s)^{1/3}$ and resumming the instanton series, we reach the residue formula
\be\label{residue_formula_GN}
\langle \CO(\sigma,\Sigma) \rangle_0 = 
-\frac{1}{2}   \sum_{m_1,m_2=0}^3  ~ 
\res_{x_a=\frac{1}{1-i^{m_a} w^{1/4}}}
\frac{(x_1-x_2)^2 (x_1 x_2)^4  \cO(1,x)}{\left[1-z (x_1 x_2)^4\right] \prod\limits_{a=1}^2 \left[(x_a-1)^4 -w x_a^4\right]} ~.
\ee

Using the notation
\be
\sigma_s \equiv \sigma~, \qquad 
\sigma_t \equiv \tr\Sigma-2\sigma~,
\ee
for the linear combinations of $\sigma$, $\tr\Sigma$ conjugate to  $\tau_0+2\tau$ and $\tau$, the correlators read 
\bea\label{correlators_GN}
\langle \sigma_s^3 \rangle &= \frac{\cN_{sss}}{\Delta}~, \quad
\langle \sigma_s^2 \sigma_t \rangle = \frac{\cN_{sst}}{\Delta} ~, \quad
\langle \sigma_s \sigma_t^2 \rangle = \frac{\cN_{stt}}{\Delta} ~, \quad
\langle \sigma_t^3 \rangle = \frac{\cN_{ttt}}{\Delta}~, \\
\langle &\sigma_s u_2(\Sigma) \rangle = \frac{\cN_{s2}}{\Delta} ~, \qquad 
\langle \sigma_t u_2(\Sigma)\rangle = \frac{\cN_{t2}}{\Delta} ~,\qquad
\langle u_3(\Sigma) \rangle = \frac{\cN_{3}}{\Delta} ~,
\eea
where the denominator
\be\label{Delta_GN}
\begin{split}
\Delta(z,w) &= \left[(1-w)^4-2 (1+6w +w^2) z+z^2\right]\cdot[(1-w)^8-4 (1-w)^4\cdot\\ 
&~\cdot(1-34w+w^2) z+2 (3+372w + 1298 w^2 +372 w^3+3w^4)) z^2+\\
&-4 (1-34w+w^2) z^3+z^4)]
\end{split}
\ee
determines the singular locus, in agreement with \cite{Jockers:2012zr,Jockers:2012dk}, and the numerators are 
\be\label{Yukawa_GN}
\begin{split}
\cN_{sss} &= 4 (1 - 2 w + w^2 - z) (1 - 2 w + w^2 + z) \cdot\\
&~~\cdot (5 - 20 w + 30 w^2 - 20 w^3 + 5 w^4 + 54 z + 212 w z + 54 w^2 z + 5 z^2)  \\
\cN_{sst} &= 4 (5-20 w+140 w^3-350 w^4+420 w^5-280 w^6+100 w^7-15 w^8+\\
&+12 z+380 w z-480 w^2 z-840 w^3 z+1420 w^4 z-372 w^5 z-120 w^6 z+\\
&-34 z^2-60 w z^2+60 w^3 z^2+34 w^4 z^2+12 z^3+212 w z^3+96 w^2 z^3+5 z^4)\\
\cN_{stt} &= 16 (1 + 2 w - 22 w^2 + 34 w^3 + 20 w^4 - 106 w^5 + 118 w^6 - 58 w^7 +  11 w^8 +\\
&- 2 z + 62 w z + 476 w^2 z - 564 w^3 z - 474 w^4 z + 438 w^5 z + 64 w^6 z +\\
&- 34 w z^2 - 94 w^2 z^2 - 94 w^3 z^2 - 34 w^4 z^2 + 2 z^3 - 30 w z^3 - 40 w^2 z^3 - z^4)\\
\cN_{ttt} &= 8 (1 + 16 w - 20 w^2 - 112 w^3 + 230 w^4 - 16 w^5 - 276 w^6 + 
   240 w^7 +\\
&-  63 w^8 - 4 z - 64 w z + 1380 w^2 z + 4224 w^3 z - 2332 w^4 z - 2944 w^5 z  +\\
&- 260 w^6 z + 6 z^2 + 80 w z^2 + 564 w^2 z^2 + 688 w^3 z^2 + 198 w^4 z^2 +\\
&- 4 z^3 - 32 w z^3 + 124 w^2 z^3 + z^4)
\end{split}
\ee
and 
\be\label{higher_Casimirs_GN}
\begin{split}
\cN_{s2} &= 4 (19 - 52 w - 88 w^2 + 556 w^3 - 970 w^4 + 836 w^5 - 368 w^6 + 
   68 w^7 +\\
& - w^8 + 136 z + 1412 w z - 604 w^2 z - 2904 w^3 z + 1296 w^4 z + 660 w^5 z +\\
&+ 4 w^6 z- 74 z^2 + 148 w z^2 + 688 w^2 z^2 + 268 w^3 z^2 - 6 w^4 z^2 +\\
&- 80 z^3 + 28 w z^3 + 4 w^2 z^3 - z^4)\\
\cN_{t2} &= 8 (9 + 16 w - 124 w^2 + 64 w^3 + 350 w^4 - 624 w^5 + 404 w^6 - 
   96 w^7 + w^8 +\\
&+ 4 z + 768 w z + 3604 w^2 z - 96 w^3 z - 3476 w^4 z - 800 w^5 z - 4 w^6 z +\\
&- 34 z^2 - 304 w z^2 - 532 w^2 z^2 - 160 w^3 z^2 + 6 w^4 z^2 +\\
&+ 20 z^3 + 32 w z^3 - 4 w^2 z^3 + z^4)\\
\cN_{3} &= 4 (21 + 132 w - 912 w^2 + 1844 w^3 - 1630 w^4 + 524 w^5 + 88 w^6 - 
   68 w^7  +\\
& + w^8 + 160 z + 3596 w z + 6668 w^2 z - 4776 w^3 z - 5160 w^4 z - 484 w^5 z +\\
&- 4 w^6 z - 126 z^2 + 220 w z^2 + 1032 w^2 z^2 + 404 w^3 z^2 + 6 w^4 z^2 +\\
&- 56 z^3 + 148 w z^3 - 4 w^2 z^3 + z^4)~.
\end{split}
\ee
The cubic correlation functions given by the first line of \eqref{correlators_GN} and \eqref{Yukawa_GN} are Yukawa couplings, which can be computed from mirror symmetry using standard techniques \cite{Candelas:1990rm, Hosono:1994av}. In order to  provide an independent check of our results, we performed that computation using the Picard-Fuchs operators in \cite{Jockers:2012dk}, and we found perfect agreement. The correlation functions given by the second line of \eqref{correlators_GN} and \eqref{higher_Casimirs_GN} are intrinsically non-abelian, and, to the best of our knowledge, their computation is a genuinely new result.




\section{Higgs branch localization and vortices}\label{section: Higgs branch loc}

In this final section, we consider an alternative localization argument. For simplicity, we only consider the special case of an abelian gauge group, 
\be
\GG= \GH= \prod_{a=1}^n U(1)_a~,
\ee
with chiral multiplets $\Phi_i$ of $R$-charges $r_i$, gauge charges $Q^a_i$, and twisted masses $m_i^F$. We shall also assume that theory has isolated Higgs vacua. None of these assumptions are strictly necessary. The generalization to a non-abelian $\GG$ could be carried out similarly, using the auxiliary Cartan theory, while the case of a continuous Higgs branch could be dealt with like in \cite{Morrison:1994fr}. 
Our main objective, here, is to  explain how the structure of the Coulomb branch formula can be understood as a more familiar sum over vortices  \cite{Morrison:1994fr}. Along the way, we introduce a simple $\epsdef$-deformation of the vortex equations, which might be of independent interest.

\subsection{Localizing on the Higgs branch}
Consider the localization Lagrangian:
\be\label{Sloc Higgs}
\SL_{\rm loc} = {1\over \e^2}(\SL_{YM} + \SL_{H}) + {1\over \g^2} \SL_{\t\Phi\Phi}~, 
\ee
where we introduced the $(\delta+\t\delta)$-exact term \cite{Benini:2012ui}:~\footnote{See  also \cite{Fujitsuka:2013fga,Benini:2013yva,Peelaers:2014ima} for further generalizations.} 
\bea
&\SL_{H} &=&\; \left(\delta+\t\delta\right) \left({\lambda_a -\t\lambda_a \over 2 i} H^a(\t\CA, \CA)\right)\cr
&&=&\; \left(D- 2 i f_{1\b 1} + i\epsdef(V_1 D_{\b 1}- V_{\b1} D_1)\t\sigma \right)_aH^a(\t\CA, \CA) + (\mathrm{fermions})~,
\eea
with $H^a(\t\CA, \CA)$ some  gauge-invariant function of the matter fields $\CA, \t\CA$. 
Note that this localizing action is invariant under $\delta+\t\delta$ instead of $\delta$ and $\t\delta$ separately, but this does not cause any difficulty.
The equation of motion for  $\t\sigma$ is simply $D_\mu D^\mu \t\sigma= O({\e^2\over \g^2})$. 
We take a double scaling limit where $\e^2, \g^2\rightarrow 0$ and ${\e^2\over \g^2} \rightarrow 0$, so that  $\t\sigma$ is constant on any saddle. The $D$ integral is Gaussian and imposes 
\be\label{eom D}
 D + i \epsdef (V_1 D_{\b1} - V_{\b1} D_1)\t\sigma = \e_0^2 H(\t\CA, \CA)~.
\ee 
We denote by $\CM_{\rm susy}$ the field configurations that satisfy the supersymmetry equations \eqref{SUSY eq for V nonAb} and \eqref{SUSY Phi}, as before. On the intersection with \eqref{eom D}, we obtain 
\bea\label{vortex equ gen}
&\t\sigma= {\rm constant}~,\qquad\quad && 2 i f_{1\b 1} =  \e_0^2 H(\t\CA, \CA)~, \cr
& (Q_i(\sigma)+ m_i^F) \CA_i = i \epsdef \CL_V^{(a)}\CA_i~, \qquad\quad &&  D_{\bz} \CA=0~,
\eea
and 
\be\label{susy higgs bis}
\CL_V \sigma=0~, \qquad  D_1 \sigma + i \epsdef V_1 \left(2 i f_{1\b 1}\right)=0~.
\ee
Here $\e^2_0$ is the ``bare'' dimensionful YM coupling appearing in \eqref{YM standard}, to be distinguished from the dimensionless $\e^2$ in \eqref{Sloc Higgs}.  One can check that any solution to \eqref{vortex equ gen}-\eqref{susy higgs bis} also solves the equation of motion of $\sigma$.
From now on, we choose 
\be\label{choice of H}
H^a(\t\CA, \CA)=   \sum_i Q_i^a \t\CA_i \CA_i 
- \t\xi^a ~.
\ee

Note that $\sigma$ is {\it not} constant on the supersymmetric saddles, which allows for contributions from nontrivial topological sectors. The parameters $\t\xi^a$ in \eqref{choice of H} are naturally identified with the $\t\tau$ couplings entering in \eqref{def tau}, with $\t\tau^a=- 2i \t\xi^a/\e^2$  like in \eqref{def txi}.
Unlike the physical FI parameters $\xi^a$, we can fix the couplings $\t\xi^a$  to our convenience. Their purpose is to localize on supersymmetric configurations similar to the Higgs branch vortices in flat space. We choose $\t\xi^a$ in a specific cone so that the vortex configurations, with $\CA \neq 0$, Higgses the gauge group to a finite subgroup.

If $\CA=\t\CA=0$, however, the localizing action \eqref{Sloc Higgs} with \eqref{choice of H} is the same as the Coulomb branch localizing action \eqref{Sloc coulomb}, and one should worry that some Coulomb branch-like configurations with the zero mode $\hs$ turned on might contribute. We expect that these additional configurations can be suppressed by sending $\t\xi \rightarrow \infty$ before taking the $\e^2 \ra 0$ limit. This scaling limit is also necessary to allow for vortices of arbitrarily large topological number, as we review below.

The fluctuation determinants of massive chiral multiplets in the vortex background can be computed by an index theorem (see appendix \ref{App: det} and references therein). Note that we are taking a double scaling limit \eqref{Sloc Higgs},  first sending  $\e^2\rightarrow 0$ to localize on the vortex saddles, and then sending $\g^2 \rightarrow 0$ to compute the fluctuations determinants.

\subsection{Vortex equations on $S^2_\Omega$}\label{subsec: vortex eq on S2}
It is instructive to first study the vortex equations \eqref{vortex equ gen} in the special case $\GG=U(1)$ with a single chiral multiplet $\Phi$ of charge $Q$, $R$-charge $r$ and twisted mass $m^F$. The vortex equations read~\footnote{Note that we take $\CA$ and its complex conjugate $\t\CA= \CA^\dagger$ in the frame basis.} 
\bea\label{vortex eq 1}
&2i f_{1\b 1} 
= \e^2_0\left( Q |\CA|^2- \t\xi\right) ~, \cr
& \left(\d_\bz -i {r\over 2}\omega_\bz- i Q a_\bz \right)\CA=0~,\cr
& \left(Q\sigma + m^F+  {r\over 2}\epsdef \right)\CA = i \epsdef V^\mu\left(\d_\mu  - i Q a_\mu\right)\CA~,
\eea
together with the remaining supersymmetry equations \eqref{susy higgs bis}.
Let us consider a given topological sector with flux:
\be\label{flux k vortex}
k ={1\over 2\pi} \int_{S^2} d^2 x \sqrt{g} (- 2 i f_{1\b1})~,
\ee
for the gauge field $a_\mu$.
The field $\CA$ can be viewed as a holomorphic section of
\be
\CK^{r\over 2}\otimes \CO(k)^Q\cong \CO( Q k -r)~,
\ee
by virtue of the second equation in \eqref{vortex eq 1}. Such sections exist if and only if $Q k - r \geq 0$, and they have $Qk -r$ simple zeros. On the $S^2_\Omega$ background,  these zeros must be partitioned between the north and south poles by rotational invariance. The section $\CA$ has a non-trivial transition function, with
\be\label{trans functions}
\CA^{(N)} = \left({z\over \bz}\right)^{\half(Q k- r)} \, \CA^{(S)}~,
\ee
between the northern and southern hemispheres.%
~\footnote{Recall that a section $\varphi$ of $\CO(n)$ transforms as $\varphi^{(N)}= z^n \varphi^{(S)}$ between patches. Thus, more precisely, $\CA$ is a section of the $U(1)$ line bundle, with first Chern class $Qk -r$, canonically associated to $\CO(Qk-r)$. Correspondingly, the gauge field $a_\mu$ can be chosen real.}
 (We write $\CA=\CA^{(N)}$, by default.)

\begin{figure}[!t]
\centering\includegraphics[width=8cm]{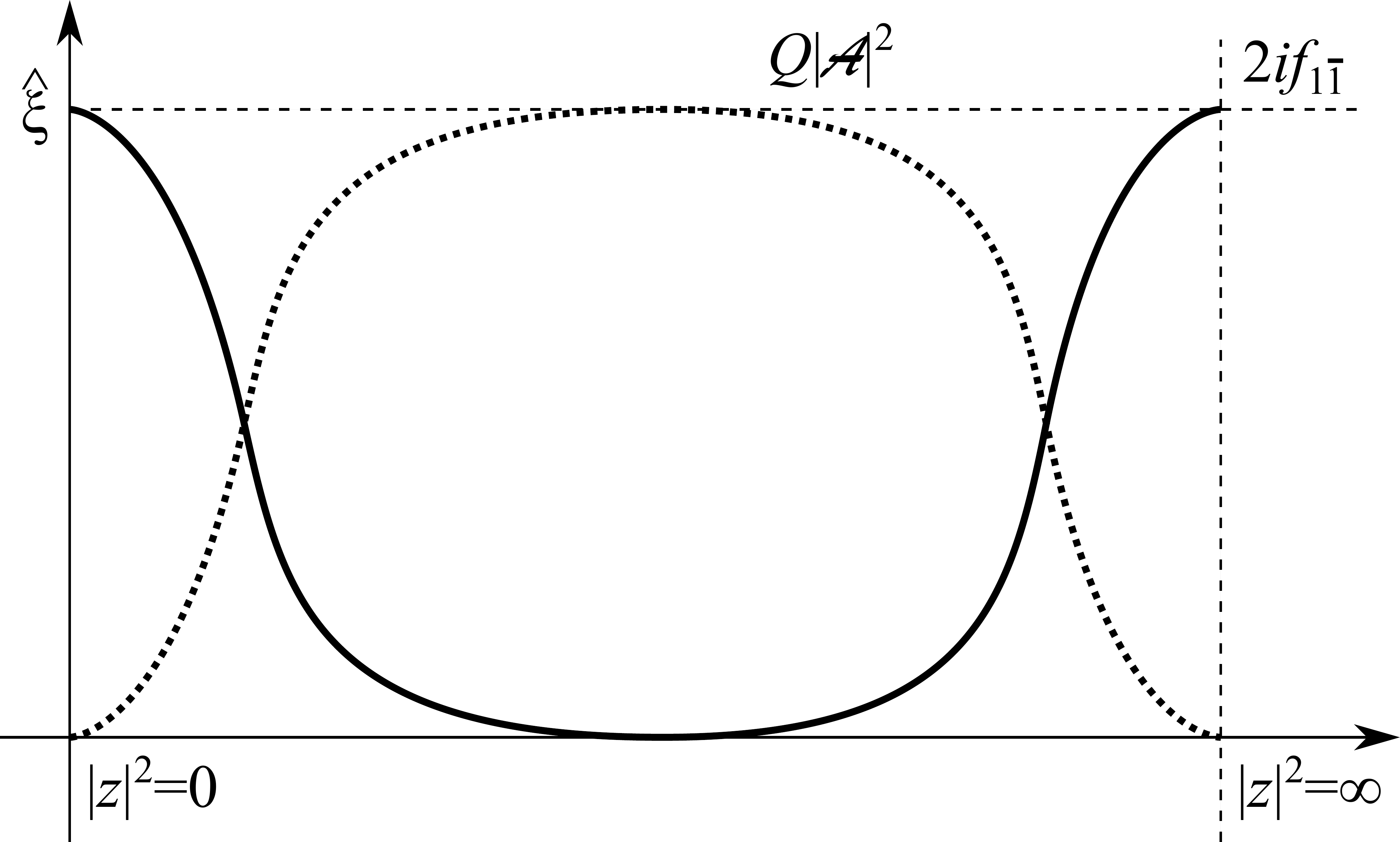}
\caption{\small  The profile of $Q|\CA|^2$ and of the gauge field flux $(-2 i f_{1\b 1})$ on the sphere (with $\e^2_0=1$). 
 The flux is localized near the poles ($|z|^2= 0$ and $|z|^2=\infty$), where $|\CA|^2$ vanishes, with size proportional to $1/(\e_0 \t\xi^{1/2})$. }
\label{f:profile}
\end{figure}

Let us choose $Q>0$ and $\t\xi > 0$, for definiteness. Then, $k$ is bounded from above according to 
$Q k - r \leq Q \e^2_0 \t\xi \, {\rm \vol}(S^2)/2\pi$. We consider a formal limit $\t\xi\rightarrow \infty$ such that all  vortex numbers are allowed. 
Consider the ansatz 
\be
\CA =  e^{f_1 + i f_2}~,
\ee 
where $f_1, f_2$ are real functions. Let us introduce the azimuthal angle $\phi$ with $z= |z| e^{i\phi}$. By rotational symmetry, we have $f_1=f_1(|z|^2)$, and $f_2$ a linear function of $\phi$. From \eqref{trans functions}, we see that $f_2^{(N)}= f_2^{(S)} + (Qk -r)\phi$.
Using the second equation in \eqref{vortex eq 1}, we can solve for a real gauge field,
\be\label{amu sol}
Q a_\mu dx^\mu  + {r\over 2}\omega_\mu dx^\mu = i (dz\d_z -d\bz \d_\bz) f_1 + df_2~.
\ee
Plugging back into \eqref{vortex eq 1}, we find an ordinary second order differential equation for $f_1(|z|^2)$,
\be
{4\over Q \e_0^2 \sqrt{g}} \d_{|z|^2}\left(|z|^2 \d_{|z|^2} \left(f_1+\frac{r}{8}\log g\right) \right)=  Q e^{2 f_1}- \t\xi~.
\ee
Solutions of this equation are known to exist \cite{Taubes:1979tm}. 
The two integration constants can be taken to be the orders of the zeros of $\CA$ at the poles. Due to the topological constraint \eqref{trans functions}, we really have a single integer parameter $p$:
\be\label{asympt A}
\CA^{(N)} \sim z^p~, \qquad \CA^{(S)}  \sim \left({1\over z}\right)^{-p+ Qk-r}~, \qquad  p=0, \cdots,  Qk-r~.
\ee
This is equivalent to 
\be\label{f asymptote}
f_1^{(N)} = {p\over 2}\log{|z|^2} + \cdots~, \qquad f_1^{(S)} = {p- Qk +r\over 2}\log{|z|^2} + \cdots~.
\ee
We therefore find a {\it vortex configuration}, schematically pictured in Figure \ref{f:profile}. 
There are $ Qk-r  +1$ distinct solutions for each flux $k$, labelled by the integer $p$ in \eqref{asympt A}.

Finally, the third equation in \eqref{vortex eq 1} determines the profile of $\sigma= \sigma(|z|^2)$ in terms of $f_1(|z|^2)$: 
\be
Q\sigma  =-m^F  - {r\over 2}\epsdef  - 2 \epsdef |z|^2 \d_{|z|^2}\left( f_1 +{r\over 8}\log{g}\right)~.
\ee
We are only interested in the values of $\sigma$ at the poles, which follow from \eqref{f asymptote}:
\be\label{sigma Higgs NS}
 Q\sigma_N+ m^F = \epsdef \left(-{r\over 2} - p \right)~, \qquad 
 Q\sigma_S+ m^F = \epsdef \left(-{r\over 2} - p + Qk\right)~.
\ee
Of course, \eqref{sigma Higgs NS}  satisfies the supersymmetry relation \eqref{rel flux to sigma},  $\sigma_S -\sigma_N= \epsdef k$.

\subsection{Higgs branch localization formula}

Consider an abelian GLSM with gauge charges $Q_i^a$, as above.  We denote by $(H)$ a solution of the equations
\be
\sum_iQ_i^a |\CA_i|^2 = \t\xi^a~\, (\forall a)~,
\qquad\quad \left(Q_i(\sigma)+ m_i^F+{r_i\over 2}\epsdef\right)\CA_i=0~\, (\forall i)~, 
\ee
such that at least $n$ distinct fields $\CA_j$, with  linearly independent charges $Q_j\in \Fh^\ast$, get a VEV, fully Higgsing the gauge group to a discrete subgroup.  We further restrict ourselves to the case when the ``Higgs vacua'' $(H)$ are isolated: that is, the VEVs of $\CA_j$ cannot be varied continuously. This can be achieved by turning on generic twisted masses $m^F$, if there are no chiral fields with the same quantum numbers. In that case, exactly $n$ fields $\CA_j$ get a VEV in a given Higgs vacuum. Let us introduce $J=\{j_1, \cdots, j_n\}$ labeling the $n$ fields $\CA_{j}$ that get a VEV in $(H)$.  For each such $(H)$, there exist $n$ distinct types of vortices behaving like in \eqref{asympt A}---that is, one has a tower of vortices indexed by $(k_H, p_H)$ for each $j\in J$, with $Q_j(k_H)- r_j \geq 0$  $\forall j\in J$. Here $k_H=(k_{H,a})_{a=1}^n$ are the fluxes of the vortex configuration, and $p_H=(p_{H,j})_{j \in J}$ with $0\leq p_{H,j} \leq Q_j(k_H)- r_j$ are the orders of zeros of $\cA_j^{(N)}$.

The values of $\sigma_N$ and $\sigma_S$ on the Higgs vacuum $(H)$ are determined by  \eqref{sigma Higgs NS}. More precisely, one needs to solve the linear system 
\bea\label{Higgs_vacuum}
& Q_j(\sigma_N|_H)+ m_j^F = \epsdef \left(-{r_j\over 2} - p_{H,j} \right) \qquad \forall j \in J~ \cr
\eea
to determine $\s_N|_H$ in terms of the twisted masses, $R$- and gauge charges, and $p_H$. Then $\sigma_S|_H= \sigma_N|_H + k_H \epsdef$  by supersymmetry.

From the previous discussions, it is clear that the Higgs branch localization formula takes the form of a sum over vortices:
\be\label{Higgs branch formula}
\left\langle  \CO^{(N)}(\sigma_N) \CO^{(S)}(\sigma_S)\right\rangle =\sum_{(H)}\sum_{k_H} q^{k_H} \sum_{p_H} Z^{\rm vortex}_{k_H, p_H}  Z^{\rm massive}_{k_H, p_H}   \CO^{(N)}(\sigma_N|_H) \CO^{(S)}(\sigma_S|_H)~.\ee
Here, we have defined  the vortex contribution 
\be\label{Z_vortex}
 Z^{\rm vortex}_{k_H, p_H}= \res_{\s_N \to \s_N|_H} \prod_{j \in J}\epsdef^{r_j- 1 - Q_j(k_H)}{ \Gamma\left({Q_j(\sigma_N )+ m_j^F\over\epsdef} +{r_j\over 2}  \right) \over  \Gamma\left({Q_j(\sigma_N + k_H \epsdef)+ m_j^F\over\epsdef} +{2-r_j\over 2}  \right) }~,
\ee
which correspond to the fluctuations of the chiral multiplets $\CA_j$, $j\in J$, with the massless modes indexed by
$p_{H,j}$  
 removed. We used the one-loop determinant \eqref{Zphi general} and the supersymmetry condition $\s_S=\s_N+k_H \epsdef$, and removed the massless modes by taking the (multi-dimensional) residue. \eqref{Z_vortex} depends on $p_H$ through the solution $\s_N|_H$ to \eqref{Higgs_vacuum}.

The remaining contribution is from all the chiral multiplets which do not participate in the vortices of $(H)$: 
\be\label{Z_massive}
 Z^{\rm massive}_{k_H, p_H}= \prod_{i \notin J}\epsdef^{r_i- 1 - Q_i(k_H)}{ \Gamma\left({Q_i(\sigma_N|_H )+ m_i^F\over\epsdef} +{r_i\over 2}  \right) \over  \Gamma\left({Q_i(\sigma_N|_H + k_H \epsdef )+ m_i^F\over\epsdef} +{2-r_i\over 2}  \right) }~.
\ee

As expected, the residues picked up by the Coulomb branch formula \eqref{formula roughly} correspond precisely to the vortices discussed here. The singularities of the integrand \eqref{formula integrand}  occur wherever some chiral multiplet  has a massless mode,  corresponding to the existence of holomorphic sections for $\CA$. By our assumption of isolated Higgs vacua, these singularities  correspond to regular hyperplane arrangements, and the JK residue  becomes an iterated residue.

\subsection{Elementary examples}
Let us illustrate \eqref{Higgs branch formula} with some elementary examples. The simplest example is the abelian Higgs model discussed in section \ref{subsec:_AHM}. In that case, there is a single field with a single Higgs vacuum with residual $\bZ_Q$ gauge symmetry, and the vortex solutions are the ones of subsection \ref{subsec: vortex eq on S2}, with $r=0$. The Higgs branch formula \eqref{Higgs branch formula} reads
\be
\left\langle  \CO^{(N)}(\sigma_N) \CO^{(S)}(\sigma_S)\right\rangle =  \sum_{k\geq 0} q^k \sum_{p=0}^{Q k} Z^{\rm vortex}_{k, p} \,  \CO^{(N)}\left(-{p\over Q}\epsdef\right) \CO^{(S)}\left(\left(-{p\over Q}+k\right)\epsdef\right)~,
\ee
with the vortex contribution 
\be\label{vortex AHM}
 Z^{\rm vortex}_{k, p} =  {1\over Q \, \epsdef^{Q k
}} {(-1)^p\over p! (Qk -p)!}~.
\ee
This obviously agrees with \eqref{correlators2_AHM}. The ``$S^2_\Omega$ vortex partition function''  \eqref{vortex AHM} can be understood as a simple gluing of flat-space vortex partition functions computed in \cite{Dimofte:2010tz}.

Another simple example is the $\mathbb{CP}^{N-1}$ model discussed in section \ref{subsec:_CP}, with generic twisted masses $m^F_i =-m_i$. We have $N$ distinct vacua $(H)$ labelled by $i$, with a single  $\CA_i$ taking VEV in each.
Therefore, \eqref{Higgs branch formula} gives a sum over $n$ vacua, where in each vacuum $i$ we have a contribution $ Z^{\rm vortex}_{k_i, p_i}$ equal to \eqref{vortex AHM} with $Q=1$, and $\sigma_N|_H= m_i -p_i \epsdef$, $\sigma_S|_H=m_i +(k_i-p_i)\epsdef$. This precisely reproduces \eqref{correlators2_CP}.

\section*{Acknowledgements} We thank Francesco Benini, Philip Candelas, Abhijit Gadde, Hans Jockers, Anton Kapustin, Ilarion Melnikov, Nikita Nekrasov, Du Pei, Antonio Sciarappa, Itamar Shamir, Mohammad Tehrani, Jie Zhou and Aleksey Zinger for interesting discussions. CC would like to thank the organizers of the FRG Workshop: Recent Progress in String Theory and Mirror Symmetry at Brandeis University, where some of these results have been presented \cite{talkBrandeis}, for their hospitality. SC thanks the Galileo Galilei Institute for Theoretical Physics and the workshop ``Holographic Methods for Strongly Coupled Systems'' for hospitality and INFN for partial support during the completion of this work. DP would like to thank the physics department at the University of North Carolina at Chapel Hill and the New High Energy Theory Center at Rutgers University for their hospitality during the completion of this work. The work of CC and DP is supported in part by DOE grant DE-FG02-92ER-40697.

\appendix
\section{Notations and conventions}\label{app: conventions}
We closely  follow the notations of \cite{Closset:2014pda}, but we shall make a few convenient field redefinitions. Moreover, we have $\epsdef$ equal to $- i \epsilon^{\rm [CC]}_\Omega$ in \cite{Closset:2014pda}, and the sign of our FI parameter $\xi$ is opposite to the one of \cite{Closset:2014pda}.

Let us consider the Riemann sphere with  complex coordinates $z, \bz$, which cover the whole $S^2$ except for the south pole at $z=\bz =\infty$. 
We consider a metric
\be
ds^2 = 2 g_{z\b z}(z, \bz) dz d\bz~, 
\ee
with a real Killing vector $V=i z\d_z - i \bz \d_\bz$, but otherwise arbitrary. We work  in the canonical frame
\be
e^1 = g^{1\over 4} dz~, \quad e^{\b 1} = g^{1\over 4} d\bz~,
\ee
with $\sqrt{g}= 2 g_{z\bz}$ by definition. Throughout the paper, we generally work with fields of definite spin, which can be obtained from geometric objects by multiplication with the vielbein. For instance, an holomorphic one-form $X_z$ will be written as a spin $1$ field $X_1 = e^z_1 X_z$, in term of the inverse vielbein $ e^z_1 = g^{-{1\over 4}}$. 
The spin connection is  given by 
\be
\omega_z=-{i\over 4}\d_z\log g~, \qquad \omega_\bz = {i\over 4}\d_\bz \log g~.
\ee
Note that, in our conventions, the Ricci scalar $R$ is negative on the round sphere. 
The covariant derivative on a field of spin $s\in \half \Z$ is
\be
D_\mu \varphi_{(s)}= (\d_\mu - i s \omega_\mu)\varphi_{(s)}~.
\ee
We generally write down derivatives in the frame basis as well: $D_1 \varphi_{(s)}= e_1^z D_z \varphi_{(s)}$ and $D_{\b 1} \varphi_{(s)}= e_{\b 1}^\bz  D_\bz\varphi_{(s)}$.
The Lie derivative  along  $V$ of a field of definite spin reads:
\be
\CL_V \varphi_{(s)} = \left[V^\mu D_\mu +2 s (D_1  V_{\b 1})\right]\varphi_{(s)} ~.
\ee
One can check that it is independent of the metric. Note that $D_1 V_{\b1} =- D_{\b1} V_1$ by the Killing equation.
We refer to appendix $A$ of \cite{Closset:2014pda} for more details on our curved-space conventions.


\subsection{$A$-twisted fields}
It is very convenient to use field variables adapted to the supersymmetries of $S^2_\Omega$. These variables are the so-called ``$A$-twisted'' variables (or rather an $\epsdef$-deformation thereof). They are given by a simple field redefinition in terms of the ``physical'' variables discussed in \cite{Closset:2014pda}.

 Let us denote by $\varphi^{\rm [CC]}$ any physical field $\varphi$ in the notation of \cite{Closset:2014pda}. 
For the bosonic components of the vector multiplet $\CV$, we define:
\bea\label{Atwisted V i}
&a_\mu= a_\mu^{\rm [CC]}+{1\over 2} \epsdef\t\sigma^{\rm [CC]}  V_\mu~, \cr
&\sigma= \sigma^{\rm [CC]} + {1\over 4} \epsdef^2 \t\sigma^{\rm [CC]}  V^\mu V_\mu~, \cr 
&\t\sigma= \t\sigma^{\rm [CC]}~, \cr
& D= D^{\rm [CC]}-i \epsdef (V_1 D_{\b 1} - V_{\b 1}D_1)\t\sigma^{\rm [CC]}~.
\eea
This $\epsdef$-dependent redefinition simplifies many formulas. For the fermionic components of $\CV$, we define~\footnote{All  these definitions of the $A$-twisted fields are written modulo powers of $\t\zeta_-\zeta_+=1$, which do not affect the discussion of the spin and  vector $R$-charge, but matter if we want to keep track of the axial $R$-charge.}
\bea\label{Atwisted V ii}
&\Lambda_1 =\t\zeta_-\left( \lambda_-^{\rm [CC]}-i \epsdef \lambda_+^{\rm [CC]}V_1\right)~,  \cr
&\t\Lambda_{\b 1} =\zeta_+\left( \t\lambda_+^{\rm [CC]}+i \epsdef \t\lambda_-^{\rm [CC]} V_{\b 1}\right)~, \cr
& \lambda=\t\zeta_- \lambda_+^{\rm [CC]}~,\cr
&  \t\lambda=\zeta_+ \t\lambda_-^{\rm [CC]}~,
\eea
in terms of the Killing spinors \eqref{KSEquivAtwist}. By construction, the $A$-twisted  fields have vanishing $R$-charge and ``twisted spin'' $s= s_0+ {r\over 2}$. (For instance, the gaugino $\lambda_+^{\rm [CC]}$  has $r=1$ and $s_0=-\half$, giving us the scalar gaugino $\lambda$.)

Similarly, for the chiral multiplet $\Phi$ of $R$-charge $r$ and the antichiral multiplet $\t\Phi$ of $R$-charge $-r$, we introduce the $A$-twisted variables
\bea\label{Atwistvar}
&\CA = (\t p_z)^{r\over 2}\, \phi^{\rm [CC]}~, 
&&\t\CA = (p_\bz)^{r\over 2} \t\phi^{\rm [CC]}~, \cr
& \CB =\sqrt2  (\t p_z)^{r\over 2}  (\zeta_+\psi_-^{\rm [CC]} -\zeta_-\psi_+^{\rm [CC]})~,\quad 
&& \t\CB =- \sqrt2  ( p_\bz)^{r\over 2}  (\t\zeta_+\t\psi_-^{\rm [CC]} -\t\zeta_-\t\psi_+^{\rm [CC]})~, \cr
&\CC ={1\over \sqrt2} (\t p_z)^{r\over 2} p_\bz \, \t\zeta_-\psi_+^{\rm [CC]}~, 
& &\t\CC ={1\over \sqrt2} (p_\bz)^{r\over 2}\t p_z \, \zeta_+ \t\psi_-^{\rm [CC]}~, \cr
&\CF = (\t p_z)^{r\over 2} p_\bz\, F^{\rm [CC]}~, 
&&\t\CF = (p_\bz)^{r\over 2}\t p_z \,\t F^{\rm [CC]}~. \cr
\eea
Here we defined
\be\label{defp}
p_\bz = - g^{1\over 4} (\zeta_+)^2~, \qquad
\t p_z = g^{1\over 4} (\t \zeta_-)^2~, 
\ee
which are nowhere vanishing sections of $\b\CK \otimes L^2$ and $\CK \otimes L^{-2}$, respectively, with $L$ the $U(1)_R$ line bundle for fields of $R$-charge $1$.

For the twisted chiral multiplet $\Omega$, we define
\be
\omega=\omega^{\rm [CC]}~, \qquad 
\CH_z = {1\over \sqrt2} g^{1\over 4}  \t\zeta_- \eta_-^{\rm [CC]}~, \qquad
\t\CH_\bz = {1\over \sqrt2} g^{1\over 4}  \zeta_+ \t\eta_+^{\rm [CC]}~, \qquad
G= G^{\rm [CC]}~,
\ee
while for the twisted antichiral multiplet $\t\Omega$:
\be
\t\omega=\t\omega^{\rm [CC]}~, \qquad 
\t h = {1\over \sqrt2}  \zeta_+ \t\eta_-^{\rm [CC]}~, \qquad 
h = {1\over \sqrt2}  \t\zeta_- \eta_+^{\rm [CC]}~, \qquad
\t G= \t G^{\rm [CC]}~.
\ee

Note that all these ``$A$-twisted fields'' are  given by a simple change of variables on a particular supersymmetric curved-space background. As emphasized  in \cite{Closset:2014uda} in a closely related context, the ``topological twist'' and ``rigid supersymmetry'' approaches to supersymmetry on curved space should be considered as two faces of the same coin.

\section{More about supersymmetry multiplets}\label{appendix: susy mult}
For completeness, let us discuss the case of a general multiplet whose lowest component is a scalar $C$ of vanishing $R$- and $Z, \t Z$-charges, $A$-in twisted notations:
\be\label{general multiplet}
\CS = \left(C~, \, \chi~,\, \t\chi~, \, \chi_{\b 1}~,\, \t\chi_1~, \, M_{\b 1}~, \, \t M_{1}~, a_\mu~,\, \sigma~, \,\t\sigma~, \,\Lambda_1~, \,\lambda~, \, \t\Lambda_{\b 1}~,\, \t\lambda~,\, D\right)~.
\ee
The lower components of  \eqref{general multiplet} are related to the ones of \cite{Closset:2014pda} by:
\bea
& C=C^{\rm [CC]}~, \qquad & \chi = \zeta_+ \chi_-^{\rm [CC]} -\zeta_- \chi_+^{\rm [CC]}~, \qquad\qquad\quad & \t\chi=\t\zeta_- \t\chi_+^{\rm [CC]}-\t\zeta_+\t\chi_-^{\rm [CC]}~,\cr
& \chi_{\b 1} = \zeta_+ \chi_+^{\rm [CC]}~, \quad & \t\chi_1 = \t\zeta_- \t\chi_-^{\rm [CC]}~, \quad  M_{\b 1} =  \zeta_+\zeta_+ M^{\rm [CC]}~, \quad &\t M_1 = \t\zeta_-\t\zeta_- M^{\rm [CC]}~,
\eea
and the higher components are defined as in \eqref{Atwisted V i}-\eqref{Atwisted V ii}.
 The supersymmetry variations are given by
\bea
&\delta C = i\chi~, \qquad && \t\delta C= - i \t\chi~,\cr
&\delta \chi= 0~,\qquad &&   \t\delta \chi= -\sigma + \epsdef V^\mu a_\mu - i \epsdef \CL_V C~,\cr
&\delta \t\chi= -\sigma + \epsdef V^\mu a_\mu + i \epsdef \CL_V C~,  
&& \t\delta \t\chi=   0~,\cr
& \delta \chi_{\b 1} = M_{\b 1}~, \qquad &&  \t\delta \chi_{\b 1}= 2D_{\b 1} C + 2 i a_{\b 1}~,\cr
& \delta  \t\chi_1=  2 D_1 C - 2 i a_1~,\qquad && \t\delta  \t\chi_1= \t M_1~,\cr
& \delta M_{\b 1}= 0~, \qquad &&   \t\delta M_{\b 1}= 2\t\Lambda_{\b 1} - 4i D_{\b 1} \chi + 2\epsdef \CL_V \chi_{\b 1}~,\cr
& \delta \t M_1= 2 \Lambda_1 + 4 i D_1 \t\chi + 2 \epsdef \CL_V \t\chi_1~,
\qquad &&\t \delta \t M_1=  0 ~,
\eea
\bea\label{susy var gen}
&\delta a_1 = D_1 \chi~, \qquad && \t\delta a_1 = - i \Lambda_1 + D_1\t\chi~,\cr
&\delta a_{\b 1} = i \t\Lambda_{\b 1}+ D_{\b 1}\chi~,  && \t\delta a_{\b 1} =D_{\b1}\t\chi~,\cr
&\delta\sigma = 2 i \epsdef V_1 \t\Lambda_{\b 1}~, 
&&  \t\delta \sigma = -2 i \epsdef V_{\b 1} \Lambda_1~,\cr
&\delta\t\sigma = -2\t \lambda~, 
&&  \t\delta \t\sigma =-2 \lambda~,\cr
& \delta \Lambda_1 = - 4 i \epsdef V_1 (D_1 a_{\b 1}- D_{\b 1}a_1) + 2 i D_1 \sigma~,
&&\t\delta \Lambda_1=0~,\cr
&\delta \t\Lambda_{\b1} =0~,   && \t\delta \t\Lambda_{\b1} = -4 i \epsdef V_{\b1}   (D_1 a_{\b 1}- D_{\b 1}a_1)\cr
&&& \qquad\;\; - 2 i D_{\b1}\sigma~,\cr
&\delta \lambda = i D +   (D_1 a_{\b 1}- D_{\b 1}a_1) - 2  \epsdef V_1 D_{\b 1} \t\sigma~,
&& \t\delta \lambda=0~,\cr
& \delta\t\lambda =0~, \qquad &&\t\delta\t\lambda = -i D-2   (D_1 a_{\b 1}- D_{\b 1}a_1) \cr
&&&\qquad\;\; - 2 \epsdef V_{\b1}D_1 \t\sigma~,\cr
&\delta D= - 2 D_1 \t\Lambda_{\b 1} + 4 i \epsdef V_1 D_{\b 1} \t\lambda \qquad && \t\delta D= - D_{\b 1} \Lambda_1 - 4 i \epsdef V_{\b 1}D_1 \lambda~,
\eea
which realize the algebra \eqref{susy algebra on fields} with vanishing central charge.

The vector multiplet $\CV$ is a particular instance of \eqref{general multiplet}, with the gauge invariance parameterized by chiral and antichiral multiplets of vanishing charges. In WZ gauge, it reduces to \eqref{V components}.
As one can readily check, the twisted chiral multiplet \eqref{Omega com} is also embedded in \eqref{general multiplet}, by the constraint $\chi=\t\chi=0$, while the twisted antichiral multiplet \eqref{t Omega com} is embedded  in \eqref{general multiplet}  by the constraint $ \chi_{\b 1}= \t\chi_1=0$ \cite{Closset:2014pda}.

A $D$-term supersymmetric Lagrangian is obtained from any neutral general multiplet \eqref{general multiplet}:
\be\label{gen Dterm Lag}
\SL_D= D- \t\sigma \CH~.
\ee
This is supersymmetric by virtue of \eqref{susy var gen}. Note also that \eqref{gen Dterm Lag} is always $\delta$-, $\t\delta$-exact, since
\be
\SL_D= \delta\t\delta \left({i\over 2} \t\sigma\right)~,
\ee
up to a total derivative. The equations \eqref{SYM Qex} and \eqref{SYM Qex} are instances of this relation. The FI parameter term in \eqref{FI Lag expanded} is {\it not} in this class because the vector multiplet is not gauge invariant.

\section{One-loop determinants}\label{App: det}
In this appendix, we collect some details on the computation of the needed one-loop determinants, for the chiral and vector multiplets in the background of a supersymmetric vector multiplet configuration.

\subsection{Chiral multiplet determinant}
Consider a chiral $\Phi$ of $R$-charge $r$ and gauge charge $Q$ under a $U(1)$ vector multiplet $\CV$. (The generalization to the general case is immediate.)  On any supersymmetric configuration \eqref{SUSY eq for V nonAb} of $\CV$ (and setting all the gaugini to zero),  the  chiral multiplet Lagrangian \eqref{Phi kinetic term} becomes
\be\label{Phi kinetic term bis}
\SL_{\t\Phi\Phi}  = \t\CA \Delta_{\rm bos} \CA  + (\t \CB, \t \CC) \Delta_{\rm fer }   \mat{\CB\cr   \CC }  -\t\CF \CF~,
\ee
\bea\label{kintops2}
&\Delta_{\rm bos} = - 4 D_1 D_{\b 1} - \, Q\t\sigma \left(-Q\sigma + i \epsdef \CL_V^{(a)}\right)~,\cr
& \Delta_{\rm fer} = - 2i \mat{ {1\over 4}Q \t\sigma &   -  D_1 \cr   D_{\b 1}   & \quad -Q\sigma + i\epsdef \CL_V^{(a)}}~.
\eea
Due to supersymmetry, the bosonic and fermionic operators are related by%
~\footnote{One needs to use $\left[D_{\b1}~,\, -Q\sigma + i\epsdef \CL_V^{(a)}\right]=0$, which can be proven using \eqref{SUSY eq for V nonAb}.}
\be\label{relbetweendets}
-2 i \,\Delta_{\rm fer}\,\mat{ - Q\sigma + i\epsdef \CL_V^{(a)} & 0 \cr  -D_{\b1}   & 1 } 
= \mat{\Delta_{\rm bos}&  4  D_1  \cr
0 & -4\left(-Q \sigma + i\epsdef \CL_V^{(a)} \right) }~.
\ee
The operator $D_{\b 1}$ naturally maps between two Hilbert spaces for the fields of spin ${{\mathbf r}\over 2}$ and ${{\mathbf r}\over 2}-1$ (with ${\mathbf r} = r-Q k$):
\be\label{Db1 op}
D_{\b 1}: \CH_{{\mathbf r}\over 2} \rightarrow \CH_{{\mathbf r}-2\over 2}~.
\ee
With some linear algebra, one finds
\be\label{Zphii}
Z^\Phi = {\det \Delta_{\rm fer} \over \det \Delta_{\rm bos}} = {\det_{{\rm coker} D_{\b1}} (- Q\sigma + i \epsdef \CL_V^{(a)})\over \det_{{\rm ker} D_{\b1}} (-Q\sigma + i\epsdef \CL_V^{(a)})}~,
\ee
up to an overall normalization. The kernel of \eqref{Db1 op} on $S^2$ is spanned by holomorphic sections of $\CO(- {\mathbf r})$. 
 Not coincidentally, this pairing between bosonic and fermionic modes works exactly like in \cite{Closset:2013sxa}.

Consider first the case of a constant supersymmetric background for $\CV$, so that $\sigma= \h\sigma$,  $\t\sigma= \t\hs$ are constant, and $a_\mu=0$ with $k=0$. Then \eqref{Zphii} can be computed very explicitly and one finds 
\be
 Z^\Phi = {\mathbf Z}^{(r)}(Q\sigma; \epsdef)~,
\ee
with the function  ${\mathbf Z}$ defined in \eqref{Zr explicit}. More generally, consider a supersymmetric background with $U(1)$ flux $k$ and a non-trivial profile of $\sigma(|z|^2)$, with values $\sigma_N$ and $\sigma_S$ at the poles. Following \cite{Pestun:2007rz, Benini:2012ui}, we can use the equivariant index theorem to show that
\be\label{Zphi general}
Z^\Phi = \prod_{n=0}^\infty {Q\sigma_S +\left({2-r\over 2} + n\right)\epsdef \over Q\sigma_N +\left({r\over 2} + n\right)\epsdef } 
= \epsdef^{r - Q k - 1}{ \Gamma\left({Q(\sigma_N)\over\epsdef} +{r\over 2}  \right) \over  \Gamma\left({Q(\sigma_S)\over\epsdef} +{2-r\over 2}  \right) }~,
\ee
where we used that $\sigma_S-\sigma_N = \epsdef k$ on a supersymmetric background. On the Coulomb branch saddle \eqref{sigma saddle}, this leads to \eqref{Zphi oneloop}.

Note that only a {\it finite} number of modes ever contribute to \eqref{Zphii}. Consequently, the one-loop determinant \eqref{Zphi general} is perfectly finite. The only  regularization ambiguity is a sign ambiguity, which we have fixed in agreement with \cite{Morrison:1994fr}.

\subsection{Chiral multiplet determinant with $\h D \neq 0$}
In the Coulomb branch localization approach with saddle \eqref{sigma saddle}, we also need to consider the case of a chiral multiplet in the background of a  zero-mode multiplet \eqref{zero mode multiplet}. In such a background, the Lagrangian \eqref{Phi kinetic term}  reads
\be\label{Lag with hD}
\SL_{\t\Phi\Phi}  = \t\CA \left(\Delta_{\rm bos} + i  Q \h D\right)\CA  + (\t \CB, \t \CC) \Delta_{\rm fer }   \mat{\CB\cr   \CC }  -\t\CF \CF+ i \CB\t\lambda\CA   + \t\CA\lambda\CB~,
\ee
with the kinetic operators defined in \eqref{kintops2}. 
 The Gaussian integral with Lagrangian \eqref{Lag with hD} is supersymmetric, and leads to a superdeterminant $Z^\Phi(\h\sigma, \t\hs, \lambda, \t\lambda, \h D)$. We are really interested in 
\be
Z^\Phi(\h\sigma, \t\hs,  \lambda, \t\lambda, \h D) = Z^\Phi(\h\sigma, \t\hs, 0, 0, \h D)~.
\ee
This has to be computed without the help of supersymmetry, unfortunately. We will thus consider a round metric of unit radius and restrict ourselves to the vanishing flux sector, ${\mathbf r}=r$, with constant background $\sigma=\hs$, $\t\sigma=\t\hs$. On the round $S^2$, we find:
\bea
&  \Delta_{\rm bos} =\Delta^\rs_{S^2} + {\rs\over 2} -Q \t\hs \left(- Q\hs + i\epsdef \CL_V\right)~, \cr
& \Delta_{\rm fer}= -i \slashed{\nabla}^\rs_{S^2} +  \mat{ -{i\over 2}Q \t\hs & 0 \cr  0  & \quad -2i\left(-Q\hs + i\epsdef \CL_V^{(a)}\right) }~.
\eea
Here $\Delta^\rs_{S^2}$ is the scalar Laplacian in a monopole background of charge $\rs$:
\be\label{S2 Lap scalar}
\Delta^\rs_{S^2} = -(1+ z \bar z)^2 \partial_z \partial_{\bar z} - \frac{\rs}{2}(1+ z \bar z) \left(z \partial_z - \bar z \partial_{\bar z} - \frac{\rs}{2} \right) - \frac{\rs^2}{4}~.
\ee
and $-i \slashed{\nabla}^\rs_{S^2}$ is the Dirac operator in that same background, acting on $(\CB, \CC)^T$:%
~\footnote{We are closely following appendix A of  \cite{Closset:2013sxa}, to which we refer for more details. Beware that the analog of our fermionic fields $\CB$ and $\CC$ is denoted there by $C$ and $B$, respectively.}
\be\label{Dirac op on S2}
-i \slashed{\nabla}^\rs_{S^2} = -i \begin{pmatrix} 0 &   (1+ z \bar z) \partial_{\bar{z}}+ \frac{1}{2}(\rs-2)z 
\\   (1+z \bar z) \partial_z - \frac{1}{2}\rs \bar{z}& 0 \end{pmatrix}.
\ee
The spectrum of these operators is well-known---see \cite{Closset:2013sxa} and references therein. Let us define 
\be
j_0(\rs) = {|\rs-1|\ov 2} -\half~.
\ee
The generic eigenvalues of $ \Delta_{\rm bos}$ are
\bea
&\Lambda_{j, m} = j(j+1) -  {\rs\ov 2}({\rs\ov 2} -1) +Q\t\hs(Q\hs + m\epsdef)~,
\eea
with $ j= j_0+1, j_0+2, \cdots$, and $m= -j, -j+1, \cdots, j$, for any $\rs$. 
Similarly, the generic eigenvalues of $\Delta_{\rm fer}$  come in pairs, $\lambda^{(+)}_{j,m}$ and $\lambda^{(-)}_{j,m}$, with $-\lambda^{(+)}_{j,m} \lambda^{(-)}_{j,m} = \Lambda_{j, m}$.  For $\rs\neq 0$, there are unpaired eigenvalues corresponding to zero modes. 
If  $\rs<1$, there is some additional bosonic mode of momentum $j=j_0$, which is only partially paired with a single fermionic zero mode. If $\rs>1$, there is an unpaired fermionic zero mode with $j=j_0$. 

The final answer for the one-loop determinant is the infinite product:
\be\label{Zphi with D formula}
Z^\Phi(\h\sigma, \t\hs, \h D) =Z^{(0)} (\hs,\t \hs, \h D) \cdot \prod_{\substack{|m| \leq j \\[2pt] j > j_0(\rs)}}
{Q\t\hs ( Q\hs + \epsdef m) + j(j+1)
- {\rs \ov 2} ({\rs\ov 2}-1)
\ov 
  i Q\h D+ Q\t \hs (Q\hs + \epsdef m) + j(j+1)
- {\rs \ov 2} ({\rs\ov 2}-1)}~.
\ee
where 
\be\label{Z0 with D}
Z^{(0)} (\hs,\t \hs, \h D) = 
\begin{cases}
\prod_{m=-\rs/2+1}^{\rs/2-1} (Q\hs + \epsdef m) &\text{if} \quad \rs>1 \,, \\
1 &\text{if} \quad \rs=1 \,, \\
\prod_{m=-|\rs|/2}^{|\rs|/2} {\t \hs\ov
\t \hs (Q\hs + \epsdef m) + i \h D} &\text{if} \quad \rs<1 \,,
\end{cases}
\ee
is  the zero-mode contribution. Note that \eqref{Zphi with D formula} holds up to a possible sign ambiguity. For $\h D=0$, this reproduces the holomorphic result \eqref{Zphi oneloop}. For $\epsdef=0$, it is easy to see that \eqref{Zphi with D formula} also holds in the presence of flux, whose only effect is to shift $\rs=r$ to $\rs = r- Qk$. From there, it is a small leap of faith to our claiming that \eqref{Zphi with D formula} is the correct answer in the general case.

\subsubsection{The large $\hs$ limit of the one-loop determinant} \label{apsub: large sigma}
In section \ref{section: derivation}, we consider the large $|\hs|$ limit of the determinant \eq{Zphi with D formula}. Let us consider the large-$|\hs|$ limit of
\be
f(\hs,i\e^2 \hD') = 
\prod_{j=0}^\infty \prod_{m=-j}^j
{|\hs|^2 + \epsdef m \b \hs + C + j(j+1)
\ov
i\e^2 \hD' + |\hs|^2 + \epsdef m \b \hs + C + j(j+1)} \,,
\ee
where $C$ is an arbitrary constant. We are interested in the limit where $|\hs|$ is taken to infinity as $\e$ is taken to zero so that
\be\label{ap limit}
\e \ra 0, \quad R = |\hs|^{1/\e^2} \ra \infty \,.
\ee
Here, we provide evidence that
\be\label{ap assumption}
\lim_{\substack{\e \ra 0 \\ R \ra \infty}}
f(\hs,i\e^2 \hD') \approx
\exp \left( 2i (1+\epsdef' \alpha) \hD' \log R \right) \,,
\ee
where $\alpha_{\epsdef'}$ is function of
\be
\epsdef' \equiv {\b\hs \ov |\hs|} \epsdef
\ee
that is well behaved in a neighborhood of $\epsdef =0$. In particular, $\alpha_{\epsdef'}$ is regular at $\epsdef=0$ and thus 
\be
\lim_{\substack{\e \ra 0 \\ R \ra \infty}}
f(\hs,i\e^2 \hD')|_{\epsdef = 0} \approx
\exp \left( 2i \hD' \log R \right) \,.
\ee
The symbol $\approx$ is used to imply that the proportionality constant of the left and the right hand side of the equation asymptotes to unity.

We use the fact that the asymptotics of $f(\hs,i\e^2 \hD')$ is well estimated by
\bea
\t f(\hs,i\e^2 \hD') &= 
\prod_{j=0}^\infty \prod_{m=-j}^j
{|\hs|^2 + \epsdef j \b \hs + C + j(j+1)
\ov
i\e^2 \hD' + |\hs|^2 + \epsdef j \b \hs + C + j(j+1)} \\
&=\prod_{j=0}^\infty \left(
{|\hs|^2 + \epsdef j \b \hs + C + j(j+1)
\ov
i\e^2 \hD' + |\hs|^2 + \epsdef j \b \hs + C + j(j+1)}
\right)^{2j+1} \,.
\eea
This follows from the fact that for $\Delta$ much smaller than $|\hs|$, and small $\epsdef$,
\be
g(-j) \leq g(m) \leq g(j)
\ee
for
\be
g(m) \equiv 
{|\hs|^2 + \epsdef m \b \hs + C + j(j+1)
\ov
\Delta + |\hs|^2 + \epsdef m \b \hs + C + j(j+1)} 
\ee
when $\epsdef \hs$ is real and $(\Delta \b \hs \epsdef)$ is a positive real number.

$\t f(\hs,i\e^2 \hD')$ can be computed using $\zeta$-function regularization. In particular, it is simple to obtain
\bea
\log \t f(\hs,i\e^2 \hD')
&= 2 \zeta'(-1,\t a_+) - 2\zeta'(-1,a_+) 
+2  \zeta'(-1,\t a_-) - 2\zeta'(-1,a_-) \\
&+  (1-2\t a_+)\zeta'(0,\t a_+) - (1-2 a_+) \zeta'(0,a_+) \\
&+(1-2\t a_-) \zeta'(0,\t a_-) - (1-2a_-) \zeta'(0,a_-) \,.
\eea
where $\zeta(s,a)$ is the Hurwitz zeta function:
\be
\zeta(s,a) = {\sum_{n=0}^\infty} {1 \ov (n+a)^s} \,, \qquad
\zeta'(s,a) = \p_s \zeta(s,a) \,.
\ee
We have defined
\bea
a_\pm &= - {\epsdef \hs +1 \ov 2} \pm
\sqrt{\left({\epsdef \hs +1 \ov 2}\right)^2 - (|\hs|^2 + C)} \\
\t a_\pm &= - {\epsdef \hs +1 \ov 2} \pm
\sqrt{\left({\epsdef \hs +1 \ov 2}\right)^2 - (|\hs|^2 + C+i\e^2\hD)} \,.
\eea
Using the asymptotic expansions of the Hurwitz zeta function,
\bea
\zeta'(0,a) &= \left(a - {1 \ov 2} \right)\log a -a + \cO(a^{-1}) \,, \\
\zeta'(-1,a) &= \left({1 \ov 2} a^2 -{1 \ov 2} a -{1 \ov 12} \right) \log a -{1 \ov 4} a^2 + {1 \ov 12} + \cO(a^{-2}) \,,
\eea
we find that
\be
\log \t f(\hs,i\e^2 \hD') \approx 2 i \e^2 \hD \left( 1 - {\epsdef' \ov \sqrt{{\epsdef'}^2-4}}\right) \ln |\hs|
\ee
up to terms that vanish in the limit \eq{ap limit}. Thus
\be
\lim_{\substack{\e \ra 0 \\ R \ra \infty}} \t f(\hs,i\e^2 \hD') \approx
\exp\left( 2 i \hD \left( 1 - {\epsdef' \ov \sqrt{{\epsdef'}^2-4}}\right) R \right) \,.
\ee
Since $({\epsdef'}^2-4)^{-1/2}$ behaves regularly in a small enough neighborhood of $\epsdef'=0$, assuming that the asymptotic behavior of $\t f(\hs,i\e^2 \hD')$ approximates that of $f(\hs,i\e^2 \hD')$ well, our assumption \eq{ap assumption} is justified.

\subsection{Gauge-fixing of the SYM Lagrangian}
Consider a non-abelian vector multiplet on $S^2_\Omega$ with the SYM Lagrangian \eqref{YM standard}. It is invariant under the gauge  group $\GG$. We can introduce BRST ghosts and auxiliary fields  $c, \t c, b$ in the adjoint of ${\mathfrak g}={\rm Lie}(\GG)$, in the standard way. The BRST transformations on ordinary fields are
\be
s a_\mu = D_\mu c~, \qquad s\varphi_b = i[c, \varphi_b]~, \qquad s\varphi_f = i\{c, \varphi_f\}~,
\ee
where $s$ denotes the BRST symmetry generator, and $\varphi_{b,f}$ stand for the bosonic and fermionic fields in the vector multiplet except $a_\mu$. We also have
\be
s c ={i\over 2}\{c, c\}~, \qquad s \t c= -b~, \qquad s b=0~.
\ee
One easily checks that $s^2=0$ using the Jacobi identity for ${\mathfrak g}$. Moreover, $s$ anticommutes with supersymmetry:
\be
\{s, \delta\}=0~, \qquad \{s, \t\delta\}=0~,
\ee
given that $\delta, \t\delta$ act trivially on $\t c, c$ and $b$.
One can then define the new supersymmetry transformations $\delta'= \delta+s$ and $\t\delta'=\t\delta+s$. All gauge invariant Lagrangians are still invariant under $\delta', \t\delta'$. 
The standard gauge-fixing action is a BRST-exact term. More precisely, we take
\be\label{gfaction}
\SL_{gf}= \half(\delta'+\t\delta')\left( \t c (G_{gf} + {\xi_{gf}\over 2} b)\right) = s\left( \t c (G_{gf} + {\xi_{gf}\over 2} b)\right)+ \half \t c(\delta+\t\delta) G_{gf}~,
\ee
for some gauge-fixing function $G_{gf}$ of the physical fields.  The additional term in the RHS of \eqref{gfaction} is needed for supersymmetry but does not affect the one-loop answer \cite{Pestun:2007rz}. Integrating out $b$, we obtain
\be\label{gfaction ii}
\SL_{gf}= {1\over 2\xi_{gf}} (G_{gf})^2 + D_\mu \t c D^\mu c + \cdots~.
\ee
On $S^2$, the ghost $c$ itself has a shift symmetry, which can be gauged-fixed as in \cite{Pestun:2007rz}. (We will be glib about it and simply remove the constant mode of $c$ by hand.)
We shall consider a convenient gauge-fixing function,
\be\label{choice of Ggf}
G_{gf}= D_\mu a^\mu + {i\over 2} \xi_{gf} [\sigma, \t\sigma]  +\sqrt{\xi_{gf}\over 2}\epsdef \CL_V^{(a)}\t\sigma~,
\ee
which is particularly adapted to the Coulomb branch.

\subsection{Vector multiplet one-loop determinant}
Let us consider the algebra ${\mathfrak g}$ in the Cartan-Weyl basis $E_\alpha, H_a$, where $a$ runs over the Cartan subalgebra and $\alpha$ denotes the non-vanishing roots. We have
\be
[H_a, E_\alpha] = \alpha_a E_\alpha~, \qquad [H_a, H_b]=0~, \qquad [E_\alpha, E_{-\alpha}]= {2\over |\alpha|^2} \alpha_a H_a~.
\ee
One can expand  the gauged-fixed Yang-Mills Lagrangian,
\be
\SL_{YM}+ \SL_{gf}~,
\ee
in this basis, with $\varphi = \varphi_a H_a + \varphi_{\alpha} E_\alpha$ for every field.
For simplicity, consider a background where $\sigma, \t\sigma$ take constant values, $\sigma_a= \hs_a$ and 
 $\t\sigma_a= \t\hs_a$. 
Expanding at second order  in the fluctuations  around $\hs, \t\hs$, a straightforward computation shows that, if we choose the gauge-fixing function \eqref{choice of Ggf} and the Feynman-like gauge $\xi_{gf}=1$, the kinetic term becomes diagonal between $a_\mu$ and $\sigma, \t\sigma$, up to terms that do not contribute to the one-loop determinant. We then obtain a simple contribution
\be
\SL_{YM}+ \SL_{gf}  \supset \half D_\mu\t\sigma D^\mu \sigma + D_\mu \t c D^\mu c~,
\ee
so that the ghost determinant completely cancels the determinant from $\sigma, \t\sigma$. The fluctuations along the Cartan of $\Fg$ can be shown to give a trivial contribution. The remaining terms come from the $W$-bosons and their fermionic partners, leading to:
\be
\SL_{YM}+ \SL_{gf}  \supset \sum_{\alpha}\Tr\left(2 a_{\b 1}^{(-\alpha)}\Delta_{\rm bos}^{(\alpha)} \,a_1^{(\alpha)} + \half   \left(2 \t \Lambda_{\b 1}^{(-\alpha)},\t \lambda^{(-\alpha)}\right) \Delta_{\rm fer }^{(\alpha)}   \mat{2 \Lambda_1^{(\alpha)}\cr  \lambda^{(\alpha)} }  \right)~,
\ee
up to terms which cannot contribute to the one-loop determinants. Here we defined
\be
 \Delta_{\rm bos}^{(\alpha)}=  - 4 D_1 D_{\b 1} - \alpha(\t\hs)\left(\alpha(\hs) + i \epsdef \CL_V\right)~,\quad \Delta_{\rm fer}^{(\alpha)} = - 2i \mat{ {\alpha(\t\hs)\over 4} &   -  D_1 \cr   D_{\b 1}   & \quad -\alpha(\hs)+i \epsdef \CL_V}~.
\ee
This shows that the supersymmetric $W$-boson contributes exactly like a chiral multiplet of $R$-charge $r=2$ on $S^2_\Omega$, with $a_1, a_{\b 1}$ playing the role of the fields $\CA, \t\CA$ in the chiral and antichiral multiplets (in particular, $\CA= a_1$ has the correct spin ${r\over 2}=1$).  
To generalize this argument, we just note that the fermionic kinetic term for the fluctuations, as computed from \eqref{YM standard}, is of the same form as $\Delta_{\rm fer}$ in  \eqref{kintops2} (with $r=2$) for any supersymmetric background $\CV$. This leads to \eqref{Zk vector}.  (More precisely, it strongly suggests it. One can also check that answer with an index theorem computation.)

\bibliographystyle{utphys}
\bibliography{bib2domega}{}

\end{document}